\documentclass[doctor,english,final]{kaist-ucs5}

\usepackage{amsmath}
\usepackage{pdfpages}
\usepackage{epigraph}
\newcommand{\RNum}[1]{\uppercase\expandafter{\romannumeral #1\relax}}
\usepackage{changepage}
\usepackage{cite}
\usepackage{hyperref}
\usepackage{bold-extra}
\usepackage[toc,page]{appendix}

\newcommand{\boldlambda}{\boldsymbol{\lambda}}
\newcommand{\boldB}{\boldsymbol{B}}
\newcommand{\boldb}{\boldsymbol{b}}
\newcommand{\boldX}{\boldsymbol{X}}
\newcommand{\boldY}{\boldsymbol{Y}}
\newcommand{\boldx}{\boldsymbol{x}}
\newcommand{\boldy}{\boldsymbol{y}}
\newcommand{\boldM}{\boldsymbol{M}}
\newcommand{\boldh}{\boldsymbol{h}}
\newcommand{\boldeps}{\boldsymbol{\epsilon}}
\newcommand{\boldW}{\boldsymbol{W}}

\newcommand{\boldomega}{\boldsymbol{\omega}}
\newcommand{\boldL}{\boldsymbol{L}}
\newcommand{\matK}{\boldsymbol{\munderbar{\munderbar{K}}}}
\newcommand{\matI}{\boldsymbol{\munderbar{\munderbar{I}}}}
\newcommand{\matLambda}{\boldsymbol{\munderbar{\munderbar{\Lambda}}}}

\expandafter\let\csname equation*\endcsname\relax
\expandafter\let\csname endequation*\endcsname\relax
\usepackage[toc,page]{appendix}
\usepackage{graphicx}
\usepackage{caption}
\usepackage{amstext}
\usepackage{amssymb}
\usepackage{amsmath}
\usepackage[utf8]{inputenc}
\usepackage{booktabs}
\usepackage{array}
\usepackage{multirow}
\usepackage{amsbsy}
\usepackage{subfig}
\usepackage{accents}
\newcommand\munderbar[1]{%
  \underaccent{\bar}{#1}}

\usepackage{color}

\newcommand{\lp}{\left(}
\newcommand{\rp}{\right)}
\newcommand{\lab}{\left<}
\newcommand{\rab}{\right>}
\newcommand{\lsb}{\left[}
\newcommand{\rsb}{\right]}

\newcommand{\labs}{\left|}
\newcommand{\rabs}{\right|}

\newcommand{\refeq}[1]{Eq. (\ref{#1})}
\newcommand{\refsec}[1]{Sec. \ref{#1}}

\newcommand{\reffig}[1]{Figure \ref{#1}}
\newcommand{\reftable}[1]{Table \ref{#1}}
\newcommand{\refmultifig}[2]{Figures \ref{#1} and \ref{#2}}

\newcommand{\abs}[1]{\labs #1 \rabs}
\newcommand{\grad}{\nabla}

\makeatletter
\newcommand*{\rom}[1]{\expandafter\@slowromancap\romannumeral #1@}
\makeatother

\usepackage{listings}
\usepackage{xcolor}

\definecolor{codegreen}{rgb}{0,0.6,0}
\definecolor{codegray}{rgb}{0.5,0.5,0.5}
\definecolor{codepurple}{rgb}{0.58,0,0.82}
\definecolor{backcolour}{rgb}{0.95,0.95,0.92}

\lstdefinestyle{mystyle}{
    backgroundcolor=\color{backcolour},   
    commentstyle=\color{codegreen},
    keywordstyle=\color{magenta},
    numberstyle=\tiny\color{codegray},
    stringstyle=\color{codepurple},
    basicstyle=\ttfamily\footnotesize,
    breakatwhitespace=false,         
    breaklines=true,                 
    captionpos=b,                    
    keepspaces=true,                 
    numbers=left,                    
    numbersep=5pt,                  
    showspaces=false,                
    showstringspaces=false,
    showtabs=false,                  
    tabsize=2
}
\title[korean] {KSTAR 플라즈마 평형을 위한\linebreak 베이즈 추론 신경망}
\title[english]{Bayesian neural network for plasma equilibria\linebreak in the Korea Superconducting Tokamak Advanced Research}

\author[korean] {정}{세 민}
\author[korean2] {정}{세민}    
\author[chinese]{丁}{世 敏}
\author[english]{Joung}{Semin}

\advisor[major]{김 영 철}{Young-chul Ghim}{signed}
\advisor[major2]{김영철}{Young-chul Ghim}{signed}    
\advisorinfo{Professor of Nuclear and Quantum Engineering} 
\department{NQE}{engineering}{a}

\referee[1]{김 영 철}
\referee[2]{김 현 석}
\referee[3]{성 충 기}
\referee[4]{윤 시 우}
\referee[5]{최 원 호}

\approvaldate{2022}{5}{31}

\refereedate{2022}{5}{31}

\gradyear{2022}

\begin{document}

   \includepdf[pages=-]{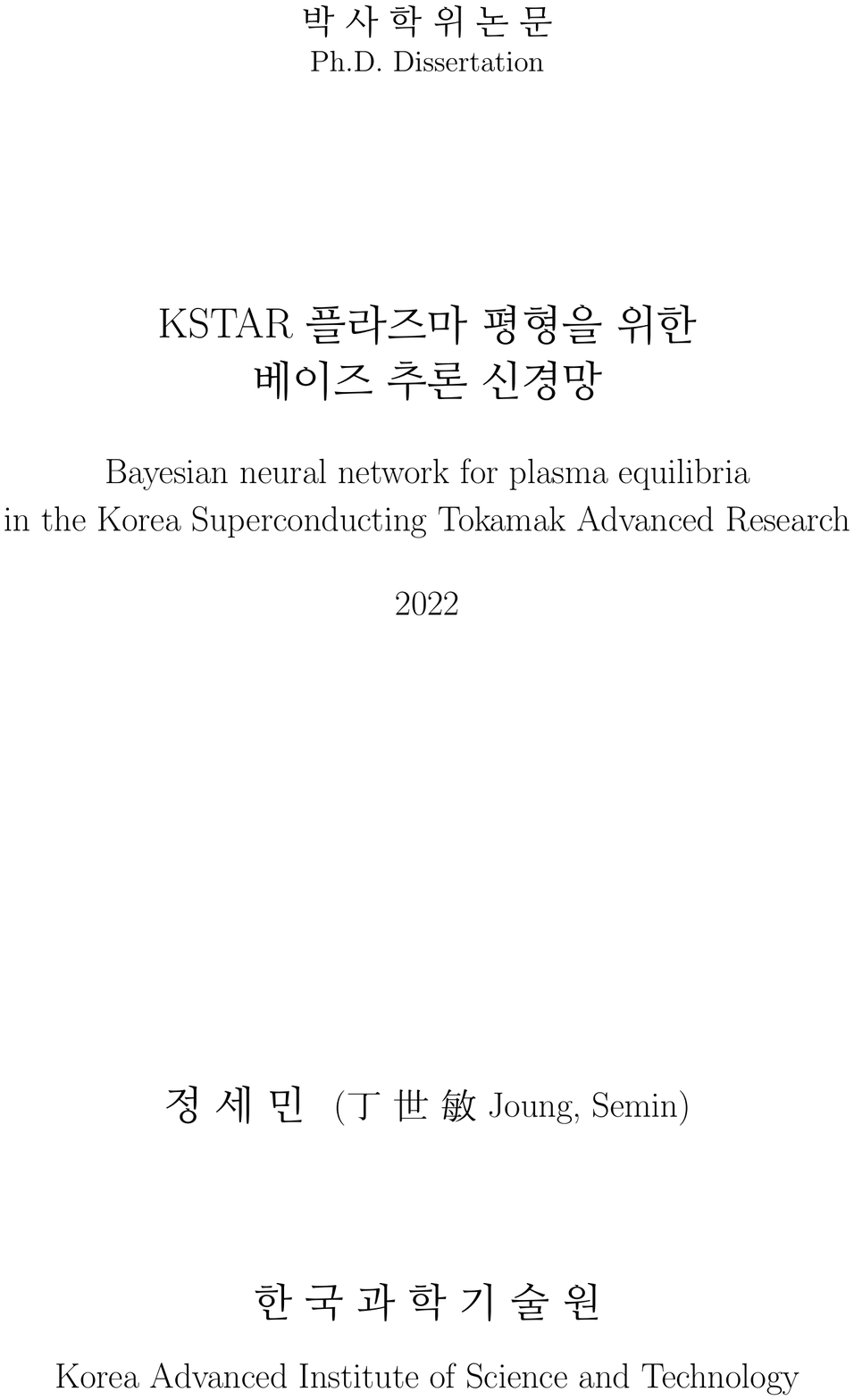}

   \thesisinfo
    \begin{summary}
    핵융합 플라즈마는 물리적으로 복잡한 시스템 중 하나이기에 핵융합로 제어를 위하여 여러 물리 현상에 대한 추가적인 이론 정비가 지속적으로 요구된다. 따라서 여러 물리 현상에 대한 심층적인 이해가 없어도 장치 제어에 도움을 일조하는 딥러닝 (심층 학습 기법) 이용은 과거 수십 년간 핵융합 플라즈마 분야에 큰 화두가 되어 왔었다. 그러나 다양한 딥러닝 기법들이 핵융합 플라즈마의 여러 분야에 연구되어 왔지만 본질적인 물리 현상에 대한 심층적 이해의 부재로 과학적인 사용에 있어서 딥러닝 기법의 신뢰성 정량화가 지속적으로 요구되어 왔었다. 이러한 요구들로 핵융합 플라즈마 분야에서 신뢰성 검증 가능하며 물리 이론까지 만족시킬 수 있는 딥러닝 개발이 새롭게 대두되었다. 우리는 본 논문에서 지배 방정식을 만족하며 관측된 물리 현상을 핵융합로 장치 제어 관련 정보로 변환시킬 수 있는 신경망 개발에 주안점을 둔다. 본 학위 논문에서는 토로이달 및 폴로이달 자기장으로 핵융합 플라즈마를 가두는 핵융합로 실험 장치 중 하나인 토카막을 활용한다. 토카막은 플라즈마를 자기적으로 가두기에 플라즈마 압력과 자기장 및 플라즈마 전류로 인한 로렌츠 힘의 균형 유지가 필수이다. 플라즈마 평형은 바로 이 균형이 유지된 상태를 이야기하는 것으로 토카막 외부 자기장 코일 전류에 의해 제어되는 플라즈마의 형상과 위치를 제공한다. 다만 약 1 억 도의 초고온 환경인 플라즈마로 인하여 플라즈마 형상을 직접 진단할 수 없기에 간접 및 국소 측정 기기로 진단된 플라즈마 물리 정보 활용으로 힘의 균형 및 맥스웰 방정식을 따르는 플라즈마 형상을 간접적으로 재구성한다. 이 재구성에 사용되는 식이 바로 힘의 균형과 맥스웰 방정식으로부터 유도된 Grad-Shafranov (GS) 식이다. 이 식은 2 차원 2 계 비선형 미분 방정식으로 수치 해석이 반드시 요구되며 수치적인 수렴을 위해 진단 데이터의 취사 선택과 같은 인간의 선택이 종종 요구된다. 또한 수치 해석의 반복 계산으로 인한 근본적인 계산 시간 한계로 실시간 토카막 운전에 활용되기 위해서는 정확도의 희생 등을 필요로 한다. 과거에 실시간 계산 한계 극복으로 인해 개발된 지도 학습 기반 신경망이 과거 제시 되었지만 결과적으로 인간의 선택을 바탕으로 한 결과로 훈련된 신경망이란 사실에는 변함이 없었다. 그러므로 이 논문은 신경망을 기반으로 하되 기존 수치 해석과 독립적이며 물리 이론을 스스로 만족함과 동시에 신뢰성의 정량화가 가능한 플라즈마 형상 재구성 방법을 제안한다. 즉 신경망을 통해 GS 식의 해를 직접 구하며 베이즈 추론 신경망을 통해 재구성된 플라즈마 평형의 신뢰성을 평가한다. 또한 자유 경계 문제 중 하나인 GS 식 해의 탐색을 위해 경계 추적이 가능한 보조 모듈을 고안하였다. 나아가 베이지안 추론과 가우시안 프로세스 및 기계 학습을 기반으로 하는 진단된 자기장 신호의 표류 현상, 신호 간의 비 일관성, 진단 신호 손실을 해결하는 방법을 소개하여 어떠한 운전 환경에서도 우리의 신경망이 사용될 수 있음을 입증한다. 그리고 신경망 훈련이 GS 식에 기반할 수 있는지를 검증하기 위해 기존 수치 해석의 데이터 및 GS 식을 신경망의 비용 함수로써 사용한 결과를 소개한다. 추가로 우리는 이 학위 논문에 활용된 원리 및 방법이 여러 딥러닝이 활용될 법한 전통적인 과학 분야에 충분히 응용될 수 있음을 밝힌다. 그리하여 단순한 딥러닝 사용을 벗어나 여러 공학 및 물리 분야에 물리적 신뢰성을 획득한 신경망 사용 방안을 제안할 수 있을 것이라고 우리는 희망한다.
    \end{summary}
   
    \begin{Korkeyword}
    케이스타, 플라즈마 평형 재구성, EFIT, 자기 진단법, 톰슨 산란 진단, 전하 교환 분광계, 가우시안 프로세스, 베이지안 추론, 베이즈 추론 신경망, 비지도 학습
    \end{Korkeyword}

    \begin{abstract}
    Fusion-graded plasmas are one of the physically complex systems, resulting in continuous establishment of plasma theories for unclarified physical phenomena in order to thoroughly control nuclear fusion reactors. Deep learning has drawn vast attention to this field of controlled fusion plasma to link physical phenomena with control-relevant parameters without a deepened understanding about plasma theories. Albeit, quantifying the uncertainty of deep learning models has been constantly requested due to their fundamental shortage of physical understanding. Thus, a concept of a reliable deep learning model to be able to present their probability distributions is raised as well as a method to inculcate physical theories in the model is also concerned. These are the main concept focused in this thesis with a tokamak experiment, one of the nuclear fusion experiments by confining the plasmas via toroidal and poloidal magnetic fields in a torus shape device. Since the tokamak confines a plasma magnetically, balancing the Lorentz force due to the magnetic field with the plasma pressure is crucial. This balanced state with equilibrium assumption is called plasma equilibrium, giving us the shape and location of the plasma determined and controlled by the external coil currents of the tokamak. However, this plasma shape cannot be directly measured due to the harsh environment caused by the plasma itself of 100 million degrees Celsius, thus the shape is indirectly reconstructed from the force balance and Maxwell's equations consistent with externally and locally measured plasma information. The Grad-Shafranov (GS) equation derived from those equations is used to reconstruct the plasma. This equation is a two-dimensional second-order differential equation, inherently requiring numerical analysis so that human decisions such as selecting some of the measured signals arbitrarily for numerical convergence are followed. Furthermore, it is likely to sacrifice accuracy of solutions of the equation for real-time tokamak controls due to multiple iterations in numerical analysis which requires intensive computations. Although there were neural network based real-time approaches via supervised learning with databases from numerical algorithms, they were inevitably under the influence of human decisions. Hence, this thesis suggests a reconstruction method based on deep neural networks which are able to not only estimate their uncertainties but also learn the governing equation themselves without depending on previous numerical algorithms. Namely, our neural networks solve the GS equation via a unsupervised learning algorithm and show probability distributions of their solutions based on Bayesian neural networks. Since solving the GS equation is a free-boundary problem, our networks are supported by an auxiliary module that detects the plasma boundary from the network outputs. Furthermore, we introduce preprocessing methods for the network inputs to address the magnetic signal drift, the flux loop inconsistency and the magnetic signal impairment based on Bayesian inference, Gaussian processes and neural networks. These methods are developed to guarantee the use of the networks in any circumstance of the tokamak experiments. In addition, we also prove that the Grad-Shafranov equation can be used as a cost function of the networks with a given equilibrium database. The principles and methods applied here are not only acceptable for fusion research but also applicable to various engineering and scientific fields. Thus, we expect that our proposal which fulfills physical reliability for the use of deep learning deserves further studies for various complex physics systems.
    \end{abstract} 
     
    \begin{Engkeyword}
    KSTAR, Grad-Shafranov equation, EFIT, Magnetic diagnostics, Thomson scattering system, Charge exchange spectroscopy, Gaussian processes, Bayesian inference, Bayesian neural networks, Unsupervised learning
    \end{Engkeyword}

    \addtocounter{pagemarker}{1}                 
    \newpage

    \tableofcontents

    \listoftables

    \listoffigures

\chapter{Preface: what would we do with a Black Box for fusion research?} \label{ch1}
\epigraph{\textit{``No one has ever proved that EL DORADO or SKYPIEA doesn't exist!!", ``Well, Let them laugh! That's what makes it A GREAT ADVENTURE!!"}}{--- Oda Eiichiro,\\ONE PIECE}

\noindent
When I started studying nuclear fusion and deep learning for the first time, I was totally absorbed in two tasks: (1) surveying the whole history of \textit{Neural Network} used in the field of nuclear fusion; (2) neural network regressions for sine functions with various \textit{signal-to-noise} ratios (SNRs). When it comes to the usage of neural networks, in brief, there were two major pedigrees such that one wished to control tokamaks accurately via neural networks, and the others tried to make neural networks predict tokamak disruptions (violent events undoubtedly forcibly terminating tokamak operations).

Starting with an idea of real-time prediction of plasma positions \cite{lister1990implementation}, mapping measurement signals of the plasma to the positions of that was extensively studied by Ref. \cite{Lister:1991gx, LAGIN19931057, Coccorese:1994jt, vanMilligen:1995dv, Bishop10.1162, windsor1997real, mc1998realtime} in the 1990s as well as a review article introducing the studies to readers with no previous knowledge of the network \cite{Bishop10}. Yet there had been no significant change in this research field \cite{Wang:JFE2016} until advanced neural network techniques (deep learning) contributed to the field again \cite{carpanese2020first, degrave2022magnetic, wai2022neural}. Regarding the disruption prediction, in 1994, there was a master's thesis \cite{markusdisruption} trying to predict disruptions in a tokamak plasma (probably for the first time) by using a neural network. Afterward, this disruption prediction field has grown rapidly over decades, being able to predict multi-tokamak disruptions based on a single neural network \cite{windsor2005cross, Tang:2019Nature} by taking advantage of previous researches that focused on disruptions occurring in a single fusion device \cite{hernandez1996neural, wroblewski1997tokamak} as a cornerstone. Of course, there is other genealogy of the neural network in fusion community which dealt with \textit{tokamak transport} \cite{tresset2002transport, wakasa2004development, wakasa2008study, Meneghini:2014ic, Citrin:2015fj, Meneghini:2017kp, van2020fast}, but I would like to spare the details of it since I believe that two main start points of the history of the neural network in fusion research are definitely \textit{tokamak control} and \textit{disruption prediction}.

In short, the neural networks have been simply used to connect control-relevant parameters of the plasma generated from \textit{plasma equilibrium} (a force-balanced state of the plasma) with given diagnostic signals as a viewpoint of tokamak control as well as they have been used to let a tokamak discharge be on a disruption alert by looking at tokamak measurements in real time in case that a disruption is about to occur. However, in either case, all the neural network did was merely to link given inputs to certain plasma physically meaningful parameters without understanding any physics behind. If the neural network doesn't know physics, how can it be used in the physics field? Thus, the idea that I have always had in my mind after finding these results is that the neural network just acts as a ``good but not great" supporting actor in nuclear fusion since the network seems to be such a magical tool that maps any inputs to any outputs, which does not require enough physics interpretation. Thus, I thought ``Isn't that a limitation of the neural network for nuclear fusion research (but also other scientific fields) because its working principle might be hard to be trusted (or scrutable)?"

\begin{figure}[t]
    \centerline{\includegraphics[width=0.575\columnwidth]{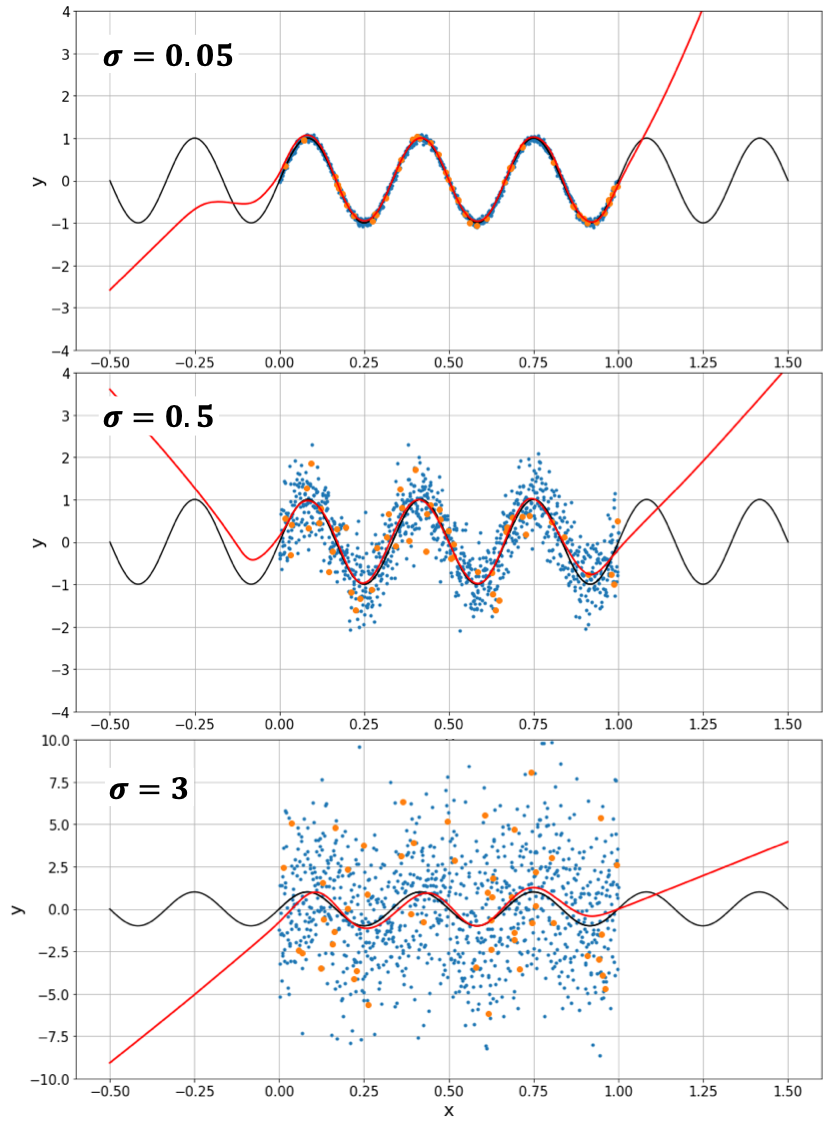}}
    \caption[Sine function regressions]{The network regression results compared to the sine functions with various noise levels, $\sigma$. The blue, the orange dots, the red lines and the black lines are the training, the validation sets, the network results and the noise-free sine functions, respectively.
    } \label{ch1-fig1}
\end{figure}

The ability of the neural network is actually impressive. The neural network has the ability to extract relations between two (or more) phenomena in terms of basic arithmetic operations. As well as two dimensional data regression in multidimensional space \cite{gauss1877theoria, bishop2006pattern, rumelhart1985learning}, deep learning is powerful to handle image processing \cite{krizhevsky2012imagenet, szegedy2015going}, which is far beyond human abilities, and able to perform natural language understanding \cite{kalchbrenner2013recurrent}, video processing \cite{sundermeyer2012lstm} and language generation \cite{mikolov2010recurrent}. These applications show that deep learning is (nearly) close to human level in various fields.

This remarkable strength of deep learning can be found even with a simple regression analysis. I would like to discuss sine function regressions here which I started figuring out when I studied neural networks for the first time. Figure \ref{ch1-fig1} shows the results of neural networks trained with data points generated from sine functions, i.e.,
\begin{equation}
t(x) = \sin{6\pi x} + \epsilon
\end{equation}
where $t$ is the observed data and $\epsilon$ stands for the level of Gaussian noise, $\epsilon=\mathcal{N}(\mu,\,\sigma^{2})$, whose mean value $\mu$ is zero, and standard deviation $\sigma$ is set arbitrarily. To train the networks, I divide the dataset into training (blue dots) and validation data (orange dots) which are used to train and validate the networks during the training procedure, respectively. The black line is the noise-free sine function, and the red line is the network results. The data points are prepared within the interval of 0 -- 1 along the $x$-axis. From top to bottom of Figure \ref{ch1-fig1}, with one significant figure, the standard deviation $\sigma$ are 0.05, 0.5 and 3 where the SNRs are approximately 200, 2 and 0.05 (20, 3 and -10 in decibel). The detailed explanation about the network training will be discussed in the following chapters.

As shown in the figure, the first two plots have the relatively less noisy observed data which is straightforward to be identified as sinusoidal functions even with our bare eyes. Thus, it is quite acceptable that the networks seem to be the sine functions since they are matched well with our intuition. However, let us take a look at the last figure. Could anyone find any features of the sine function from the blue (and orange) dots with the naked eye only? I swear that no one can do that. It means that even if the networks tell us the blue dots are possibly produced from a sinusoidal function, because our intuition (we indeed have our intuition in this case although the intuition is not proper anymore) refuses to accept the argument provided by the network since it is difficult to determine that the network fits to our intuition or not, therefore we cannot believe and trust what the network insists on. 

The result above is quite surprising since the network actually ``did not know" what the observed data is made from, but ``infer" the sine function. This process is seemingly ``magical" like a black box to those who were not involved in the process of generation of the training data, which could be the fact that makes us doubt the network's capabilities. In other words, I would like to argue that the sine function can be regarded as physical parameters of interest, and similarly the intuition can turn out to be physical theories. Namely, although the neural network is able to generate physically reliable outputs, we would not depend on the results since we can think the network never fully follows laws of physics of interest. 

Then how can the neural network be counted on for scientific uses (or nuclear fusion research)? To answer this question, I shall start from defining \textit{uncertainty} in deep learning. First of all, there are situations such that observed (or collected) data is too noisy or too sparse to sufficiently cover possible phenomena, which leads to the first type of uncertainty, i.e., \textit{aleatoric uncertainty}. This is yielded by the poor quality of observed (or collected) data. The other type of uncertainty is caused by a network model itself such that the model is not as complex as collected data, or free-parameters of the model is poorly determined.\footnote{The etymology of the word \textit{aleatoric} is the Latin ``aleator”, or ``dice player”, meaning that aleatoric uncertainty is the ``dice player's uncertainty". The etymology of the word \textit{epistemic} is the Greek ``episteme" known as ``knowledge", meaning that epistemic uncertainty is ``knowledge uncertainty". Epistemic uncertainty can be often reducible through having more knowledge, while aleatoric uncertainty sometimes cannot be reduced due to the measurement noise or the inherent stochasticity.} These yield \textit{epistemic uncertainty} (also referred to as model uncertainty). Both uncertainties result in \textit{predictive uncertainty} which quantifies how the network is sure about its prediction.

Perhaps, if we have a lot of data which can sufficiently cover all possible physical phenomena, then we can possibly give credence to whatever a neural network outputs. This is unfortunately almost impossible, and does not always guarantee that the neural network follows physics theories. Then, what if we train a neural network with physics theories? Does this way quantify (or reduce) the network's epistemic uncertainty (knowledge uncertainty) related to the ``theories"? As a result, does this mean the network truly follows the theories and shows how it is confident with respect to given inputs (and given theories)? In particular, unlike the past ``magic" approaches, can't that lead us to trust the neural network a little more (or further)? Answers to these questions are the main gist of this thesis, which will be provided in the following chapters from the perspective of tokamak control.

With the Korea Superconducting Tokamak Advanced Research (KSTAR), Article \RNum{4} has been developed to show that a neural network can learn a plasma `theory' with the support of a database prepared from a numerical algorithm by reconstructing plasma equilibria based on the Grad-Shafranov (GS) equation. This is a preliminary application for the network to provide the possibility of a complete unsupervised learning for the reconstruction such that the neural network can understand the GS equation itself, and the database from the numerical algorithm is no longer required. This is described in Article \RNum{5}, providing how a principle of the unsupervised learning works, and why this kind of network is required for tokamak control. Article \RNum{1}, Article \RNum{2} and Article \RNum{3} have been developed to preprocess KSTAR measurements used to inputs of our networks since baseline increases of measured signals in time (signal drift), missing signals due to mechanical issues and inconsistency between signals should be handled to use our networks in any experimental circumstances. From \textit{Bayesian neural networks}, our applications are able to quantify the epistemic uncertainty related to the plasma theory by obtaining inference results of the GS equation as well as plasma information such as positions and locations of the plasmas (which are hard to be measured directly) in the KSTAR. One can find the detailed fundamentals and analyses in the next chapters.

I have recognized that the neural network is treated as a black box, inscrutable as well as unbelievable. How can this prejudice be resolved? Let me leave this here: solving differential equations numerically also faced a tough proof back then around 1950. I would say, we are simply taking a look at a novel method whose detail is not fully figured out yet, and I hope this thesis corroborates it is fine to use a neural network in nuclear fusion research, including the controls of the tokamak plasmas in real time.

\chapter{Nuclear Fusion} \label{ch2}
\epigraph{\textit{나가와 도깨비, 인간, 레콘이 살고 있는 집에서 누군가가 바닥에 바늘을 떨어뜨렸다. 잘 보이지 않는 바늘을 찾아내는 방법은?\\ 답 : 도깨비가 바늘이 뜨거워질 정도의 도깨비불을 퍼뜨리고 나가가 뜨겁게 달아오른 바늘을 눈으로 확인하여 집어 올린다. 그리고 인간은 온 힘을 다해 레콘을 말려야 한다. 설득력이 충분하다면 레콘이 집을 들어 흔드는 것은 막을 수 있을 것이다.}}{--- 이영도,\\피를 마시는 새}

\noindent
What is nuclear fusion? It is the morning and the evening star. I slightly transform a Sinclair Lewis (\textbf{Harry Sinclair Lewis}, February 7, 1885 -- January 10, 1951) quote to start explaining what nuclear fusion is in the less heavy mood. The energy from the nucleus can be obtained by combining light nuclei into heavier ones. This is called nuclear fusion energy \cite{cite-keyVerberk}, which is a foundation of energy generated from the Sun and stars in outer space. Before deepening our sight about nuclear fusion, let me consider three very different time scales which are involved in climate change or energy sources if one considers either of them. The first view is a few months -- a few years scale that is a short time scale to be required to take temporary solutions such as making an agreement like the Kyoto Protocol, issuing carbon credits, limiting the speed limit of automobiles, offering tax credits for renewable energy plants, etc. As a relatively longer time scale, 10 -- 50 years, we can use this to take such solutions like developing new clean (or carbon-free) energy sources. The last perspective about the time scales is the longest time scale, 100 -- 5000 years, which is the faraway future, so  we barely know what will happen in the future. The problem is we are already facing global warming and sea level rise as well as rising fuel prices. A more vicious truth is that we do not have much time to prepare them and, especially, the intermediate time scale. (I largely take this information from Ref. \cite{Chen:2011wc}).

\begin{figure}[t]
    \centerline{\includegraphics[width=0.65\columnwidth]{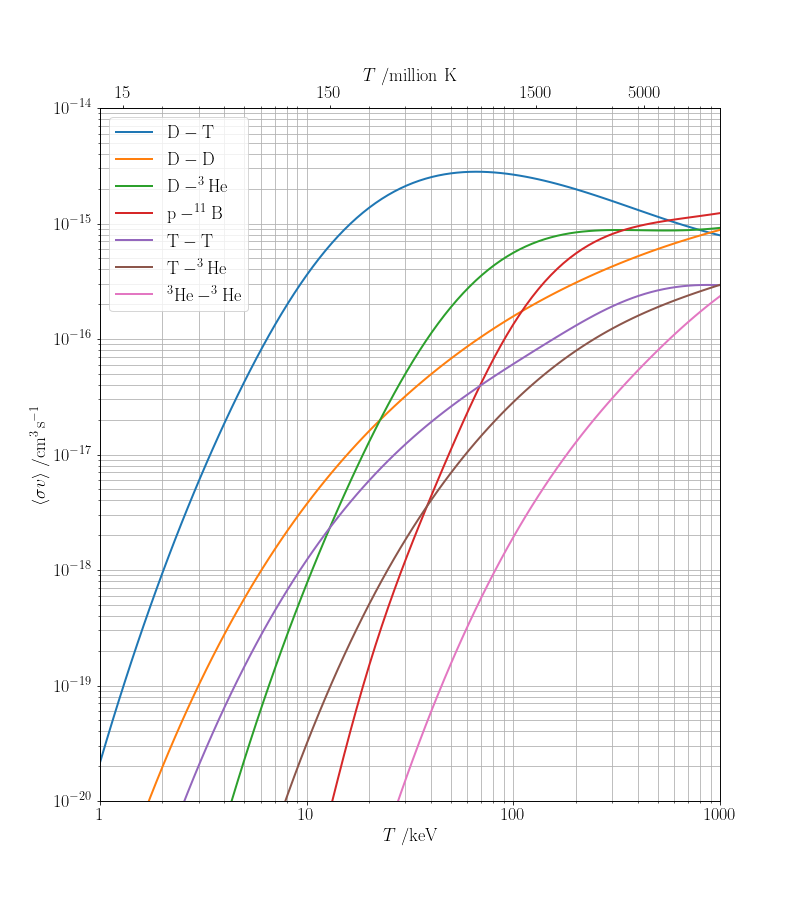}}
    \caption[Fusion reaction rates]{Fusion reaction rates of deuterium and tritium (D-T) and deuterium and deuterium (D-D) with some well-known cross sections. The D-T fusion reaction rate is remarkably higher than the other reaction rates at the temperature of the order of 10--100 keV.
    } \label{ch3-fig1}
\end{figure}

In these circumstances, nuclear fusion is a solution as a clean energy source which has ideally no blemish to be worried about generating any carbon-like byproducts and a vast resource of fusion fuels available from the sea. Although fusion power will take time (and money also) to be realized, however, we are living in the land and era of taking photos of Pluto, trying to reuse space rockets and build AI technology in our daily life, thus, they can be affordable. Over decades, we have been trying to build nuclear fusion power plants which are able to contain and sustain fusion reactions occurring only in really extreme conditions on Earth. The sequence of the fusion reaction of interest is following: two small nuclei are given enough kinetic energy to pass over a potential (Coulomb) barrier due to their charges, then they fuse together and are transformed into another nuclei, and then they produce large energy which is often called fusion energy. We often tap deuterium and tritium as the two nuclei since their reaction rate is extraordinarily superior to the other famous reaction rates shown in Figure \ref{ch3-fig1}, which is
\begin{equation}
\label{eq:nf1}
{}_{1}^{2}\text{D}+{}_{1}^{3}\text{T} \longrightarrow {}_{2}^{4}\text{He}\text{ (3.5 MeV)}+{}_{0}^{1}\text{n}\text{ (14.1 MeV)}
\end{equation}
where the goal to crash through the Coulomb barrier is to heat them up to the temperature of the order of 10 keV ($\approx 10^{8}$ K), which makes the deuterium and the tritium to become \textbf{plasma}, the fourth state of matter. In this state, the dynamics of the plasma is governed by a collective behaviour and sensitive to externally applied electromagnetic fields. Sensitivity to external fields can be interpreted as a way to control the plasma through the fields, allowing us to have a fusion reactor confining the plasma magnetically for a sufficiently long time to acquire enough fusion reactions. The alpha particles (${}_{2}^{4}\text{He}$) confined in the magnetic cage can heat the plasma through collisions, while the neutrons (${}_{0}^{1}\text{n}$) ignoring magnetic fields can be used for a blanket \cite{federici2019overview} to capture the neutrons and convert their energy into heat.

\begin{figure}[t]
    \centerline{\includegraphics[width=0.65\columnwidth]{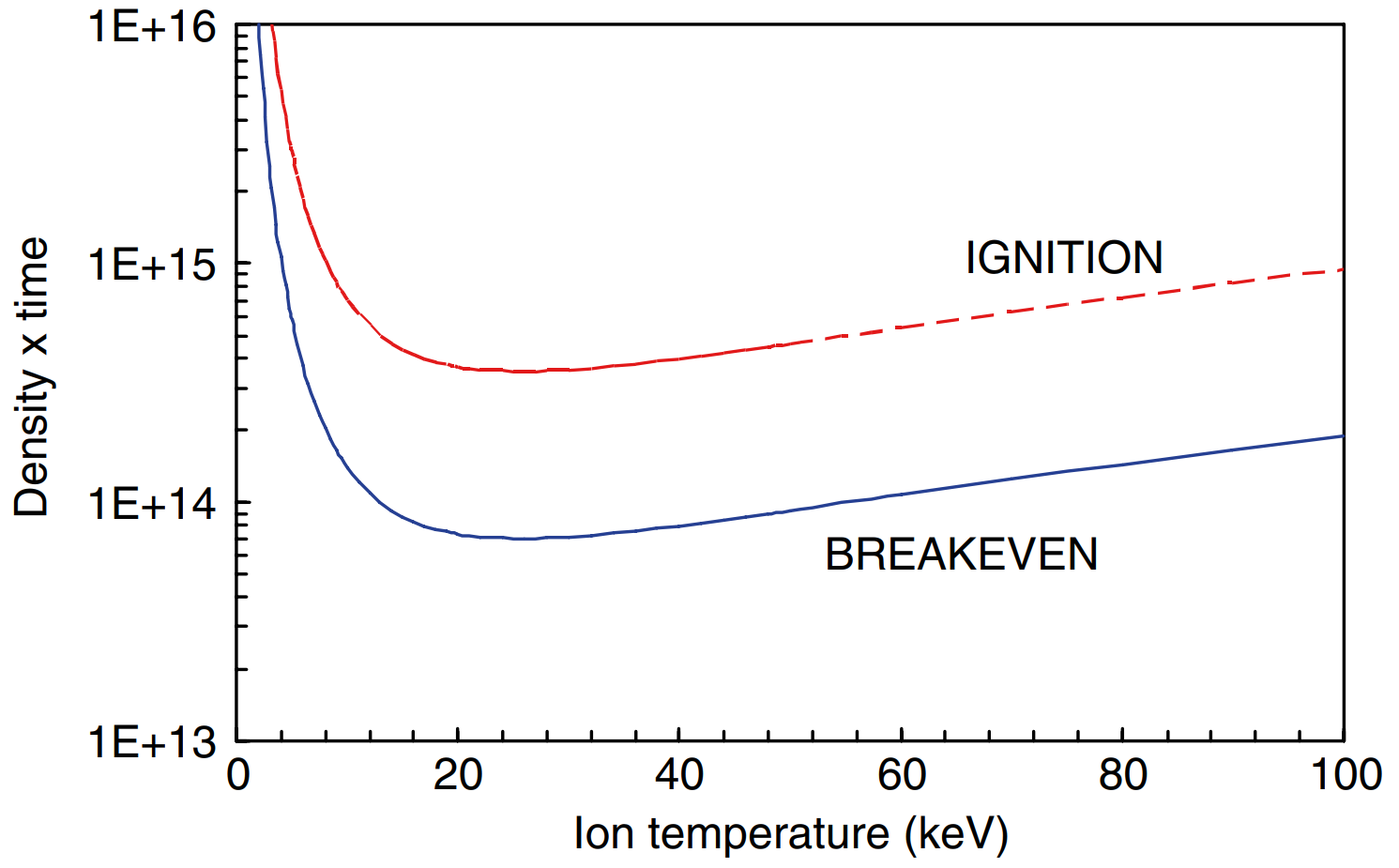}}
    \caption[Lawson criterion]{Lawson criterion for D-T fusion. The ordinate stands for $n\tau_{E}$. This figure is taken from \cite{Chen:2011wc}.
    } \label{ch3-fig2}
\end{figure}

From now on, let me consider an actual time scale that we need to have in order to see enough fusion reactions in the fusion reactors. There is a relation portraying how much time do the reactions require when a certain amount of plasma and a certain plasma temperature are given. This is known as the \textbf{Lawson Criterion} which describes the relation between plasma density $n$, ion temperature $T$ and confinement time $\tau_{E}$ as shown below
\begin{equation}
\label{eq:nf2}
nT\tau_{E} \geq 3\times10^{21} \text{ keV s/}m^{3}
\end{equation}
where the confinement time $\tau_{E}$ is the ratio of the plasma thermal energy density $W$ to the power $P_{heat}$ that is needed to keep the plasma at a certain temperature as shown below
\begin{equation}
\label{eq:nf3}
\tau_{E} = W/P_{heat}.
\end{equation}

A modified Lawson Criterion is shown in Figure \ref{ch3-fig2} where BREAKEVEN stands for balances between the fusion energy and the energy needed to sustain the plasma, and IGNITION is a condition to ignite a self-sustaining plasma. The figure says that we need at least $n\tau_{E}$ of the order of $10^{20} \text{s/}m^{3}$ to achieve the breakeven condition with $\tau_{E}$ of the order of 1 sec if we expect that a reasonable plasma density is $\sim 10^{20} \#/m^{3}$. Dramatically speaking, we must hold the plasma for 1 sec inside of the fusion reactors, and I develop a neural network to control the plasma for the time scale of 1 sec really precisely by reconstructing a better plasma equilibrium in real time, which will be introduced out of this thesis. Thus, we can probably say that nuclear fusion is figuratively the morning and the evening star of which awakens us to reach a future of using the clean and carbon-free energy invented by humankind's knowledge.


\section{Tokamak} \label{ch2-1}

\begin{figure}[t]
    \centerline{\includegraphics[width=0.95\columnwidth]{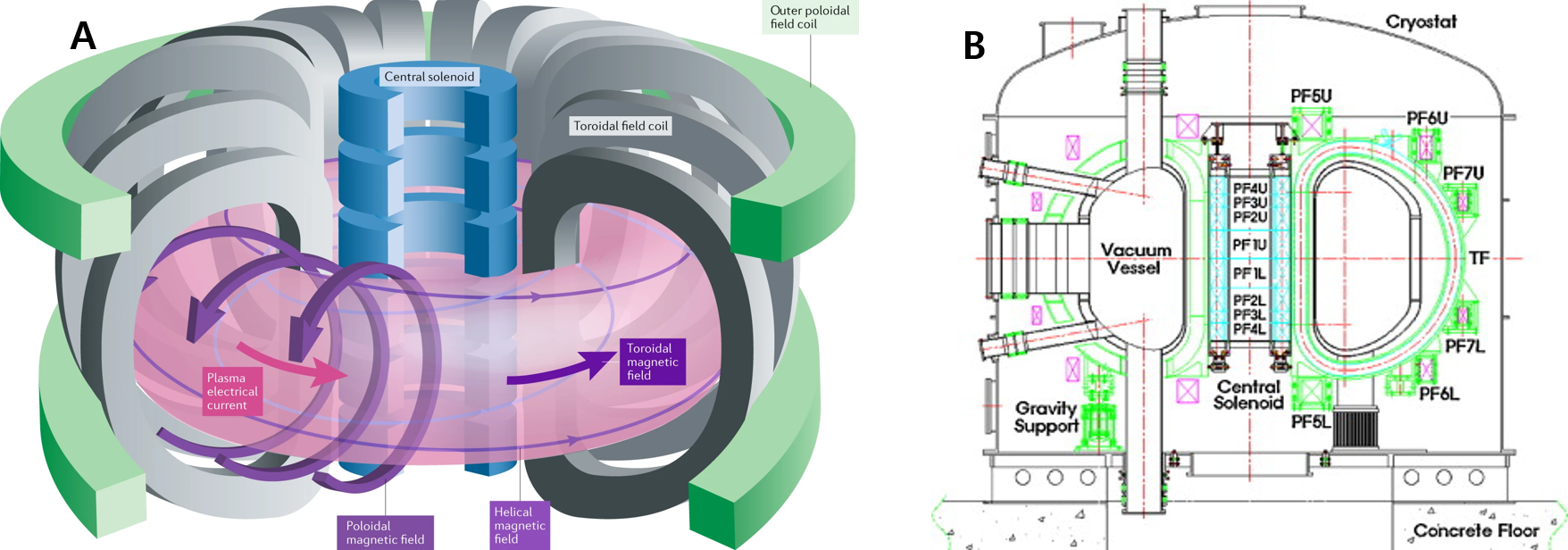}}
    \caption[Tokamak configuration]{(a) A typical tokamak configuration. Image courtesy of EUROfusion. (b) Elevation view of ths KSTAR tokamak \cite{oh2009commissioning}.
    } \label{ch3-fig3}
\end{figure}

\noindent
Previously, I mentioned that the plasma responds to the external electromagnetic fields sensitively, and this leads us to build the fusion reactor generating the magnetic fields to confine and control the plasma. One of the reactors working such a way is \textit{tokamak} whose name comes from the Russian words \textit{\textbf{to}roidalnaya \textbf{ka}mera \textbf{ma}gnitnaya \textbf{k}atushka}. These words mean toroidal chamber magnetic coils. Although there are various concepts of magnetic confinement devices such as stellarator and reversed field pinch device, I would like to consider the tokamak solely in this thesis since this thesis is mainly based on tokamak experimental results. As shown in Figure \ref{ch3-fig3} (a), the tokamak generates two major directions of the magnetic fields; toroidal magnetic field and poloidal magnetic field. The toroidal field coils represented as the gray structures generate the toroidal magnetic field in the direction of the red arrow in the figure. Similarly, The poloidal field coils (green) and the central solenoids (blue) produce the poloidal magnetic field in the direction of the purple arrows, while purposes of those coils are slightly different: the poloidal field coils are generally used to control the plasma position; the object of the central solenoids is to induce a current to the plasma in order to generate a plasma current in toroidal direction. Therefore, the plasma current is the main source of the poloidal magnetic field.

I would like to note that the tokamak experiments done by the \textbf{K}orea \textbf{S}uperconducting \textbf{T}okamak \textbf{A}dvanced \textbf{R}esearch (KSTAR) which is one of the first research tokamaks with fully superconducting magnets in the world located at South Korea have been dedicated to developments of this thesis. The elevation view of KSTAR is shown in Figure \ref{ch3-fig3} (b). I outline briefly below KSTAR and its specifications.

\begin{table}[]
\centering
\caption{Designed ranges of major paramters of KSTAR \cite{lee2001design}}
\label{tab:nf1}
\begin{tabular}{clllclc}
\hline
\multicolumn{1}{l}{Symbol} &  & Parameter &  & \multicolumn{1}{l}{Baseline} &  & \multicolumn{1}{l}{Upgrade} \\ \hline
$B_{T}$ &  & Toroidal field {[}T{]} &  & 3.5 &  &  \\
$I_{p}$ &  & Plasma current {[}MA{]} &  & 2.0 &  &  \\
$R_{0}$ &  & Major radius {[}m{]} &  & 1.8 &  &  \\
$a$ &  & Minor raidus {[}m{]} &  & 0.5 &  &  \\
$\kappa$ &  & Elongation &  & 2.0 &  &  \\
$\delta$ &  & Triangularity &  & 0.8 &  &  \\
- &  & Pulse length {[}s{]} &  & 20 &  & 300 \\
- &  & Neutral beam {[}MW{]} &  & 8.0 &  & 16.0 \\
- &  & Ion cyclotron {[}MW{]} &  & 6.0 &  & 6.0 \\
- &  & Lower hybrid {[}MW{]} &  & 1.5 &  & 3.0 \\
- &  & Electron cyclotron {[}MW{]} &  & 0.5 &  & 1.0 \\ \hline
\end{tabular}
\end{table}

KSTAR is one of the first fusion experimental reactors using superconducting magnets in the world. The typical and designed ranges of the major specifications of KSTAR are shown in Table \ref{tab:nf1}. It is worth mentioning that KSTAR recently sustained the ion (deuterium) temperature up to $\sim$100 million degree Kelvin at the center of the plasma for $\sim$20 sec for the first time \cite{han2022sustained}. KSTAR has a major radius of 1.8 m and a minor radius of 0.5 m. The central solenoids of KSTAR are designed to induce the plasma current of 2.0 MA, and the toroidal field coils are capable of generating the toroidal magnetic field of 3.5 T. The application in this thesis provides a self-sustained deep learning approach for the plasma equilibrium from KSTAR plasma diagnostic data which has been supported from developed preprocessors based on Bayesian inference and deep learning respectively. I will introduce what I mean by the plasma equilibrium in the next section.

\section{Equilibrium}  \label{ch2-2}

\noindent
In the field of nuclear fusion, \textit{Equilibrium}, \textit{Tokamak Equilibrium}, \textit{Plasma Equilibrium} and \textit{Magnetic Equilibrium} all mean the same phenomenon that the Lorentz force exerted on the plasma balances the plasma pressure (or pressure gradient) in a macroscopic equilibrium state inside the tokamak. The basic condition for the plasma equilibrium suggest that the force on the plasma be zero at all plasma regions. This equilibrium comes from the single fluid magnetohydrodynamic (MHD) equations \cite{Freidberg:1987} which explain fluid-like, macroscopic behaviors of ionized ions and electrons.

Before explaining equations of the plasma equilibrium, I would like to state two fundamental aspects of the equilibrium: (1) the internal balance between the two forces as introduced above; (2) there is the shape and position of the plasma determined and controlled by the external coil currents. 

From now on, Let us take a look at the MHD equation to arrive at the force balance equation and beyond. The MHD momentum is given by,
\begin{equation}
\label{eq:gs1}
\rho \left[\frac{\partial \vec{v}}{\partial t} + \Big(\vec{v}\cdot\nabla \Big)\vec{v} \right] = \vec{J}\times\vec{B} - \nabla p
\end{equation}
where $\rho$ is the mass density, $\vec{v}$ is the bulk plasma velocity field, $\vec{J}$ is the (plasma and external) current density, $\vec{B}$ is the magnetic field, and $p$ is the plasma pressure. If the static equilibrium conditions ($\vec{v}=0$ and $\partial/\partial t=0$) are assumed, the equation turns out to be the force balance equation which is
\begin{equation}
\label{eq:gs2}
\vec{J}\times\vec{B} = \nabla p.
\end{equation}
From this equation, it is obvious that there is no pressure gradient along the magnetic field lines, which is
\begin{equation}
\label{eq:gs3}
\vec{B} \cdot \nabla p = 0
\end{equation}
where it also means that the plasma forms magnetic surfaces of constant pressure. Likewise, the force balance equation tells us a relation:
\begin{equation}
\label{eq:gs4}
\vec{J} \cdot \nabla p = 0
\end{equation}
which also imply that the current lines lie in the magnetic surfaces. Furthermore, it is convenient to define the poloidal magnetic function $\psi$. This is a constant quantity on each magnetic surface acting as the poloidal flux lying within that surface. Thus, there is another relation with the magnetic field:
\begin{equation}
\label{eq:gs5}
\vec{B} \cdot \nabla \psi = 0.
\end{equation}

Through the force balance equation, based on a cylindrical coordinate and an axisymmetric systems with Maxwell's equations ($\nabla\cdot\vec{B}=0$ and $\nabla\times\vec{B}=\mu_{0}\vec{J}$), we can now derive a differential equation for the poloidal flux function $\psi$ which is called the Grad-Shafranov (GS) equation \cite{grad1958hydromagnetic, shafranov1966plasma} as shown below 
\begin{equation}
\label{eq:gs6}
\begin{split}
\Delta^{*}\psi &= R \frac{\partial}{\partial R}\frac{1}{R}\frac{\partial \psi}{\partial R} + \frac{\partial^{2} \psi}{\partial Z^{2}} \\
&= - \mu_{0}R J_{t} \\
&= -R^{2}\mu_{0}\frac{\partial p \big(\psi \big)}{\partial \psi} - F \big(\psi \big)\frac{\partial F \big(\psi \big)}{\partial \psi}
\end{split}
\end{equation}
where $F \big(\psi \big)$ is the poloidal current function as a function of $\psi$, which is related with the toroidal magnetic field $B_{T}$ as $F \big(\psi \big) = R\: B_{T}$. The first two lines of the equation include effects of Maxwell's equations, and the second to third lines are influenced by the force-balance equation. 

To obtain the solution of the equation, $\psi(R,Z)$, it is required to observe $J_{t}$ over the whole plasma volume. Unfortunately, magnetic measurements externally installed from the plasma are generally available, together with local temperature and density data of the plasma. Furthermore, the plasma exists in a certain region inside the tokamak. A boundary dividing the plasma with a vacuum region is called \textit{plasma boundary}, \textit{last closed flux surface} and \textit{separatrix}. This boundary is determined after the solution $\psi$ is found. Thus, this leads us to solve a free boundary and inverse problem.

Usually, a Green function's formulation is carried out to solve the GS equation as follows:
\begin{equation}
\label{eq:gs7}
\psi(R,Z) = \int\!\!\int G(R,Z;R',Z') J_{\phi}(R',Z') dR'dZ'
\end{equation}
where $G$ is the free-space Green's function, and $(R',Z')$ is the position of a current source. But, one can raise a question like ``Does the GS equation just give us a relation between given current densities and structures of the poloidal magnetic field (or flux surfaces) after all?", ``If that is true, can we use the Biot-Savart law instead of such complex differential equations?" Well, as a conclusion, it turns out that the formulation of the Green function and using the Biot-Savart law are the same eventually, meaning that using the law is viable. Therefore, I would like to introduce how to use the Biot-Savart law in our case and what is insufficient if we use the law only in the following subsection.

\subsection{Biot-Savart Law} \label{ch2-2-1}

\begin{figure}[t]
    \centerline{\includegraphics[width=1.00\columnwidth]{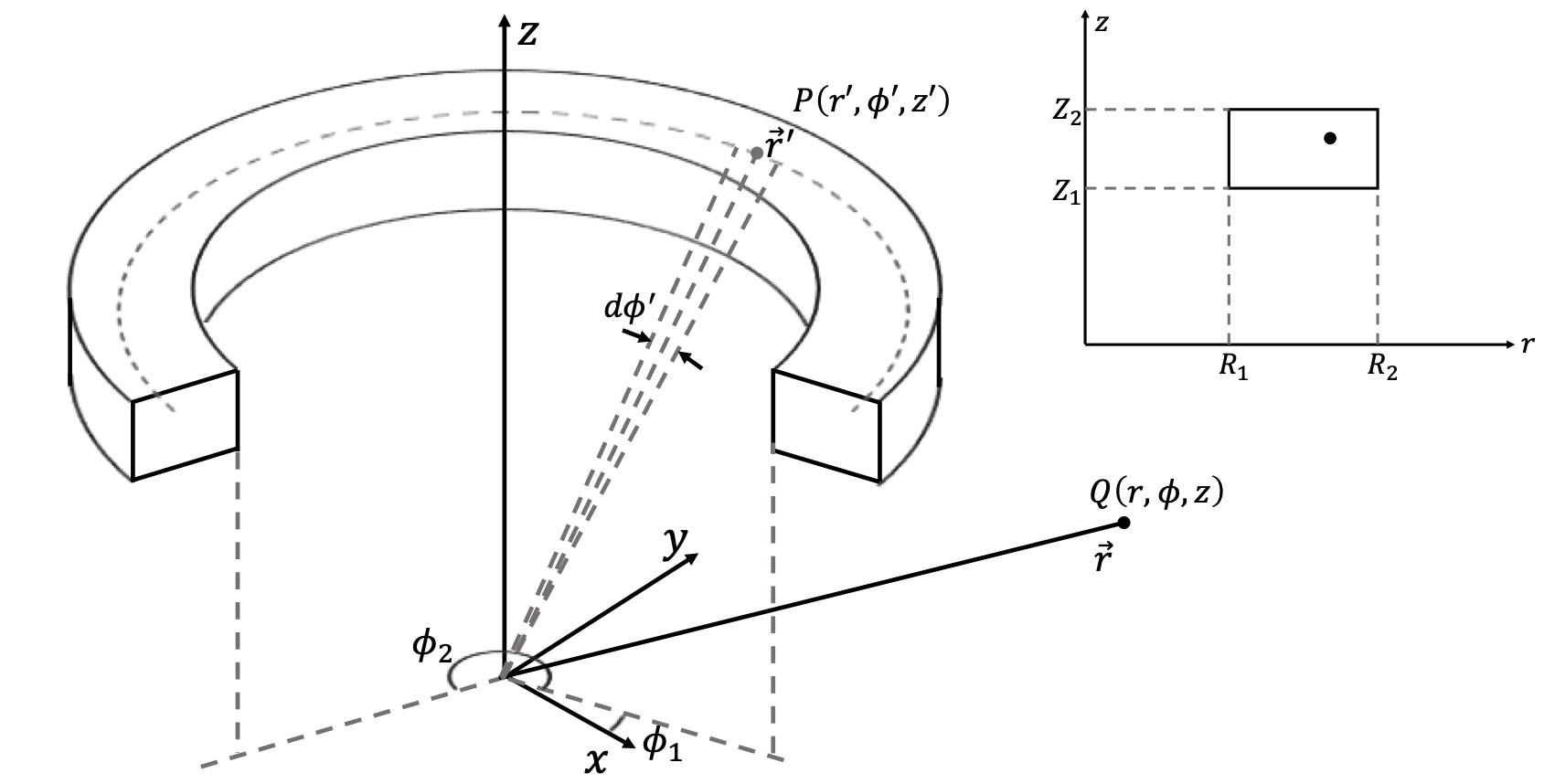}}
    \caption[Circular wire]{A segment of the circular wire showing rectangular cross-section and carrying toroidal current in cylinder coordinates.
    } \label{ch3-fig4}
\end{figure}

\noindent
In the tokamak, there are four major sources of the poloidal magnetic field, i.e., the poloidal field coils, the central solenoids, in-vessel coils \cite{kim2009design} and the plasma (current). Eddy currents (vessel currents) which are currents induced on tokamak vessel structures are ignored for simplicity. As one can find in Ref. \cite{kim2005status} and Ref.\cite{kim2009design}, all the external current coils have rectangular cross-sections, and they carry constant currents at a certain time. This means that if we assume the plasma current as a collection of wires whose cross-sections look rectangular shapes, we can use the Biot-Savart law for the vector potential and the magnetic field due to an arbitrary three-dimensional volume current with a rectangular cross-section $(R_{2}-R_{1})\times(Z_{2}-Z_{1})$ shown in Figure \ref{ch3-fig4} in cylindrical coordinates. It is worth to mention that derivations introduced in this subsection are mainly based on Ref. \cite{urankar1982vector}.

Let me consider the Biot-Savart law for the vector potential and the magnetic field by a volume current:
\begin{equation}
\label{eq:bs1}
\vec{A}\left(\vec{r}\right) = \frac{1}{4\pi} \int_{l} \int_{s} j d\vec{l} \left|\vec{r}-\vec{r}' \right|^{-1} ds,
\end{equation}
\begin{equation}
\label{eq:bs2}
\vec{B}\left(\vec{r}\right) = \frac{\mu_{0}}{4\pi} \int_{l} \int_{s} j \left[d\vec{l} \times \left(\vec{r} - \vec{r}' \right) \right] d\vec{l} \left|\vec{r}-\vec{r}' \right|^{-3} ds,
\end{equation}
where $\vec{r}$ and $\vec{r}'$ are positions of the field and source respectively, $j$ is a constant current, $ds$ is a differential element of cross-sectional area perpendicular to $d\vec{l}$ which is a segmented line element along the current. Then, as shown in Figure \ref{ch3-fig4}, set the field and source positions as $\vec{r}=(r,\phi,z)$ and $\vec{r}'=(r',\phi',z')$, respectively, where a current-carrying arc segment has properties such as $R_{1}\leq r'\leq R_{2}$, $Z_{1}\leq z'\leq Z_{2}$ and $(\phi_{2}-\phi_{1})$ as the azimuthal length of the arc segment.

I can rewrite the above equations as the following simple expressions 
\begin{equation}
\label{eq:bs3}
\vec{A}\left(\vec{r}\right) = \frac{J}{4\pi} \int_{R_{1}}^{R_{2}} dr' \int_{Z_{1}}^{Z_{2}} dz' \vec{\hat{A}} \left(\vec{r} \right),
\end{equation}
\begin{equation}
\label{eq:bs4}
\vec{B}\left(\vec{r}\right) = \frac{\mu_{0}J}{4\pi} \int_{R_{1}}^{R_{2}} dr' \int_{Z_{1}}^{Z_{2}} dz' \vec{\hat{H}} \left(\vec{r} \right),
\end{equation}
where $J$ is the azimuthal constant current density, and $A_{z}=0$ due to the conductor structure. Now, I can define the relevant forms as follows:
\begin{equation}
\label{eq:bs5}
\begin{split}
\hat{A}_{j} \left(\vec{r} \right) = \frac{1}{2} \int_{\phi_{1}}^{\phi_{2}} d\Phi& \Big[\gamma D(\Phi) + 2\gamma r \cos{\Phi}\sinh^{-1}{\beta_{1} (\Phi)} \\
&+ \big(r'^{2} - r^{2}\cos{2\Phi}\sinh^{-1}{\beta_{2}(\Phi)} \big) \\
&- r^{2}\sin{2\Phi}\tan^{-1}{\beta_{3}(\Phi)}\Big]
    \begin{cases}
      &-\sin{\Phi},\\
      &\cos{\Phi},
    \end{cases}\\
&j = r, \phi,
\end{split}
\end{equation}
where the first and the second terms inside the bracket correspond to $r$ and $\phi$ directions, respectively, and
\begin{equation}
\label{eq:bs6}
\begin{split}
\hat{H}_{l} \left(\vec{r} \right) =& \int_{\phi_{1}}^{\phi_{2}} d\Phi
    \begin{cases}
      & \cos{\Phi} \left[D(\Phi) + r \cos{\Phi} \sinh^{-1}{\beta_{1} (\Phi)} \right], \\
      & \sin{\Phi} \left[D(\Phi) + r \cos{\Phi} \sinh^{-1}{\beta_{1} (\Phi)} \right], \\
      & \gamma \sinh^{-1}{\beta_{1}(\Phi)} - r \cos{\Phi}\sinh^{-1}{\beta_{2}(\Phi)} - r\sin{\Phi}\tan^{-1}{\beta_{3}(\Phi)},
    \end{cases}\\
&l=r,\phi,z,
\end{split}
\end{equation}
where the components inside the brackets correspond to $r$, $\phi$ and $z$ directions from the top. I also define the following expressions as:
\begin{equation}
\label{eq:bs7}
\begin{split}
& \gamma = z' - z,\\
& \Phi = \phi' - \phi, \\
& D^{2}(\Phi) = \gamma^{2} + B^{2}(\Phi),\\
& B^{2}(\Phi) = r'^{2} + r^{2} - 2rr' \cos{\Phi},\\
& G^{2}(\Phi) = \gamma^{2}+r^{2}\sin^{2}{(\Phi)},\\
& \beta_{1}(\Phi) = (r'-r\cos{\Phi})/G(\Phi),\\
& \beta_{2}(\Phi) = \gamma/B(\Phi),\\
& \beta_{3}(\Phi) = \gamma(r'-r\cos{\phi})/[r\sin{\phi}D(\Phi)].
\end{split}
\end{equation}
Therefore, for all the conductors in the tokamak, the only task left is calculating the equations above for each conductor with a condition of $(\phi_{2}-\phi_{1})=2\pi$. Now, we finally have the poloidal flux function $\psi$ related to the Biot-Savart law given by $\psi(R,Z)=2\pi RA_{\phi}$.

At this moment, it seems that we have a complete formula for a solution of the tokamak equilibrium. However, we should remember that our problem is a free-boundary and inverse problem. For simplicity, let us consider the inverse problem only. Thus, if we assume an arbitrary distribution of the plasma current density, then we can obtain a distribution of $\psi$ from the equations above and update the plasma current density again using the obtained $\psi$ based on $\nabla\times\vec{B}=\mu_{0}\vec{J}$ or the first and second lines of the GS equation to be consistent with the external magnetic measurements. If we keep carrying out those sequences repeatedly, we may end up with a converged distribution of $\psi$, i.e., flux surfaces; seemingly we finish solving the tokamak equilibrium! However, what we must never forget during the sequences is whether or not the result satisfies the force balance. In other words, if our result does not meet the second and third lines of the GS equation, i.e., $J_{\phi}=J_{\phi}(R,\boldsymbol{\psi})$, then our result is totally meaningless. We fully contemplate this in order to design a neural network that solves the GS equation without any support of solutions of the GS equation, which will be presented soon.

\section{Plasma diagnostics} \label{ch2-3}

\noindent
In this section, I briefly introduce plasma diagnostics used in this thesis. The GS equation shows that the spatial variation of $\psi$ is related to the current density. Namely, if the current density is exactly known, then the solution $\psi$ would be obtained through the GS equation. Unfortunately, knowing internal information of the plasma is barely straightforward due to the temperature of the plasma ($\sim10^8$ K), therefore the current density should be inferred as well by taking advantage of externally and locally measured data. From the inferred current, the GS equation is iteratively solved until an estimated equilibrium fits the measured data reasonably. Here, the external and the local measurements are magnetic measurements and plasma pressure measurements, respectively, which are essential to observe the plasma equilibrium and energy transport in nuclear fusion experiments.

\subsection{Magnetic diagnostics} \label{ch2-3-1}

\begin{figure}[t]
    \centerline{\includegraphics[width=0.60\columnwidth]{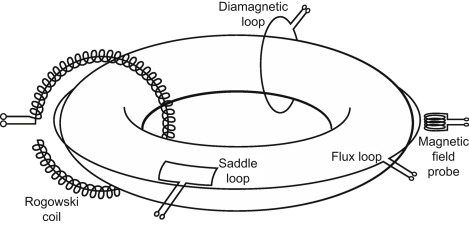}}
    \caption[Magnetic diagnostics]{A schematic of the magnetic diagnostics. Image courtesy of Ref. \cite{Strait:2008}.
    } \label{ch3-fig5}
\end{figure}

\noindent
Here, I would like to deal with magnetic diagnostics relevant to the poloidal magnetic field since solving the GS equation is highlighted. KSTAR has installed the magnetic measurements \cite{Lee:2001cp} on the KSTAR vessel wall far away from the plasma as induction coil-type measurements with analogue integrators \cite{Strait:2008}. Among them, I take advantage of 84 magnetic pick-up coils (magnetic probes) which measure the poloidal magnetic fields of the normal and the tangential directions to the vessel wall and 45 flux loops (FLs) measuring the poloidal magnetic fluxes, respectively \cite{Lee:2001cp, Lee:2008cl}. I also use Rogowski coils measuring the total plasma current, the poloidal field coil currents and the in-vessel coil currents. Basic forms of the diagnostics are shown in Figure \ref{ch3-fig5}.


The brief history of the KSTAR magnetic diagnostic system is as follows. A schematic design of the diagnostics and how to install it to the KSTAR were suggested \cite{Lee:2001cp}, and performances of the designed magnetic pick-up coils were tested in a vacuum chamber \cite{bak2001performance}. The designs were improved and analyzed in a radio-frequency environment \cite{bak2004performance}. After that, it was reported that some of the fabricated magnetic measurements were installed at the KSTAR vessel \cite{lee2006fabrication} as well as the design of the integrators for the systems was reported \cite{Bak:2007jj}. In 2008, only some of the poloidal field coils were driven to analyze operation performances of the measurements \cite{Lee:2008cl} for the first plasma on KSTAR. Diamagnetic loops were also designed and installed \cite{bak2008diamagnetic}, and their performances were reported in 2011 \cite{bak2011diamagnetic}. An outline of the first plasma of the KSTAR was published in 2010 \cite{lee2010diagnostics} with measured magnetic data from various measurements. In 2017, a study about how to improve a plasma operation based on the magnetic diagnostics was introduced \cite{kim2017improvements}.

When it comes to the integrator, several hardware designs were introduced \cite{ali1993long, strait1997hybrid, werner2006w7, Bak:2007jj, seo2010development, wang2014digital} to correct a phenomenon called \textit{signal drift}, i.e., a baseline of a measured signal drifting in time, and software correction approaches were suggested \cite{ali1993long, strait1997hybrid, seo2009plasma, Moreau:2009dg}.

\subsection{Pressure measurements} \label{ch2-3-2}

\begin{figure}[t]
    \centerline{\includegraphics[width=0.70\columnwidth]{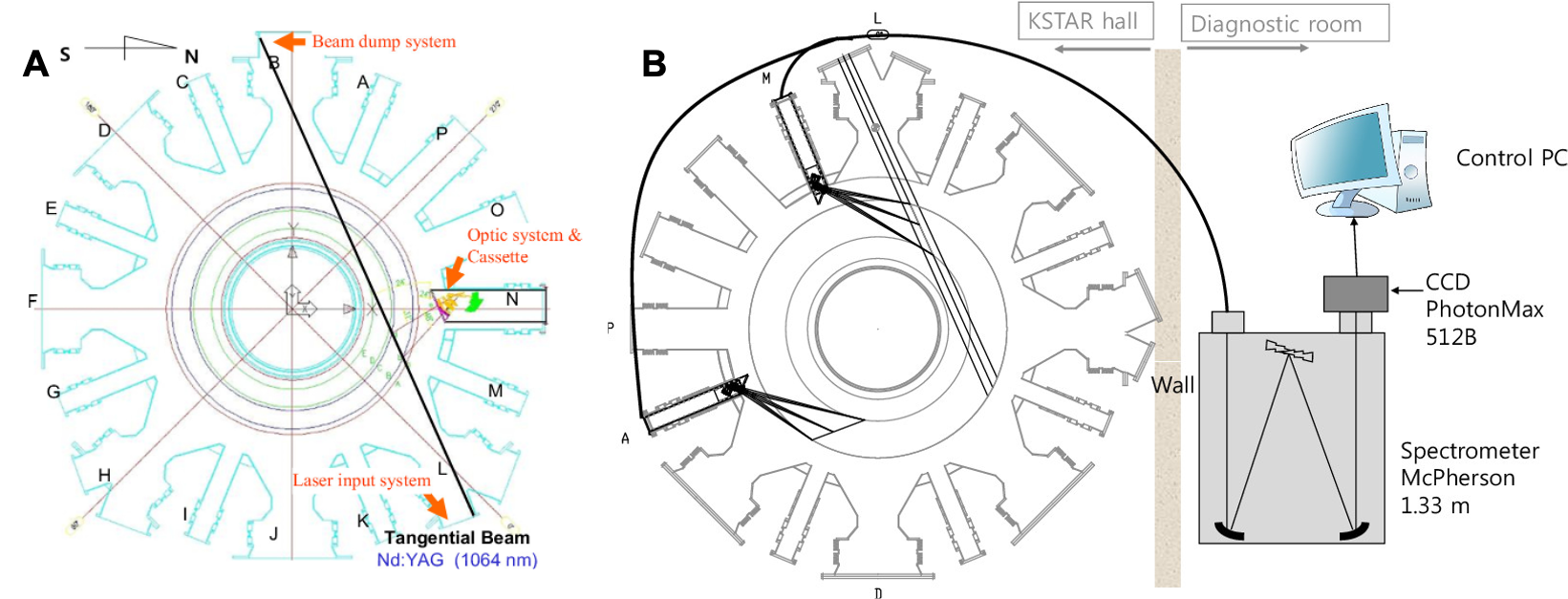}}
    \caption[KSTAR TS and CES systems]{(a) Port allocation of KSTAR TS system. Image courtesy of Ref. \cite{doi:10.1063/1.3494275}. (b) KSTAR CES layout composed of a window, mirror, collection lens, a spectrometer and a CCD camera. Image courtesy of Ref. \cite{doi:10.1063/1.3496991}.
    } \label{ch3-fig6}
\end{figure}

\noindent
To solve the GS equation, the plasma pressure on the $\psi$ coordinate is necessary over the whole tokamak region. Currently, all we can do is measuring local \textit{thermal} pressures of the plasma. These pressures are estimated from the ideal gas law, i.e., the pressure $p=nT$ where $n$ and $T$ are the density and the temperature, respectively. The reason why I stress the term ``thermal" is there is the fast ion pressure, $p_{fast}$, which is contributed by the fast ion (more energetic than a thermal ion) populations coming from the Neutral Beam Injection (NBI) system \cite{bae2012commissioning} in the KSTAR. Profile measurement systems for this pressure still need to be developed further \cite{sarwar2018effective, kim2012initial, yoo2021fast}. 

To estimate the plasma pressure, we need to measure $n$ and $T$ of the electron and the ion, respectively. The Thomson Scattering (TS) system is used detect the electron density $n_{e}$ and the electron temperature $T_{e}$ simultaneously, which is one of the major methods to obtain those information by measuring the scattered and Doppler-shifted photons from an interaction between high-power laser photons and the plasma electrons. The KSTAR TS system provides 31 local measurements with a spatial resolution of 6 mm to 13 mm, together with a temporal resolution of 50 Hz, as shown in Figure \ref{ch3-fig6} (a).

Regarding the ion density, quasi-neutrality works here, i.e., $n_{i}\approx n_{e}$. For the ion temperature, the Charge Exchange Spectroscopy (CES) system, which obtain local carbon density and flow velocity measurements along the NBI beam path by capturing the Doppler line width and deviation of a spectrum emitted from an interaction between the neutral beam ions and the carbon ions in the tokamak. The KSTAR CES system has 32 local measurements with a spatial resolution $<$ 5 mm, together with a temporal resolution $<$ 100 Hz, as shown in Figure \ref{ch3-fig6} (b).

Therefore, based on the equation below, we can relate the density and temperature measurements with the plasma pressure, i.e., 
\begin{equation}
\label{eq:pm1}
p_{tot} = n_{e}T_{e} + n_{i}T_{i} + p_{fast} + p_{rest}
\end{equation}
where $p_{rest}$ is the pressure of impurities in the tokamak whose profile is not sufficiently available yet. Thus, we can use this for $p(\psi)$ term in the GS equation, which will be dealt with in Article \RNum{5}.

One can have curiosity about the $F(\psi)$ term in the GS equation. In fact, the Motional Stark Effect (MSE) system \cite{levinton1989magnetic} which measures local magnetic field pitch angles along the NBI path from the polarization of the motional Stark effect emission signals by the NBI beam. In this thesis, I would like to prove the fact whether deep learning can solve the GS equation by only using the equation itself or not, and a way of using the MSE measurements is considerably similar with that of the plasma pressure. Thus, I would like to leave this as a future work. It is worth to mention that the KSTAR MSE system has 25 local measurements with a spatial resolution of 1 cm to 3 cm, together with a temporal resolution of 100 Hz \cite{ko2013polarimetric}.

\section{Deep learning for tokamak equilibrium} \label{ch2-4}

\noindent
This thesis addresses reconstructing ``tokamak equilibria" in real time in the field of magnetically controlled nuclear fusion, from the perspective of solving the GS equation by an application of deep learning. As explained before, the GS equation gives us two facts: (1) balancing the plasma pressure and the Lorentz force and (2) information of plasma positions in the tokamak. Although there were previous approaches \cite{Lao:1985hn, Lao:2005kd, Ferron:2002fw, Yue:2013cj, berkery2021kinetic} to find a solution of the GS equation, sacrificing accuracy or depending on human subjectivity (manually determined complexity of the solutions) is still left to be resolved. Thus, we present a deep learning method which solves the GS equation by itself with no guess of the GS equation.

How can we believe that the networks truly understand the GS equation? Are they soluble if the network's outputs are trained with well-calculated tokamak equilibria given by the previous methods? Unfortunately, I do not think so. The networks may not be able to capture the force balance behind the dataset \cite{joung2019deep}, and there is still the human decision regarding numerical convergence such that some of the measured signal are arbitrarily selected for the reconstruction, although the network can produce the equilibria outstandingly. To answer those questions, I make the networks find the equilibria after they fully understand the GS equation without any human selection. Eventually, they can generate the equilibria consistent with given measurements. How this is possible will be served in Article \RNum{5} in chapter \ref{ch4}. Article \RNum{4} provides a prototype of Article \RNum{5} by trying to learn the GS equation by means of the KSTAR EFIT database \cite{Park:2011go}. Article \RNum{1}, Article \RNum{2} and Article \RNum{3} provides how to preprocess inputs of the networks, which guarantees that the networks can be used in any circumstance.

\chapter{Deep learning and Bayesian Inference} \label{ch3}
\epigraph{\textit{``황새의 울음을 듣겠느냐?"\\ 정우는 놀란 얼굴로 새장을 바라보다가 말했다. ``동백꽃의 향기요?"\\ 회의장의 사람들은 자신들의 이해력에 도대체  무슨 문제가 있나 고민했다.}}{--- 이영도,\\피를 마시는 새}

\noindent
The human brain is a extremely \textit{complex}, \textit{non-linear} and \textit{parallel} information-processing system, which is constituted with \textit{neurons}, the brain's structural elements, performing logical, cognitive and unconscious reasoning relatively efficiently compared to contemporary digital computers. This ability is purportedly built up over time with certain rules called \textit{experience}. This keeps continuing the development of our brain constantly. Artificial neural networks (simply referred to as \text{neural networks}) are designed to mimic our brain's functioning by using a massive interconnection of simple digital units called \textit{neurons}, \textit{perceptrons} or \textit{nodes}. This intuitive structure, i.e., \textit{plasticity} is an inception of \textit{deep learning}, an enormous pile of the interconnection to perform human-like or \textit{superhuman} capabilities.

In 1943--1958, the formation of neural networks begins with McCulloch and Pitts (1943)\cite{mcculloch1943logical} suggesting the idea of neural networks as computing machines, Hebb (1949)\cite{hebb1949organization} underlying self-organizing learning, and Rosenblatt (1958)\cite{rosenblatt1958perceptron} introducing the \textit{perceptron} as the first model for supervised learning. Although there were critics pointed out by Minsky and Selfridge (1961, 1969, 1988)\cite{minsky1960learning, minsky1969perceptrons, minsky1988perceptrons} that the perceptron is not essentially capable of being globally generalized based on locally learned examples, it would not be an overstatement that we are living in an era of neural networks and deep learning.

Then, how can we implement physically reliable deep learning? The methods that I propose in this thesis are to develop a \textit{Bayesian neural network} which is capable of perceiving physics theories and quantifying its confidence level with given input information. Although the ways are not restricted to the field of nuclear fusion, I would like to propose a neural network available for tokamak control based on learning with plasma magnetohydrodynamic (MHD) theory \cite{Freidberg:1987} and Maxwell's equations. Thus, I would like to prove that there is definitely a physically reliable neural network based on the arguments in this thesis.

From the next section, I will describe a structure of a feedforward neural network and a basic notion of supervised learning on the basis of the simple sine regression introduced in Chapter \ref{ch1}. Then I will introduce uncertainty in neural networks in light of Bayesian neural networks. How networks are taught with physics theories will be also given later. Finally, I will convey a short discussion about the usage of Generative Adversarial Networks (GANs) in light of tokamak control.

\section{Feedforward Neural Network} \label{ch3-1}

\begin{figure}[t]
    \centerline{\includegraphics[width=0.35\columnwidth]{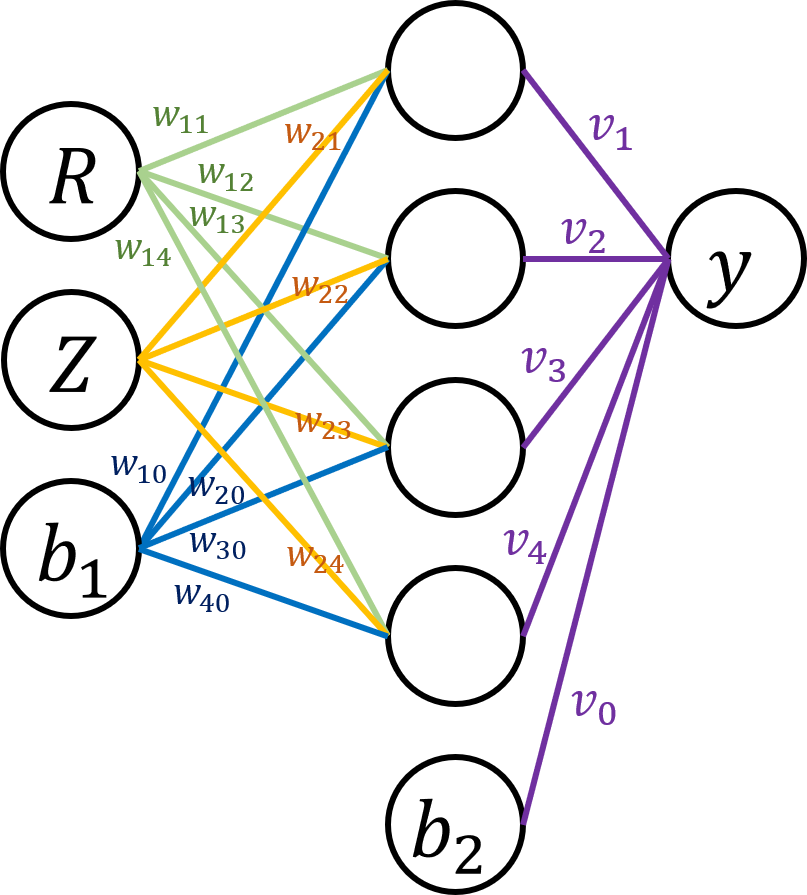}}
    \caption[Simple neural network]{A simple neural network having two input nodes (except the input bias), a single hidden layer and an output node.
    } \label{ch2-fig1}
\end{figure}

\noindent
Feedforward neural networks basically have an input layer, hidden layers and an output layer, and Figure \ref{ch2-fig1} shows an example of them with a simple architecture. Given arbitrary inputs $R$ and $Z$, the input information flows through the hidden layer toward the output node after going through an activation function in each layer. This process is non-linear, which lends the network to resemble any forms of data distributions. Each layer has its own bias as well.

The formula for the network in Figure \ref{ch2-fig1} can be expressed as follows:
\begin{equation} 
\label{eq:fn1}
\begin{split}
\hat{y} = v_{0} &+ v_{1} f\big(w_{10}+w_{11}R+w_{21}Z\big) \\
&+ v_{2} f\big(w_{20}+w_{12}R+w_{22}Z\big) \\
&+ v_{3} f\big(w_{30}+w_{13}R+w_{23}Z\big) \\
&+ v_{4} f\big(w_{40}+w_{14}R+w_{24}Z\big)
\end{split}
\end{equation}
where $w$ and $v$ are the weights between the input and the hidden layers, and between the hidden and the output layers, respectively. $f$ is an activation function originated from the biological activation function where \textit{sigmoid}, $\tanh$, and \textit{ReLU} functions are popular to be used.

Among the training methods for the network, here I would like to introduce supervised learning whose cost function is a function of observed quantities $t$ and the network outputs $\hat{y}$ as shown below:
\begin{equation} 
\label{eq:fn2}
\epsilon = \large(t_i - \hat{y}_i \large)^{2}
\end{equation}
where $i$ is the feature of a database for training the network. With the sine function regression mentioned in Chapter \ref{ch1}, we can define $t_{i}\sim0$ at $x_{i}=0$ as an instance. This makes the network be almost zero when the input is zero if the network has a single input node by taking advantage of the well-known \textit{gradient descent method}.

\begin{figure}[t]
    \centerline{\includegraphics[width=0.40\columnwidth]{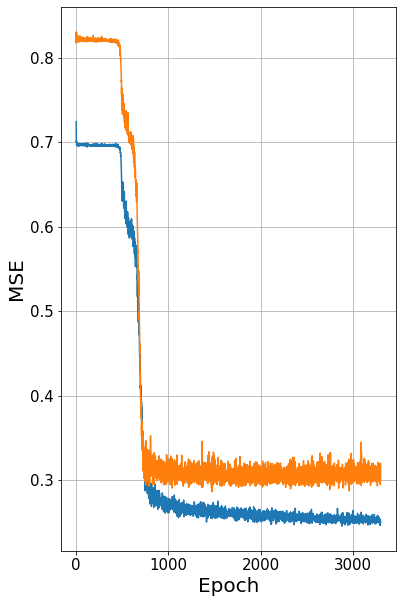}}
    \caption[Learning curve]{Training (blue) and validation (orange) costs versus epochs.
    } \label{ch2-fig2}
\end{figure}

In general, there are features involved in \textit{training} set and \textit{validation} set separately. With the sine example again, I create (observe) a total of 1024 features (data points) in interval $(0, 1)$ along x-axis. 90 percent of these features are used to build the training set, while the remaining features form the validation set. The training set indeed takes part in the training process, i.e., updating weights of the network. Conversely, the validation set does not contribute to the update process, while it is used to stop the training \textit{early} to avoid over-fitting issue. The over-fitting is a problem that the network try to follow all the sporadic data points exactly, deteriorating the network's prediction ability, i.e., \textit{network generality}.

Figure \ref{ch2-fig2} shows the training (blue) and validation (orange) costs calculated based on Equation \ref{eq:fn2} over epochs (a single loop of the update procedure). The epoch is identical to an iteration if we use whole features to update the weights at once in the single iteration. This figure is from $\sigma=0.5$ case in Figure \ref{ch1-fig1}. The training cost keeps decreasing in the figure, while the validation cost shows a stagnation (or increase) after decreasing. This shows the network is over-fitted to the data points at $\text{epoch}=3000$ since the network follows the training data points well compared to the validation data points. In other words, the network generality is degenerated. Therefore, the network at $\text{epoch}\sim2000$ is possibly optimal where the validation cost is about to increase. It is worth mentioning that I construct the neural network having four hidden layers and 300 nodes each. The activation used is \textit{swish} function.

The generalized equational forms for the neural network are as follows:
\begin{equation} 
\label{eq:fn3}
\begin{split}
&\widehat{\boldy} = f(\boldx\boldW_{1}+\boldb)\boldW_{2}, \\
&E^{\boldW_{1},\boldW_{2},\boldb}(\boldX,\boldY) = \frac{1}{2N}\sum_{i=1}^{N} ||\boldy_{i}-\widehat{\boldy}_{i}||^{2}, \\
&\mathcal{L}(\boldW_{1}, \boldW_{2}, \boldb) \equiv E^{\boldW_{1},\boldW_{2},\boldb}(\boldX,\boldY) + \lambda_{1}||\boldW_{1}||^{2} + \lambda_{2}||\boldW_{2}||^{2} + \lambda_{3}||\boldb||^{2}
\end{split}
\end{equation}
where we slightly change a notation such that $\boldy$ is the observed data, the boldface of the uppercase letters are the matrices, the boldfaces of the lowercase letters are the vectors, and the lowercase letters are the scalars. $\boldX$ and $\boldY$ are the observed input and output data. Based on these notations, I will relate the neural network to Bayesian inference to quantify the predictive uncertainty, i.e., \textit{Bayesian neural network}. Furthermore, the training method is related to the principle of Occam's razor (if no evidence, avoid over-fitting), which will be also discussed in the following sections.

\subsection{Bayesian Inference} \label{ch3-1-1}

\noindent
Before discussing \textit{Bayesian neural network}, I would like to explain Bayesian inference. The power of Bayes' theorem is the fact that the probability of \textit{hypothesis} being true given \textit{data} is linked to the probability of the data being able to be observed if the hypothesis was true:
\begin{equation} 
\text{prob}(hypothesis|data,I) \propto \text{prob}(data|hypothesis,I)\times\text{prob}(hypothesis|I).
\end{equation}
where $\text{prob}(hypothesis|I)$ is the $prior$ probability (representing our state of ignorance before the data have been measured regarding the truth of the hypothesis), $\text{prob}(data|hypothesis,I)$ is the $likelihood$ probability (modifying the prior by the measurements), and $\text{prob}(hypothesis|data,I)$ is the $posterior$ probability (illustrating our state of knowledge in the data point of view regarding the truth of the hypothesis). $\text{prob}(data|I)$ that is not shown in the equation is called $evidence$ which plays an important role in some situations like $model selection$. The quantities on the right hand side can be denoted as $\text{prob}(data, hypothesis|I)=\text{prob}(data|hypothesis,I)\times \text{prob}(hypothesis|I)$, which means that \textit{if we first specify how much we believe that the hypothesis is true, and then state how much we believe that the data is true given that the hypothesis is true, then we must implicitly have specified how much we believe that both the data and the hypothesis are true} \cite{Sivia:2006}.

In light of the neural network, the quantities of interest to be found through Bayesian inference are the weights. Given training inputs $\boldX$ and their corresponding outputs $\boldY$, the posterior probability of the network weights is:
\begin{equation} 
\label{eq:bi1}
p\big(\boldomega|\boldX,\boldY \big) = \frac{p\big(\boldY|\boldX,\boldomega \big)p(\boldomega)}{p\big(\boldY|\boldX \big)}
\end{equation}
where $\boldomega$ is the weight matrix of the network. The posterior represents the most probable weights given the training data. 


Like the evidence that I describe above, we can perform an integration of the posterior over the space of the weights which is called $marginalization$ as shown below:
\begin{equation} 
\label{eq:bi3}
p\big(\boldY|\boldX \big) = \int p\big(\boldY|\boldX, \boldomega \big) p\big(\boldomega) d\boldomega,
\end{equation}
which is in other words we marginalize over all unknown parameters, i.e., an weighted average of $\boldomega$ with respect to its prior distribution.

In light of real world observations, inferring $p\big(\boldomega|\boldX,\boldY \big)$ analytically is often unavailable. Therefore, an arbitrary $variational$ distribution whose parameter is $\theta$, $q_{\theta}(\boldomega)$, is defined to be used for the inference straightforwardly. $q_{\theta}(\boldomega)$ is suggested to be closer to the original posterior distribution, driving us to use the Kullback-Leibler (KL) divergence over $\theta$. This tells us how similar both two distributions are:
\begin{equation} 
\label{eq:bi4}
KL\big(q_{\theta}(\boldomega) \big|\big| p\big(\boldomega|\boldX,\boldY \big) \big) = \int q_{\theta}(\boldomega) \log \frac{q_{\theta}(\boldomega)}{p\big(\boldomega|\boldX,\boldY \big)} d\boldomega.
\end{equation}


Minimizing the KL divergence is identical to maximization of the \textit{evidence lower bound} (ELBO) with respect to $q_{\theta}(\boldomega)$ also known as \textit{variational lower bound}, i.e.,
\begin{equation} 
\label{eq:bi6}
\begin{split}
\mathcal{L}_{VI}(\theta) &\equiv \int q_{\theta}(\boldomega) \log p\big(\boldY|\boldX,\boldomega \big) d\boldomega - KL\big(q_{\theta}(\boldomega) \big|\big| p(\boldomega) \big) \leq \log p\big(\boldY|\boldX \big)\\
&= \int q_{\theta}(\boldomega) \log p\big(\boldY|\boldX,\boldomega \big) d\boldomega \\
&- \int q_{\theta}(\boldomega) \log q_{\theta}(\boldomega) d\boldomega + \int q_{\theta}(\boldomega) \log p\big(\boldomega \big) d\boldomega \\
&- \int q_{\theta}(\boldomega) \log p\big(\boldY|\boldX \big) d\boldomega + \int q_{\theta}(\boldomega) \log p\big(\boldY|\boldX \big) d\boldomega
\end{split}
\end{equation}
where we can find the evidence of the posterior on the far right in the first line. This plays a role of ``Occam's razor" which penalize $q_{\theta}(\boldomega)$ since the first term in the middle of the first line increases the degree of freedom of $q_{\theta}(\boldomega)$, while the second term in the same line let $q_{\theta}(\boldomega)$ be as close as the prior $p(\boldomega)$. This will show up in Appendix \ref{ap1} to explain that this also governs the degree of freedom of the network.

The procedure above is known as \textit{variational inference} (VI) which results in capturing model uncertainty, and allows us to replace the marginalization with the optimization. The Bayes' theorem and the marginalization have enormously attracted attention in nuclear fusion in light of Bayesian forward models \cite{kwak2017bayesian,kwak2020bayesian,kwak2021bayesian} and physical parameter regressions \cite{chilenski2015improved, chilenski2017experimentally, ho2019application, linder2018flux, leddy2022single}.

\subsection{Sine function Regression: Part 1} \label{ch3-1-2}

\begin{figure}[t]
    \centerline{\includegraphics[width=0.575\columnwidth]{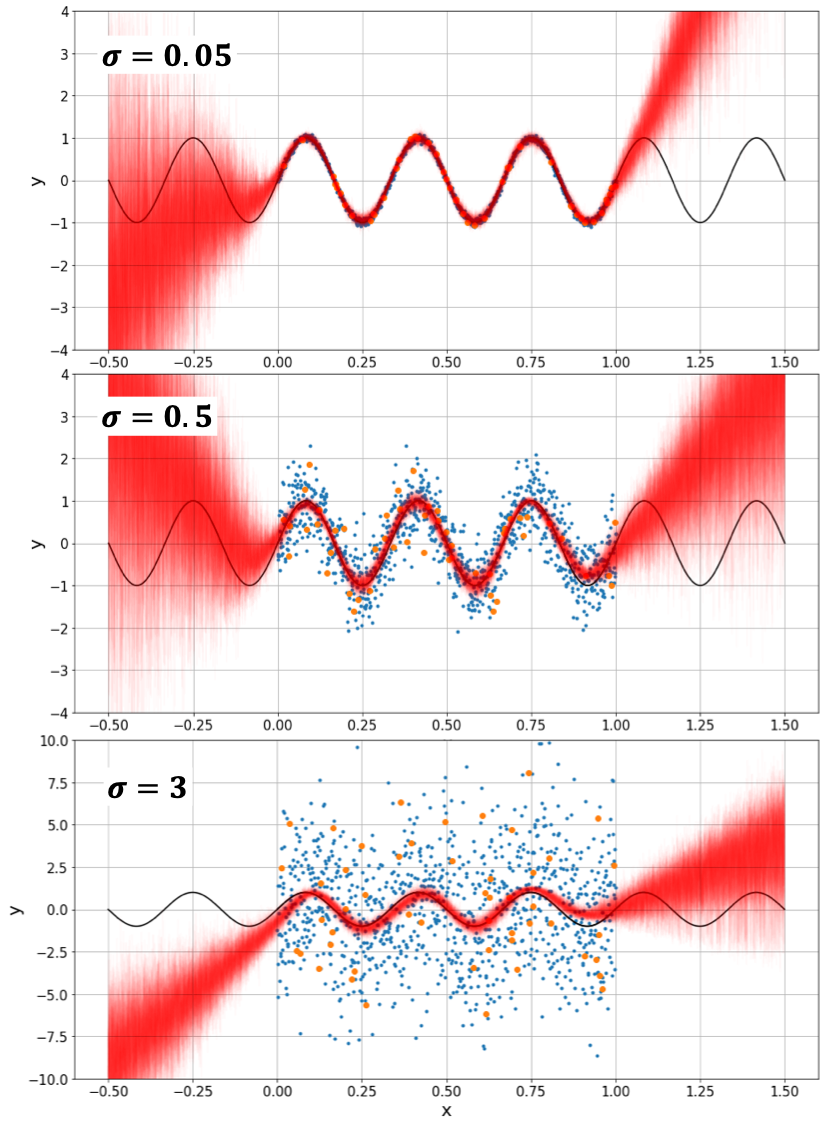}}
    \caption[Bayesian neural network regressions]{A Bayesian neural network posterior with various SNR of the observed sine functions. The networks have the dropout probability $p=0.2$. This figure is identical to Figure \ref{ch1-fig1} except the predictive uncertainty expressed in the red area.
    } \label{ch2-fig3}
\end{figure}

\noindent
With Appendix \ref{ap1} explaining what \textit{Bayesian deep learning} is, we have explored dropout in terms of Bayesian neural networks and predictive uncertainty. To implement the uncertainty mentioned in Chapter \ref{ch1} from dropout, we simply need to go through the stochastic process of dropout. In other words, we can obtain a Bayesian neural network posterior if dropout is applied during the training, naturally giving us the network's uncertainty over the network's parameter space. I apply this process for the sine function regressions, which I have covered previously. As a coincidence, I already used dropout for the problem, and let me confirm the predictive uncertainty of the network with the scattered data points of the sine functions.

Figure \ref{ch2-fig3} shows the Bayesian neural network posteriors with their uncertainty expressed in the red areas. Same with Figure \ref{ch1-fig1}, the black line is the noise-free sine functions, the blue and orange dots are the training and validation sets, and the red areas represent 1 $\sigma$ standard deviation. Since I synthesized a total of 1024 data points, I used the dropout probability of 0.2 following Figure 6.14 in Ref. \cite{gal2016uncertainty}. Without being caught in \textit{over-fitting} (following all the data points exactly), the network results are close to the correct answers (black line) with reasonable uncertainty even though the observed data are quite sporadic. Furthermore, the thickness of the red area is gradually noticeably increased when SNR of the sine data is increased. This means that the \textbf{magic approach} mentioned in Chapter \ref{ch1} is no longer magic, rather is expressed in the quantified uncertainty through Bayesian inference, convincing us it is reliable. 

Now, I would like to mention that we partially prove the neural networks are out of the \textit{black box} except that we yet prove the networks can understand physics explained in Chapter \ref{ch1}. To make this a total belief, we extend our result to show neural networks learning physical theories.


\subsection{Sine function Regression: Part 2} \label{ch3-1-3}

\begin{figure}[t]
    \centerline{\includegraphics[width=0.90\columnwidth]{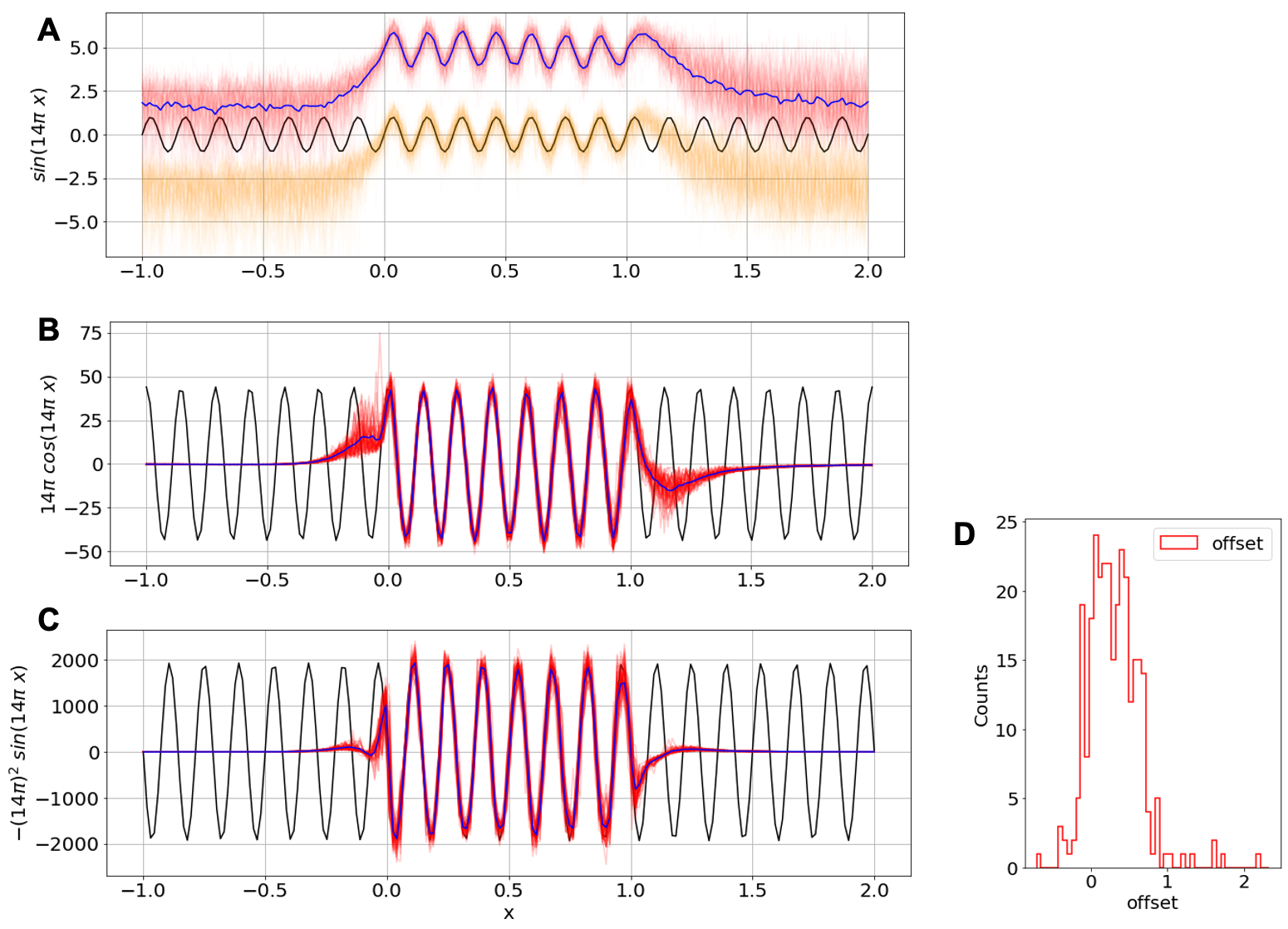}}
    \caption[Bayesian neural network differentiation]{(a) The black line is $\sin{(14\pi x)}$ which is the solution of Equation. (b) The first order derivative of the solution with respect to $x$. (c) The second order derivative. The red areas are the network results with their uncertainty, while the blue lines are the means of them. (d) The distribution of the random offset from 300 different networks.
    } \label{ch2-fig6}
\end{figure}

\noindent
With Appendix \ref{ap2} describing neural network differentiation, I, here, wish to train a neural network by Equation \ref{eq:rn1} whose solution is $t(x)=\sin{(14\pi x)} + \text{const}$ where we simply set const to zero. Following the procedure explained before, I train the neural network which has four hidden layers with 100 neurons and a bias each by using Equation \ref{eq:rn1} as a cost function. The training data is generated from the first order derivative of the solution, i.e., $14\pi\cos{(14\pi x)}$ between zero and one on the x-axis without adding noises. Figure \ref{ch2-fig6} (b) shows the first order derivative as the black line. Similarly, Figure \ref{ch2-fig6} (a) shows the solution as well as I prepare the second order derivative as well in Figure \ref{ch2-fig6} (c). In the figure, the blue lines are the network results.

As one can see, the network is capable of generating its first order derivative with respect to the input $x$ corresponding to our differential equation. Furthermore, its own output and the second order derivative (with respect to $x$) are truly matched with the solution and the second order derivative of Equation \ref{eq:rn1} as long as we shift the blue line in Figure \ref{ch2-fig6} (a) to the origin. This shift results from the fact that I do not explicitly control an offset (bias) of the network output from the cost function. Instead, I provide Figure \ref{ch2-fig6} (d), i.e., a distribution of the random offset from 300 different networks which seemingly follows a normal distribution. Lastly, Bayesian neural network posterior is also applied here where the red (and orange) areas indicate the network uncertainties analyzed by dropout. Thus, this fulfills the concept of the total belief such that we believe not only the network is able to quantify its confidence but also it can grasp a differential equation or a certain physical theory.

So far, how to teach a network physics has been introduced with the simple first order differential equation. It is worth to mention that this training method is somewhat close to \textit{supervised learning} since the network can be taught with the data generated from the cosine function although it never notices how the sine function looks like. Then what if we would like to teach high order differential equations or what if we cannot prepare not only solutions of differential equations but also their corresponding derivatives at all? Could it be called supervised learning as well? In fact, these are raised when I deal with applying a network to the purpose of tokamak control based on a plasma governing equation. I teach a second order (elliptical) partial differential equation without having a dataset for its derivatives through a neural network. Therefore, one can find answers to the questions in the following chapters.

\section{Advanced topic: GAN} \label{ch3-2}

\begin{figure}[t]
    \centerline{\includegraphics[width=0.30\columnwidth]{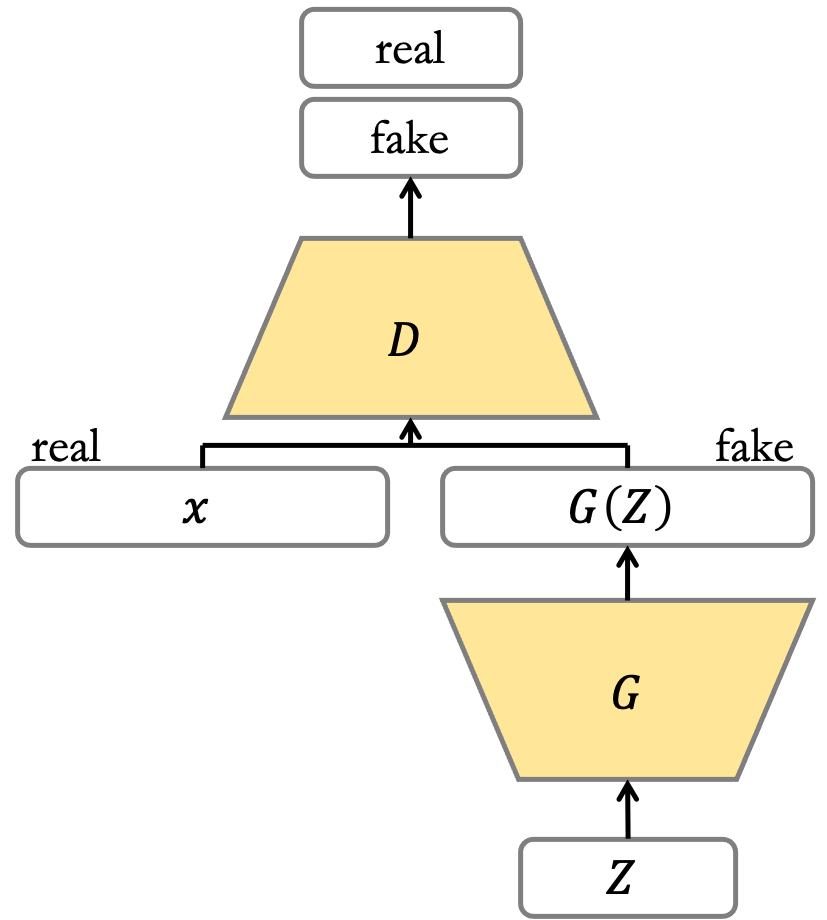}}
    \caption[GAN architecture]{A fundamental GAN architectures
    } \label{ch2-fig7}
\end{figure}

\noindent
This section is prepared to find other research fields where the method we have looked at can be helpful. Generative Adversarial Networks (GANs) \cite{goodfellow2014generative, creswell2018generative} have emerged as a type of unsupervised learning to generate network representations from distributions of data without experiencing them explicitly. Figure \ref{ch2-fig7} shows a basic structure of GAN.

In this figure, there are the generator $G$ and the discriminator $D$ which act as the forger and the expert. The forger falsifies a network output to be realistic data from a random noise, while the expert identifies real data from the forgeries. The equation below is a realization of the forger-expert relation as a cost function of GAN:
\begin{equation} 
\label{eq:nn1}
\begin{split}
\mathop{\min}_{G} \mathop{\max}_{D} V\large(D,G\large) =& E_{x\sim p_{data}(x)} [\log D(\boldx)] \\
&+ E_{z\sim p_{z}(z)} [\log (1 - D(G(\boldsymbol{z})))]
\end{split}
\end{equation}
where the first line on the right hand side is a cost for the discriminator, the second line is for the generator, $\boldx$ is the real data, and $\boldsymbol{z}$ is the random noise. 

This is powerful for data generation even not being contained in a prepared dataset. Thus, there was studies to use GANs to replace typical numerical simulations in the field of physics such as accelerator \cite{de2017learning, paganini2018calogan, paganini2018accelerating} and materials \cite{kim2020generative, kim2020inverse}. I would like to briefly introduce the use of GAN in the field of tokamak control by using \textit{plasma equilibria} database. Plasma equilibrium is a reconstructed magnetic topology of the plasma which will be discussed in the next chapter. Below are the relevant python codes using TensorFlow \cite{tensorflow2015-whitepaper}.

\begin{lstlisting}[language=Python, caption=The use of GAN with plasma equilibria: Load essential libraries.]
import h5py
import matplotlib.pyplot as plt
import numpy as np

import tensorflow as tf
from tensorflow.keras.layers import Activation, BatchNormalization, Dense, Dropout, Flatten, Reshape
from tensorflow.keras.layers import LeakyReLU, ZeroPadding2D
from tensorflow.keras.layers import Conv2D, Conv2DTranspose
from tensorflow.keras import Input
from tensorflow.keras.layers import InputLayer
from tensorflow.keras.models import Sequential
from tensorflow.keras.optimizers import Adam

from sklearn.model_selection import train_test_split
from collections import defaultdict
import argparse

\end{lstlisting}

\begin{lstlisting}[language=Python, caption=The use of GAN with plasma equilibria: Load equilibria.]
X = f3['psi'][:].transpose(2, 0, 1)

ix = list(range(X.shape[0]))
np.random.shuffle(ix)
# ix = ix[:results.nb_points]

X = X[ix]
X_train, X_test = train_test_split(X, train_size=0.9)
X_train = np.expand_dims(X_train, axis=-1)
X_test = np.expand_dims(X_test, axis=-1)
X_train = X_train.astype(np.float32)
X_test = X_test.astype(np.float32)

nb_train, nb_test = X_train.shape[0], X_test.shape[0]
\end{lstlisting}

\begin{lstlisting}[language=Python, caption=The use of GAN with plasma equilibria: Define the GAN architecture.]
img_rows = 64
img_cols = 64
channels = 1

# input image dimension
img_shape = (img_rows, img_cols, channels)

# latent space dimension
z_dim = 100

def build_generator(z_dim):

    model = Sequential()
    # From Dense 8x8x256
    model.add(Dense(256 * 8 * 8, input_dim=z_dim))
    model.add(Reshape((8, 8, 256)))
    # 8x8x256 => 16x16x128 
    model.add(Conv2DTranspose(128, kernel_size=3, strides=2, padding='same'))
    model.add(BatchNormalization())
    model.add(LeakyReLU(alpha=0.01))
    # 16x16x128 => 32x32x64
    model.add(Conv2DTranspose(64, kernel_size=3, strides=2, padding='same'))
    model.add(BatchNormalization())
    model.add(LeakyReLU(alpha=0.01))
    # 32x32x64 => 32x32x32
    model.add(Conv2DTranspose(32, kernel_size=3, strides=1, padding='same'))
    model.add(BatchNormalization())
    model.add(LeakyReLU(alpha=0.01))
    # 32x32x32 => 64x64x1
    model.add(Conv2DTranspose(1, kernel_size=3, strides=2, padding='same'))
    model.add(Activation(tf.nn.leaky_relu))

    return model

def build_discriminator(img_shape):

    model = Sequential()
    model.add(Input(shape=img_shape))

    model.add(
        Conv2D(32,
               kernel_size=3,
               strides=2,
               padding='same'))
    model.add(LeakyReLU(alpha=0.01))
    # 32x32x32 -> 16x16x64 
    model.add(
        Conv2D(64,
               kernel_size=3,
               strides=2,
               padding='same'))
    model.add(LeakyReLU(alpha=0.01))
    # 16x16x64 -> 8x8x128
    model.add(
        Conv2D(128,
               kernel_size=3,
               strides=2,
               padding='same'))
    model.add(LeakyReLU(alpha=0.01))
    # 8x8x128 -> 4x4x256
    model.add(
        Conv2D(256,
               kernel_size=3,
               strides=2,
               padding='same'))
    model.add(LeakyReLU(alpha=0.01))    
    model.add(Flatten())
    model.add(Dense(1, activation='sigmoid'))

    return model

def build_gan(generator, discriminator):

    model = Sequential()
    model.add(generator)
    model.add(discriminator)

    return model

\end{lstlisting}

\begin{lstlisting}[language=Python, caption=The use of GAN with plasma equilibria: Compile the GAN defined.]
# d model
discriminator = build_discriminator(img_shape)
discriminator.compile(loss='binary_crossentropy',
                      optimizer=Adam(),
                      metrics=['accuracy'])
# g model
generator = build_generator(z_dim)
# let d be non-trainable
discriminator.trainable = False
# g + d compile
gan = build_gan(generator, discriminator)
gan.compile(loss='binary_crossentropy', optimizer=Adam())
\end{lstlisting}

\begin{lstlisting}[language=Python, caption=The use of GAN with plasma equilibria: Train the GAN.]
train_history = defaultdict(list)
test_history = defaultdict(list)
nb_epochs = 100
batch_size = 100
latent_size = 100

real = np.ones((batch_size, 1))
fake = np.zeros((batch_size, 1))

for epoch in range(nb_epochs):
    print('Epoch {} of {}'.format(epoch + 1, nb_epochs))

    nb_batches = int(X_train.shape[0] / batch_size)
    epoch_gen_loss = []
    epoch_disc_loss = []

    for index in range(nb_batches):

        if index % 100 == 0:
            print('processed {}/{} batches'.format(index + 1, nb_batches))

        # generate a new batch of noise
        noise = np.random.normal(0, 1, (batch_size, latent_size))
        # get a batch of real images
        image_batch = X_train[index * batch_size:(index + 1) * batch_size]

        # generate a batch of fake images, 
        # using the generated labels as a
        # conditioner. We reshape the sampled labels to be
        # (batch_size, 1) so that we can feed them 
        # into the embedding
        # layer as a length one sequence
        generated_images = generator.predict(noise)

        # see if the discriminator can figure itself out...
        real_batch_loss = discriminator.train_on_batch(
            image_batch, real
        )

        # note that a given batch should have 
        # either *only* real or *only* fake,
        # as we have both minibatch discrimination 
        # and batch normalization, both
        # of which rely on batch level stats
        fake_batch_loss = discriminator.train_on_batch(
            generated_images, fake
        )
        d_loss, accuracy =  0.5 * np.add(real_batch_loss, fake_batch_loss)

        epoch_disc_loss.append(d_loss)

        # we want to train the genrator to trick
        # the discriminator
        # For the generator, we want all the {fake, real} labels 
        # to say real trick = np.ones(batch_size)
        gen_losses = []

        # we do this twice simply to match the number of batches
        # per epoch used to
        # train the discriminator
        for _ in range(2):
            noise = np.random.normal(0, 1, (batch_size, latent_size))

            gen_losses.append(gan.train_on_batch(
                noise,
                real
            ))

\end{lstlisting}

\begin{figure}[t]
    \centerline{\includegraphics[width=1.00\columnwidth]{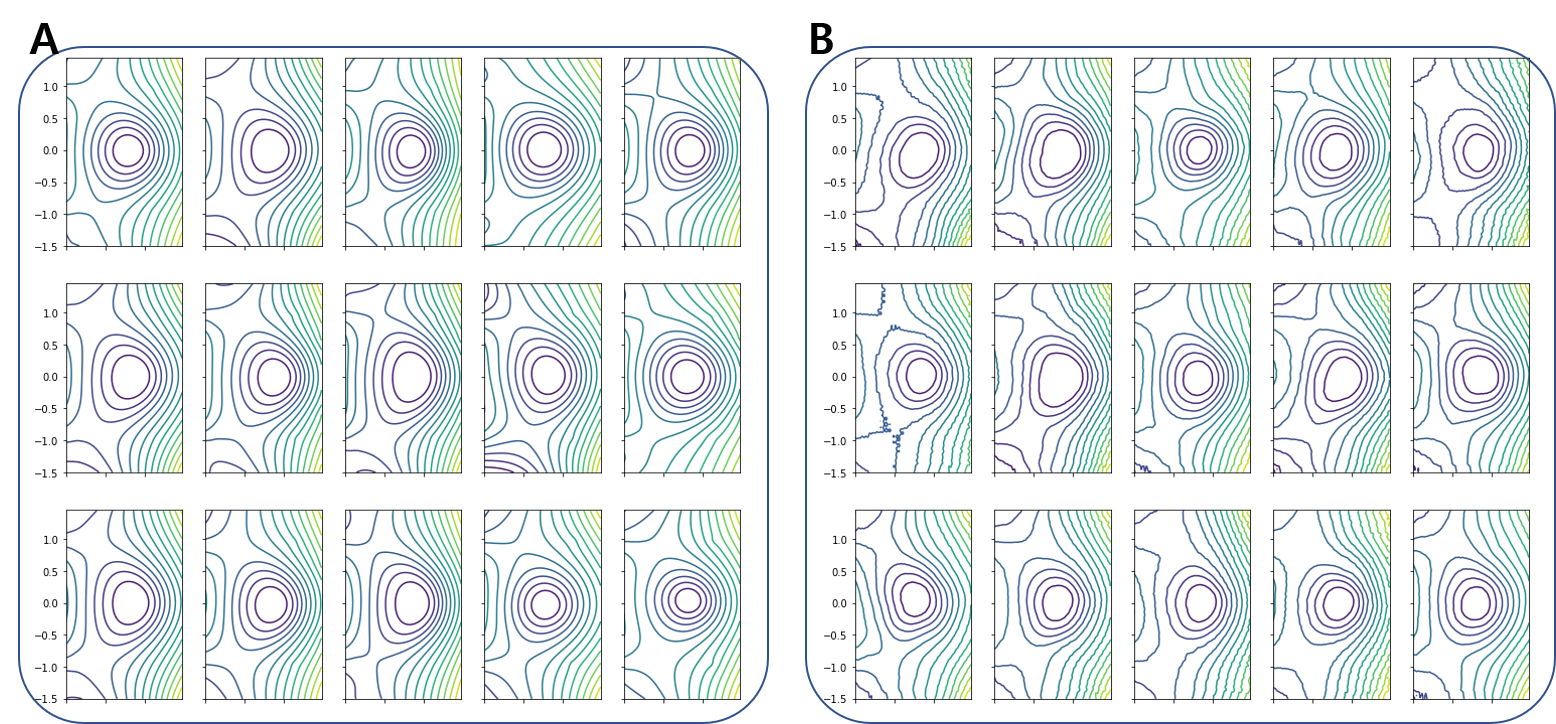}}
    \caption[GAN results]{(a) Examples from the equilibrium database, (b) Examples of the GAN results.
    } \label{ch2-fig8}
\end{figure}

As one may have noticed, there are no physical constraints in this training procedure although Figure \ref{ch2-fig8} shows a great similarity between the prepared database and the GAN results except wrinkled features in the GAN. Of course, this is a simple example but the cost function of GAN does not contain any physical restrictions. Therefore, applying our approach in the previous section to a GAN may result in a physically constrained GAN result which might be helpful to be used for simulations instead.

\section{Outlook} \label{ch3-3}

\noindent
I have reviewed a part of constituents for learning physics via neural networks in this thesis. I explain how the neural networks can be trained with not only their output but also derivatives. From this perspective, we can gain insight into an interesting paradigm that the networks can learn a physical system if the system is governed by physical theories, then the networks can use the theories as their cost functions even if simulated data for the phenomenon is not prepared yet to train the networks. It can be asserted that the network results can be more reliable than a usual training procedure since the networks literally understand the physics theories based on the novel paradigm. If the network results become more credible, humans can trust the networks and entrust them to more tasks related to the physical system (especially, tokamak operations). Perhaps, this paradigm might be regarded as a cornerstone of a pure autonomous tokamak control via deep learning. Anyhow, as a test bed, reconstructing plasma equilibria in the field of magnetic confinement fusion is chosen to prove this idea. Reconstructing plasma equilibria requires solving a second-order partial differential equation, which will be introduced in the next chapter.

\chapter{Bayesian neural network in fusion research} \label{ch4}

\noindent
Here, Chapter \ref{ch4} constitutes the main outcome of this thesis. The findings listed in this chapter are applications of the principles and methods that are described in the previous chapters in order to reconstruct plasma equilibria as scientific and practical usages of deep learning in nuclear fusion research. 

With the Korea Superconducting Tokamak Advanced Research (KSTAR), Article \RNum{4} has been developed to show that a neural network can learn a plasma `theory' with the support of a database prepared from a numerical algorithm by reconstructing plasma equilibria based on the Grad-Shafranov (GS) equation. This is a preliminary application for the network to provide the possibility of a complete unsupervised learning for the reconstruction such that the neural network can understand the GS equation itself, and the database from the numerical algorithm is no longer required. This is described in Article \RNum{5}, providing how a principle of the unsupervised learning works, and why this kind of network is required for tokamak control. Article \RNum{1}, Article \RNum{2} and Article \RNum{3} have been developed to preprocess KSTAR measurements used to inputs of our networks since baseline increases of measured signals in time (signal drift), missing signals due to mechanical issues and inconsistency between signals should be handled to use our networks in any experimental circumstances. From \textit{Bayesian neural networks}, our applications are able to quantify the epistemic uncertainty related to the plasma theory by obtaining inference results of the GS equation as well as plasma information such as positions and locations of the plasmas (which are hard to be measured directly) in the KSTAR.

Again, the principles and methods that I have used for the applications are explained in the previous chapters, thus the reader who wants to take a look at these is recommended to read chapter \ref{ch2} and chapter \ref{ch3}.

\section{Article \RNum{1}: Signal drift correction}

\noindent
This approach deals with Bayesian based numerical method for real-time correction of signal drifts in magnetic measurements from tokamaks\footnote{Reproduced from \textsc{S. J{\protect\small oung}} et al. the Appendix in 'Deep neural network Grad–Shafranov solver constrained with measured magnetic signals'. In: \textit{Nuclear Fusion}, Vol.60.1 (3$^{rd}$ Dec. 2019), page  \textsc{{\protect\small 016034}}, \textsc{{\protect\small DOI:}}\href{https://doi.org/10.1088/1741-4326/ab555f}{10.1088/1741-4326/ab555f}}, which is largely taken from Ref \cite{joung2019deep}, as a part of \textit{preprocessing magnetic measurements via Bayesian inference and neural networks}.

This article is to model signal drift which is a phenomenon that baselines of measured signals increase or decrease in time by using Bayesian inference. Magnetic signals such as magnetic fields and fluxes typically measured from inductive coils with analogue integrators can be obtained by integrating voltages induced in the coils by time-varying magnetic fields from current sources. KSTAR usually has the poloidal field coils and the plasma as the current sources, and vessel currents (induced current in KSTAR vessel structure) are often regarded as significant current sources as well. Besides, KSTAR measures the poloidal magnetic fields and fluxes from magnetic pick-up coils and flux loops installed on the vacuum vessel wall. When the voltages induced from the sources are integrated, spurious offsets are also often accumulated, causing the magnetic signals to tend to be increased (or decreased) over time. This phenomenon should be compensated properly to be used for various plasma analyses based on the magnetic signals such as EFIT (plasma equilibrium fitting).

Thus, the Bayesian model for the KSTAR pick-up coils and flux loops are suggested with information of initial magnetization stage which is a step that all the poloidal field coils are being charged to be ready for tokamak discharges. In this stage, currents of the poloidal field coils become a steady state after being fully charged, meaning that the magnetic signals also have ideally no variance in time. Thus, any variances in this phase can be considered as the signal drift which is alleviated by our Bayesian model. To model the signal drift, linearly increased drift model is assumed. This can reasonably handle the measured signals from KSTAR short-pulse discharges ($\leq$ 20 sec), while being required to be improved for applications of KSTAR long-pulse discharges. Nevertheless, this method is quite effective in KSTAR discharges where the short-pulse discharges account for the majority. Thus, Article \RNum{2}--\RNum{5} employs this development in order to preprocess the signal drifts in the magnetic fields and fluxes. Note that this article \RNum{1} is a long version of an appendix in Article \RNum{4}.

\subsection{Introduction}
\label{sec:intro}

 \begin{figure}[t]
    \centerline{\includegraphics[width=0.55\textwidth]{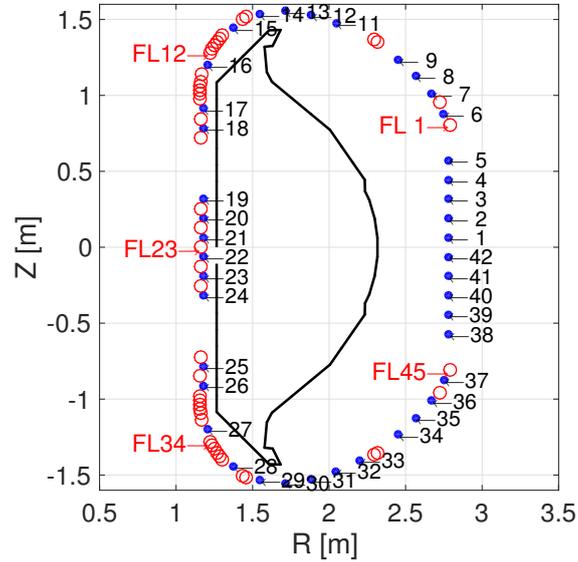}}
    \caption{Configuration of magnetic diagnostics on a poloidal cross-section of KSTAR at a certain toroidal position. Blue dots show the positions of both MP$_n$ and MP$_t$, and red open circles for the positions of the FLs. The black thick line shows the first wall. Note that we only show five FL sensor numbers out of 45 of them for simplicity.} 
    \label{MD-config}
\end{figure}

\noindent
Magnetic diagnostics (MDs) are one of the most fundamental and widely used sensors installed in almost all (if not all) magnetic-confinement fusion devices, for instance LHD \cite{Sakakibar2010}, MAST \cite{Edlington:2001jg}, D\rom{3}-D \cite{Strait:2006kh}, TCV \cite{Moret:1998fh}, EAST \cite{Liu:2013dn}, JET \cite{Peruzzo:2009et}, ITER \cite{Chavan:2009gk} and KSTAR \cite{Lee:2001cp, Lee:2008cl}. Reference \cite{Orlinskij:191982} also discusses magnetic diagnostics on TFTR, JET, JT-60 and  D\rom{3}-D. Various magnetic signals from MDs play significant roles in real-time plasma controls, detecting MHD (magnetohydrodynamics) events \cite{Shiraki:2015jp, Snipes:1988cq, Gerasimov:2014dp, Buttery:2002im, Park:2013ex} as well as reconstructing magnetic equilibria \cite{Svensson:2008in, Lazerson:2015bb, Romero:2013eg, Qian:2017gj}, e.g., EFIT \cite{Lao:1985hn}. Albeit such important roles, baselines of the measured magnetic signals often suffer from drifts in time mainly due to capacitor leakage in analogue integrators \cite{Xia:2015cv} and possibly radiations \cite{Shikama:2003jn}. This phenomenon is typically called `signal drift' whose error must be eliminated in order to conform with required accuracy for EFIT \cite{Lao:1985hn}, magnetic control \cite{Ariola:2008} and neural network applications \cite{Lister:1991gx, Bishop:1994kr, Coccorese:1994jt}. In this paper, we propose a novel algorithm that removes the signal drifts in real-time only based on the experimentally measured data.

Most of previous researches resolve the signal drifts by modifying hardware systems \cite{Arshad:1993ev, Strait:1997ds, Werner:2006hz, Bak:2007jj, seo2010development, Wang:2014el} which is a good solution but more cumbersome than having a simple numerical solution. We develop a novel numerical method capable of inferring how much magnetic signals drift and correcting the signal drifts in real-time that can work with existing MDs without any modification of the hardware systems. 

The method is based on Bayesian probability theory \cite{Sivia:2006}, and finds the slope and the offset of the drift sequentially, thus a `two-step drift correction method,' during the initial magnetization stage, i.e., before the plasma initiation. This allows one to have not only more accurate magnetic signals for post-discharge analyses but also to improve real-time monitoring and control systems such as real-time EFIT \cite{Ferron:2002fw}. We note that existing numerical algorithms to correct such drifts require post discharge information \cite{seo2009plasma, Arshad:1993ev, Strait:1997ds, Moreau:2009dg} which inhibits real-time application.

In this work, we first present a detailed description of the Bayesian based real-time two-step correction method in Sec. \ref{sec:drift_correction}. We, then, provide how the method is applied to existing KSTAR experimental data and how effectively the method removes the signal drifts from the magnetic measurements in Sec. \ref{sec:results}, followed by discussions of our proposed method on the short pulse discharge ($< 40$ sec in terms of poloidal field (PF) coil operation time) and long pulse discharge ($> 40$ sec) as well as abnormal magnetic signals in Sec. \ref{sec:discussion}. Our conclusions are stated in Sec. \ref{sec:conclusions}.

 \begin{figure}[t]
    \centerline{\includegraphics[width=0.70\textwidth]{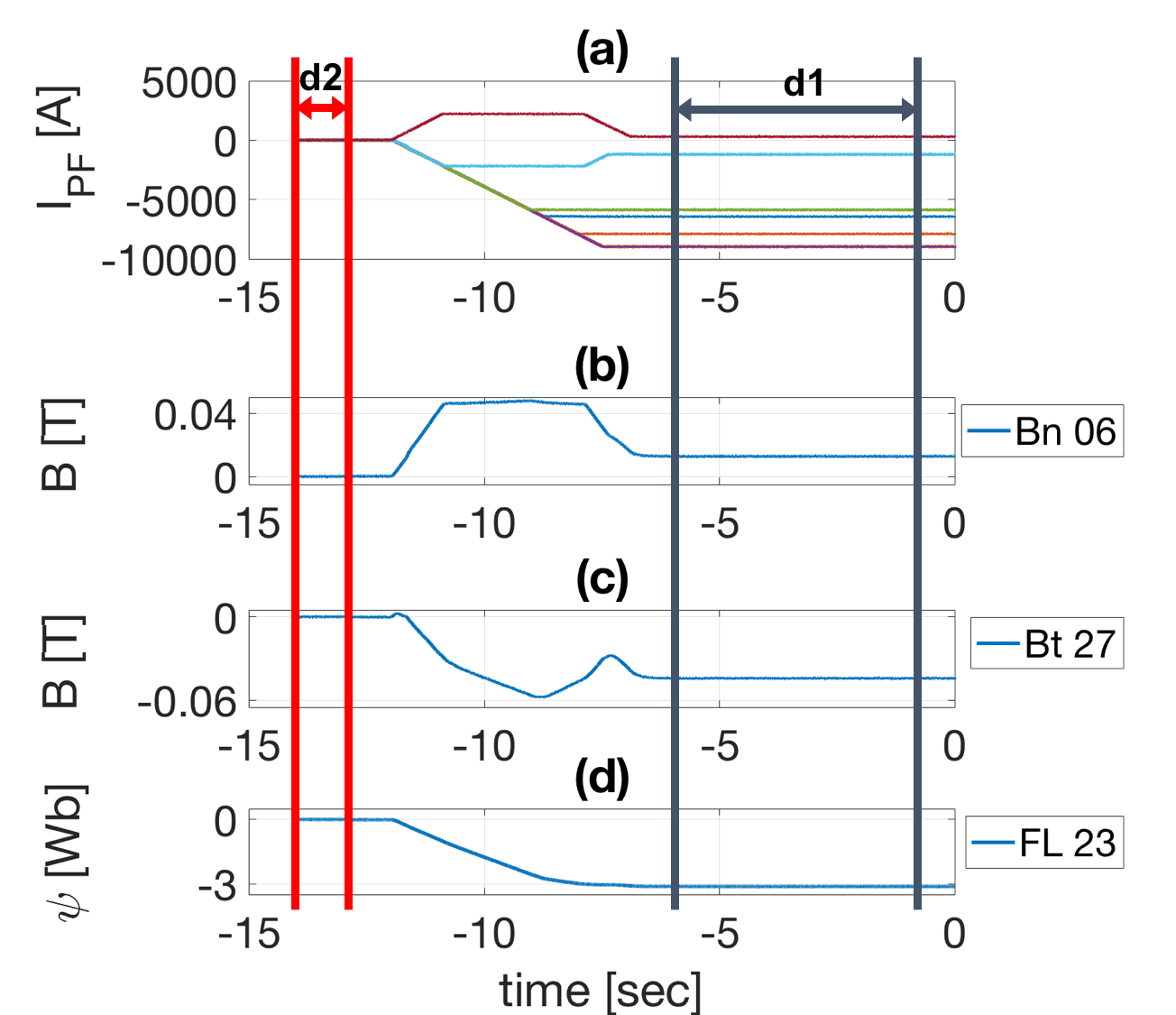}}
    \caption{An example of temporal evolutions of (a) currents in the PF coils, (b) normal and (c) tangential components of magnetic fields measured by an MP$_n$ and an MP$_t$, respectively, and (d) magnetic flux measured by an FL during the initial magnetization stage, i.e., $t<0$, for a typical KSTAR discharge. Information from the time interval d1 (d2) is used to estimate $a_i^m$ ($b_i^m$).} 
    \label{PF-fig}
\end{figure}

\subsection{Real-time drift correction based on Bayesian inference}
\label{sec:drift_correction}

\noindent
Fig. \ref{MD-config} shows the locations of the magnetic diagnostics (MDs) with the sensor numbers \cite{Lee:2008cl} at a certain toroidal position of KSTAR \cite{Kwon:2011dq}. The blue dots are the magnetic probes (MPs) measuring both normal ($B_n$ measured by MP$_n$) and tangential ($B_t$ measrued by MP$_t$) components of the magnetic fields. Note that MP \#$10$ does not exist at this toroidal position. The red circles are the flux loops (FLs) measuring magnetic fluxes. There are total of $45$ FLs on KSTAR, but we only show five sensor numbers out of $45$ of them in the figure for simplicity. In this work, we focus on correcting the signal drifts in real-time for the total number of $127$ magnetic signals, i.e., $2 \times 41$ MPs for both MP$_n$ and MP$_t$ and $45$ FLs.

\begin{figure}[t]
  \centerline{\includegraphics[width=0.70\textwidth]{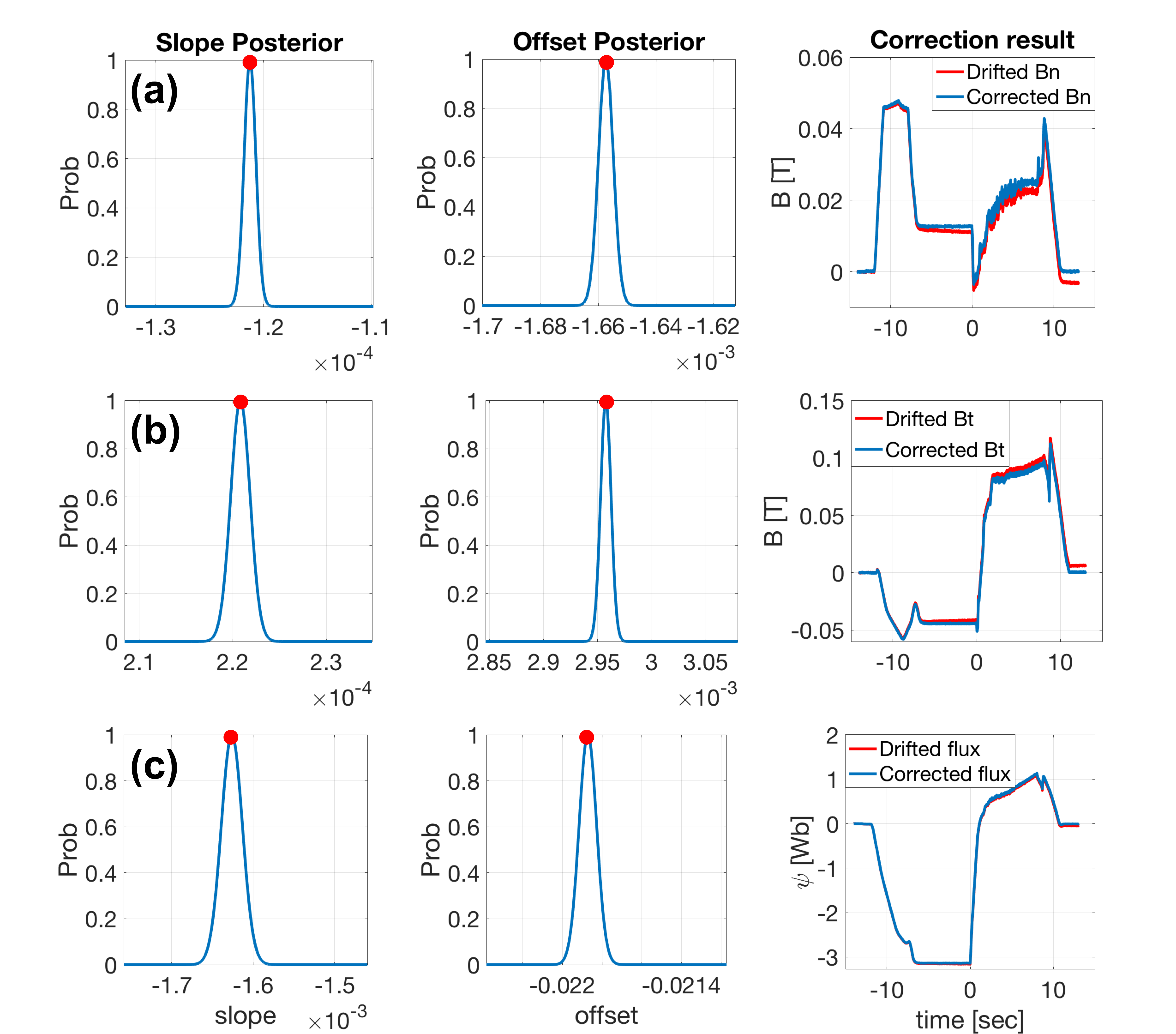}}
  \caption{Examples of the proposed two-step drift correction method for the MDs of (a) MP$_n$~\#6, (b) MP$_t$~\#27 and (c) FL~\#45. Left and middle panels show the \textit{posteriors} of the slope and the offset for each MD where the red dots depict the \textit{maximum a posterior}. Right panel shows both the original magnetic signals with the signal drifts (red) and the drift corrected signals (blue).} \label{Correc-fig}
\end{figure}

\subsection{Two-step drift correction method}
\label{subsec:two_step}

 \begin{figure}[h]
    \centerline{\includegraphics[width=0.50\textwidth]{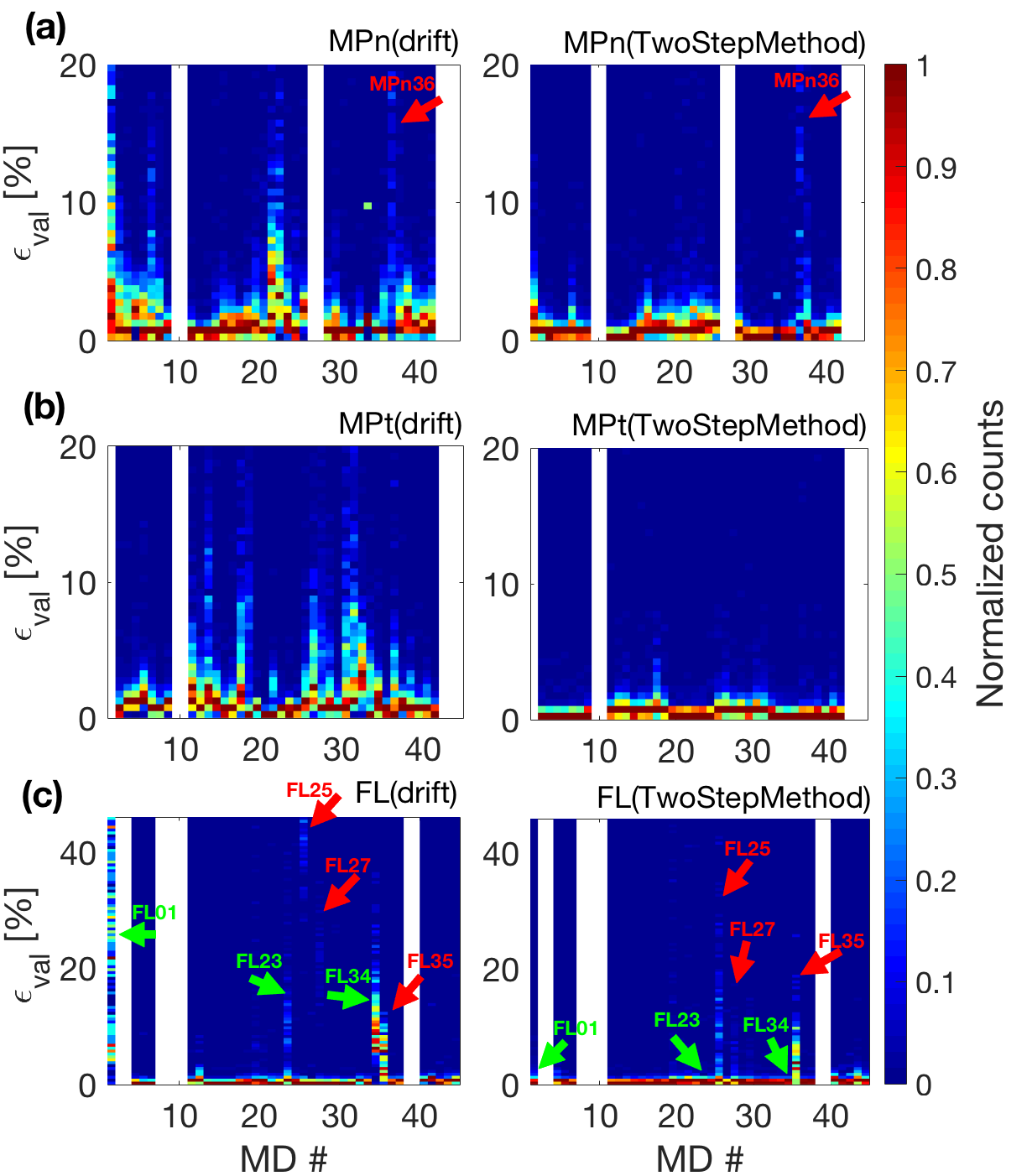}}
    \caption{Histograms of the validation errors for randomly selected 297 KSTAR discharges before (left panel) and after (right panel) the two-step drift correction for (a) $m=$MP$_n$ measuring $B_n$, (b) $m=$MP$_t$ measuring $B_t$, and (c) $m=$FL measuring magnetic fluxes. MD~\# in horizontal axes denote the MD sensor numbers, i.e., subscript $i$ in $\epsilon_{i, s}^m$. Colors represent the relative occurrence normalized to a unity for every sensor. Non-existing magnetic signals are displayed as white streaks.} 
    \label{valid-fig2}
\end{figure}

\noindent
To remove the signal drifts, we deem {\it a priori} that the signals drift linearly in time \cite{Strait:1997ds, Xia:2015cv, Ka:2008jg}, which we substantiate our assumption based on the measured data obtained during actual plasma operations in Sec. \ref{sec:results}. Therefore, we take the drifting components of the signals ($y_i^m$) from various types of MDs to follow:
\begin{equation} \label{eq:lineardrift}
y_i^m = a_i^m t + b_i^m,
\end{equation}
where $t$ is the time. $a_i^m$ and $b_i^m$ are the slope and the offset, respectively, of a drift signal for the $i^\text{th}$ magnetic sensor of a type $m$ (MP$_n$, MP$_t$ or FL). Then, our goal simply becomes finding $a_i^m$ and $b_i^m$ for all $i$ and $m$ of interests before a plasma starts or the blip time ($t=0$) so that $y_i^m$ can be subtracted from the measured magnetic signals in real-time. Here, we assume that $a_i^m$ and $b_i^m$ do not change over one plasma discharge. One can consider such linearization in time as taking up to the first order of Taylor expanded drifting signals. Therefore, we have to examine carefully our proposed method for long pulsed discharges with large nonlinearities, which is discussed in Sec. \ref{sec:discussion}.

We use two different time intervals during the initial magnetization stage for every plasma discharge to find $a_i^m$ and $b_i^m$, sequentially, thus the name `two-step drift correction method.' Fig. \ref{PF-fig} shows an example of temporal evolutions of currents in the poloidal field (PF) coils, $B_n$ and $B_t$ measured by an MP$_n$ and an MP$_t$, respectively, and magnetic flux by an FL up to the blip time ($t=0$) of a typical KSTAR discharge. 

During the time interval d1 in Fig. \ref{PF-fig}, all the magnetic signals must be constant in time because there are no changes in currents of all the PF coils as well as there are no plasmas yet that can change the magnetic signals. Therefore, any temporal changes in a magnetic signal during d1 can be regarded as due to a non-zero $a_i^m$. With the knowledge of $a_i^m$ from d1 time interval, we obtain the value of $b_i^m$ using the fact that all the magnetic signals must be zeros during the time interval d2 because there are no sources of magnetic fields, i.e., all the currents in the PF coils are zeros. 

Summarizing our procedure, (1) we first obtain the slopes $a_i^m$ based on the fact that all the magnetic signals must be constant in time during d1 time interval, and then (2) find the offsets $b_i^m$ based on the fact that all the magnetic signals, after the linear drifts in time are removed based on the knowledge of $a_i^m$, must be zeros during d2 time interval.

\subsubsection{Bayesian inference}

\begin{figure}[h]
    \centerline{\includegraphics[width=0.45\textwidth]{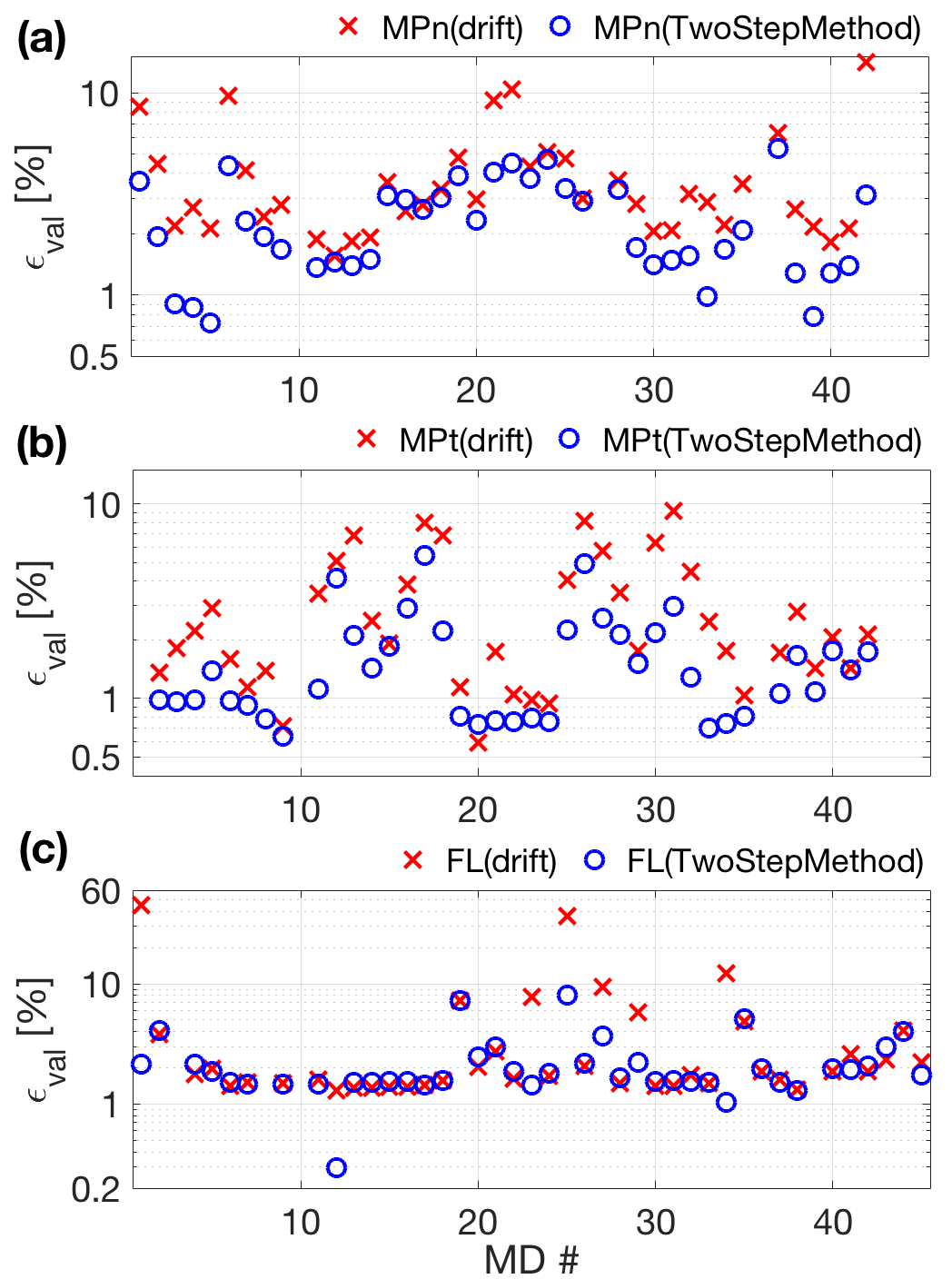}}
    \caption{Averaged validation errors $\lab\epsilon_{i}^m\rab$ for $297$ KSTAR discharges for (a) the normal (MP$_n$) and (b) the tangential (MP$_t$) components of magnetic signals, and for (c) the flux loop (FL) measurements. Blue circles indicate the validation errors after the two-step drift correction method, and red crosses mean the validation errors before applying our correction method.} 
    \label{pclog_val}
\end{figure}

\noindent
Bayesian probability theory \cite{Sivia:2006} has a general form of
\begin{equation}
p\lp\mathcal{W}|\mathcal{D}\rp=\frac{p\lp\mathcal{D}|\mathcal{W}\rp p\lp\mathcal{W}\rp}{p\lp\mathcal{D}\rp},
\label{eq:Bayes}
\end{equation}
where $\mathcal{W}$ is a (set of) parameter(s) we wish to infer, i.e., $a_i^m$ and $b_i^m$ for our case, and $\mathcal{D}$ is the measured data, i.e., measured magnetic signals during the time intervals of d1 and d2 in Fig. \ref{PF-fig}. The \textit{posterior} $p\lp\mathcal{W}|\mathcal{D}\rp$ provides us probability of having a certain value for $\mathcal{W}$ given the measured data $\mathcal{D}$ which is proportional to a product of \textit{likelihood} $p\lp\mathcal{D}|\mathcal{W}\rp$ and \textit{prior} $p\lp\mathcal{W}\rp$. Then, we use the \textit{maximum a posterior (MAP)} to select the value of $\mathcal{W}$. The \textit{evidence} $p\lp\mathcal{D}\rp$ (or marginalized \textit{likelihood}) is typically used for a model selection and is irrelevant in this study as we are only interested in estimating the parameter $\mathcal{W}$, i.e., $a_i^m$ and $b_i^m$.

We estimate values of the slope $a_i^m$ and the offset $b_i^m$ based on Eq. (\ref{eq:Bayes}) in two steps as described in Sec. \ref{subsec:two_step}: 
\begin{equation} \label{eq:baye-slope}
\text{Step (1)}\: : \: p(a_i^{m}|\mathcal{\vec D}_{i, d1}^m) \propto p(\vec D_{i, d1}^m|a_i^{m})p(a_i^{m}),
\end{equation}
\begin{equation} \label{eq:baye-off}
\text{Step (2)}\: : \: p(b_i^{m}|\mathcal{\vec D}_{i, d2}^m, a_i^{m*}) \propto p(\vec D_{i, d2}^m|b_i^{m}, a_i^{m*})p(b_i^{m}),
\end{equation}
where $\mathcal{\vec D}_{i, d1}^m$ ($\mathcal{\vec D}_{i, d2}^m$) are the time series data from the $i^\text{th}$ magnetic sensor of a type $m$ (MP$_n$, MP$_t$ or FL) during the time intervals of d1 (d2) as shown in Fig. \ref{PF-fig}. $a_i^{m*}$ is the \textit{MAP}, i.e., the value of $a_i^m$ maximizing the \textit{posterior} $p(a_i^{m}|\mathcal{\vec D}_{i, d1}^m)$. Since we have no prior knowledge on $a_i^m$ and $b_i^m$, we take \textit{priors}, $p(a_i^{m})$ and $p(b_i^{m})$, to be uniform allowing all the real numbers. We mention that the \textit{posterior} for $b_i^m$ should, rigorously speaking, be obtained by marginalizing over all possible $a_i^m$, i.e., $p(b_i^{m}|\mathcal{\vec D}_{i, d2}^m)=\int p(b_i^{m}|\mathcal{\vec D}_{i, d2}^m, a_i^m) p(a_i^{m}|\mathcal{\vec D}_{i, d1}^m) da_i^m$. However, as we are only interested in \textit{MAP} rather than obtaining full probability distribution of $b_i^m$, we omit the marginalization procedure and simply use $a_i^{m*}$. Furthermore, as we are interested in real-time application, we must consider the computation time as well.

With Eq. (\ref{eq:lineardrift}), we model \textit{likelihoods}, $p(\vec D_{i, d1}^m|a_i^{m})$ and $p(\vec D_{i, d2}^m|b_i^{m}, a_i^{m*})$, as Gaussian: 
\begin{align}
& p(\vec D_{i, d1}^{m}|a_i^m) = \frac{1}{\sqrt{(2\pi)^L} |\sigma_{i, d1}^{m}| } \cr
&\! \times\! \exp \! \left(\!- \frac{\sum\limits_{t_l \in d1}^L \left[ a_i^m(t_{l} - t_0) - \left( D_{i, d1}^m(t_l) - \left< D_{i, d1}^m(t_0)\right> \right)\right]^2}{2(\sigma_{i, d1}^m)^2} ) \! \right), \cr
\label{eq:like-slope}
\end{align}
\begin{align}
p(\vec D_{i, d2}^m | b_i^m, & a_i^{m*}) = \frac{1}{\sqrt{(2\pi)^K} |\sigma_{i, d2}^m| } \cr
& \times\! \exp \! \left(\!- \frac{\sum\limits_{t_k\in d2}^K \left[ b_i^m - \left( D_{i, d2}^m(t_k) -a_i^{m*} t_k\right)\right]^2}{2(\sigma_{i, d2}^m)^2} \! \right), \cr
\label{eq:like-offset}
\end{align}
which simply state that noises in the measured signals follow Gaussian distributions. Here, $\sigma_{i,d1}^m$ and $\sigma_{i,d2}^m$ are the experimentally obtained noise levels for the $i^\text{th}$ magnetic sensor of a type $m$ (MP$_n$, MP$_t$ or FL) during the time intervals of d1 and d2 in Fig. \ref{PF-fig}, respectively. $t_l$ and $t_k$ define the actual time intervals of d1 and d2, i.e., $t_l \in [-6, -1]$ sec and $t_k \in [-14, -13]$ sec with $L$ and $K$ being the numbers of the data points in each time interval, respectively. $t_0$ can be any value within the d1 time interval, and we set $t_0=-2$~sec in this work. $\left< D_{i, d1}^m(t_0)\right>$, removing the offset effect to obtain only the slope, is the time averaged value of $D_{i, d1}^m(t)$ for $t\in [t_0-0.5, t_0+0.5]$~sec. We use the time averaged value to minimize the effect of the noise in $D_{i, d1}^m(t)$ at $t=t_0$.

With our choice of uniform distributions for \textit{priors} in Eqs. (\ref{eq:baye-slope}) and (\ref{eq:baye-off}), \textit{MAPs} for $a_i^m$ and $b_i^m$, which we denote them as $a_i^{m*}$ and $b_i^{m*}$, coincide with the maximum \textit{likelihoods} which can be analytically obtained by maximizing Eqs. (\ref{eq:like-slope}) and (\ref{eq:like-offset}) with respect to $a_i^m$ and $b_i^m$, respectively: 
\begin{equation} \label{eq:direct-slope}
a_i^{m*} = \frac{\sum\limits_{t_l \in d1}^L \left[ \left(D_{i, d1}^m(t_l) - \left< D_{i, d1}^m(t_0)\right>\right)\left(t_l - t_0 \right) \right]}{\sum\limits_{t_l \in d1}^L  \left[ t_l - t_0 \right]^2},
\end{equation}
\begin{equation} \label{eq:direct-off}
b_i^{m*} = \frac{1}{K} \sum\limits_{t_k \in d2}^K  \left[ D_{i, d2}^m(t_k) - a_i^{m*} t_k \right].
\end{equation}
Now, we have attained simple algebraic equations based on Bayesian probability theory which can provide us values of the slope $a_i^m$ and the offset $b_i^m$ before the blip time, i.e., before $t=0$. 

As will be discussed in Sec. \ref{sec:results}, we find slopes and offsets for all $127$ MDs shown in Fig. \ref{MD-config} within $\sim0.2$~sec (before the plasma starts for each shot) using MATLAB on a typical laptop within of the order of $1$\% average validation errors, except few abnormal events which are also discussed in Sec. \ref{sec:discussion}. This means that we can correct the drifts of magnetic signals in real-time. 

\begin{figure}[t]
    \centerline{\includegraphics[width=0.605\textwidth]{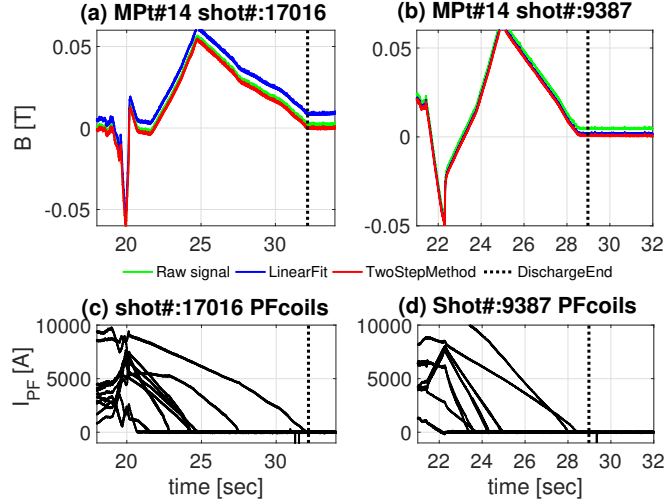}}
    \caption{Qualitative comparisons between a typical chi-square linear fitting method (blue line) and our proposed two-step method (red line) with the raw (before correction) signal (green line) in (a) KSTAR shot \#17016 and (b) \#9387 for the tangential component of magnetic signal MP$_t$ \#14. (c) and (d) show temporal evolutions of currents through KSTAR PF coils, and vertical dotted lines indicate the time where we expect all the magnetic signals return to zeros if there were no signal drifts. Note that the blue line in (b) is almost overlapped with the red line, but it is slightly more off from the zero compared to the red line.} 
    \label{comp-linear}
\end{figure}

\subsection{Results with KSTAR experimental data} \label{sec:results}

\noindent
As examples of the results of the proposed two-step drift correction method described in Sec. \ref{sec:drift_correction}, Fig. \ref{Correc-fig} shows \textit{posteriors} of the slopes (left panel) and the offsets (middle panel) of a few MDs: (a) MP$_n$~\#6 for the normal component of the magnetic field $B_n$, (b) MP$_t$~\#27 for the tangential component of the magnetic field $B_t$ and (c) FL~\#45 for the poloidal magnetic flux from KSTAR shot \#8775. The right panel shows results of removing the signal drifts (blue lines) from the original magnetic signals (red lines), where the slopes and the offsets are selected as the values corresponding to the \textit{maximum a posterior (MAP)}, i.e., the values with the maximum probabilities depicted as red dots in the left and middle panels of Fig. \ref{Correc-fig}. It is indisputable how effectively the proposed method removes the signal drifts. 

\begin{figure}[t]
    \centerline{\includegraphics[width=0.505\textwidth]{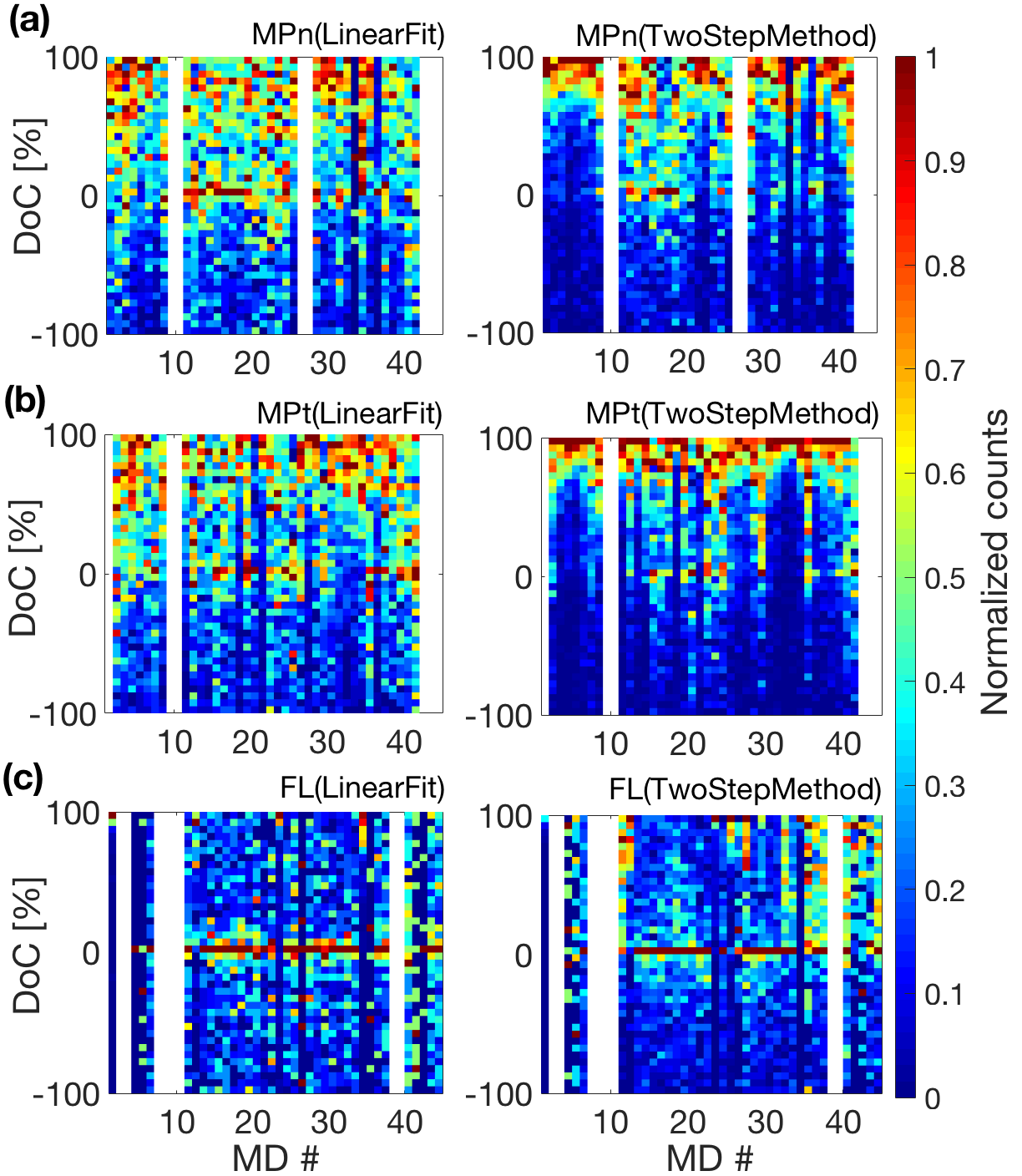}}
    \caption{Histograms of the degree of corrections (DoC's), where signal drift corrections are performed based on a typical chi-square linear fitting method (left panel) and our proposed two-step drift correction method (right panel) for (a) MP$_n$ measuring $B_n$, (b) MP$_t$ measuring $B_t$, and (c) FL measuring magnetic fluxes. MD~\# in horizontal axes denote the MD sensor numbers. Colors represent the relative occurrence normalized to a unity for every sensor. Same sets of magnetic signals used to generate Fig. \ref{valid-fig2} are used. Non-existing magnetic signals are displayed as white streaks.} 
    \label{comp-doc}
\end{figure}

For the purpose of real-time correction during a plasma operation, it is not necessary to generate full \textit{posteriors} based on Eqs. (\ref{eq:baye-slope})-(\ref{eq:like-offset}), rather we can simply calculate the \textit{MAPs} of the slope and the offset using Eqs. (\ref{eq:direct-slope}) and (\ref{eq:direct-off}). For a post-discharge analysis, having full \textit{posteriors} is beneficial as they provide quantitative uncertainties of the estimated slopes and offsets which are required information to perform a proper error propagation.  

It is worthwhile to mention that `drift signals' in this work are actually ``corrected" signals in some degrees. KSTAR executes a $60$-sec-long shot with the predefined waveforms on the PF coils to calibrate (to obtain the slopes and the offsets of) magnetic signals without plasmas every morning during a campaign. As right panel of Fig. \ref{Correc-fig} shows such calibration retains observable non-zero values in correcting drift signals. Our two-step drift correction method is applied in these `corrected' drift signals.

\subsubsection{Validation error: How good is the two-step drift correction method?}\label{subsec:val_err}

\begin{figure}[t]
    \centerline{\includegraphics[width=0.605\textwidth]{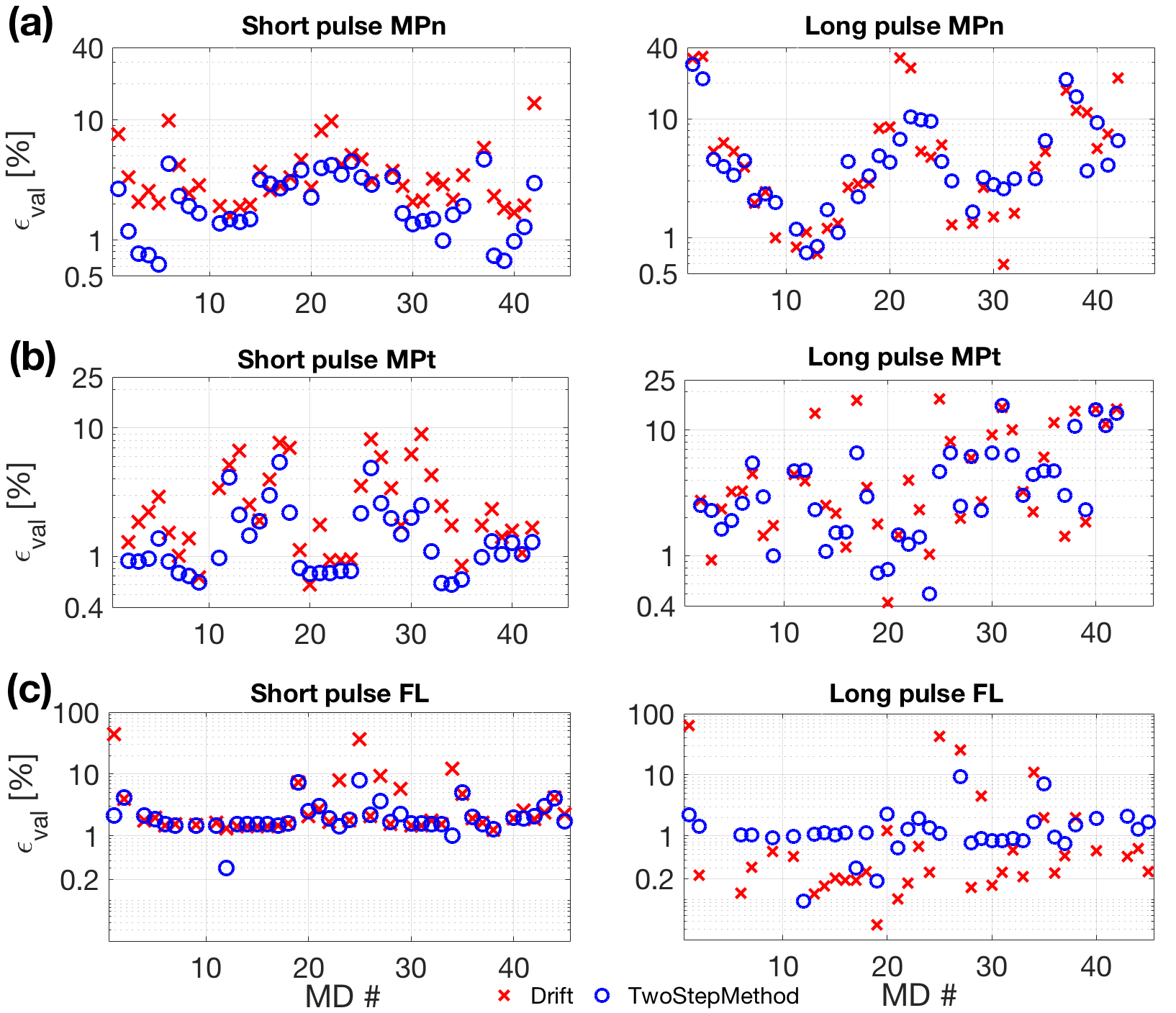}}
    \caption{Averaged validation errors as in Fig. \ref{pclog_val} before (red crosses) and after the correction (blue circle) for (a) the normal component (MP$_n$) and (b) the tangential component (MP$_t$) of magnetic signals and for (c) flux loop measurements. Left panels show the results for the 286 short pulse discharges ($< 40$ sec), while the right panels show the results for the 11 long pulse discharges ($> 40$ sec). Note the different scales for y-axis.} 
    \label{short_vs_long-fig}
\end{figure}
\begin{figure}[t]
    \centerline{\includegraphics[width=0.64\textwidth]{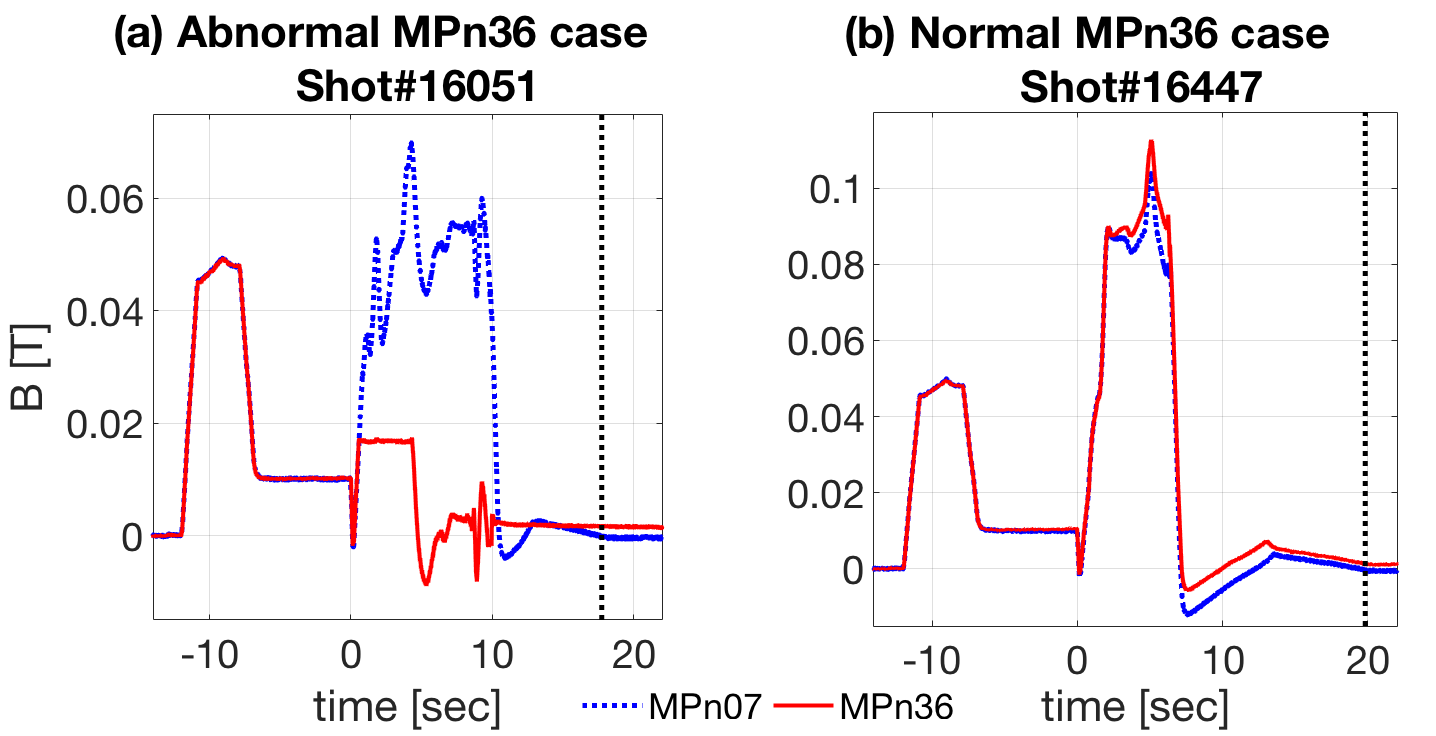}}
    \caption{Temporal evolutions of the magnetic signals measured by MP$_n$ \#07 (blue) and MP$_n$ \#36 (red) for (a) KSTAR shot \#16051 (abnormal MP$_n$ \#36) and (b) \#16447 (normal MP$_n$ \#36). These two magnetic sensors are located at the up-down symmetric positions as shown in Fig. \ref{MD-config}, and the discrepancy between MP$_n$ \#07 and \#36 in (a) are too large compared to (b) to be explained by the slight up-down asymmetry of the KSTAR plasmas. Vertical dotted lines indicate where all the currents through the PF coils are returned to zeros.} 
   \label{fault_Mpn-fig}
\end{figure}
\begin{figure}[t]
    \centerline{\includegraphics[width=0.64\textwidth]{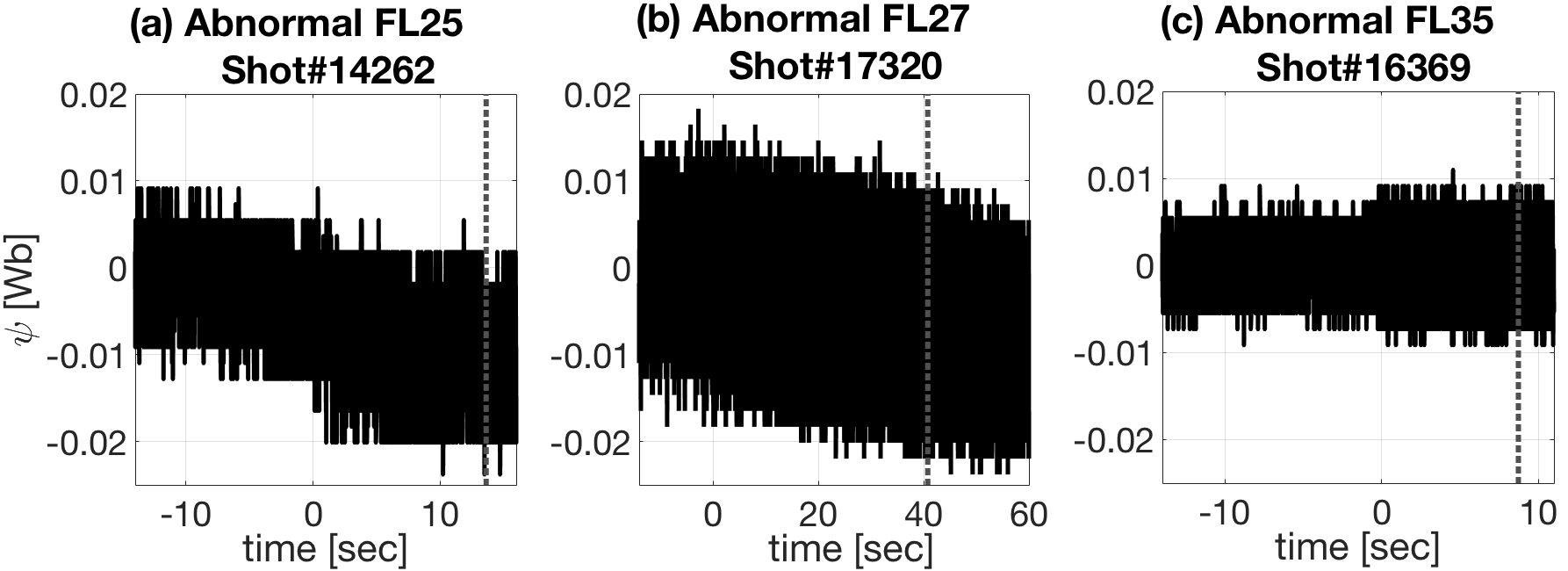}}
    \caption{Temporal evolutions of the magnetic signals measured by (a) FL \#25 (KSTAR shot \#14262), (b) FL \#27 (KSTAR shot \#17320) and (c) FL \#35 (KSTAR shot \#16369). These signals are basically noises (see Fig. \ref{Correc-fig}(c) as an example of working FL signal). Vertical dotted lines indicate where all the currents through the PF coils are returned to zeros.} 
   \label{fault_FL-fig}
\end{figure}

\noindent
As a measure of merit of the proposed two-step drift correction method, we define a validation error $\epsilon_{i, s}^m$ of a KSTAR shot number $s$ for the $i^\text{th}$ sensor of a type $m$ (MP$_n$, MP$_t$ or FL) as follow:
\begin{equation} \label{eq:linear-valid}
\epsilon_{i, s}^m = 100 \times \frac{\delta_{i, s}^m}{\max \left[\abs{f_{i, s}^m (t)}\right]_\text{flat-top}}\;[\%],
\end{equation}
\begin{equation} \label{eq:linear-valid_avg}
\lab\epsilon_{i}^m\rab =  \frac{1}{N} \sum_{s} \epsilon_{i, s}^m,
\end{equation}
where $f_{i, s}^m(t)$ is a magnetic signal of the KSTAR shot \#$s$, and a $\max\left[\abs{\cdot}\right]_\text{flat-top}$ operator selects the maximum absolute value of the argument during the flat-top phase. $\delta_{i, s}^m$ is the mean value of the $f_{i, s}^m (t)$ after all the currents of the KSTAR PF coils are returned to zeros, i.e., we expect $\delta_{i, s}^m$ to be zero if signal drifts are correctly removed (or if there were no signal drifts). $N$ is the total number of KSTAR shots we have used to estimate the average values of the validation errors.

The validation error defined in Eq. (\ref{eq:linear-valid}) quantifies how close $\delta_{i, s}^m$ is to zero relative to the maximum magnetic signal during a flat-top plasma operation. We normalize $\delta_{i, s}^m$ because its absolute value near zero is arbitrary, i.e., we cannot quantify how close to zero is close enough in absolute sense. Therefore, this validation error provides us quantitative measure of effectiveness of the proposed two-step drift correction method as well as goodness of the assumptions that signal drifts are linear in time, and the slopes and the offsets do not change significantly over one plasma discharge \footnote{If we have a large validation error, then we do not know whether the estimated slope and offset are inaccurate, or the assumptions are not valid. On the other hand, if we have a small validation error, then it is likely that the assumptions are valid, AND the slope and the offset are accurately estimated.}.

Fig. \ref{valid-fig2} shows histograms of the validation errors for (a) the normal (MP$_n$) and (b) the tangential (MP$_t$) components of magnetic signals, and for (c) the flux loop (FL) measurements; while left and right panels show before and after the two-step drift correction, respectively. MD~\# (horizontal axes) indicate the MD sensor numbers, i.e., subscript $i$ in $\epsilon_{i, s}^m$, and vertical axes are the validation errors. Colors representing the number of relative occurrence within a magnetic sensor are normalized to a unity for every sensor. We have randomly selected 297 KSTAR discharges from the 2013 KSTAR campaign to the 2017 campaign with a constraint that magnetic signals must exist after all the currents of the PF coils are returned to zeros \footnote{Existence of the data after all the currents of the PF coils are returned to zeros is necessary to estimate the validation error, but it is not required for real-time application of our two-step drift correction method.} so that we can estimate the validation errors. Non-existing magnetic signals are displayed as white streaks. It is evident that large validation errors are suppressed by our proposed method as the widths of the histograms are reduced to in the range of smaller values of the validation errors.

There are a few notable magnetic sensors indicated by red(MP$_n$ \#$36$, FL \#$25$, FL \#$27$ and FL \#$35$) and green(FL \#$01$, FL \#$23$ and FL \#$34$) arrows in Fig. \ref{valid-fig2}. As we discuss in more detail regarding these magnetic sensors in the discussion section (Sec. \ref{subsec:disc_abnormal}), we just briefly mention that red arrowed magnetic sensors correspond to a case where the two-step drift correction method does not work, i.e., large validation errors (large drift) before the correction is not improved by the proposed method, due to abnormal magnetic signals. Contrarily, we assert that our proposed method works well even on the sensors with large drifts as long as magnetic signals are not abnormal as indicated by green arrows. Note that there are many similar cases, i.e., large validation errors before the correction and small validation errors after the correction, for MP$_n$ and MP$_t$ signals as attested by the data in Fig. \ref{valid-fig2}.

Fig. \ref{pclog_val} shows the averaged validation errors $\lab\epsilon_{i}^m\rab$ for $N=297$ (same data sets used to generate Fig. \ref{valid-fig2}), showing that the validation errors are indeed reduced for MP$_n$ and MP$_t$ signals. Note that the drift corrected signal of MP$_t$ \#$20$ is worse than the value before the correction, but we argue that this is not so much a problem since the validation error is still less than the others. MP$_n$ \#$16$ is also worse after the correction, but the difference in the validation error is negligibly small. Our method is less effective for the FL measurements. However, large errors such as FL \#01, \#23, \#25, \#27, \#29 and \#34 are certainly reduced by our proposed method. 

Note that Figs. \ref{valid-fig2} and \ref{pclog_val}  summarize all the 297 KSTAR discharges whose pulse length (in terms of PF coil operation time) is less than $90$~sec. In Sec. \ref{sec:discussion}, we break down our results into short ($< 40$ sec) and long ($> 40$ sec) pulses, and discuss the appropriateness and limitations of our proposed method.

\subsubsection{Degree of Correction: Is the two-step drift correction method better than a typical linear fitting method?}\label{subsec:DoC}

\noindent
We now turn our attention to show how good our proposed method is compared to a typical chi-square linear fitting method. The slopes ($a_i^m$) and the offsets ($b_i^m$) in Eq. (\ref{eq:lineardrift}) are estimated simultaneously (rather than the two-step method as proposed) using the magnetic data from the time interval of d$2$ in Fig. \ref{PF-fig}. Since our method is proposed for a real-time control purpose, we must compare with the existing method that can be applied to a real-time control as well.

As qualitative comparisons, we show two cases in Fig. \ref{comp-linear}: (a) and (c) for KSTAR shot \#17016, and (b) and (d) for KSTAR shot \#9387. (a) and (b) show tangential component of magnetic signal MP$_t$ \#$14$; while (c) and (d) show temporal evolutions of currents through the KSTAR PF coils. Vertical dotted lines indicate the time where we expect all the magnetic signals return to zeros if there were no signal drifts. Fig. \ref{comp-linear}(a) shows a case where a typical chi-square linear fitting method (blue line) makes the error worse compared to the raw data (green line), i.e., before any correction, while our two-step method (red line) makes the error much smaller, i.e., closer to zero. Fig. \ref{comp-linear}(b) shows a case where a typical chi-square linear fitting method (blue line) works well bringing the magnetic signal closer to zero, but our proposed method (red line) is  even better.  

For more thorough and quantitative comparison, we define a degree of correction (DoC) as
\begin{equation}\label{eq:DoC}
\text{DoC\:[\%]} = \frac{\lsb\delta_{i, s}^m\rsb_\text{raw}-\lsb\delta_{i, s}^m\rsb_\text{corr}}{\lsb\delta_{i, s}^m\rsb_\text{raw}} \times 100,
\end{equation}
where $\lsb\delta_{i, s}^m\rsb$ has the same meaning as in Eq.(\ref{eq:linear-valid}). The subscript `raw' means before the correction, and `corr' stands for after the correction using either a typical chi-square linear fitting method (denoted as Linear Fit) or our proposed two-step drift correction method. If the DoC is 100\:\%, then the correction is perfect, i.e., $\lsb\delta_{i, s}^m\rsb_\text{corr}=0$ bringing the magnetic signal back to zero after the correction; while $0<\text{DoC\:[\%]}<100$ means that the applied method has corrected the signal in finite degrees. However, $\text{DoC\:[\%]}\le 0$ corresponds to a case where applied method makes no correction or even worse than a before-correction case.

\begin{figure}[t]
    \centerline{\includegraphics[width=0.6700\textwidth]{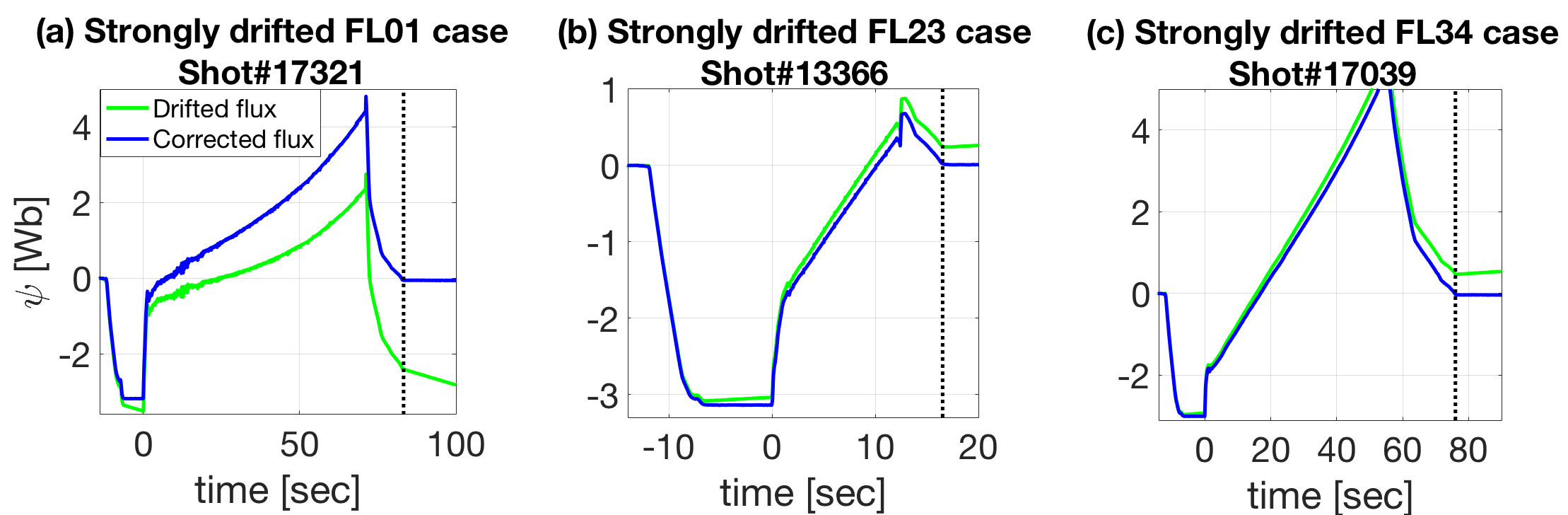}}
    \caption{Temporal evolutions of the magnetic signals measured by (a) FL \#01 (KSTAR shot \#17321), (b) FL \#23 (KSTAR shot \#13366) and (c) FL \#34 (KSTAR shot \#17039). Blue (green) line is the signal after (before) the correction. Vertical dotted lines indicate where all the currents through the PF coils are returned to zeros.} 
   \label{normal_FL-fig}
\end{figure}

Fig. \ref{comp-doc} shows histograms of the DoC's for (a) the normal (MP$_n$) and (b) the tangential (MP$_t$) components of magnetic signals, and for (c) the flux loop (FL) measurements; while the right (left) panels show the results based on our proposed two-step method (a linear fitting method). Here, same sets of magnetic signals as in Fig. \ref{valid-fig2} are used, i.e., 297 KSTAR discharges, and horizontal axes and color contours (normalized counts) represent same as in Fig. \ref{valid-fig2}. 

The calculated DoC's support that our proposed method, in general, brings the magnetic signals closer to zeros, i.e., the values of DoC's are typically within 0 to 100\:\%. In addition, it works better than a typical chi-square linear fitting method. Again, corrections on the FL signals are less effective (but still effective), which can be anticipated from the results of the validation errors shown in Fig. \ref{pclog_val}(c). 

As we do not claim that our proposed method is a perfect solution (rather we claim that it is an easy and better solution), we argue that our proposed method is a meaningful and valid solution based on the measures of the validation error and the degree of correction.

\subsection{Discussions} \label{sec:discussion}

\noindent
There exist at least two questions that need to be addressed. First, how good or applicable is the assumption of \textit{linear} model as in Eq. (\ref{eq:lineardrift})? Mathematically, this is a Taylor expansion of the \textit{true} drift signals neglecting higher order (nonlinear) terms. Therefore, it is evident that our proposed method will not work for long pulse discharges. This is also true even if we are able to obtain the second order term, i.e., nonlinear term forcing the drift model to be nonlinear, since it is still an approximation neglecting even higher order terms. Thus, a valid question we need to raise is: how long a discharge can be without a failure of our proposed method?

Another question is on the issue of abnormal magnetic signals which is briefly mentioned in Sec. \ref{subsec:val_err}. How do abnormal magnetic signals affect our proposed method? What good does our proposed method do for abnormal magnetic signals? 

\subsubsection{Short pulse vs. Long pulse} \label{subsec:disc_shot_long}

\noindent
We define a discharge pulse length based on the PF coil operation time in this work rather than a usual plasma operation time since we need to be able to estimate the validation error for quantitative judgement, and validation error can only be estimated after all the currents through the PF coils are returned to zeros. We have examined 286 short pulse ($< 40$ sec) discharges and 11 long pulse ($> 40$ sec) discharges, summing up to 297 KSTAR discharges from 2013 to 2017 campaigns. Notice that the number of long pulse discharges are much smaller than the short pulse ones because smaller number of long pulse experiments have been carried out, and we have randomly selected KSTAR discharges.

Fig. \ref{short_vs_long-fig} shows the averaged validation errors $\lab\epsilon_{i}^m\rab$ (defined in Eq. (\ref{eq:linear-valid_avg})) as in Fig. \ref{pclog_val} for (a) the normal component (MP$_n$) and (b) the tangential component (MP$_t$) of magnetic signals and for (c) flux loop measurements. Left and right panels show the results for the short and long pulse discharges, respectively. It is clear that our proposed method performs reasonably good corrections on the drift signals if a discharge pulse length is less than $40$ sec. We also have scrutinized the validation errors with a ten-second step, such as $0-10$ sec, $10-20$ sec, $20-30$ sec, etc., and we have confirmed that $40$ sec is an impartial and justifiable pulse length limitation for our proposed method to work properly. 

For the investigated long pulse discharges, the two-step drift correction method is modestly working for MP$_n$ and MP$_t$ signals, whereas the corrections on the flux loop measurements make the results worse. If we take a careful look on the results of the flux loop measurements in Fig. \ref{short_vs_long-fig}(c), we notice that the validation errors before the correction (red crosses) are smaller for the long pulse discharges compared to the short pulse discharges, while levels of the validation errors after the correction (blue circles) are similar for short and long pulse discharges. Although not conclusive, such results can be explained if a slope of the drift signal, i.e., $a_i^m$ in Eq. (\ref{eq:lineardrift}), changes its sign over a long pulse discharge, certainly a nonlinear effect.

Having discussed on the limitation of our proposed method due to a finite nonlinear effect, we enunciate that our two-step drift correction method do work properly (Sec. \ref{subsec:val_err}) and better than a typical linear fitting method (Sec. \ref{subsec:DoC}) at least for the pulse length less than $40$ sec. Existing magnetic confinement devices such as tokamaks, stellarators, linear machines with non-superconducting magnetic coils, in general, would not suffer from such a limitation as pulse lengths tend to be shorter. Therefore, our proposed method can readily be used for those existing devices without modifying any hardware systems. 

For future machines such as ITER, DEMO or even fusion power plants, this limitation must be overcome for steady-state long pulse discharges. One possible numerical solution is based on the laws of physics. We can conceive correcting the drift signals using physics constrained Bayesian probability theory \cite{Sivia:2006} and Gaussian processes \cite{Rasmussen:2006} since the measured magnetic signals must conform with Amp\`{e}re's law ($\nabla\times\vec B=\mu_0 \vec J$ ignoring the displacement current term as usual) and Gauss's law for magnetism ($\nabla\cdot\vec B = 0$) as has been done for imputation of faulty magnetic sensors \cite{Joung:2018ju}. Needless to say, the best solution will be based on the development of new kinds of hardwares. 

\subsubsection{Abnormal magnetic signals}\label{subsec:disc_abnormal}

\noindent
As mentioned before (Sec. \ref{subsec:val_err}) some of the magnetic signals are not corrected by our proposed method as indicated by the red arrows in Fig. \ref{valid-fig2}, i.e., MP$_n$ \#36, FL \#25, FL \#27 and FL \#35. This is due to malfunction of the magnetic sensors for the investigated KSTAR discharges. 

Fig. \ref{fault_Mpn-fig} shows the temporal evolutions of MP$_n$ \#7 and \#36, where these sensors are located at the up-down symmetric positions as shown in Fig. \ref{MD-config}. Slight up-down asymmetry of KSTAR plasmas cannot explain such large discrepancy in these two signals shown in Fig. \ref{fault_Mpn-fig}(a). If the sensor were working properly, we would expect that these two magnetic sensors output similar temporal behaviour as shown in Fig. \ref{fault_Mpn-fig}(b) which is a case with normal MP$_n$ \#36 signal. Therefore, we conclude that abnormal MP$_n$ \#36 signal has contributed such a large validation error even after applying our proposed method to the signal.

Fig. \ref{fault_FL-fig} shows examples of FL \#25, FL \#27 and FL \#35 signals selected from three different KSTAR discharges to substantiate that these sensors are not working properly. They just show features of random noises compared to a proper flux loop measurement as shown in the right panel of Fig. \ref{Correc-fig}(c).  Notice the scale difference of y-axes in Figs. \ref{Correc-fig} and \ref{fault_FL-fig}. Again, we conclude that abnormal FL \#25, FL \#27 and FL \#35 signals have resulted in large validation errors even after applying our proposed method to the signals.

Contrarily, we have many cases for MP$_n$, MP$_t$ and FL signals where the large validation errors before the corrections become noticeably smaller after the corrections. Such examples are indicated by the green arrows in Fig. \ref{valid-fig2}(c), i.e., FL \#01, FL \#23 and FL \#34. This means that large drifts are well corrected by our proposed method as shown in Fig. \ref{normal_FL-fig}.

With these observations that abnormal signals have the large validation errors both before and after the correction while normal signals have the small validation errors after the correction even if the validation errors are large before the correction, we argue that the estimated average validation errors can be used to detect flawed magnetic sensors automatically without scrutinizing hundreds of magnetic signals.

\subsection{Conclusions} \label{sec:conclusions}

\noindent
Magnetic measurements with many kinds of magnetic probes and flux loops are indispensable for preparing, operating and analyzing magnetically confined plasmas. Yet, they suffer from the drifts in many cases, and many engineers and scientists are required to provide non-trivial efforts to correct the obtained signals.

We have proposed the two-step drift correction method which resolves the drift problem in real-time. The method is based on Bayesian probability theory and obtains necessary information to correct the drifts before each plasma discharge initiates. This means that we can correct the drifts in real-time and provide more accurate information for real-time control of plasmas such as for real-time EFIT reconstruction. Our method is capable of correcting the drifts within of the order of 1\% average validation errors at least for the pulse length (in terms of PF coil operation time) less than $40$ sec. Furthermore, the average validation errors can be used to automatically detect defected magnetic sensors without going through hundreds of magnetic signals one-by-one to find such flawed ones.

Many real-time applications are developed or proposed based on neural networks these days. If one attempts to utilize neural networks with magnetic signals as inputs to the networks, then our method can also be heavily used for such applications.

\newpage
\section{Article \RNum{2}: Imputation}

\noindent
This approach deals with the imputation of faulty magnetic sensors with coupled Bayesian and Gaussian processes to reconstruct the magnetic equilibrium in real time\footnote{\textsc{S. J{\small oung}, J. K{\small im}, S. K{\small wak}, K. P{\small ark}, S.H. H{\small ahn}, H.S. H{\small an}, H.S. K{\small im}, J.G. B{\small ak}, S.G. L{\small ee}} and \textsc{Y.{\small -c}. G{\small him}} \\
\textit{Review of Scientific Instruments}, Vol.89.10 (7$^{th}$ May 2018), \textsc{{\small DOI:}}\href{https://doi.org/10.1063/1.5038938}{10.1063/\linebreak1.5038938}}, which is largely taken from Ref \cite{Joung:2018ju}, as a part of \textit{preprocessing magnetic measurements via Bayesian inference and neural networks}.

This article describes a Bayesian modelling of a magnetic diagnostic system to infer one (or more) missing magnetic signals based on Maxwell's equations. This Bayesian model has been applied to normal and tangential magnetic pick-up coils at the Korea Superconducting Tokamak Advanced Research (KSTAR). These pick-up coils measure the poloidal magnetic field, respectively, normal and tangential to the vacuum vessel wall where the coils are installed. As the pick-up coils are subject to impairment during plasma operations, faulty plasma operations and incorrect data analyses can be caused by the missing magnetic fields. Thus, the normal pick-up coils are forward-modelled together by Gauss's law for magnetism, and Amp\`{e}re's law is used to build a forward model for the tangential pick-up coils. 

We divide the missing magnetic signals from the measured signals while forward-modelling the signals in order that the missing signals can be inferred consistently with the measured magnetic signals as long as they satisfy Maxwell's equation. These models reasonably work when the number of missing signals is only one although we obtain an infinite number of solutions from the \textit{maximum a posteriori} method if there are more than one unknown of the missing components. Thus, Gaussian processes assisted forward models are introduced to restrict the solution spaces by arbitrarily relating the missing and the measured signals with each other based on non-parametric Gaussian process regressions. Therefore, all of the signals can be represented as a function of one arbitrary missing signal, and after the selected missing signal is determined from the forward model, then multiple signals can be subsequently inferred with the Gaussian processes. This approach, therefore, can infer the missing fields even if they are more than one. Note that the results of Article \RNum{1} have been used here in order to preprocess the signal drift. 

\subsection{Introduction}

\noindent
Magnetic pick-up coils installed on magnetic confinement devices such as tokamaks and stellarators in addition to Rogowski and flux loop coils provide magnetic information such that high temperature fusion-grade plasmas can be controlled in real time and that magnetic equilibria can be reconstructed for data analyses. Neural networks also have been developed to provide the positions of X-point and plasma boundary in real time\cite{Lister:1991gx, Coccorese:1994jt} where input signals to the networks are magnetic signals. Therefore, integrity of the magnetic signals is of paramount importance. As magnetic probes are subject to impairment during plasma operations, faulty plasma operations and incorrect data analyses can be caused by missing magnetic signals. For the case of neural networks trained with full sets of magnetic signals, even a single missing signal may cause the networks not to work properly.\cite{vanLint:2005cn}

\begin{figure}[t]
    \centering
    \includegraphics[width=0.699\linewidth]{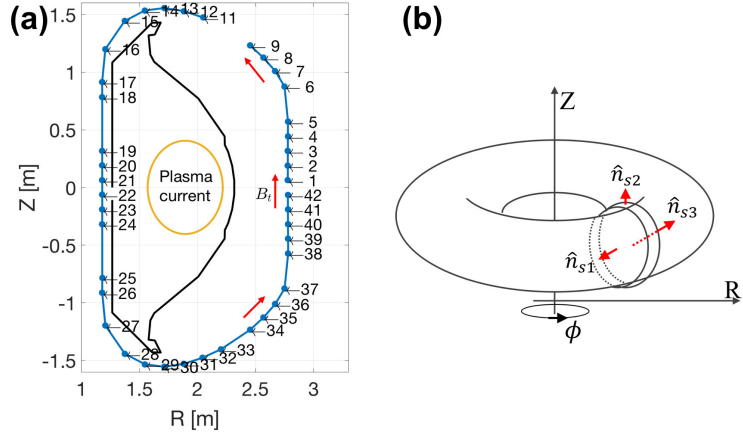}
    \caption{Schematics of (a) the Amperian loop (blue line connecting blue dots) for $\grad\times\vec B = \mu_0\vec J$ and (b) the pancake-shaped Gaussian surface with three surfaces $s_1$, $s_2$ and $s_3$ for $\grad\cdot\vec B = 0$. Blue dots with the numbers in (a) indicate the magnetic probes.\cite{Lee:2008cl}}
    \label{fig:MS-modelling}
\end{figure}

We present how one can numerically infer, thus impute, missing magnetic signals in real time based on a Bayes' model\cite{Sivia:2006} coupled with the Gaussian Process\cite{Rasmussen:2006} (GP). Likelihood is constructed using Gauss's law for magnetism and Amp\`{e}re's law and ensuring the consistency with the measured data. A couple of algorithms to detect faulty magnetic sensors in real time have been developed,\cite{Neto:2014cm, Nouailletas:2012ho} and an inference method for just one faulty signal has also been proposed.\cite{Nouailletas:2012ho} Our proposed method in this work is tested with up to nine non-consecutive missing magnetic probe signals installed on KSTAR.\cite{Kwon:2011dq} We find that the method infers the correct values in less than $1$ msec on a typical personal computer. Then, a full set of raw data, i.e., inferred ones together with measured ones, can be passed for real-time EFIT reconstruction\cite{Lao:1985hn, Ferron:2002fw} and neural networks. 

Detailed descriptions on how we generate the likelihood and estimate the maximum a posteriori of the Bayes' model and how well the model infers the missing values as well as its limitation are provided in section \ref{subsec:bayes}. The limitation on the Bayes' model motivates us to use the GP discussed in section \ref{subsec:GP} which also has a certain drawback. In section \ref{subsec:bayes_GP}, we present improved performance, i.e., resolving the defects of the Bayes' model and the GP while retaining their advantages, achieved by coupling the Bayes' model with the GP. To test our proposed method we assume that the intact magnetic signals are missing and compare the measured signals with the inferred values. Our conclusion is presented in section \ref{sec:conclusion:imp}.

\subsection{Imputation Scheme: Based on Bayes' model}
\label{subsec:bayes}

\begin{figure}[t]
    \centering
    \includegraphics[width=0.595\linewidth]{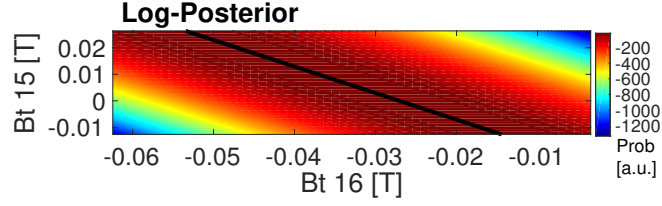}
    \caption{Log-posterior, $\ln[p(\boldB^*_\oplus | \boldB_\oplus, \Omega_\oplus)]$, of the missing magnetic signals inferred by the Bayes' model with the Maxwell's equations, when two tangential components ($B_t$ from MPs \#15 and \#16) of the magnetic signals are missing. Thick black line marks where the posterior is maximum indicating that infinite number of solutions are possible. Data are inferred for KSTAR shot \#9010 at $0.1$\:sec.}
    \label{fig:Inference-limit}
\end{figure}

\noindent
Magnetic probes,\cite{Lee:2008cl} depicted in Fig. \ref{fig:MS-modelling}(a) as the blue dots with the probe numbers, installed on KSTAR at a certain toroidal location measure tangential ($B_t$) and normal ($B_n$) components of the magnetic fields with respect to the wall. Missing tangential components are inferred using Amp\`{e}re's law with the measured plasma currents by Rogowski coils, i.e., $\grad\times\vec B=\mu_0\vec J$ neglecting $\partial\vec E/\partial t$ term based on a usual magnetohydrodynamic assumption,\cite{Freidberg:2014dt} and missing normal components using Gauss's law for magnetism, i.e., $\grad\cdot\vec B=0$.

With the Amperian loop, the blue line connecting the blue dots shown in Fig. \ref{fig:MS-modelling}(a), the tangential components of the magnetic signals $B_t$ must approximately satisfy
\begin{eqnarray}
\label{eq:ampere}
\mu_0 I_p &=& \oint_L \vec B \cdot d\vec{l} \approx \oint_L \left( B_t^\text{MP} - B_t^\text{PF} \right) \: dl \nonumber \\
&\approx& \left[ \sum_{m=1}^{N_m} \Delta l_{m}^* \left( B_{t, m}^{*\text{MP}} - B_{t, m}^{*\text{PF}} \right) + \right. \nonumber \\
&& \qquad\qquad\qquad\qquad \left. \sum_{i=1}^{N_i} \Delta l_i \left( B_{t, i}^\text{MP} - B_{t, i}^\text{PF} \right) \right] \nonumber \\
&=&\boldlambda^{*T}\left(\boldB_t^{*\text{MP}} - \boldB_t^{*\text{PF}} \right) + \nonumber \\
&& \qquad\qquad\qquad\qquad \boldlambda^T \left(\boldB_t^{\text{MP}} - \boldB_t^{\text{PF}} \right), 
\end{eqnarray}
where $I_p$ is the total plasma current assuming that the effect of transient eddy currents is negligible\cite{Tsaun:2007il} which, in general, is acceptable at least during a flat-top phase. $B_t^\text{MP}$ and $B_t^\text{PF}$ are the tangential components of the magnetic fields measured by the magnetic probes (MP) and induced by the poloidal field (PF) coils, respectively. Note that KSTAR has 14 PF coils, and their contributions are not perfectly canceled out due to change of integral form to a summation. Therefore, we remove the PF coil contributions. $m$ and $i$ are the indices for the missing and the intact magnetic signals; whereas $N_m$ and $N_i$ are the total numbers of the missing and the intact signals, respectively. $\Delta l$, an approximation of $dl$, denotes the segment distance between the magnetic probes, i.e., the distance between the consecutive probe numbers in Fig. \ref{fig:MS-modelling}(a). $\Delta l$ is different for different probes as can be seen in Fig. \ref{fig:MS-modelling}(a). Superscripted asterisk means the missing magnetic signal. The last line in Eq. (\ref{eq:ampere}) is just a reformulation of the second line using the vector notations, i.e., $\boldlambda^{(*)}=\{\Delta l^{(*)}_{i(m)}\}$ and $\boldB_t^{(*)}=\{B_{t,i(m)}^{(*)}\}$. Moret \textit{et al.}\cite{Moret:1998fh} has used Eq. (\ref{eq:ampere}) to obtain plasma currents in TCV tokamak; whereas we apply the same idea to obtain the missing magnetic signals based on the plasma currents measured by Rogowski coils.

For the normal components of the magnetic signals, we utilize the pancake-shaped Gaussian surface as depicted in Fig. \ref{fig:MS-modelling}(b) consisting of three surfaces $s_1$, $s_2$ and $s_3$. We force the Gaussian surface to be flat enough, so that the magnetic fluxes through the surfaces of $\hat n_{s_1}$ and $\hat n_{s_3}$ cancel each other as $\hat n_{s_1}\cdot\hat n_{s_3}=-1$, where $\hat n$ is a unit normal vector. Then, $\grad\cdot\vec B=0$ can be written as
\begin{equation}
\label{eq:gauss_1}
0=\oint_{s_1+s_2+s_3} \!\!\!\!\!\!\!\!\!\!\!\!\!\!\!\! \vec B \cdot d\vec S \approx \int_{s_2} B_n\:dA\approx\Delta w \int_L B_n\: dl,
\end{equation}
where $dA$ ($= \Delta w\:dl$) is the differential area normal to the surface $s_2$ (parallel to $B_n$) with $\Delta w$ being the thickness of the Gaussian surface. $dl$ is the differential length encompassing the minor radius (or the poloidal cross-section) and essentially same as the blue line in Fig. \ref{fig:MS-modelling}(a). Since $\Delta w\ne 0$, we have, again with the vector notations, 
\begin{eqnarray}
\label{eq:gauss}
0&=&\int_L B_n\: dl\approx\boldlambda^{*T}\boldB_n^{*}  + \boldlambda^T \boldB_n.
\end{eqnarray}

Assuming that the noise in magnetic signals is Gaussian, the likelihood is
\begin{eqnarray}
\label{eq:likebi}
p( \boldB_\oplus | \boldB_\oplus^*, \Omega_\oplus) &=& \frac{1}{\sqrt{2 \pi} \sigma} \times \nonumber \\
&&\!\!\!\!\!\!\!\!\!\!\!\!\!\!\!\!\!\!\!\!\!\!\!\!\!\! \exp \left[-\frac{\left(\boldlambda^{*T} \boldB_\oplus^* - \left(\Omega_\oplus - \boldlambda^{T} \boldB_\oplus \right) \right)^2}{2 \sigma^2} \right],
\end{eqnarray}
where $\boldB_\oplus^{(*)}$ is either $\boldB_t^{(*)\text{MP}}-\boldB_t^{(*)\text{PF}}$ or $\boldB_n^{(*)}$ depending on whether we are interested in the tangential or normal component, respectively. Likewise, the value of $\Omega_\oplus$ is $\mu_0 I_p$ for the tangential component or simply $0$ for the normal component. $\sigma$ is the noise standard deviation based on the measured magnetic signals with the uncertainty propagation, and measured to be $\mathcal{O}(10^{-4})$. 

Finally, we obtain posterior as
\begin{equation} \label{eq:msbayeinf}
p(\boldB^*_\oplus | \boldB_\oplus, \Omega_\oplus) \propto p(\boldB_\oplus | \boldB_\oplus^* , \Omega_\oplus)\:p(\boldB^*_\oplus | \Omega_\oplus),
\end{equation}
providing us inferred values of the missing magnetic signals ($\boldB_\oplus^*$) consistent with the measured signals ($\boldB_\oplus$ and $\sigma$) and the Maxwell's equations ($\Omega_\oplus$) assuming that $p(\Omega_\oplus)=1$. With a uniform prior $p(\boldB^*_\oplus|\Omega_\oplus)$, it is obvious that we obtain infinite number of solutions from \textit{maximum a posteriori} (MAP) method if we have more than one unknown of the same component. In simpler words, we have only one equation for the tangential (Amp\`ere's law) or the normal (Gauss's law for magnetism) component; thus, more than one unknown of the same component results in infinite number of solutions. Fig. \ref{fig:Inference-limit} shows an estimated log-posterior distribution, $\ln[p(\boldB^*_\oplus | \boldB_\oplus, \Omega_\oplus)]$, where we have removed two $B_t$ measurements, i.e., probe numbers \#15 and \#16, and confirms this effect clearly as depicted by the thick black line corresponding to the MAPs. This is the limitation of the imputation scheme solely based on the Bayes' model consistent with the Maxwell's equations.

\subsection{Based on Gaussian Process}
\label{subsec:GP}

\begin{figure}[t]
    \centering
    \includegraphics[width=.85\linewidth]{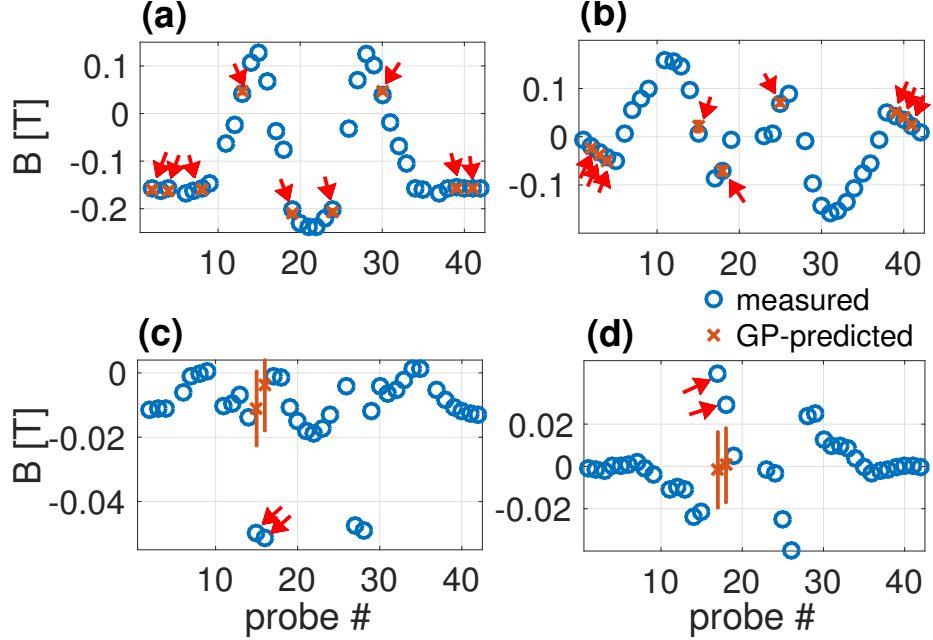}
    \caption{Successful GP predictions (red crosses) compared with the actual data (blue circles) for (a) $B_t$ and (b) $B_n$ at $3.70$\:sec of KSTAR shot \#9010 where we remove nine non-consecutive signals (indicated by red arrows) simultaneously to examine the proposed GP imputation scheme. On the other hand, if the magnetic signals are spatially varying fast such as (c) $B_t$ of MPs \#15 and \#16 and (d) $B_n$ of MPs \#17 and \#18 at $0.10$\:sec of the same shot, the GP imputation scheme fails to infer the correct values.}
    \label{fig:MS-gp}
\end{figure}

\noindent
Motivated by the limitation of the Baye's model with the Maxwell's equations, we introduce Gaussian Process\cite{Rasmussen:2006} (GP) in our imputation scheme. We express the probability distribution of $\boldB^*$ ($N_m \times 1$ column vector) given the measured data $\boldB$ ($N_i \times 1$ column vector) without any analytic expression of the data \textit{a priori} as described elsewhere\cite{Rasmussen:2006, Mises:1964}
\begin{equation}
\label{eq:gp}
p\left(\boldB^* | \boldB \right) = \mathcal{N}\left( \matK^*\matK^{-1}\boldB,\; \matK^{**}-\matK^*\matK^{-1}\matK^{*T}\right),
\end{equation}
with
\begin{eqnarray}
\label{eq:gpcond}
&& \quad \matK \equiv \matK\left(\boldX, \boldX \right)+\sigma_n^2\matI,  \quad (N_i\times N_i\text{ matrix}) \nonumber \\
&&\quad \matK^* \equiv \matK\left(\boldX^*, \boldX \right), \quad\quad\quad (N_m\times N_i\text{ matrix}) \nonumber \\
 &&\quad\matK^{**} \equiv \matK\left(\boldX^*, \boldX^* \right), \quad\quad (N_m\times N_m\text{ matrix}), \nonumber
\end{eqnarray}
where $\mathcal{N}(\:,\:)$ is the usual notation for a normal distribution, and $\matI$ the identity matrix. $\sigma_n^2\sim\mathcal{O}(10^{-4})$ is determined by treating it as a hyperparameter for the numerical stability during matrix inversion.\cite{Kwak:2017gy, Foster:2009rr, Rasmussen:2006} Recall that $N_i (N_m)$ is the total number of intact (missing) magnetic signals. Here, $\boldX^{(*)}$ is the $2\times N_i (N_m)$ matrix containing the physical positions of all the intact (missing) magnetic probes in two dimensional space, i.e., physical $R$ and $Z$ positions at a fixed toroidal location. 

The $i^{\text{th}}$ and $j^{\text{th}}$ component of a covariance matrix $\matK^{(*\text{ or }**) }$ is defined as
\begin{eqnarray}
&& K_{ij}^{(*\text{ or }**)}\left(\boldx_i^{(*)}, \boldx_j^{(*)}\right) = \nonumber \\
&& \sigma_f^2 \exp\left[-\frac{1}{2} \left(\boldx_i^{(*)}-\boldx_j^{(*)}\right)^T  \begin{bmatrix} \ell_{R}^2 & 0 \\0& \ell_{Z}^2 \end{bmatrix}^{-1} \left(\boldx_i^{(*)}-\boldx_j^{(*)}\right)  \right], \nonumber \\
\end{eqnarray}
where $\boldx_i^{(*)}$ is the $i^{\text{th}}$ column vector of the $\boldX^{(*)}$, i.e., $2\times1$ column vector containing the physical positions of the $i^{\text{th}}$ magnetic probe in $R$ and $Z$ coordinate. Hyperparameters $\sigma_f^2$, $\ell_R$ and $\ell_Z$ are the signal variance and the length scales in $R$ and $Z$ directions, respectively. These hyperparameters govern the characteristic of the Gaussian process, i.e., Eq. (\ref{eq:gp}), and we select the hyperparameters such that the evidence $p(\boldB)$ is maximized\cite{Kwak:2016kv} with an assumption\cite{Li:2013apa} of $\ell_R=\ell_Z$ for simplicity. As searching for the hyperparameters may become time consuming, thus not applicable for real-time control, one can obtain these values beforehand using many existing plasma discharges as for the case of density reconstruction.\cite{Kwak:2017gy} Once we have values for the hyperparameters, i.e., $\sigma_f\sim\mathcal{O}(10^{-2})$ and $\ell_R=\ell_Z\sim\mathcal{O}(10^{-1})$ in this study, we use Eq. (\ref{eq:gp}) to obtain the values of the missing magnetic signals $\boldB^*$, i.e., $\boldB^*=\matK^*\matK^{-1}\boldB$.

Fig. \ref{fig:MS-gp}(a) and (b) show that our proposed GP imputation scheme successfully infers the missing magnetic signals both for (a) $B_t$ and (b) $B_n$ where the red crosses are the inferred values and the blue circles are the measured (actual) values. We have examined our scheme with up to nine non-consecutive missing signals indicated by the red arrows. 

We have also found that the GP imputation scheme fails to infer the correct values if the magnetic signals are varying fast in space as shown in Fig. \ref{fig:MS-gp}(c) for $B_t$ and (d) for $B_n$. This is the limitation of the GP-only imputation scheme.

\subsection{Based on Bayes' model coupled with Gaussian Process}
\label{subsec:bayes_GP}

\begin{figure}[t]
    \centering
    \includegraphics[width=0.7485\textwidth]{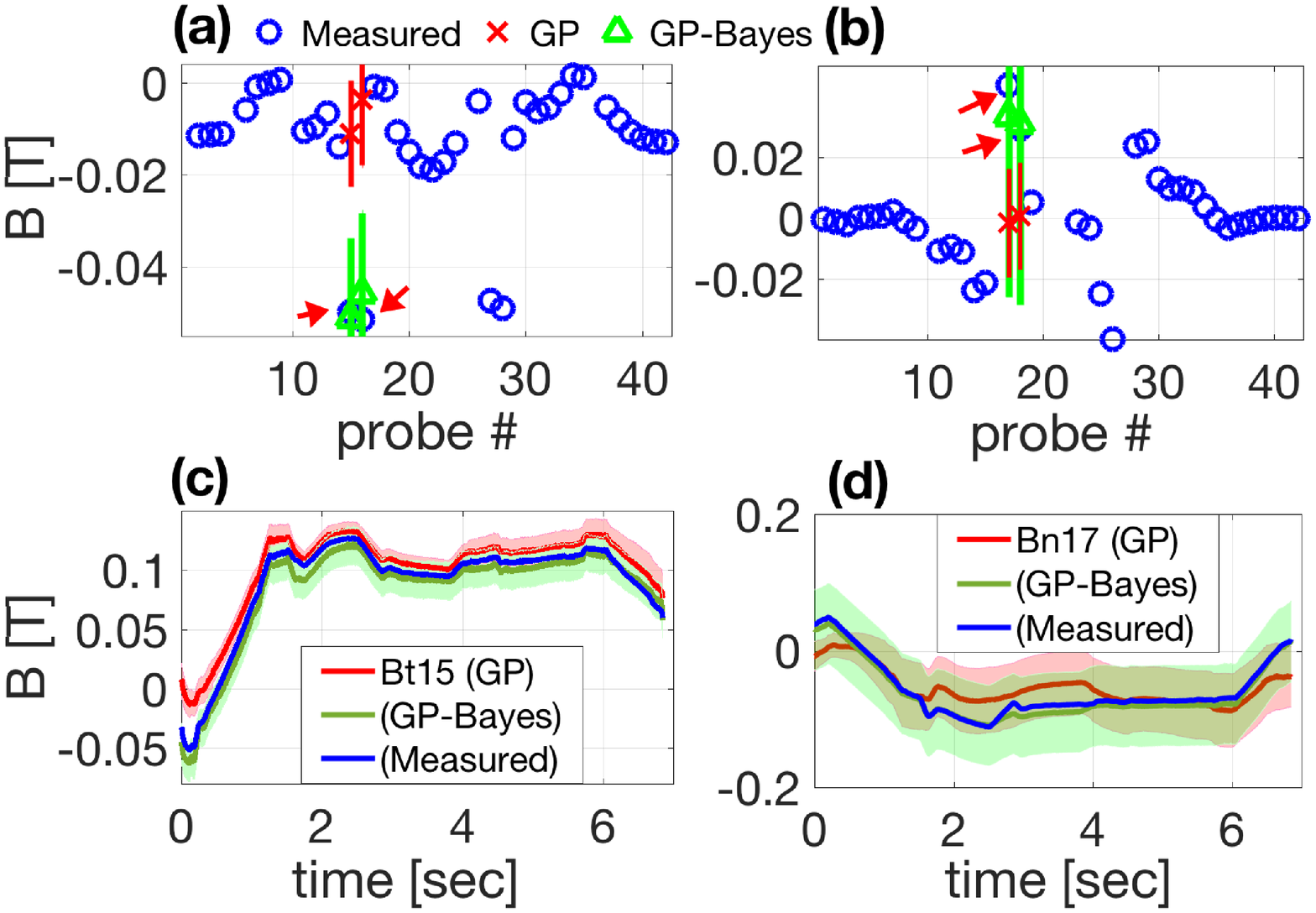}
    \caption{(a) $B_t$ from MPs \#15 and \#16 and (b) $B_n$ from MPs \#17 and \#18 from KSTAR shot \#9010 at $0.1$\:sec as shown in Fig. \ref{fig:MS-gp}(c) and (d). Green triangles obtained by the Bayes' mode with the GP match the measured values (blue circles) well, while the GP-only method (red crosses) fails to do so as has been discussed in Sec. \ref{subsec:GP}.  Comparisons of temporal evolutions for (c) $B_t$ from MP \#15 and (d) $B_n$ from MP \#17 from KSTAR shot \#9427 where blue line is the measured values, red line for the GP-only and green line for the Baye's model with the GP. Green lines agree well with blue lines well throughout the whole discharge including ramp-up and ramp-down phases.} 
    \label{fig:MS-gpwithbayes}
\end{figure}

\noindent
As we find the limitations of the Bayes' model (infinite number of solutions for more than one missing magnetic signal) and the GP (incorrect inference for spatially fast-varying missing magnetic signals), we resolve such weaknesses by combining the two schemes: for instance, if we have seven missing signals, we select one missing signal among the seven. Then, we use the GP to infer the non-selected six missing ones based on the intact signals together with the selected missing one which is inferred based on the Bayes' model.

Let us denote the selected missing magnetic signal as $B_k^*$, and define $\check\boldX^*$ to contain the positions of $R$ and $Z$ for all the missing magnetic signals except the ones corresponding to $B_k^*$ resulting in $2\times (N_m -1)$ matrix; while $\check\boldX$ containing those of $B_k^*$ in addition to intact magnetic signals becoming $2\times (N_i +1)$ matrix, i.e., concatenate those of $B_k^*$ at the last column of $\boldX$. With $\check\boldX^*$ and $\check\boldX$ our covariance matrices become
\begin{eqnarray}
\label{eq:gpcond_bayes}
&& \check\matK \equiv \check\matK\left(\check\boldX, \check\boldX \right)+\sigma_n^2\matI, ( (N_i+1)\times (N_i+1)\text{ matrix} ) \nonumber \\
&& \check\matK^* \equiv \check\matK\left(\check\boldX^*, \check\boldX \right), \quad\quad ( (N_m-1)\times (N_i+1)\text{ matrix} ) \nonumber \\
 && \check\matK^{**} \equiv \check\matK\left(\check\boldX^*, \check\boldX^* \right), \quad ( (N_m-1)\times (N_m-1)\text{ matrix} ). \nonumber
\end{eqnarray}
We separate out the last column of the $(N_m-1)\times (N_i+1)$ matrix of $\check\matK^*\check\matK^{-1}$ containing $B_k^*$ information and denote this column vector as $\boldL$ and the rest of the matrix, i.e., without the last column of $\check\matK^*\check\matK^{-1}$, be $\matLambda$. Since we have found that $\boldB^*=\matK^*\matK^{-1}\boldB$ in Sec. \ref{subsec:GP}, we obtain
\begin{equation}
\label{eq:gp_reduction}
\check\boldB^*_\oplus = \matLambda\boldB_\oplus + \boldL B_{k\oplus}^*,
\end{equation}
stating that once $B_k^*$ is determined, then all the other missing magnetic signals $\check\boldB^*$ are determined by the GP. We find the unknown $B_k^*$ using the Bayes' model where it is perfectly applicable since we have only one missing signal as discussed in Sec. \ref{subsec:bayes}. Thus, $\boldlambda^{*T}\boldB_\oplus^*$ in Eq. (\ref{eq:likebi}) is
\begin{eqnarray}
\boldlambda^{*T}\boldB_\oplus^* &=& \lambda_k^*B_{k\oplus}^* + \check\boldlambda^{*T}\check\boldB_\oplus^* \nonumber \\ 
&=&\left(\lambda_k^* + \check\boldlambda^{*T}\boldL \right) B_{k\oplus}^* + \check\boldlambda^{*T}\matLambda\boldB_\oplus,
\end{eqnarray}
where $\lambda_k^*$ and $\check\boldlambda^*$ are the segment distances for the selected missing one $B_{k\oplus}^*$ and for the rest of the missing signals, respectively.

Slightly modifying Eq. (\ref{eq:likebi}) to include the GP scheme, our likelihood for the Bayes' model, then, becomes
\begin{eqnarray}
\label{eq:likebi_gp}
p( \boldB_\oplus | B_{k\oplus}^*, \Omega_\oplus) &=& \frac{1}{\sqrt{2 \pi} \sigma} \exp \left[ \frac{1}{2\sigma^2} \left(  \left(\lambda_k^* + \check\boldlambda^{*T}\boldL \right) B_{k\oplus}^*  \right.\right.  \nonumber \\
& - & \left.\left. \left( \Omega_\oplus - \boldlambda^T\boldB_\oplus -  \check\boldlambda^{*T}\matLambda\boldB_\oplus  \right) \right) ^2 \right].
\end{eqnarray}
The likelihood now contains only one unknown $B_k^*$, and all the rest of the missing signals are treated as known ones using the GP, i.e,. Eq. (\ref{eq:gp_reduction}). 

We construct the prior $p(B_{k\oplus}^*|\Omega_\oplus)$ to follow a Gaussian distribution with the mean of $B_{k\oplus}^\text{pair}$ and the variance of $\sigma^2_{\text{prior}}$. $B_{k\oplus}^\text{pair}$ is the signal of the magnetic probe from the up-down symmetric position of the missing signal $B_{k\oplus}^*$. MPs \#6 and \#37, MPs \#12 and \#31, and MPs \#19 and \#24 in Fig. \ref{fig:MS-modelling}(a) are examples. We use such a paired magnetic signal as a prior mean of the missing signal because KSTAR discharges are quite up-down symmetric, so that a typical correlation between the paired signals is about $0.9$. Regarding the prior variance $\sigma_\text{prior}^2$, to minimize possible biases we set it to be $500$ which means that the prior distribution is largely uniform since the actual values of the magnetic signals are not much larger than $0.1$\:T as shown in Fig. \ref{fig:MS-gp}.  

We finally obtain the posterior following Eq. (\ref{eq:msbayeinf}) as
\begin{eqnarray}
\label{eq:posterior_gp}
p(B_{k\oplus}^* | \boldB_\oplus, \Omega_\oplus) &\propto&   \nonumber \\
&&\!\!\!\!\!\!\!\!\!\!\!\!\!\!\!\!\!\!\!\!\!\!\!\!\!\!\!\!\!\!\!\!\!\!\!\!\!\!\!\!\!\! \exp \left[ - \frac{\left( B_{k\oplus}^* - B_{k\oplus}^\bigstar \right)^2 }{2\sigma_\text{GP}^2} - \frac{\left(B_{k\oplus}^* - B_{k\oplus}^\text{pair}  \right)^2}{2\sigma^2_{\text{prior}}}  \right],
\end{eqnarray}
where 
\begin{eqnarray}
B_{k\oplus}^\bigstar &=&\frac{ \Omega_\oplus - \boldlambda^T\boldB_\oplus -  \check\boldlambda^{*T}\matLambda\boldB_\oplus }{\lambda_k^* + \check\boldlambda^{*T}\boldL }   \nonumber \\
\sigma_\text{GP}^2&=&\left[\frac{\sigma} {\lambda_k^* + \check\boldlambda^{*T}\boldL }\right]^2. \nonumber
\end{eqnarray}
Thus, \textit{maximum a posteriori} (MAP) denoted as $B_{k\oplus}^\text{MAP}$ can be analytically estimated and is
\begin{equation}
\label{eq:bkmap}
B_{k\oplus}^\text{MAP}=\left[\frac{B_{k\oplus}^\bigstar}{\sigma_\text{GP}^2} + \frac{B_{k\oplus}^\text{pair}}{\sigma_\text{prior}^2} \right]\left[\frac{1}{\sigma_\text{GP}^2} + \frac{1}{\sigma_\text{prior}^2} \right]^{-1}
\end{equation}
with the posterior variance $\sigma^2_\text{post} = (1/\sigma_\text{GP}^2 + 1/\sigma_\text{prior}^2 )^{-1}$. Once $B_{k\oplus}^\text{MAP}$ is found, then all the other missing signals are determined by Eq. (\ref{eq:gp_reduction}). This completes the imputation process.

To validate our proposed imputation scheme based on the Bayes' model coupled with the GP, we take the same examples shown in Fig. \ref{fig:MS-gp}(c) and (d). Fig. \ref{fig:MS-gpwithbayes}(a) $B_t$ from MPs \#15 and \#16 and (b) $B_n$ from MPs \#17 and \#18 show considerable improvements where the green triangles inferred by the Bayes' model coupled with the GP are very close to the blue circles which are the measured values. Again, the red crosses obtained only by the GP fails to do so. 

Fig \ref{fig:MS-gpwithbayes}(c) $B_t$ from MP \#15 and (d) $B_n$ from MP \#17 from KSTAR shot \#9427 show temporal evolutions of the inferred values where the blue line is the measured values, the red line for the GP only and the green line for the Bayes' model with the GP. Typically, the GP-only method fails largely during ramp-up and ramp-down phases while it is not too bad during the flat-top phase; whereas the Bayes' model with the GP finds the correct values throughout the whole discharge.

Eq. (\ref{eq:bkmap}) contains no unknowns which means that $B_{k\oplus}^\text{MAP}$ can be estimated in real-time. In fact, our proposed method takes less than $1$\:msec on a typical personal computer. The hyperparameters are prepared beforehand based on many previous discharges, and missing or faulty signals can be identified\cite{Neto:2014cm, Nouailletas:2012ho} in real-time. What one requires to do is simply to perform the following three steps in real-time: (1) select a missing signal ($B_{k\oplus}^*$) among all the missing ones ($\boldB_\oplus^*$), (2) estimate noise levels ($\sigma$) of the measured signals and (3) apply Eq. (\ref{eq:bkmap}) and Eq. (\ref{eq:gp_reduction}) to impute more than one missing magnetic signals. Good choice of a missing signal ($B_{k\oplus}^*$) is from the ones that spatially vary fast if they exist. In KSTAR such signals are $B_t$ from MPs \#15 and \#16, and $B_n$ from MP \#17 and \#18 in almost all cases, if not all.

\subsection{Discussion and Conclusion}
\label{sec:conclusion:imp}

\noindent
We have developed and presented a real-time inference scheme, thus imputation scheme, for missing or faulty magnetic signals. Our method, Bayes' model with the likelihood constructed based on Gauss's law for magnetism and Amp\`ere's law, coupled with the Gaussian process, allows one to infer the correct values even if more than one missing signal that is spatially varying fast exists. The coupled method outperforms the Baye's-only and the GP-only methods without losing their own advantages. We have examined our method up to nine non-consecutive missing magnetic signals.

The proposed method takes less than $1$\:msec on a typical personal computer, so that the method can be applied to fusion-grade plasma operations where real-time reconstruction of magnetic equilibria is crucial. It can also be used for a neural network trained with a complete set of magnetic signals without fearing the possible loss of magnetic signals during plasma operations.

As a possible future work, developing a real-time searching algorithm for the hyperparameters in the Gaussian process that optimizes the evidence will be beneficial. Although results with the predetermined hyperparameters based on many previous discharges can be satisfying, the hyperparameters specific to a current discharge may provide much better plasma controls especially for those discharges that we have not yet explored much. In addition, including the effect of eddy currents can improve the performance of our method especially during the ramp-up and down phases and disruptions.

\newpage
\section{Article \RNum{3}: Preprocessing flux loop}

\noindent
This approach deals with A deep learning approach to recover hidden consistency of KSTAR flux loop signals\footnote{\textsc{S. J{\small oung}, J. K{\small im}, H.S. H{\small an}, J.G. B{\small ak}} and \textsc{Y.{\small -c}. G{\small him}} \\
\textit{Scientific Reports}, (2022), in preparation}, which is largely taken from Ref \cite{joung2022fl}, as a part of \textit{preprocessing magnetic measurements via Bayesian inference and neural networks}.

This article describes a deep neural network applied to the KSTAR poloidal flux loops to recover consistency between the measured flux signals. The poloidal magnetic signals are typically utilized to reconstruct a plasma equilibrium which is a state when a plasma is assumed to be in an ideal magnetohydrodynamic equilibrium state. The plasma equilibrium can be obtained by iteratively solving the Grad-Shafranov equation, together with the measured poloidal magnetic fields and fluxes: first estimate toroidal current density from the Grad-Shafranov equation by using assumed equilibrium; second, calculate magnetic signals from the current density, and then update the current by comparing the calculated signals with the measured signals based on the Grad-Shafranov equation; finally, update the equilibrium, and repeat these procedure until the calculated signals are close enough to the measured signals within a criterion. 

However, the reconstruction procedure is often abortive when all of the measured magnetic signals (the fields and the fluxes) are used simultaneously although impaired signals are expelled from the reconstruction. Especially, the intact signals measured from flux loops cannot be utilized at once since they yield unreasonable reconstructions, making humans pick only a few of them meticulously. The deep neural network is developed to compensate for the inconsistency between the flux loops and generate cleaned fluxes as well as the missing fluxes from its output. Thus, the network generated fluxes can be used for the reconstruction procedure at the same time, and the reconstruction results are quite reasonable compared to existing equilibrium databases reconstructed by humans with their careful decisions about selections of the magnetic signals. Note that Article \RNum{5} employs this approach in order to reconstruct plasma equilibria solely based on neural networks without depending on humans.

\subsection{Introduction}

\noindent
Plasma is an ionized gas which is a fuel of thermonuclear fusion\cite{cite-keyVerberk, Chen:2011wc}, a sustainable and clean energy source. Tokamak is a device that confines the plasma in the magnetic fields to help the fusion reactions continue. Thus, it is important to measure magnetic signals inside the tokamak generated from the magnet coils as well as the plasma. To this end, an induction coil-type diagnostics with the integrator\cite{Strait:2008} are widely used for measuring the poloidal magnetic field, the magnetic flux, and the plasma current in various magnetic confinement devices\cite{Sakakibar2010, Edlington:2001jg, Strait:2006kh, Moret:1998fh, Liu:2013dn, Peruzzo:2009et,Lee:2001cp, Lee:2008cl, Orlinskij:191982} including ITER, the International Thermonuclear Experimental Reactor\cite{Chavan:2009gk}, which is built to prove the possibility for constructing the fusion power plant.

Basically, magnetic diagnostics are installed on the vacuum vessel wall of the tokamak far away from the plasma since the plasma temperature can reach about 100 million degrees. Thus, the magnetic data solely contains indirect information about the internal properties of the plasma related to the plasma geometry or large-scale magnetohydrodynamic (MHD) activities\cite{Shiraki:2015jp, Snipes:1988cq, Gerasimov:2014dp, Buttery:2002im, Park:2013ex, Svensson:2008in, Lazerson:2015bb, Romero:2013eg, Qian:2017gj, Lao:1985hn}. Here we focus on real-time control of the plasma shape and position which are obtained from the plasma reconstruction by solving the Grad-Shafranov (GS) equation\cite{grad1958hydromagnetic, shafranov1966plasma} where the GS equation is derived based on the MHD equation in equilibrium state\cite{Freidberg:1987}. To this end, KSTAR\cite{lee2001design} has 84 magnetic probes and 45 flux loops (FLs) which measure the poloidal magnetic fields and fluxes, respectively\cite{Lee:2001cp, Lee:2008cl}. 

However, compared with the probes, there are inconsistencies between the magnetic fluxes hindering the plasma from being reconstructed reasonably, although we can compensate impaired probes\cite{Joung:2018ju} as well as signal drifts (signals tending to unintentionally increase or decrease over time) in magnetic measurements\cite{joung2019deep} through Bayesian inference\cite{Sivia:2006}. These inconsistencies give rise to the dependency on human expert knowledge which possibly causes biases to select the specific signals for the reconstruction process.

Thus, with tensorflow\cite{tensorflow2015-whitepaper}, we propose a method to recover the consistency based on the capability of deep neural networks which are able to learn differential equations by themselves. Recently, this capability was used for various research fields such as fluid dynamics\cite{worswick2018deep, bar2019learning, raissi2020hidden}, quantum mechanics\cite{pfau2020ab, mills2017deep, torlai2018neural, li2021kohn}, and plasma physics\cite{joung2022deep}. By means of differentiating neural networks analytically with respect to their inputs, we present that the network can produce the consistent magnetic fluxes based on the measured magnetic fields. Using the network outputs, we also prove that reconstructing plasma equilibrium at the level of the expert is plausible without relying on the expert knowledge.

\subsection{Magnetic data collection}

\noindent
We collect 701 shots among the KSTAR 2020 year campaign experiments whose discharge length is greater than or equal to 10 $sec$. In each shot, the magnetic signals from 200 $msec$ to the half of the whole discharge length are collected at intervals of $\sim$10 $msec$. Thus, we extract a total of 369,610 time slices for the magnetic probes, the flux loops, the plasma current, and the poloidal field coil currents. To train the neural network, we have a total of 13,675,570 $(=37\times369,610)$ magnetic fields normal to the wall, and 14,414,790 $(=39\times369,610)$ magnetic fields tangential to the wall, and we also have a total of 11,827,520 magnetic fluxes obtained from 32 FLs. 90$\%$ of these collected signals are used for the network training, 5$\%$ for the network validation, and the remaining 5$\%$ for the test dataset. 

\begin{figure}[t]
\centering
\includegraphics[width=0.88\columnwidth]{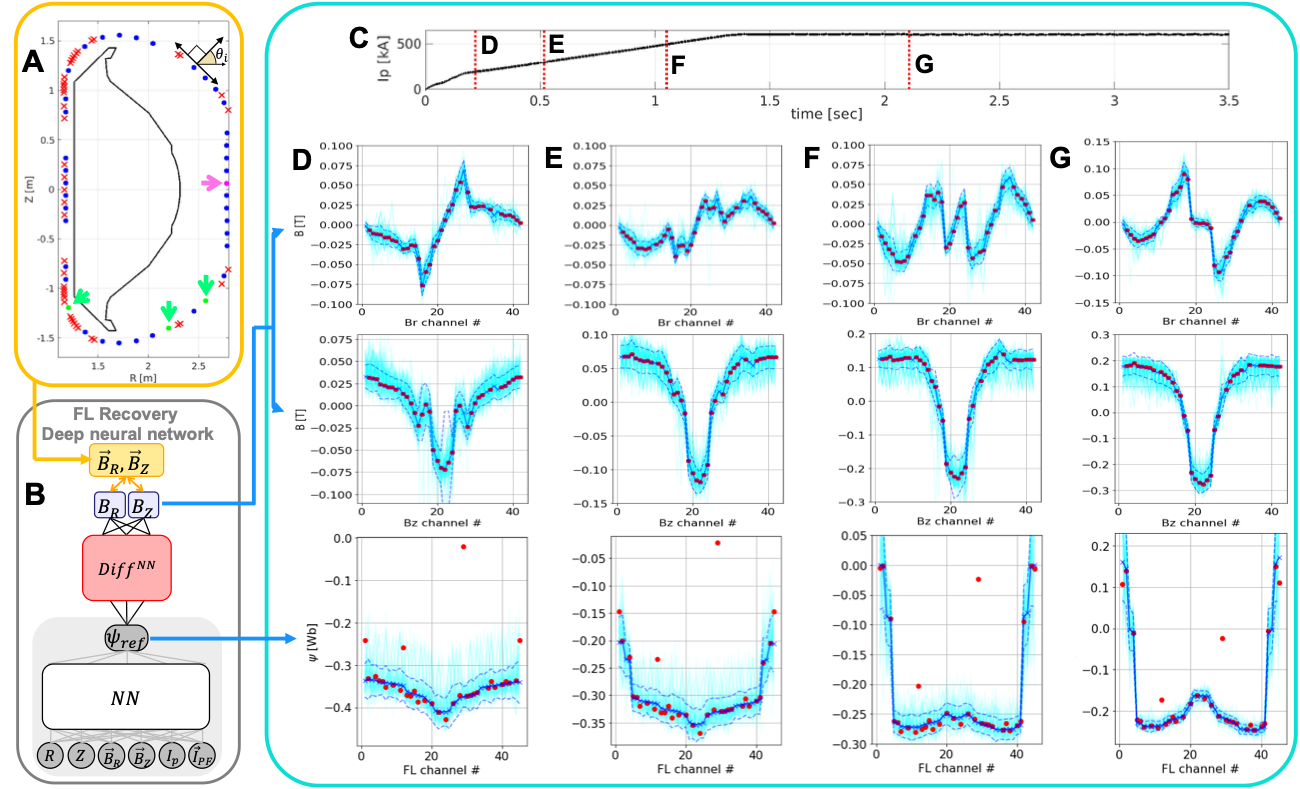}
\caption{Schematic diagram of FL recovery via a deep neural network. (A) The positions of magnetic measurements on the KSTAR poloidal cross-section where both $B_n$ and $B_t$ exist (blue dots), only $B_n$ exists (pink dot), only $B_t$ exists (green dots), and FLs exist (red crosses). (B) Schematics of the deep neural network whose output stands for the flux function converted into $B_R$ and $B_Z$ through analytic differentiation. (C) Temporal evolution of the plasma current for KSTAR 24445 shot. (D--G) The network results (blue area) at the red dotted lines in (C) are presented compared with the measured signals (red dots). First row is for $B_R$, second row is for $B_Z$, and the last row is for FL.}
\label{fig:1}
\end{figure}

\subsection{Domain knowledge regarding the poloidal magnetic field}

\noindent
The magnetic fields normal and tangential to the wall where the probes are installed can be converted into the $R$ and $Z$ components of the fields based on the angle (Fig. 1A, upper right) between the normal direction to the wall and the $R$ direction on the $R$--$Z$ plane. Thus, we compute $B_R$ and $B_Z$ as shown below.
\begin{equation}
\begin{split}
B_{R,i} = B_{t,i} cos(\theta_i) - B_{n,i} sin(\theta_i) \\
B_{Z,i} = B_{t,i} sin(\theta_i) + B_{n,i} cos(\theta_i)&
\end{split}
\end{equation}
where the subscript $i$ denotes the channel number of the probes, $B_t$ is the tangential magnetic field, and $B_n$ is the normal magnetic field. From the Gauss’s law for the magnetism, we can relate the $B_R$ and $B_Z$ with the poloidal flux function, $\psi$, based on the vector potential on the cylindrical coordinates, i.e.,
\begin{equation}
\begin{split}
B_{R,i} = -\frac{1}{R} \frac{d\psi}{dZ_i} \\
B_{Z,i} = \frac{1}{R} \frac{d\psi}{dR_i}&
\end{split}
\end{equation}
where the subscript $i$ denotes the channel number of the probes, $R_i$ and $Z_i$ are the position where the probe is on the $R$--$Z$ domain, and $\psi$ is the poloidal flux function which is equal to the poloidal flux measured from the flux loop divided by $2\pi$. 

\begin{figure}[t]
\centering
\includegraphics[width=0.78\columnwidth]{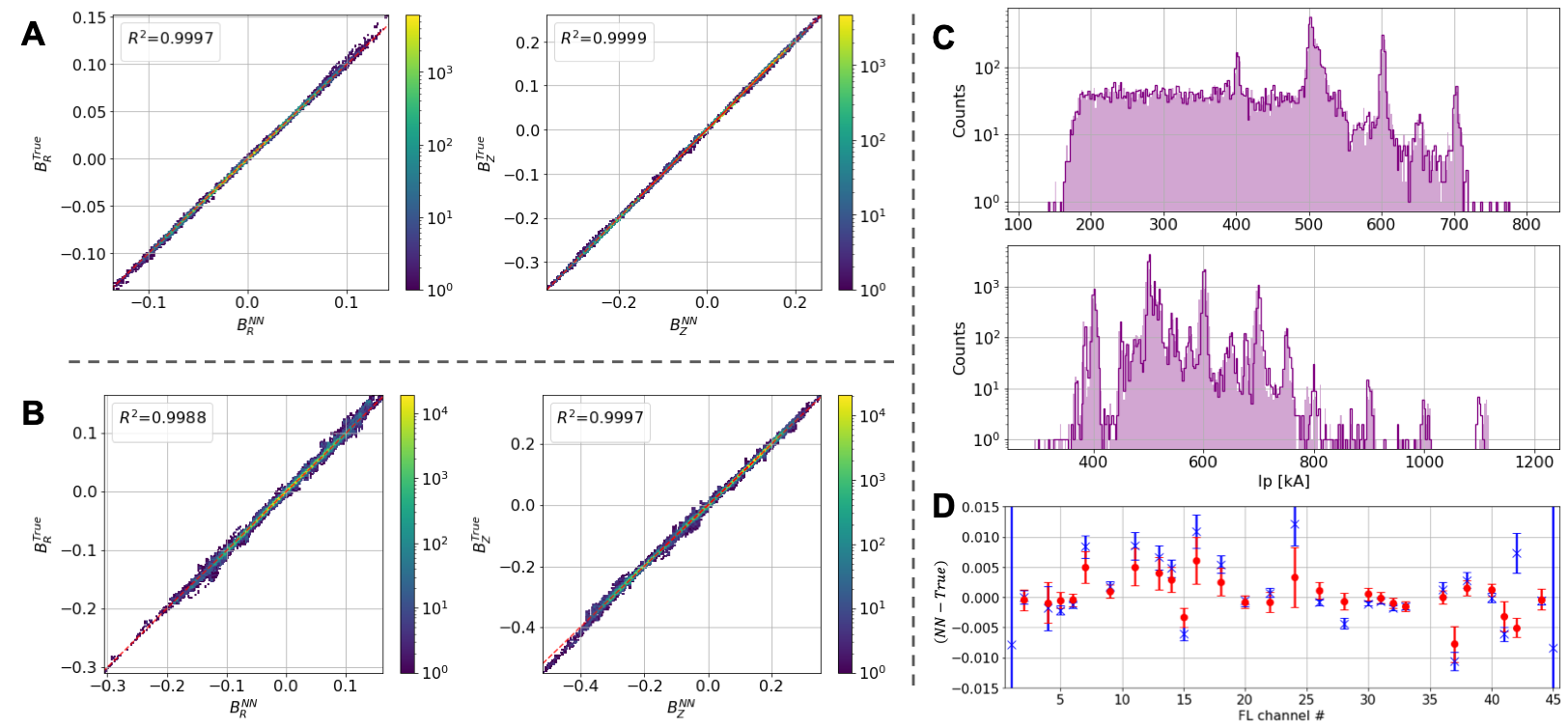}
\caption{Statistical analysis of the trained network with test dataset. (A) Statistics of the coefficient of determination between the measured and the network $B_R$ (left) and $B_Z$ (right) from the ramp-up phase, and (B) from the flat-top phase. (C) Distributions of the plasma current measured from the Rogowski coil (purple line) and the network (purple area) of the ramp-up phase (top) and the flat-top phase (bottom). (D) Using 24445 shot, Statistics on the difference between the measured $\psi$ and the network $\psi_{rel}^a$ of the ramp-up phase (blue) and the flat-top phase (red).}
\label{fig:2}
\end{figure}

\subsection{Modelling the network architecture}

\noindent
Based on the relationship between the magnetic field and the flux, we construct the network architecture whose output is the flux function turning out to be $B_R$ and $B_Z$ through the analytic differentiation of the network (Fig. \ref{fig:1}B). The network is a 4-layer fully-connected network with 100 hidden neurons and one bias per hidden layer. The network weights are randomly initialized\cite{pmlr-v9-glorot10a}, and the activation function is the swish function\cite{ramachandran2017searching}, i.e.,
\begin{equation}
swish(x) = x \times sigmoid(x) = \frac{x}{1-\exp{(-x)}}
\end{equation}
The cost function for training the network is $e = e_1 + e_2$ where $e_1$ is built with $B_R$ and $B_Z$ from the probe positions having both the tangential and the normal fields (Fig. \ref{fig:1}A, blue dots)
\begin{equation}
e_1 = \frac{1}{N} \sum_{i=1}^{N} \left(B_{R, i}^{NN} - B_{R, i}^{MD} \right)^2 + \frac{1}{N} \sum_{i=1}^{N} \left(B_{Z,i}^{NN} - B_{Z,i}^{MD} \right)^2
\end{equation}
where $N$ is 11,985,084 which is the total number of $B_R$ and $B_Z$ signals in the training set. On the other hand, $e_2$ is for the normal (Fig. \ref{fig:1}A, pink dot) or the tangential probes (Fig. \ref{fig:1}A, green dots) where there is no signals being paired, i.e.,
\begin{equation}
e_2 = \frac{1}{N_{2}} \sum_{i=1}^{N_{2}} \left( \left(\alpha_{t(n), i} B_{R,i} + \beta_{t(n), i} B_{Z,i}\right)^{NN} - B_{t(n), i}^{MD} \right)^2
\end{equation}
where the subscript $i$ is the no-pair channel number of the probes, $N_2$ is 1,330,596 which is the number of the no-pair signals in the training dataset, $\alpha_{t(n)}$ is $sin(\theta)$ (or $cos (\theta)$), and $\beta_{t(n)}$ is $cos(\theta)$ (or $-sin(\theta)$). The superscript $NN$ stands for the network results, and the superscript $MD$ refers to the measurements. Additionally, since the network output only learns the relative value for $\psi$, we process $\psi_{rel}$ to be the absolute value $\psi_{rel}^a$ as follows,
\begin{equation}
\psi_{rel, i}^a = \psi_{rel, i} - \frac{1}{45} \sum_{i=1}^{45} \psi_{ref, i} + \frac{1}{32} \sum_{i=1}^{32} \frac{\psi_{FL, i}}{2\pi} 
\end{equation}
where the subscript $i$ means the channel number of the flux loops, and $\psi_{FL}$ is the poloidal flux measured from the flux loops.

\begin{figure}[t]
\centering
\includegraphics[width=0.74\columnwidth]{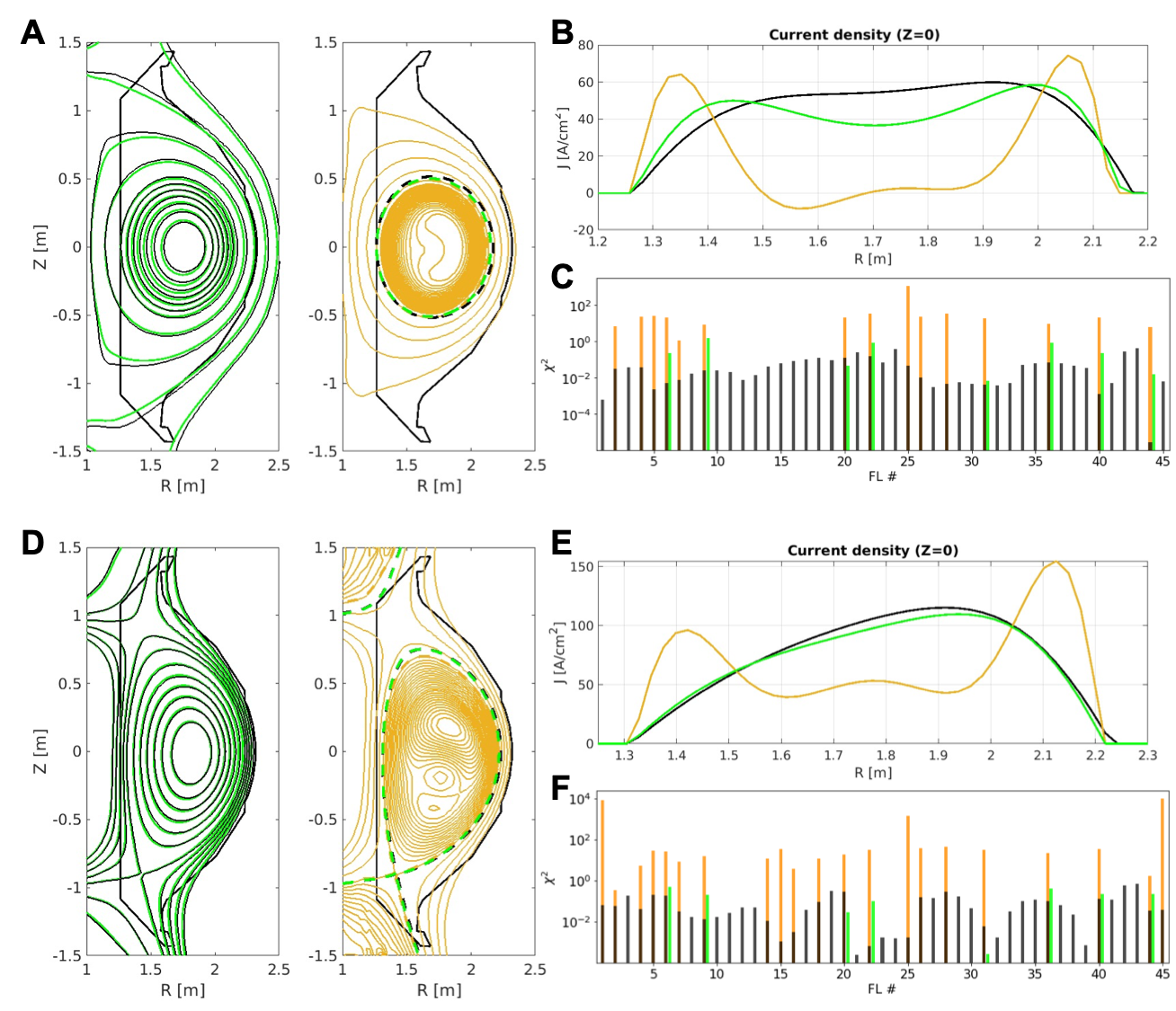}
\caption{Comparison of equilibrium reconstructions based on the network (black), the expert (green) and the novice (orange). (A) Left: comparison of the equilibrium results at 498.5 msec of 24445 shot. Right: a novice producing equilibrium overlaid with the plasma boundaries from the network and the expert (dotted lines). (B) Spatial profiles of the plasma current density at Z=0 location. (C) Chi-square results of the $\psi$. (D-F) Same results with (A-C) at 2704.3 msec.}
\label{fig:3}
\end{figure}

\subsection{Recovering flux loop consistency via the deep neural network}

\noindent
To restore the consistency between magnetic fluxes, we apply the method using the deep neural network in which the network output itself is learned by training the derivatives of it\cite{joung2022deep}. From the relationship between the magnetic fields and the magnetic fluxes based on Maxwell's equations, we can model the network architecture (Fig. \ref{fig:1}B) whose output is the poloidal magnetic flux function, $\psi$, which can be transformed into the poloidal magnetic fields in the $R$ and $Z$ direction, $B_R$ and $B_Z$, via the analytic differentiation of the network $\psi$, while the network is fed with the spatial positions $R$ and $Z$, the magnetic fields, the plasma current, and the poloidal field coil currents. Since the network $\psi_{rel}$ trained from its derivatives has no information about the absolute value, we turn $\psi_{rel}$ into $\psi_{rel}^a$ to have the information by using the measured FL signals (see Methods subsection ‘Domain knowledge regarding the poloidal magnetic field’). 

Using 36 $B_R$ and 36 $B_Z$ signals at the blue dots in Fig. \ref{fig:1}A which are computed based on the normal and tangential components of the poloidal magnetic fields with respect to the vessel wall where the probes are installed, we generate the network results qualitatively based on KSTAR 24445 shot. Typically, the plasma current which is the induced current in the plasma starts being discharged at 0 $sec$, reaches the flat-top phase where the current is approximately in equilibrium state after passing through the ramp-up phase where the current increases in time (Fig. \ref{fig:1}C, black line). During the discharge, we choose four different time slices (Fig. \ref{fig:1}C, red dotted lines) to generate the network results with its uncertainties\cite{gal2016dropout} (Fig. \ref{fig:1}D--G, blue areas) in comparison with the measured magnetic data (Fig. \ref{fig:1}D--G, red dots).

The first and second rows in Fig. \ref{fig:1}D--G shows the spatial profiles of $B_R$ and $B_Z$ where the x-axes stand for the probe channel numbers, while the figures in the last row represent the spatial profiles of $\psi$. We can find that the measured $\psi$ are sporadically scattered, being laid within the network uncertainties. We presume that this scatteredness is not due to the mechanical uncertainties of the FLs since the uncertainty scale of the FL is about $10^{-3}$. Thus, this makes us confirm that the scatteredness is elminated in the network results, meaning that the network can recover the inconsistency between the FL signals.

\subsection{Quantitative assessment of the network}

\noindent
So far, we have qualitatively demonstrated that the consistency between FLs is reasonably recovered through the deep neural network. Here we show that statistical approaches to evaluate the network quality by using the test dataset. First, we use the $R^2$ metric, the coefficient of determination\cite{rao1973linear}, for validating the network generality based on the training targets, i.e., $B_R$ and $B_Z$. We present that the network $B_R$ and $B_Z$ signals from the 36 probes each show a fairly linear relation with those of the measurements for the ramp-up phase (Fig. \ref{fig:2}A; Left: $R^2$ of $B_R$ = 0.9997, Right: $R^2$ of $B_Z$ = 0.9999). Similarly, the $R^2$ results for the flat-top phase also show considerable linearity (Fig. \ref{fig:2}B; Left: $R^2$ of $B_R$ = 0.9988, Right: $R^2$ of $B_Z$ = 0.9997), indicating that our network can fully imitate the test dataset.

To validate the network quality including the excluded $B_R$ and $B_Z$ signals out of the 42 probes each, we estimate the plasma current from the poloidal magnetic fields tangential to the vessel wall by computing the network $B_R$ and $B_Z$ along with the Ampere’s law\cite{Joung:2018ju}. Compared with the distributions of the measured plasma current at the ramp-up (top) and the flat-top phase (bottom) (Fig. \ref{fig:2}C, purple lines), the network quite well picks up the features of the plasma currents from various KSTAR discharges (Fig. \ref{fig:2}C, purple areas). In addition, we calculate the differences between the measured $\psi$ and the network $\psi_{rel}^a$ over KSTAR 24445 shot for the ramp-up (Fig. \ref{fig:2}D, blue bars) and the flat-top (Fig. \ref{fig:2}D, red bars) phases in order to indicate that the network can produce a similar level of the $\psi_{rel}^a$ with respect to the measurements over the discharge.

\subsection{Equilibrium reconstruction using the network magnetic flux}

\noindent
We now demonstrate that the use of the network can help the plasma be reconstructed at the level of the expert without depending on the expert decisions. Thus, we reconstruct the plasma equilibria using KSTAR 24445 shot based on the network results, the expert selection, and a novice’s trial. Note that, in Fig. \ref{fig:3}, all the black colors refer to the network results, while the green colors are for the expert, and the orange colors are related to the novice. To reconstruct the plasma, we use an algorithm called EFIT\cite{Lao:1985hn} which is a reconstruction code widely used in various tokamaks\cite{Park:2011go, Sabbagh:2001hh, Jinping:2009jn, Lao:2005kd, OBrien:1992gm}, where we use all 45 recovered $\psi$ signals for the network case, but, the expert only uses 8 selected FL signals. For mimicking the novice’s trial, we randomly choose 15 FL signals for the ramp-up phase, and 21 FL signals for the flat-top phase which account for half of the total number of the FL.

Overall, we can find that the network helps to pick up the flux surfaces (Fig. \ref{fig:3}A and D) and the current density profiles (Fig. \ref{fig:3}B and E) comparable to the expert’s reconstructions, while the novice creates unreasonable and significantly different reconstruction results although the plasma boundaries are similar with those of the network and the expert (Fig. \ref{fig:3}A and D, dotted lines). This indicates that the plasma reconstruction can be highly distorted by the reconstruction attempts of the unskilled, which possibly gives an adverse impact to the real-time control of the plasma.

Furthermore, as a result of $\chi^2 = ((\psi_{EFIT} - \psi_{used})/\sigma)^2$ (Fig. \ref{fig:3}C and F), we can clearly distinguish the reconstruction qualities such that the $\chi^2$ for the novice's trial is significantly higher than those of the expert as well as the network. It also would be noted that the total summation of the network $\chi^2$ for the ramp-up case (Fig. \ref{fig:3}C) is 2.78 which is lower than the expert’s case, i.e., the total sum of $\chi^2=3.77$ where only 8 signals participate in the reconstruction. This informs that the expert-level of the reconstruction can be achieved by even non-experts if the network results are utilized. Thus, we can expect that we can exclude the expert decision during the reconstruction so that a complete automation of the process potentially comes true.

\subsection{Discussion}

\noindent
We have developed a method that recovers the flux loop consistency by means of the network ability to be able to understand differential equations. Based on how the network output is learned by itself from its derivatives, we have applied this approach to produce the poloidal flux function in order to remove its innate scatteredness based on the poloidal magnetic fields. Through the neural network, the consistent flux signals are available to be used for the plasma reconstruction where the expert-quality of the reconstruction can be attainable with no needs to become proficient about magnetic diagnostics. In conclusion, we can expect the fully automated reconstruction process can possibly be realized.

As future works, we deal with the other magnetic measurements installed in KSTAR such as saddle loops measuring radial magnetic fluxes, Mirnov coils detecting MHD activities, and a diamagnetic loop measuring a diamagnetic flux by means of the deep neural network. Additionally, we expand our technique to the long-pulse discharge sustaining the plasma over 300 $sec$ where deterioration in the magnetic signals are likely to occur, e.g., due to the signal drift with assistance of Bayesian statistical analysis.

\newpage
\section{Article \RNum{4}: Preliminary result under supervised learning}

\noindent
This approach deals with deep neural network Grad–Shafranov solver constrained with measured magnetic signals\footnote{\textsc{S. J{\small oung}, J. K{\small im}, S. K{\small wak}, J.G. B{\small ak}, S.G. L{\small ee}, H.S. H{\small an}, H.S. K{\small im}, G. L{\small ee}, D. K{\small won}} and \textsc{Y.{\small -c}. G{\small him}} \\
\textit{Nuclear Fusion}, Vol.60.1 (3$^{rd}$ Dec. 2019), \textsc{{\small DOI:}}\href{https://doi.org/10.1088/1741-4326/ab555f}{10.1088/1741-4326/ab555f}}, which is largely taken from Ref \cite{Joung:2019}, as a part of \textit{reconstruction of plasma equilibria via deep neural networks}.

This article describes a deep neural network approach to reconstruct plasma equilibria in real time by using the existing equilibrium database. This also proves that the solution of the Grad-Shafranov equation can be achieved by the neural network although the network is trained based on a supervised learning manner. As discussed in the synopsis of Article \RNum{3}, the equilibrium reconstruction requires the iterative estimation which takes time not to be suitable for a real-time application. The reconstruction informs tokamak controllers of positions of the plasma, and thus the controllers regulate powers of the poloidal field coils to manage the magnetic fields inside the tokamak. Therefore, to control tokamak plasmas precisely, the reconstruction should be done in real time. 

However, due to the iterative scheme, simplifying the original reconstruction algorithm is applied such as limiting the number of iterations or reusing equilibria reconstructed previously. In this sense, the network trained based on the database estimated by the original algorithm is suggested in order to take advantage of the original-like (or off-line-EFIT-like) equilibrium in real time. This network is fed with the poloidal magnetic fields and fluxes as its input, and produces a solution of the Grad-Shafranov equation as its output. To make the network generate rigorous equilibria satisfying the Grad-Shafranov equation, the equation itself is used as a cost function for the network training. Furthermore, This adopts the results of Article \RNum{2} in case of inferring missing inputs, which guarantees the use of the network in any circumstances. Since this Article proves that a neural network is able to encode rigorous plasma equilibria to its architecture, Article \RNum{5} is developed based on this approach.

\subsection{Introduction} \label{S1:intro}

\noindent
Magnetic equilibrium is one of the most important information to understand the basic behavior of plasmas in magnetically confined plasmas, and the off-line EFIT \cite{Lao:1985hn} code has been extensively used to reconstruct such equilibria in tokamaks.  Its fundamentals are basically finding a solution to an ideal magnetohydrodynamic equilibrium with toroidal axisymmetry, known as the Grad-Shafranov (GS) equation \cite{Freidberg:1987}:
\begin{equation}
\begin{split} \label{eq:gseq}
\Delta^*\psi &\equiv \lp R\frac{\partial}{\partial R} \frac{1}{R} \frac{\partial}{\partial R} + \frac{\partial^2}{\partial Z^2} \rp \psi \\
& =  -\mu_{0}R j_\phi \\
& = -\mu_0 R^2 \frac{d p(\psi)}{d \psi} - F(\psi)\frac{d F(\psi)}{d \psi},
\end{split}
\end{equation}
where $\psi=\psi\lp R, Z\rp$ is the poloidal flux function, $j_\phi=j_\phi\lp R, Z\rp$ the toroidal current density function, $p(\psi)$ the plasma pressure. $F(\psi)$ is related to the net poloidal current. Here, $R$, $\phi$ and $Z$ denote the usual cylindrical coordinate system.  As the $\Delta^*$ is a two-dimensional nonlinear partial differential operator, the off-line EFIT \cite{Lao:1985hn} finds a solution with many numerical iterations and has been implemented in many tokamaks such as D\rom{3}-D \cite{Lao:2005kd}, JET \cite{OBrien:1992gm}, NSTX \cite{Sabbagh:2001hh}, EAST \cite{Jinping:2009jn} and KSTAR \cite{Park:2011go} to name some as examples.

With an aim of real-time control of tokamak plasmas, real-time EFIT (rt-EFIT) \cite{Ferron:2002fw} code is developed to provide a magnetic equilibrium fast enough whose results are different from the off-line EFIT results. As pulse lengths of tokamak discharges become longer \cite{VanHoutte:1993, Ekedahl_2010, Itoh_1999, Zushi_2003, Saoutic_2002, Park:2019, Wan:2019}, demand on more elaborate plasma control is ever increased. Furthermore, some of the ITER relevant issues such as ELM (edge localized mode) suppression with RMP (resonant magnetic perturbation) coils \cite{Park_NP:2018} and the detached plasma scenarios \cite{Reimold:2015, Jaervinen:2016} require sophisticated plasma controls, meaning that the more accurate magnetic equilibria we have in real time, the better performance we can achieve.

There has been an attempt to satisfy such a requirement of acquiring a more accurate, i.e., closer to the off-line EFIT results compared to the rt-EFIT results, magnetic equilibrium in real-time using graphics processing units (GPUs) \cite{Yue:2013cj} by parallelizing equilibrium reconstruction algorithms. The GPU based EFIT (P-EFIT)  \cite{Yue:2013cj} enabled one to calculate a well-converged equilibrium in much less time; however, the benchmark test showed similar results to the rt-EFIT rather than the off-line results \cite{Huang:2016gz}.

\begin{figure}[t]
    \centering
    \includegraphics[width=0.344\linewidth]{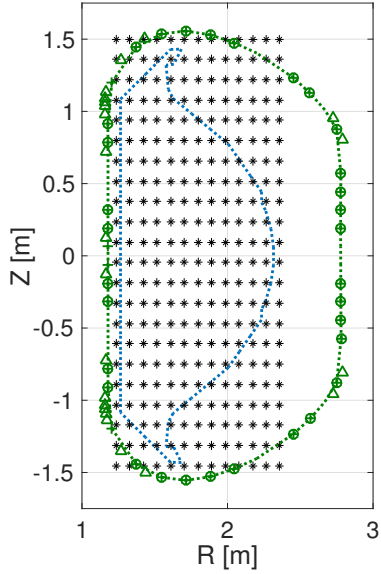}
    \caption{A poloidal cross-section of KSTAR with the first wall (blue dotted line). Green dotted line indicates a Rogowski coil measuring the plasma current ($I_\textrm{p}$). Green open circles and crosses depict locations of the magnetic pick-up coils measuring $32$ normal ($B_\textrm{n}$) and $36$ tangential ($B_\textrm{t}$) magnetic fields, respectively, whereas green triangles represent $22$ flux loops measuring poloidal magnetic fluxes ($\Psi_\textrm{FL}$). Black asterisks ($22\times 13$ spatial positions) show locations where we obtain the values of $\psi$ from the off-line EFIT results.}
    \label{fig:kstarConfig}
\end{figure}

Thus, we propose a reconstruction algorithm based on a neural network that satisfies the GS equation as well as the measured magnetic signals to obtain accurate magnetic equilibrium in real time. We note that usage of neural networks in fusion community is increasing rapidly, and examples are radiated power estimation \cite{Barana:2002RSI}, identifying instabilities \cite{Murari:2013cm}, estimating neutral beam effects \cite{Boyer_2019}, classifying confinement regimes \cite{Murari:2012fl}, determination of scaling laws \cite{MURARI2010850, Gaudio_2014}, disruption prediction \cite{Tang:2019Nature, Cannas:NF2010, Pau:NF2019}, turbulent transport modelling \cite{Meneghini:2014ic, Meneghini:2017kp, Citrin:2015fj, Felici:2018db}, plasma tomography with the bolometer system \cite{Matos:2017kl, Ferreira:2018}, coil current prediction with the heat load pattern in W7-X \cite{Bockenhoff:2018hl}, filament detection on MAST-U \cite{Cannas:2019gr}, electron temperature profile estimation via SXR with Thomson scattering \cite{Clayton:2013dx} and equilibrium reconstruction \cite{Lister:1991gx, Coccorese:1994jt, Bishop:1994kr, Cacciola:2006bq, Jeon:2001bd, Wang:JFE2016} together with an equilibrium solver \cite{vanMilligen:1995dv}. Most of previous works on the equilibrium reconstruction with neural networks have paid attention to finding the poloidal beta, the plasma elongation, positions of the X-points and plasma boundaries, i.e., last closed flux surface, and gaps between plasmas and plasma facing components, rather than reconstructing the whole internal magnetic structures we present in this work.

The inputs to our developed neural networks consist of plasma current measured by a Rogowski coil, normal and tangential components of magnetic fields by magnetic pick-up coils, poloidal magnetic fluxes by flux loops and a position in $\lp R, Z\rp$ coordinate system, where $R$ is the major radius, and $Z$ is the height as shown in \reffig{fig:kstarConfig}. The output of the neural networks is a value of poloidal flux $\psi$ at the specified $\lp R, Z\rp$ position. To train and validate the neural networks, we have collected a total of $1,118$ KSTAR discharges from two consecutive campaigns, i.e., $2017$ and $2018$ campaigns. We, in fact, generate three separate neural networks which are NN$_\textrm{2017}$, NN$_\textrm{2018}$ and NN$_\textrm{2017, 2018}$ where subscripts indicate the year(s) of KSTAR campaign(s) that the training data sets are obtained from. Additional $163$ KSTAR discharges (from the same two campaigns) are collected to test the performance of the developed neural networks.

We train the neural networks with the KSTAR off-line EFIT results, and let them be \textit{accurate} magnetic equilibria. Note that disputing on whether the off-line EFIT results we use to train the networks are accurate or not is beyond the scope of this work. If we find more accurate EFIT results, e.g., MSE(Motional Stark Effect)-constrained EFIT or more sophisticated equilibrium reconstruction algorithms that can cope with current-hole configurations (current reversal in the core) \cite{Rodrigues:2005PRL, Rodrigues:2007PRL, Ludwig:2013NF}, then we can always re-train the networks with new sets of data as long as the networks follow the trained EFIT data with larger similarity than the rt-EFIT results do. This is because supervised neural networks are limited to follow the training data. Hence, as a part of the training sets we use the KSTAR off-line EFIT results  as possible examples of \textit{accurate} magnetic equilibria to corroborate our developed neural networks. 

\begin{figure}[t]
    \centering
    \includegraphics[width=0.7\linewidth]{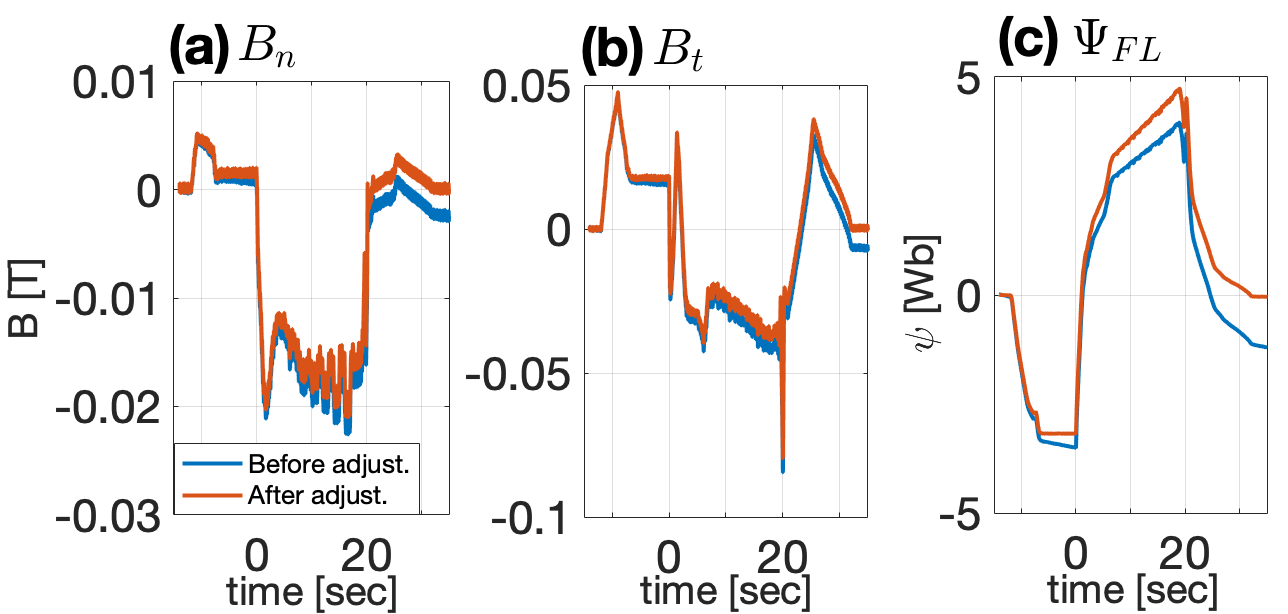}
    \caption{Before (blue) and after (red) the magnetic signal adjustments for (a) normal and (b) tangential components of magnetic fields measured by the magnetic pick-up coils, and (c) poloidal magnetic flux measured by one of the flux loops. The signals return closer to zeros after the adjustment when all the external magnetic coils (except the toroidal field coils) are turned off at around $30$ sec in this KSTAR discharge.}
    \label{fig:inputDrift}
\end{figure}

To calculate the output data a typical neural network requires the same set of input data as it has been trained. Therefore, even a single missing input (out of input data set) can result in a flawed output  \cite{vanLint:2005cn}. Such a case can be circumvented by training the network with possible combinations of missing inputs. As a part of input data, we have $32$ normal and $36$ tangential magnetic fields measured by the magnetic pick-up coils. If we wish to cover a case with one missing input data, then we will need to repeat the whole training procedure with $68$ ($32+36$) different cases. If we wish to cover a case with two or three missing input data, then we will need additional $2,278$ and $50,116$ different cases to be trained on, respectively. This number becomes large rapidly, and it becomes formidable, if not impossible, to train the networks with reasonable computational resources. Since the magnetic pick-up coils are susceptible to damages, we have developed our networks to be capable of inferring a few missing signals of the magnetic pick-up coils in real-time by invoking an imputation scheme \cite{Joung:2018ju} based on Bayesian probability \cite{Sivia:2006} and Gaussian processes \cite{Rasmussen:2006}.

In addition to reconstructing \textit{accurate} magnetic equilibria in real-time, the expected improvements with our neural networks compared to the previous studies are at least fourfold: (1) the network is capable of providing whole internal magnetic topology, not limited to boundaries and locations of X-points and/or magnetic axis; (2) spatial resolution of reconstructed equilibria is arbitrarily adjustable within the first wall of KSTAR since $\lp R, Z\rp$ position is a part of the input data; (3) the required training time and computational resources for the networks are reduced by generating a coarse grid points also owing to $\lp R, Z\rp$ position being an input, and (4) the networks can handle a few missing signals of the magnetic pick-up coils using the imputation method.

We, first, present how the data are collected to train the neural networks and briefly discuss real-time preprocessing of the measured magnetic signals in \refsec{S2:collection}. For the readers who are interested in thorough description of the real-time preprocessing, Article \RNum{1} provides the details. Then, we explain the structure of our neural networks and how we train them in \refsec{S3:nn}. In \refsec{S4:trainefit}, we present the results of the developed neural network EFIT (nn-EFIT) in four aspects. First, we discuss how well the NN$_\textrm{2017, 2018}$ network reproduces the off-line EFIT results. Then, we make comparisons among the three networks, NN$_\textrm{2017}$, NN$_\textrm{2018}$ and NN$_\textrm{2017, 2018}$, by examining in-campaign and cross-campaign performance. Once the absolute performance qualities of the networks are established, we compare relative performance qualities between nn-EFIT and rt-EFIT. Finally, we show how the imputation method support the networks when there exist missing inputs. Our conclusions are presented in \refsec{S6:con}.

\begin{table}[t]
\centering
\caption{Summary of the data samples to train and validate the networks}
\label{tab:NNdata}
\resizebox{0.78\textwidth}{!}{%
\begin{tabular}{cccc}
\hline\hline
Parameter & Definition & \multicolumn{1}{l}{Data size} & \multicolumn{1}{l}{No. of samples} \\ \hline
$I_\textrm{p}$ & Plasma current & 1 &  \\
& (Rogowski coil) & & \\
 &  & \multicolumn{1}{l}{} & \multicolumn{1}{l}{} \\
$B_\textrm{n}$ & Normal magnetic field & 32 &  \\
& (Magnetic pick-up coils) & & \\
 &  &  & 217,820 \\
$B_\textrm{t}$ & Tangential magnetic field & 36 & (time slices) \\
& (Magnetic pick-up coils) & & \\
 &  &  &  \\
$\Psi_\textrm{FL}$ & Poloidal magnetic flux & 22 &  \\
& (Flux loops) & & \\
 &  &  &  \\ \hline
$R$ & Position in major radius & 1 & 286 \\
 &  &  & ($22\times13$ grids) \\
$Z$ & Position in height & 1 &  \\
 &  &  &  \\ \hline
Network Input size &  & 93 (+1 for bias) &  \\
 &  &  &  \\
Total no. of samples &  &  & 62,296,520 \\ \hline\hline
\end{tabular}%
}
\end{table}

\subsection{Collection and real-time preprocessing of data}
\label{S2:collection}

\noindent
\reffig{fig:kstarConfig} shows locations where we obtain the input and the output data with the first wall (blue dotted line) on a poloidal cross-section of KSTAR. The green dotted line indicates a Rogowski coil measuring the plasma current ($I_\textrm{p}$).  The green open circles and crosses show locations of the magnetic pick-up coils measuring $32$ normal ($B_\textrm{n}$) and $36$ tangential ($B_\textrm{t}$) components of magnetic fields, respectively, whereas the green triangles show $22$ flux loops measuring the poloidal magnetic fluxes ($\Psi_\textrm{FL}$). These magnetic signals are selectively chosen out of all the magnetic sensors in KSTAR \cite{Lee:2008cl} whose performance has been demonstrated for many years, i.e., less susceptible to damages.

Although KSTAR calibrates the magnetic sensors (magnetic pick-up coils and flux loops) regularly during a campaign to remove drifts in the magnetic signals, it does not guarantee to fully eliminate such drifts. Thus, we preprocess the signals to adjust the drifts. \reffig{fig:inputDrift} shows examples of before (blue) and after (red) the drift adjustment for (a) normal and (b) tangential components of magnetic fields measured by the magnetic pick-up coils and (c) poloidal magnetic flux measured by one of the flux loops. Here, a KSTAR discharge is sustained until about $20$ sec, and all the external magnetic coils (except the toroidal field coils) are turned off at about $30$ sec. Therefore, we expect all the magnetic signals to return to zeros at around $30$ sec. If not, we envisage that there has been residual drifts. This means that we need to be able to preprocess the magnetic signals in real-time so that the input signal characteristics for predictions are similar to the trained ones. Article \RNum{1} describes in detail how we preprocess the magnetic signals in real-time.

The black asterisks in \reffig{fig:kstarConfig} show the $22\times 13$ grid points where we obtain the values of $\psi$ from the off-line EFIT results as outputs of the networks. We note that the original off-line EFIT provides the values of $\psi$ with $65\times 65$ grid points. The $22\times 13$ grid points are selected such that the distances between the neighboring channels in $R$ and $Z$ directions are as similar as possible while covering whole region within the first wall.  By generating such coarse grid points we can decrease the number of samples to train the network, thus consuming less amount of computational resources. Nevertheless, we do not lose the spatial resolution since $\lp R, Z\rp$ position is an input, i.e., the network can obtain the value of $\psi$ at any position within the first wall (see \refsec{S4:trainefit}).

\begin{figure}[t]
    \centering
    \includegraphics[width=0.425\linewidth]{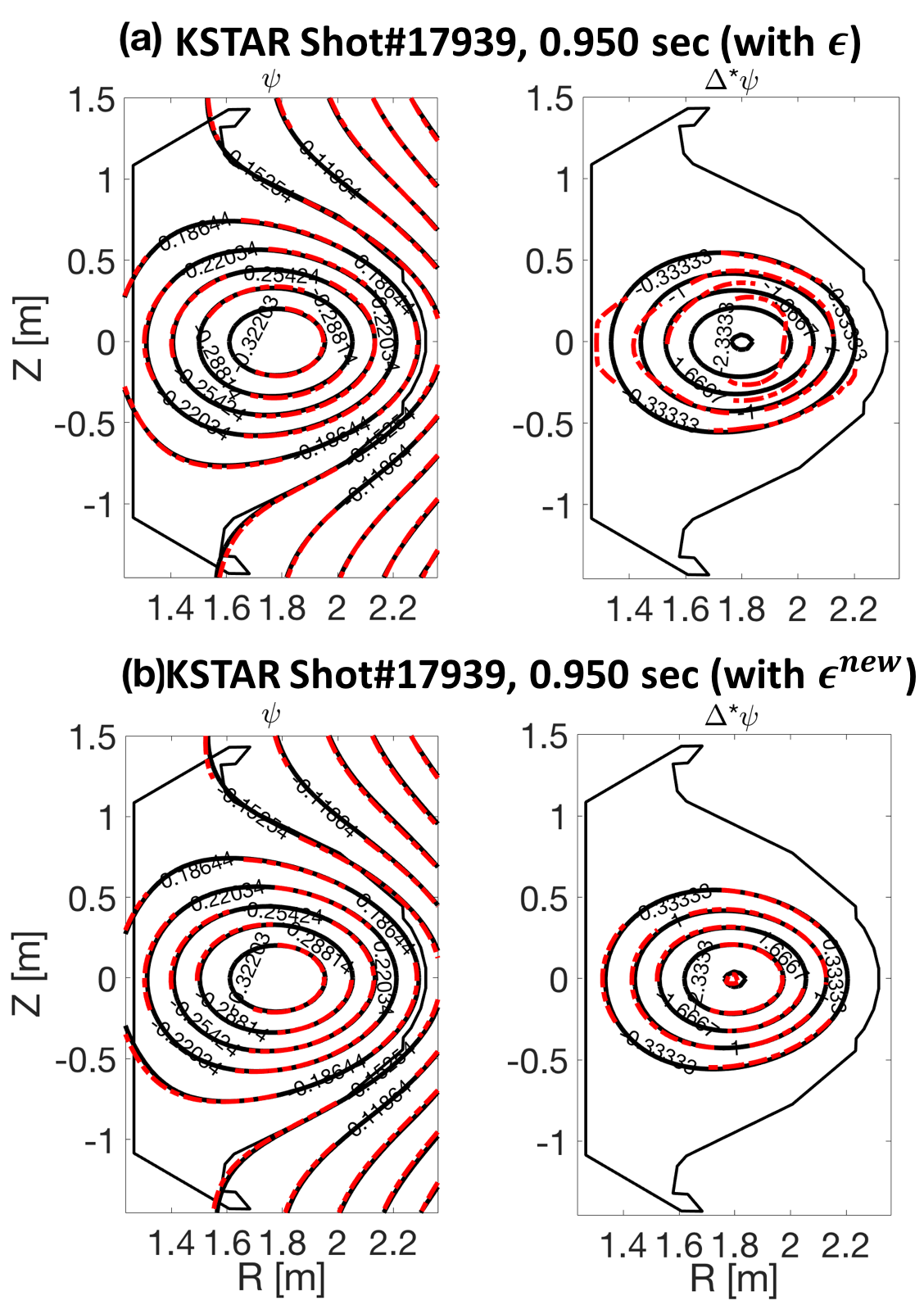}
    \caption{An example of the two networks' results trained with the cost function (a) $\epsilon$ and (b) $\epsilon^\textrm{new}$ for KSTAR shot\# $17939$ at $0.950$ sec. Both networks (red dashed line) reproduce the $\psi^\textrm{Target}$ (black line) well (left panels), but only the network trained with $\epsilon^\textrm{new}$ reproduces $\Delta^*\psi^\textrm{Target}$ (right panels).}
    \label{fig:costco}
\end{figure}

With an additional input for the spatial position $R$ and $Z$, each data sample contains $93$ inputs (and yet another input for bias) and one output which is a value of $\psi$ at the specified $\lp R, Z\rp$ location. We randomly collect a total of $1,118$ KSTAR discharges from $2017$ and $2018$ campaigns. Since each discharge can be further broken into many time slices, i.e., every $50$ msec following the temporal resolution of the off-line EFIT, we obtain $217,820$ time slices. With a total of $286$ value of $\psi$ from $22\times 13$ spatial points, we have a total of $62,296,520\lp =217,820\times286\rp$ samples to train and validate the networks. $90$\% of the samples are used to train the networks, while the other $10$\% are used to validate the networks to avoid overfitting problems. Note that an overfitting problem can occur if a network is overly well trained to the \textit{training data} following the very details of them. This inhibits generalization of the trained network to predict \textit{unseen data}, and such a problem can be minimized with the validation data set. All the inputs except $R$ and $Z$ are normalized such that the maximum and minimum values within the whole samples become $1$ and $-1$, respectively. We use the actual values of $R$ and $Z$ in the unit of meters.

\reftable{tab:NNdata} summarizes the training and validation samples discussed in this section. Additionally, we also have randomly collected another $163$ KSTAR discharges in the same way discussed here which are different from the $1,118$ KSTAR discharges to test the performance of the networks.

\subsection{Neural network model and training} \label{S3:nn}

\subsubsection{Neural network model}

\noindent
We develop the neural networks that not only output a value of $\psi$ but also satisfies \refeq{eq:gseq}, the GS equation. With the total of $94$ input nodes ($91$ for a plasma current and magnetic signals, two for $R$ and $Z$ position, one for the bias) and one output node for a value of $\psi$, each network has three fully connected hidden layers with an additional bias node at each hidden layer. Each layer contains $61$ nodes including the bias node. The structure of our networks is selected by examining several different structures by error and trials.

Denoting the value of $\psi$ calculated by the networks as $\psi^\textrm{NN}$, we have
\begin{equation}
\label{eq:nn_structure}
\begin{split}
&\psi^\textrm{NN}\! = \! s_{0} + \sum_{l=1}^{60}\! s_{l} \\
&\times f\!\! \left(\! u_{l0}\! + \sum_{k=1}^{60}\! u_{lk} f\!\! \left(\! v_{k0}\! +\! \sum_{j=1}^{60}\! v_{kj} f\!\! \left(\! w_{j0}\! +\! \sum_{i=1}^{93}\! w_{ji} x_{i}\! \right)\!\!\! \right)\!\!\! \right),
\end{split}
\end{equation}
where $x_i$ is the $i^\textrm{th}$ input value with $i=1,\dots,93$, i.e., $91$ measured values with the various magnetic diagnostics and two for $R$ and $Z$ positions. $w_{ji}$ is an element in a $61\times 94$ matrix, whereas $v_{kj}$ and $u_{lk}$ are elements in $61\times 61$ matrices. $s_l$ connects the $l^\textrm{th}$ node of the third (last) hidden layer to the output node. $w, v, u$ and $s$ are the weighting factors that need to be trained to achieve our goal of obtaining accurate $\psi$. $w_{j0}, v_{k0}, u_{l0}$ and $s_0$ are the weighting factors connecting the biases, where values of all the biases are fixed to be unity. We use a hyperbolic tangent function as the activation function $f$ giving the network non-linearity \cite{Haykin:2008}:
\begin{equation} \label{eq:NN-actfcn}
f\! \left(t \right) = \tanh(t) = \frac{2}{1 + e^{-2t}} - 1.
\end{equation}

The weighting factors are initialized as described in \cite{pmlr-v9-glorot10a} so that a good training can be achieved. They are randomly selected from a normal distribution whose mean is zero with the variance set to be an inverse of total number of connecting nodes. For instance, our weighting factor $w$ connects the input layer (94 nodes with bias) and the first hidden layer (61 nodes with bias), therefore the variance is set to be $1/(94+61)$. Likewise, the variances for $v$, $u$ and $s$ are $1/(61+61)$, $1/(61+61)$ and $1/(61+1)$, respectively.

\begin{figure}[t]
    \centering
    \includegraphics[width=0.495\linewidth]{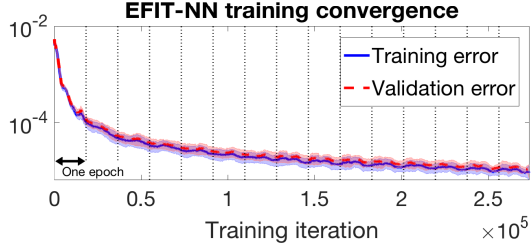}
    \caption{The descending feature of training (blue line) and validation (red dashed line) errors as a function of iterations. Shaded areas represent standard deviation of the errors.}
    \label{fig:conv}
\end{figure}

\subsubsection{Training}

\noindent
With the aforementioned network structure, training (or optimizing) the weighting factors to predict the correct value of $\psi$ highly depends on a choice of the cost function. A typical choice of such cost function would be:
\begin{equation} \label{eq:costfcn}
\epsilon = \frac{1}{N} \sum_{i=1}^{N} \left( \psi_{i}^\textrm{NN} - \psi_{i}^\textrm{Target} \right)^{2},
\end{equation}
where $\psi^\textrm{Target}$ is the target value, i.e., the value of $\psi$ from the off-line EFIT results in our case, and $N$ the number of data sets.

As will be shown shortly, minimizing the cost function $\epsilon$ does not guarantee to satisfy the GS equation (\refeq{eq:gseq}) even if $\psi^\textrm{NN}$ and $\psi^\textrm{Target}$ matches well, i.e., the network is well trained with the given optimization rule. Since $\Delta^*\psi$ provides information on the toroidal current density directly, it is important that $\Delta^*\psi^\textrm{NN}$ matches $\Delta^*\psi^\textrm{Target}$ as well. We have an analytic form representing $\psi^\textrm{NN}$ as in \refeq{eq:nn_structure}; therefore, we can analytically differentiate $\psi^\textrm{NN}$ with respect to $R$ and $Z$, meaning that we can calculate $\Delta^*\psi^\textrm{NN}$ during the training stage. Thus, we introduce another cost function:
\begin{equation} \label{eq:costfcn_plusdiff}
\begin{split}
\epsilon^\textrm{new}& = \frac{1}{N} \sum_{i=1}^{N} \left( \psi_{i}^\textrm{NN} - \psi_{i}^\textrm{Target} \right)^{2} \\
&+ \frac{1}{N} \sum_{i=1}^{N} \left( \Delta^*\psi_{i}^\textrm{NN} - \Delta^*\psi_{i}^\textrm{Target} \right)^{2},
\end{split}
\end{equation}
where we obtain the value of $\Delta^*\psi^\textrm{Target}$ from the off-line EFIT results as well. 

To acknowledge difference between the two cost functions $\epsilon$ and $\epsilon^\textrm{new}$, we first discuss the results. \reffig{fig:costco} shows the outputs of the two trained networks with the cost function (a) $\epsilon$ and (b) $\epsilon^\textrm{new}$. It is evident that in both cases the network output $\psi^\textrm{NN}$ (red dashed line) reproduces the off-line EFIT $\psi^\textrm{Target}$ (black line). However, only the network trained with the cost function $\epsilon^\textrm{new}$ reproduces the off-line EFIT $\Delta^*\psi^\textrm{Target}$. Both networks are trained well, but the network with the cost function $\epsilon$ does not achieve our goal, that is correctly predicting $\psi^\textrm{Target}$ \textit{and} $\Delta^*\psi^\textrm{Target}$.

Since our goal is to develop a neural network that solves the GS equation, we choose the cost function to be $\epsilon^\textrm{new}$ to train the networks. We optimize the weighting factors by minimizing $\epsilon^\textrm{new}$ with the Adam \cite{DBLP:journals/corr/KingmaB14} which is one of the gradient-based optimization algorithms. With $90$\% and $10$\% of the total data samples for training and validation of the networks, respectively, we stop training the networks with a fixed number of iterations that is large enough but not too large such that the validation errors do not increase, i.e., to avoid overfitting problems. The whole workflow is carried out with Python and Tensorflow \cite{tensorflow2015-whitepaper}.

With the selected cost function we create three different networks that differ only by the training data sets. NN$_\textrm{2017}$, NN$_\textrm{2018}$ and NN$_\textrm{2017, 2018}$ refer to the three networks trained with the data sets from only $2017$ ($744$ discharges), from only $2018$ ($374$ discharges) and from both $2017$ and $2018$ ($744+374$ discharges) campaigns, respectively. 

\begin{figure}[t]
    \centering
    \includegraphics[width=0.495\linewidth]{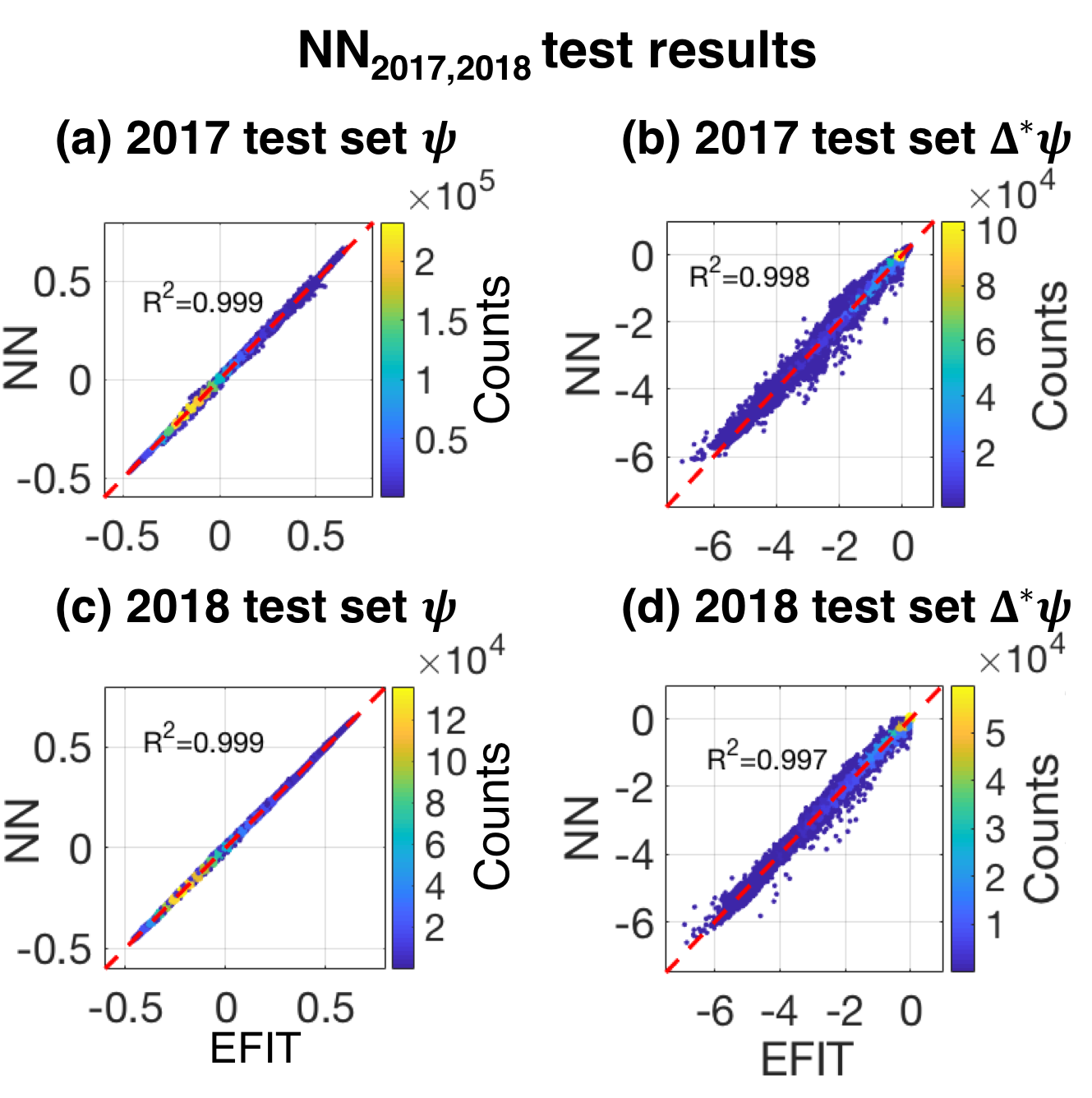}
    \caption{Performance tests of the NN$_{2017, 2018}$ network on the unseen KSTAR discharges from (a)(b) 2017 campaign and (c)(d) 2018 campaign. The values of R$^2$ and histograms of (a)(c) $\psi^\textrm{NN}$ vs. $\psi^\textrm{Target}$ and (b)(d) $\Delta^*\psi^\textrm{NN}$ vs. $\Delta^*\psi^\textrm{Target}$ with colors representing number of counts manifest goodness of the NN$_{2017, 2018}$ network. Red dashed line is the $y=x$ line.}
    \label{fig:teLinear1718}
\end{figure}

The descending feature of the cost function $\epsilon^\textrm{new}$ as a function of the training iteration for NN$_{2017, 2018}$ network is shown in \reffig{fig:conv}. Both the training errors (blue line) and validation errors (red dashed line) decrease together with similar values which means that the network is well generalized. Furthermore, since the validation errors do not increase, the network does not have an overfitting problem. Note that fluctuations in the errors, i.e., standard deviation of the errors, are represented as shaded areas. 

Small undulations repeated over the iterations in \reffig{fig:conv} are due to the mini-batch learning. Contrary to the batch learning, i.e., optimizing the network with the entire training set in one iteration, the mini-batch learning divides the training set into some number of small subsets ($1,000$ subsets for our case) to optimize the networks sequentially. One cycle that goes through all the subsets once is called an epoch. The mini-batch learning helps to escape from local minima in the weighting factor space \cite{DBLP:journals/corr/GeHJY15} via the stochastic gradient descent scheme \cite{Bottou10large-scalemachine}.

\begin{figure}[t]
    \centering
    \includegraphics[width=0.695\linewidth]{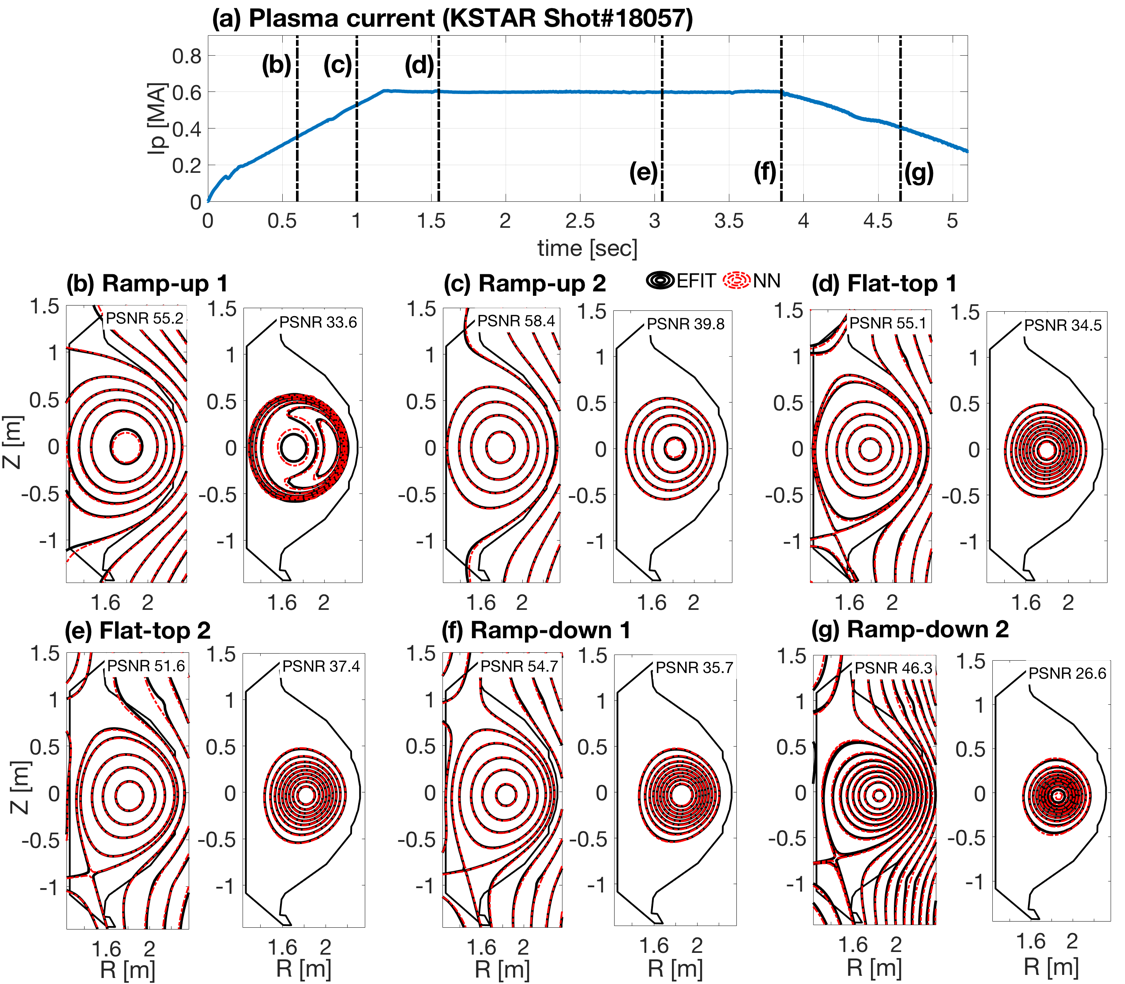}
    \caption{The actual reconstruction results for the KSTAR shot\#18057, comparing the network results and off-line EFIT reconstructions for ramp-up ((b) and (c)), flat-top ((d) and (e)), ramp-down ((f) and (g)) phases following (a) the plasma current evolution. Black lines indicate the flux surfaces from the off-line EFIT, overlaid with the red dotted lines which stand for the NN reconstructions. As a figure of merit, magnitudes of PSNR metric are written on each figure.}
    \label{fig:teFSshot}
\end{figure}

\subsection{Performance of the developed neural networks: Benchmark tests} \label{S4:trainefit}

\noindent
In this section, we present how well the developed networks perform. Main figures of merit we use are peak signal-to-noise ratio (PSNR) and mean structural similarity (MSSIM) as have been used perviously \cite{Matos:2017kl} in addition to the usual statistical quantity R$^2$, coefficient of determination. We note that obtaining full flux surface information $\psi\lp R, Z\rp$ on $22\times 13$ or $65\times 65$ spatial grids with our networks takes less than $1$ msec on a typical personal computer. 

First, we discuss the benchmark results of the NN$_{2017, 2018}$ network. Then, we compare the performance of NN$_{2017}$, NN$_{2018}$ and NN$_{2017, 2018}$ networks. Here, we also investigate cross-year performance, for instance, applying the NN$_{2017}$ network to predict the discharges obtained from 2018 campaign and vice versa. Then, we evaluate the performance of the networks against the rt-EFIT results to examine possibility of supplementing or even replacing the rt-EFIT with the networks. Finally, we show how the imputation scheme supports the networks' performance. Here, all the tests are performed with the unseen (to all three networks, i.e., NN$_{2017}$, NN$_{2018}$ and NN$_{2017, 2018}$) KSTAR discharges which are $88$ and $75$ KSTAR discharges from 2017 and 2018 campaigns, respectively.

\subsubsection{Benchmark results of the NN$_{2017, 2018}$ network} \label{S4-1:benchmark}

\begin{figure}[t]
    \centering
    \includegraphics[width=0.585\linewidth]{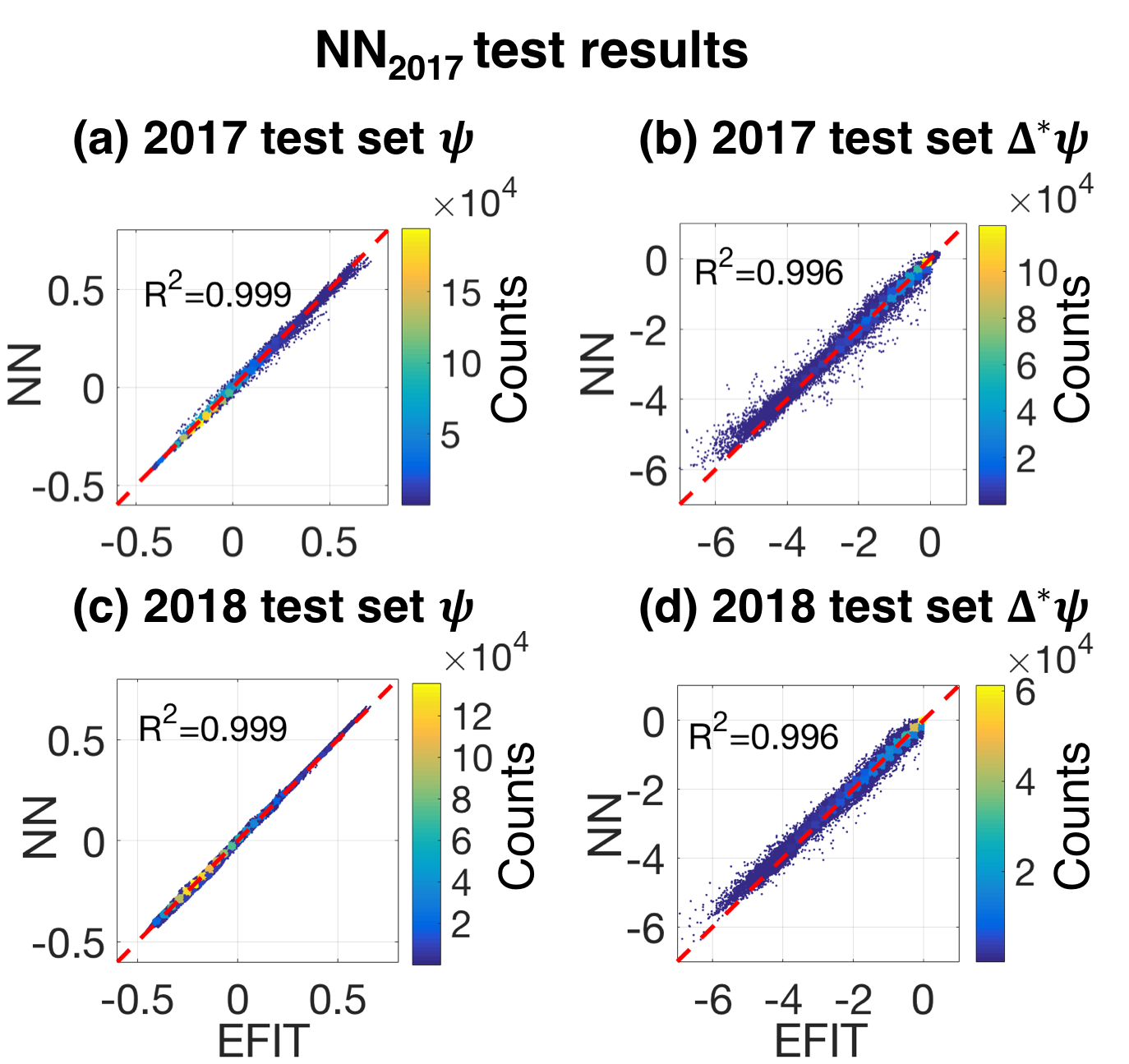}
    \caption{Same as \reffig{fig:teLinear1718} for the the NN$_{2017}$ network, i.e., trained with the data sets from 2017 campaign.}
    \label{fig:teLinear17}
\end{figure}

\begin{figure}[t]
    \centering
    \includegraphics[width=0.585\linewidth]{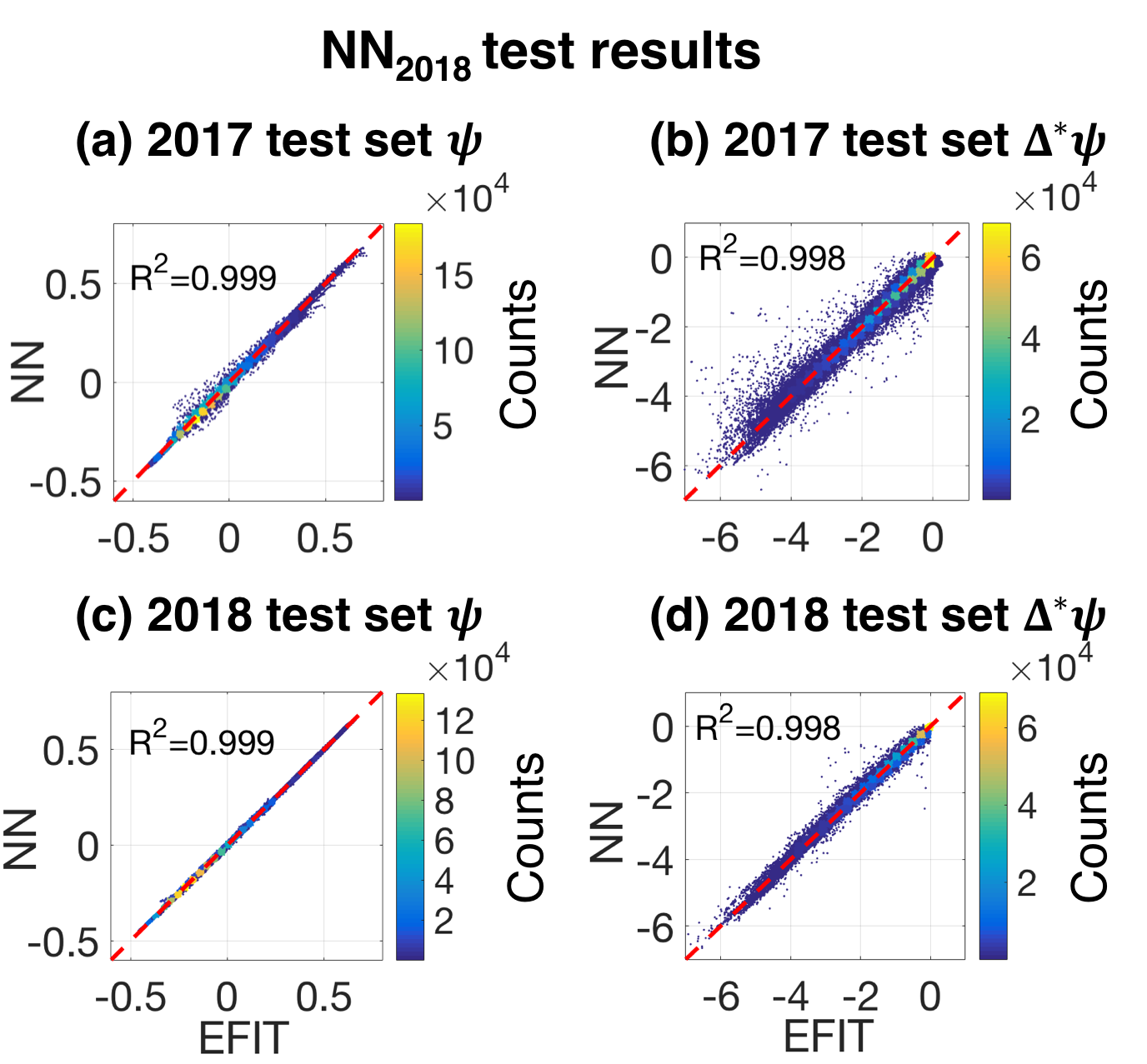}
    \caption{Same as \reffig{fig:teLinear1718} for the the NN$_{2018}$ network, i.e., trained with the data sets from 2018 campaign.}
    \label{fig:teLinear18}
\end{figure}

\noindent
\reffig{fig:teLinear1718} show the benchmark results of the NN$_{2017, 2018}$ network, i.e., network trained with the data sets from both 2017 and 2018 campaigns. (a) and (b) show the results with the test discharges from 2017 campaign; while (c) and (d) present the results with the test discharges from 2018 campaign. Histograms of (a)(c) $\psi^\textrm{NN}$ vs. $\psi^\textrm{Target}$ and (b)(d) $\Delta^*\psi^\textrm{NN}$ vs. $\Delta^*\psi^\textrm{Target}$ are shown with colors representing the number of counts. For instance, there is a yellow colored point in \reffig{fig:teLinear1718}(a) around $(-0.1, -0.1)\pm\varepsilon$, where $\varepsilon$ is a bin size for the histogram. Since yellow represents about $2\times 10^5$ counts, there are approximately $2\times 10^5$ data whose neural network values and EFIT values are $-0.1\pm\varepsilon$ simultaneously within our test data set. Note that each KSTAR discharge contains numerous time slices whose number depends on the actual pulse length of a discharge, and each time slice generates the total of $22\times 13=286$ data points. The values of $\psi^\textrm{Target}$ and $\Delta^*\psi^\textrm{Target}$ are obtained from the off-line EFIT results. It is clear that the network predicts the target values well. 

As a figure of merit, we introduce the R$^2$ metric (coefficient of determination) defined as 
\begin{equation}
\textrm{R}^2 = 1 - \frac{\sum_{i=1}^{L} \left(y_i^\textrm{Target} - y_i^\textrm{NN} \right)^2}{\sum_{i=1}^{L} \left(y_i^\textrm{Target} - \frac{1}{L}\sum_{j=1}^{L} y_{j}^\textrm{Target} \right)^2},
\end{equation}
where $y$ takes either $\psi$ or $\Delta^*\psi$, and $L$ is the number of test data sets. The calculated values are written in \reffig{fig:teLinear1718}, and they are indeed close to unity, implying the existence of very strong linear correlations between the predicted (from the network) and target (from the off-line EFIT) values. Note that R$^2=1$ means the perfect prediction. The red dashed lines on the figures are the $y=x$ lines.

\reffig{fig:teFSshot} is an example of reconstructed magnetic equilibria using KSTAR shot \#18057 from 2017 campaign. (a) shows the evolution of the plasma current. The vertical dashed lines indicate the time points where we show and compare the equilibria obtained from the network (red) and the off-line EFIT (black) which is our target. (b) and (c) are taken during the ramp-up phase, (d) and (e) during the flat-top phase, and (f) and (g) during the ramp-down phase. In each sub-figure from (b) to (g), left panels compare $\psi$, and right panels are for $\Delta^*\psi$. We mention that the equilibria in \reffig{fig:teFSshot} are reconstructed with $65 \times 65$ grid points even though the network is trained with $22 \times 13$ grid points demonstrating how spatial resolution is flexible in our networks.

For a quantitative assessment of the network, we use an image relevant figure of merit that is peak signal-to-noise ratio (PSNR) \cite{HuynhThu:2008fm} originally developed to estimate a degree of artifacts due to an image compression compared to an original image. Typical PSNR range for the JPEG image, which preserves the original quality with a reasonable degree, is generally in 30--50 dB \cite{Matos:2017kl, Ebrahimi:2004fz}. For our case, the networks errors relative to the off-line EFIT results can be treated as artifacts. As listed on \reffig{fig:teFSshot}(b)-(g), PSNR for $\psi$ is very good, while we achieve acceptable values for $\Delta^*\psi$.

\subsubsection{The NN$_{2017}$, NN$_{2018}$ and NN$_{2017, 2018}$ networks}

\begin{figure}[!]
    \centering
    \includegraphics[width=0.50\linewidth]{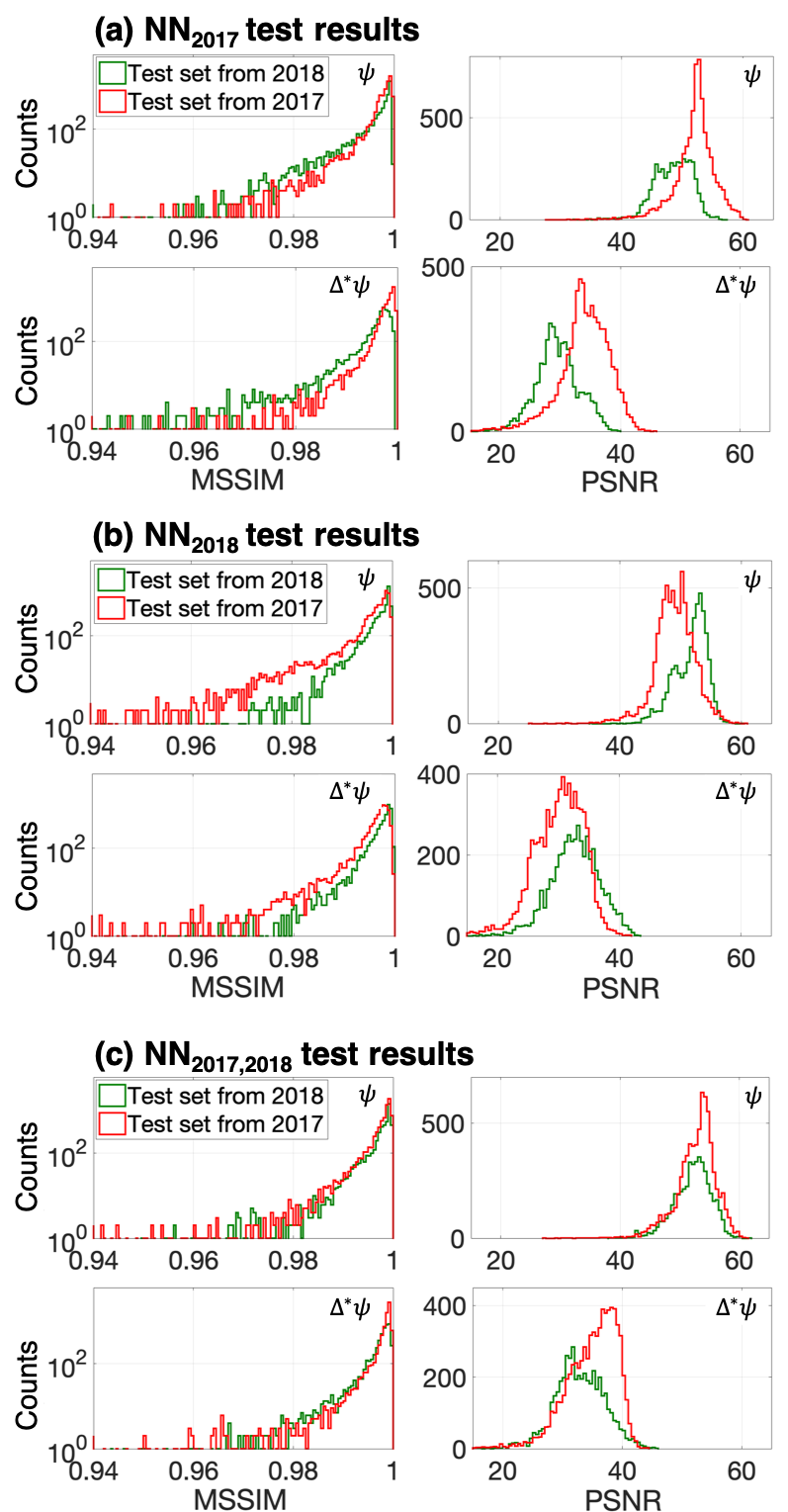}
    \caption{Histograms of MSSIM (left panel) and PSNR (right panel) for (a) NN$_{2017}$, (b) NN$_{2018}$ and (c) NN$_{2017, 2018}$. Red (green) line indicates the test results on the data sets from 2017 (2018) campaign. In each sub-figure, top (bottom) panel show the results for $\psi$ ($\Delta^*\psi$). The off-line EFIT results are used as reference.}
    \label{fig:tedist}
\end{figure}

\noindent
Similar to shown in \reffig{fig:teLinear1718}, we show the benchmark results of the NN$_{2017}$ (trained with the data sets from 2017 campaign) and the NN$_{2018}$ (trained with the data sets from 2018 campaign) in \refmultifig{fig:teLinear17}{fig:teLinear18}, respectively. R$^2$ metric is also provided on the figures. Again, overall performance of the networks are good. 

The NN$_{2017}$ and NN$_{2018}$ networks are trained with only in-campaign data sets, e.g., NN$_{2018}$ with the data sets from only 2018 campaign, and we find slightly worse results, but still good, on predicting cross-campaign magnetic equilibria, e.g. NN$_{2018}$ predicting equilibria for 2017 campaign. Notice that the NN$_{2017}$ seems to predict cross-campaign equilibria better than in-campaign ones by comparing \reffig{fig:teLinear17}(a) and (c) which contradicts our intuition. Although the histogram in \reffig{fig:teLinear17}(c) seems tightly aligned with the $y=x$ line (red dashed line), close inspection reveals that the NN$_{2017}$ network, in general, underestimates the off-line EFIT results from 2018 campaign marginally. This will be evident when we compare image qualities.

Mean structural similarity (MSSIM) \cite{Wang:2004gj} is another image relevant figure of merit used to estimate perceptual  similarity (or perceived differences) between the true and reproduced images based on inter-dependence of adjacent spatial pixels in the images. MSSIM ranges from zero to one, where the closer to unity the better the reproduced image is.

Together with PSNR, \reffig{fig:tedist} shows MSSIM for (a) NN$_{2017}$, (b) NN$_{2018}$ and (c) NN$_{2017, 2018}$ where the off-line EFIT results are used as reference. Notice that counts in all the histograms of MSSIM and PSNR in this work correspond to the number of reconstructed magnetic equilibria (or a number of time slices) since we obtain a single value of MSSIM and PSNR from one equilibrium; whereas counts in Figures \ref{fig:teLinear1718}, \ref{fig:teLinear17} and \ref{fig:teLinear18} are much bigger since $286(=22\times 13)$ data points are generated from each time slice. Red (green) line indicates the test results on the data sets from 2017 (2018) campaign. In general, whether the test data sets are in-campaign or cross-campaign, image reproducibility of all three networks, i.e., predicting the off-line EFIT results, is good as attested by the fact that MSSIM is quite close to unity and PSNR for $\psi$ ($\Delta^*\psi$) ranges approximately $40$ to $60$ ($20$ to $40$). It is easily discernible that in-campaign results are better for both NN$_{2017}$ and NN$_{2018}$ unlike what we noted in \reffig{fig:teLinear17}(a) and (c). Not necessarily guaranteed, we find that the NN$_{2017, 2018}$ network works equally well for both campaigns as shown in \reffig{fig:tedist}(c).

\subsubsection{Comparisons among nn-EFIT, rt-EFIT and off-line EFIT} \label{S4-2:rtefit}

\begin{figure}[t]
    \centering
    \includegraphics[width=0.320\linewidth]{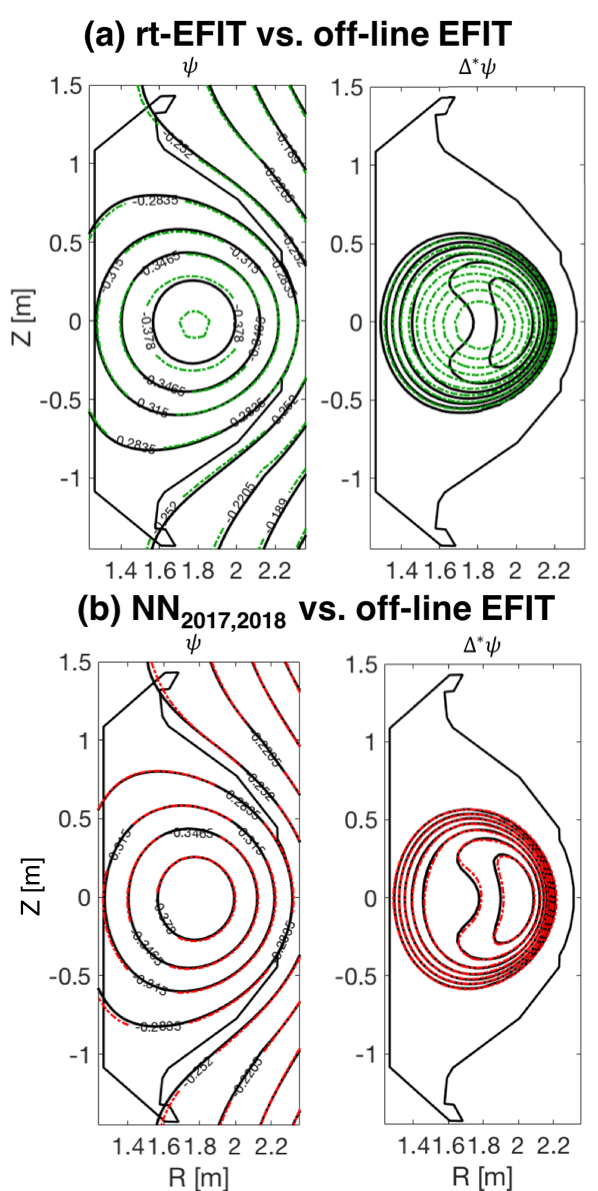}
    \caption{An example of reconstructed $\psi\lp R, Z\rp$ (left panel) and $\Delta^*\psi\lp R, Z\rp$ (right panel) for KSTAR shot \#$17975$ at $0.7$ sec comparing (a) rt-EFIT (green) and off-line EFIT (black) and (b) nn-EFIT (NN$_{2017, 2018}$) (red) and off-line EFIT (black).}
    \label{fig:rtFSshot}
\end{figure}

\noindent
It is widely recognized that rt-EFIT results and off-line results are different from each other. If we allow the off-line EFIT results used to train the networks to be accurate ones, then the reconstruction of equilibria with the neural networks (nn-EFIT) must satisfy the following criterion: nn-EFIT results must be more similar to the off-line EFIT results than rt-EFIT results are to the off-line EFIT as mentioned in \refsec{S1:intro}. Once this criterion is satisfied, then we can always improve the nn-EFIT as genuinely more accurate EFIT results are collected. For this reason, we make comparisons among the nn-EFIT, rt-EFIT and off-line EFIT results. 

\reffig{fig:rtFSshot} shows an example of reconstructed magnetic equilibria for (a) rt-EFIT vs. off-line EFIT and (b) nn-EFIT (the NN$_{2017, 2018}$ network) vs. off-line EFIT for KSTAR shot \#$17975$ at $0.7$ sec with $\psi$ (left panel) and $\Delta^*\psi$ (right panel). Green, red and black lines indicate rt-EFIT, nn-EFIT and off-line EFIT results, respectively. This simple example shows that the nn-EFIT is more similar to the off-line EFIT than the rt-EFIT is to the off-line EFIT, satisfying the aforementioned criterion.

\begin{figure}[t]
    \centering
    \includegraphics[width=0.48\linewidth]{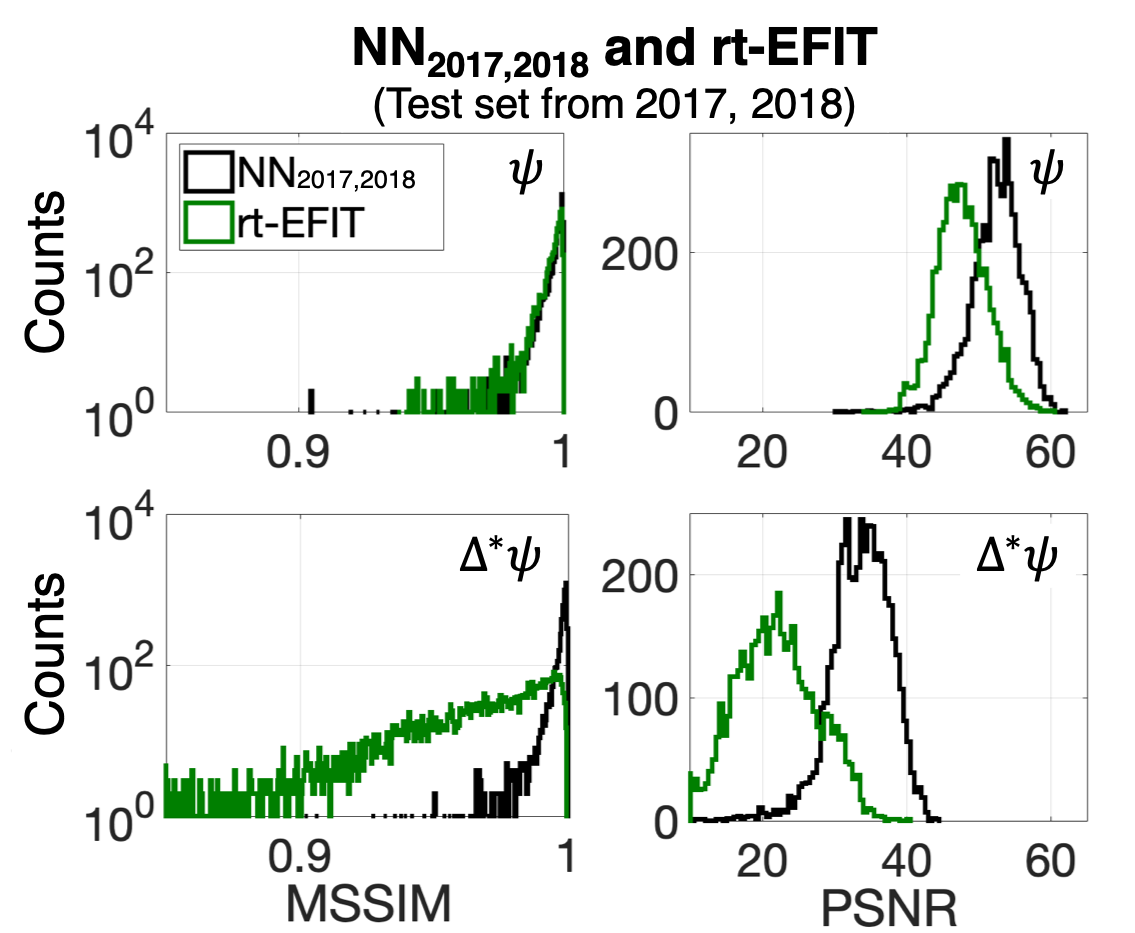}
    \caption{Histograms of MSSIM (left panel) and PSNR (right panel) of $\psi$ (top) and $\Delta^*\psi$ (bottom) calculated by the nn-EFIT (black) and the rt-EFIT (green), where the nn-EFIT is the NN$_{2017, 2018}$. For both the nn-EFIT and the rt-EFIT, the off-line EFIT is treated as reference.}
    \label{fig:rtEFIT_nn_2017_2018}
\end{figure}

To validate the criterion statistically, we generate histograms of MSSIM and PSNR for the nn-EFIT and the rt-EFIT with reference to the off-line EFIT. This is shown in \reffig{fig:rtEFIT_nn_2017_2018} as histograms, where MSSIM (left panel) and PSNR (right panel) of $\psi$ (top) and $\Delta^*\psi$ (bottom) are compared between the nn-EFIT (black) and the rt-EFIT (green). Here, the nn-EFIT results are obtained with the NN$_{2017, 2018}$ network on the test data sets. We confirm that the criterion is satisfied with the NN$_{2017, 2018}$ network as the histograms in \reffig{fig:rtEFIT_nn_2017_2018} are in favour of the nn-EFIT, i.e., larger MSSIM and PSNR are obtained by the nn-EFIT. This is more conspicuous for $\Delta^*\psi$ than $\psi$.

\begin{figure}[t]
    \centering
    \includegraphics[width=0.44\linewidth]{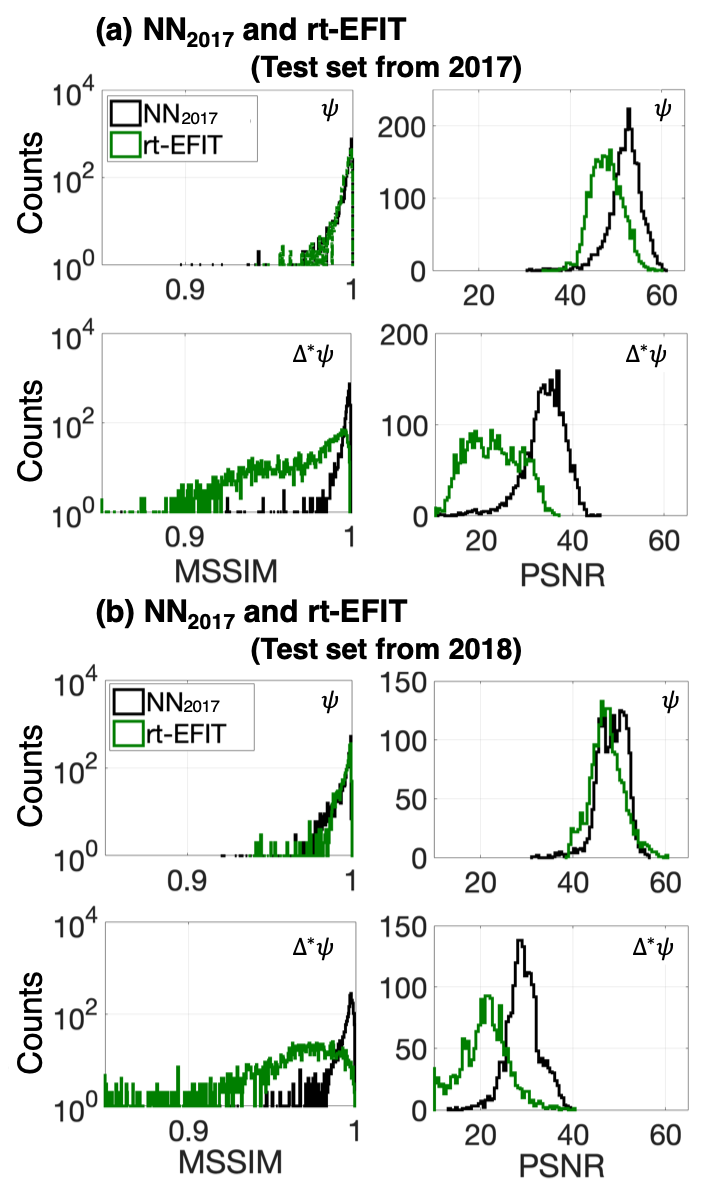}
    \caption{Same as \reffig{fig:rtEFIT_nn_2017_2018} with the NN$_{2017}$ as the nn-EFIT where the test data sets are obtained from (a) 2017 campaign and (b) 2018 campaign.}
    \label{fig:rtEFIT_nn_2017}
\end{figure}

\begin{figure}[t]
    \centering
    \includegraphics[width=0.44\linewidth]{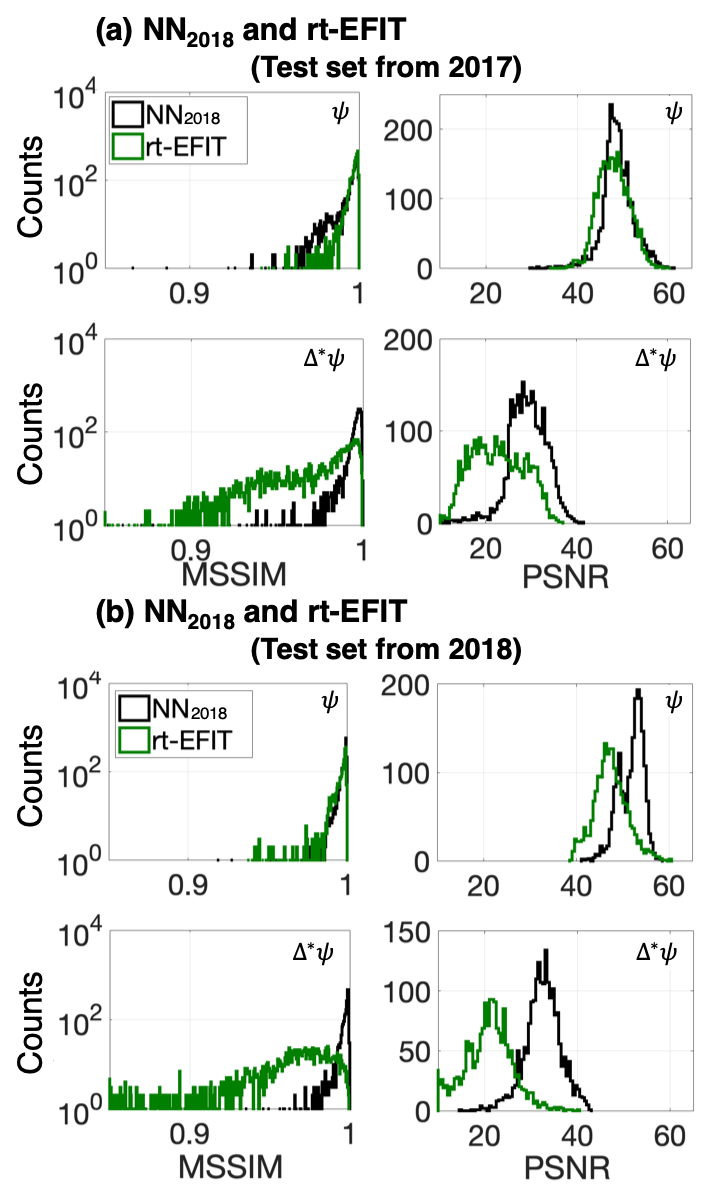}
    \caption{Same as \reffig{fig:rtEFIT_nn_2017_2018} with the NN$_{2018}$ as the nn-EFIT where the test data sets are obtained from (a) 2017 campaign and (b) 2018 campaign.}
    \label{fig:rtEFIT_nn_2018}
\end{figure}

We perform the similar statistical analyses for the other two networks, NN$_{2017}$ and NN$_{2018}$, which are shown in Fig. \ref{fig:rtEFIT_nn_2017} and \ref{fig:rtEFIT_nn_2018}. Since these two networks are trained with the data sets from only one campaign, we show the results where the test data sets are prepared from (a) 2017 campaign or (b) 2018 campaign so that in-campaign and cross-campaign effects can be assessed separately. We find that whether in- or cross-campaign, the criterion is fulfilled for both $\psi$ and $\Delta^*\psi$.

\subsubsection{The NN$_{2017, 2018}$ network with the imputation scheme} \label{S5:imputation}

\noindent
If one or a few magnetic pick-up coils which are a part of the inputs to the nn-EFIT are impaired, then we will have to re-train the network without the damaged ones or hope that the network will reconstruct equilibria correctly by padding a fixed value, e.g., zero-padding, to the broken ones. Of course, one can anticipate training the network by considering possible combinations of impaired magnetic pick-up coils. With the total number of $68$ signals from the magnetic pick-up coils being inputs to the network in our case, we immediately find that the number of possible combinations increases too quickly to consider it as a solution. 

Since inferring the missing values is better than the null replacement \cite{vanLint:2005cn}, we resolve the issue by using the recently proposed imputation method \cite{Joung:2018ju} based on Gaussian processes (GP) \cite{Rasmussen:2006} and Bayesian inference \cite{Sivia:2006}, where the likelihood is constructed based on Maxwell's equations. The imputation method infers the missing values fast enough, i.e., less than $1$ msec to infer at least up to nine missing values on a typical personal computer; thus, we can apply the method during a plasma discharge by replacing the missing values with the real-time inferred values. 

\begin{figure}[t]
    \centering
    \includegraphics[width=0.47\linewidth]{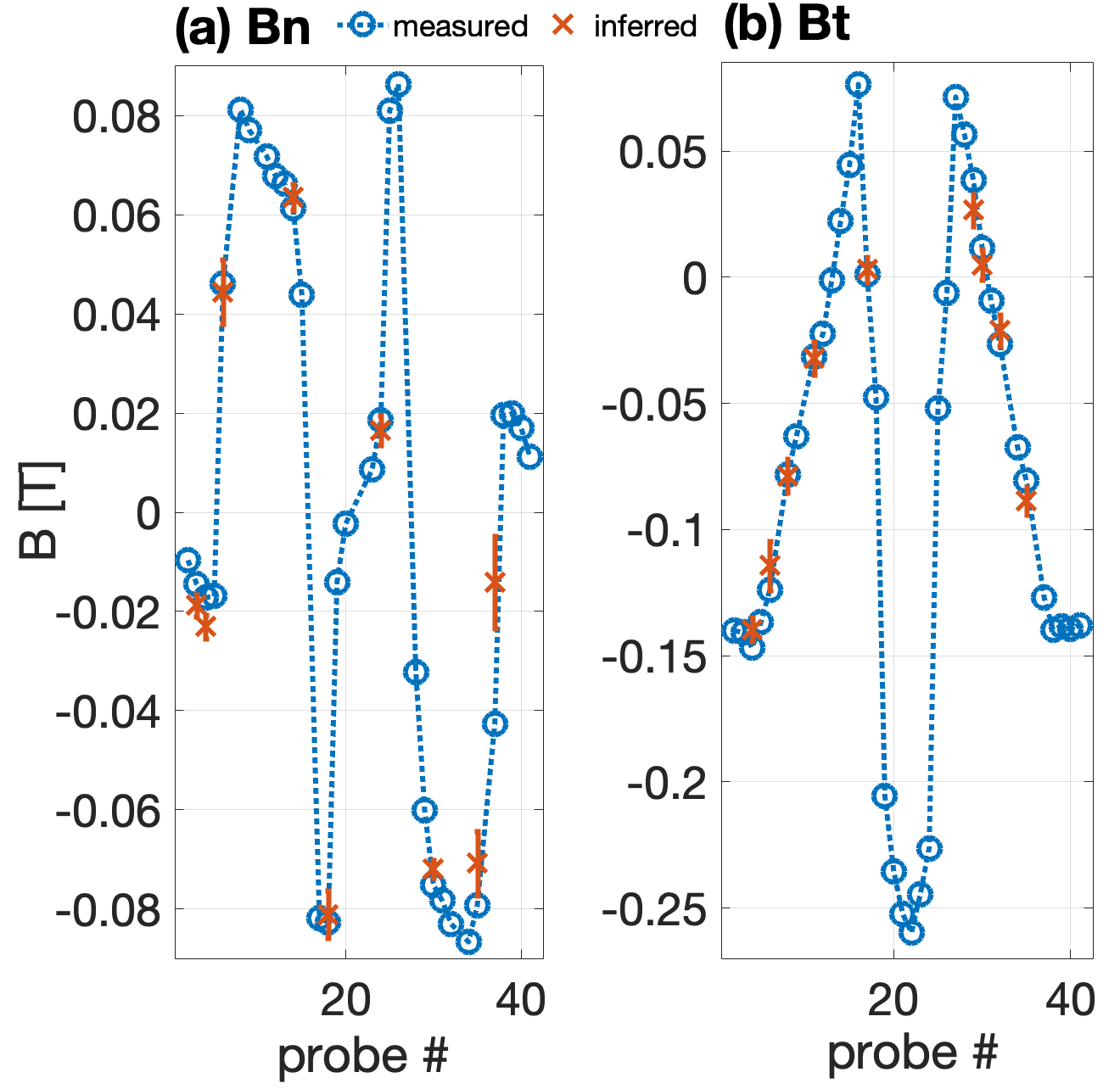}
    \caption{Measured (blue open circles) and inferred with the imputation method \cite{Joung:2018ju} (red crosses with their uncertainties) values for (a) $B_\textrm{n}$ and (b) $B_\textrm{t}$. Probe \# on the horizontal axis is used as an identification index of magnetic pick-up coils. Inferred probes are Probe \#3, 4, 6, 14, 18, 24, 30, 35, 37 for $B_\textrm{n}$ and Probe \#4, 6, 8, 11, 17, 29, 30, 32, 35 for $B_\textrm{t}$.}
    \label{fig:imp_example}
\end{figure}

\begin{table}[h]
\centering
\caption{The imputation results shown in Figure \ref{fig:imp_example} with KSTAR shot \#20341 at 2.1 sec.}
\label{tab:imputation}
\resizebox{0.48\textwidth}{!}{%
\begin{tabular}{crrcrr}
\hline \hline
\multicolumn{3}{c}{$B{_n}$ {[}T{]} $\times 10^{-2}$} & \multicolumn{3}{c}{$B{_t}$ {[}T{]} $\times 10^{-2}$} \\ 
\hline
No. & \multicolumn{1}{c}{Measured} & \multicolumn{1}{c}{Inferred} & No. & \multicolumn{1}{c}{Measured} & \multicolumn{1}{c}{Inferred} \\ \hline
3 & -1.45 & -1.88$\pm$0.22 & 4 & -14.69 & -13.97$\pm$0.47 \\
4 & -1.72 & -2.31$\pm$0.24 & 6 & -12.38 & -11.42$\pm$0.97 \\
6 & 4.62 & 4.45$\pm$0.65 & 8 & -7.82 & -7.88$\pm$0.67 \\
14 & 6.13 & 6.36$\pm$0.27 & 11 & -3.15 & -3.22$\pm$0.65 \\
18 & -8.27 & -8.11$\pm$0.48 & 17 & 0.10 & 0.30$\pm$0.52 \\
24 & 1.86 & 1.65$\pm$0.30 & 29 & 3.84 & 2.65$\pm$0.64 \\
30 & -7.52 & -7.19$\pm$0.18 & 30 & 1.15 & 0.49$\pm$0.61 \\
35 & -7.93 & -7.08$\pm$0.65 & 32 & -2.65 & -2.11$\pm$0.62 \\
37 & -4.27 & -1.41$\pm$0.93 & 35 & -8.07 & -8.87$\pm$0.55 \\ \hline \hline
\end{tabular}%
}
\end{table}

\begin{figure}
    \centering
    \includegraphics[width=0.46\linewidth]{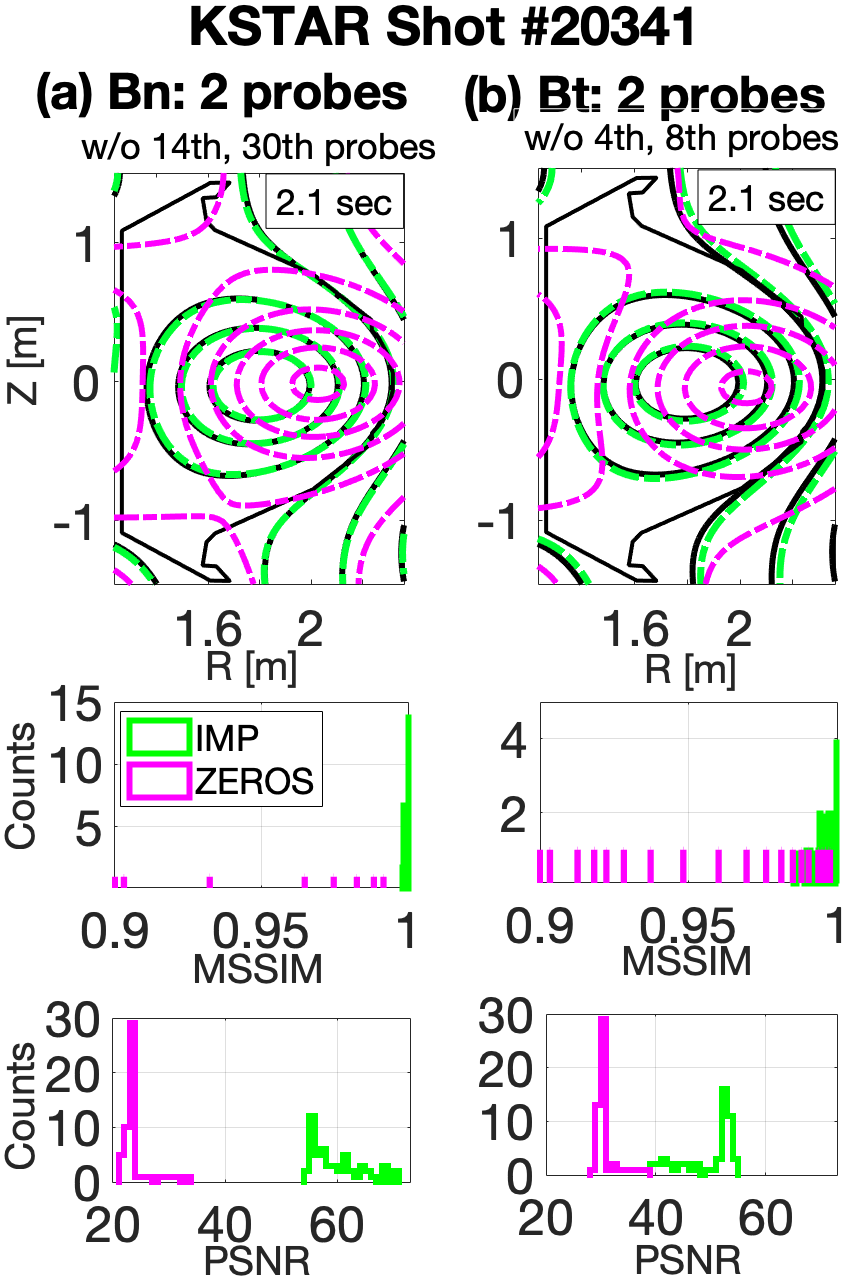}
    \caption{Top panel: nn-EFIT (NN$_{2017, 2018}$ network) reconstructed equilibria without any missing values (black line), and with two missing values replaced with the inferred values using the imputation method (green line) or with the zeros using the zero-padding method (pink dashed line), where the missing values are (a) $B_\textrm{n}$ Probe \#$14$ and $30$ (left panel) and (b) $B_\textrm{t}$ Probe \#$4$ and $8$ (right panel). Bottom panels: histograms of MSSIM and PSNR using the imputation method (green) and the zero-padding method (pink) for all the equilibria obtained from KSTAR shot \#20341, where the reference values are those obtained using nn-EFIT without any missing values. Note that there are many more counts less than $0.9$ for MSSIM with the zero-padding method. }
    \label{fig:imp_two}
\end{figure}

We have applied the imputation method to KSTAR shot \#$20341$ at $2.1$ sec for the normal ($B_\textrm{n}$) and tangential ($B_\textrm{t}$) components of the magnetic pick-up coils as an example. We have randomly chosen nine signals from the $32$ $B_\textrm{n}$ measurements and another nine from the $36$ $B_\textrm{t}$ measurements and pretended that all of them ($9+9$) are missing simultaneously. \reffig{fig:imp_example} shows the measured (blue open circles) and the inferred (red crosses with their uncertainties) values for (a) $B_\textrm{n}$ and (b) $B_\textrm{t}$. Probe \# on the horizontal axis is used as an identification index of the magnetic pick-up coils. \reftable{tab:imputation} provides the actual values of the measured and inferred ones for better comparisons. We find that the imputation method infers the correct (measured) values very well except Probe \#$37$ of $B_\textrm{n}$. Inferred (missing) probes are Probe \#3, 4, 6, 14, 18, 24, 30, 35, 37 for $B_\textrm{n}$ and Probe \#4, 6, 8, 11, 17, 29, 30, 32, 35 for $B_\textrm{t}$. Here, we provide all the Probe \#'s used for the neural network: $B_\textrm{n}$ Probe \#[2, $\dots$, 6, 8, 9, 11, $\dots$, 15, 17, $\dots$, 20, 23, $\dots$, 26, 28, $\dots$, 32, 34, 35, 37, $\dots$, 41] (a total of $32$) and $B_\textrm{t}$ Probe \#[2, $\dots$, 6, 8, 9, 11, $\dots$, 32, 34, 35, 37, $\dots$, 41] (a total of $36$).

\begin{figure}[t]
    \centering
    \includegraphics[width=0.46\linewidth]{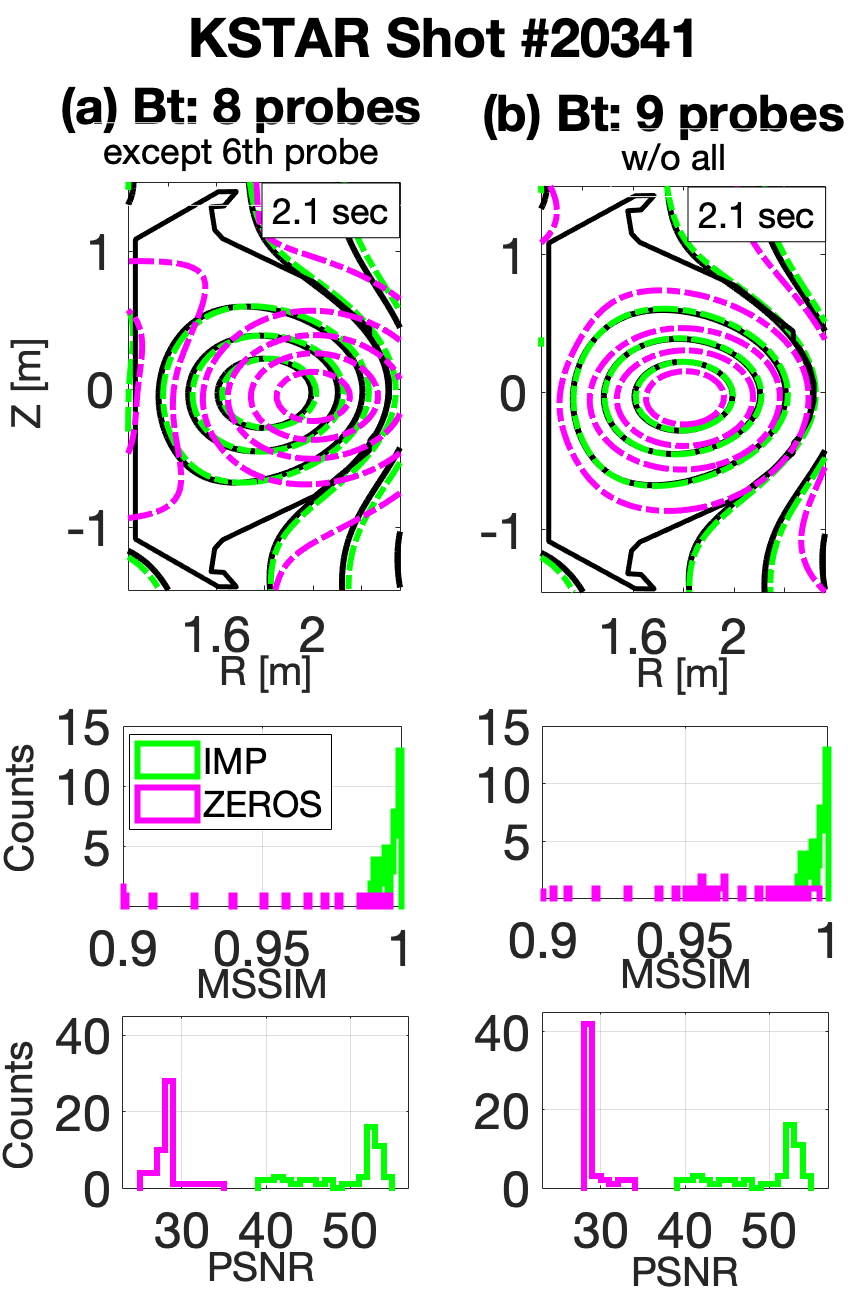}
    \caption{Same color code as in \reffig{fig:imp_two}. Missing values are (a) eight $B_\textrm{t}$ (except only Probe \#$6$) and (b) all nine $B_\textrm{t}$.}
    \label{fig:imp_many_Bt}
\end{figure}

Comparisons between the nn-EFIT without any missing values, which we treat as reference values, and the nn-EFIT with the imputation method or with the zero-padding method are made. Here, nn-EFIT results are obtained using the NN$_{2017, 2018}$ network. Top panel of \reffig{fig:imp_two} shows $\psi\lp R, Z\rp$ obtained from the nn-EFIT without any missing values (black line) and from the nn-EFIT with the two missing values replaced with the inferred values (green line), i.e., imputation method, or with zeros (pink dashed line), i.e., zero-padding method for (a) $B_\textrm{n}$ (left panel) and (b) $B_\textrm{t}$ (right panel) at $2.1$ sec of KSTAR shot \#20341. Probe \#14 and 30 for $B_\textrm{n}$ and Probe \#4 and 8 for $B_\textrm{t}$ are treated as the missing ones. Bottom panels compare histograms of MSSIM and PSNR using the imputation method (green) and the zero-padding method (pink) for all the equilibria obtained from KSTAR shot \#20341.

It is clear that nn-EFIT with the imputation method (green line) is not only much better than that with the zero-padding method (pink dashed line) but it also reconstructs the equilibrium close to the reference (black). In fact, the zero-padding method is too far off from the reference (black line) to be relied on for plasma controls.

\begin{figure}[t]
    \centering
    \includegraphics[width=0.46\linewidth]{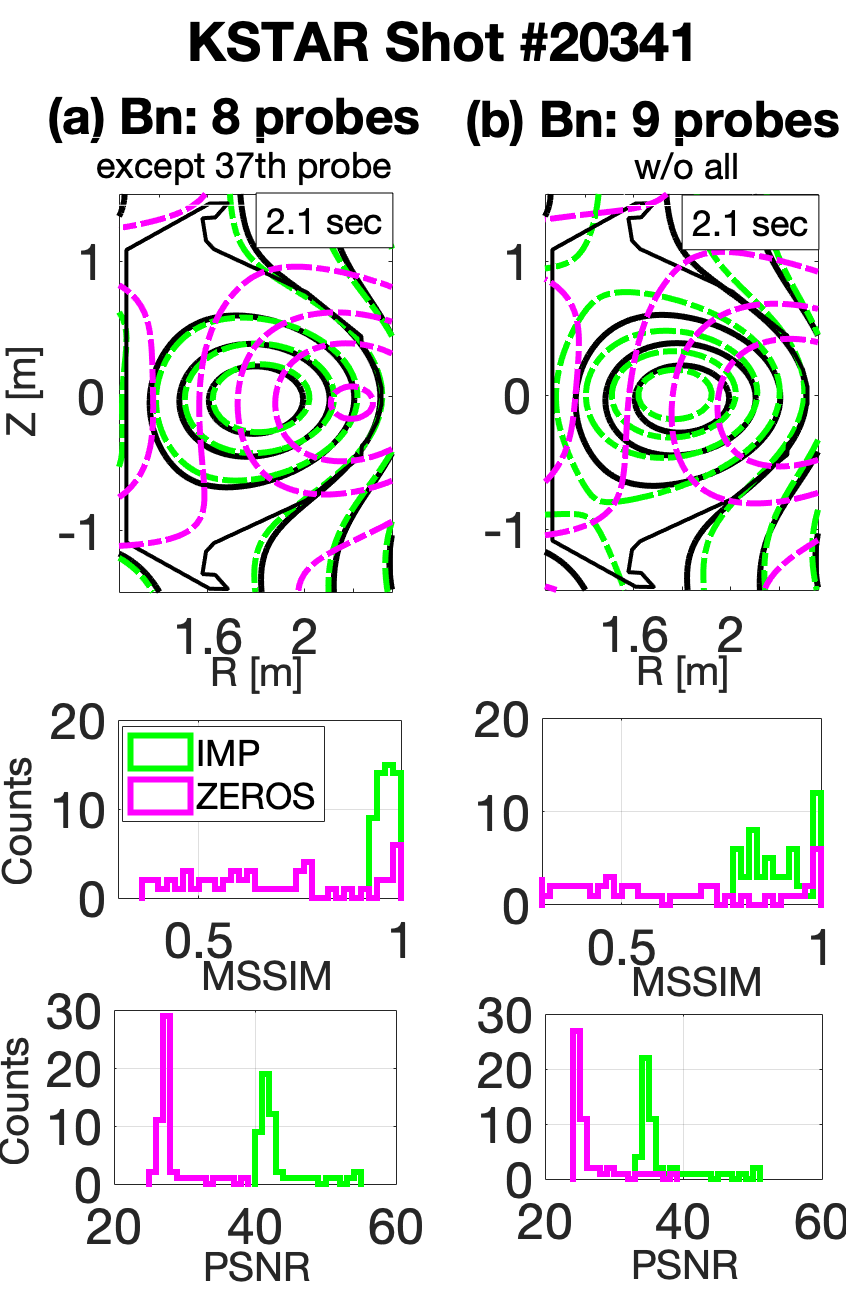}
    \caption{Same color code as in \reffig{fig:imp_two}. Missing values are (a) eight $B_\textrm{n}$ (except only Probe \#$37)$, (b) all nine $B_\textrm{n}$.}
    \label{fig:imp_many_Bn}
\end{figure}

Motivated by such a successful result of the nn-EFIT with the imputation method on the two missing values, we have increased number of missing values as shown in Figures \ref{fig:imp_many_Bt} and \ref{fig:imp_many_Bn} for the same KSTAR discharge, i.e., KSTAR shot \#20341. Let us first discuss \reffig{fig:imp_many_Bt} which are with (a) the eight (except only Probe \#6) and (b) nine (all) missing values of $B_\textrm{t}$. Color codes are same as in \reffig{fig:imp_two}, i.e., the reference is black, and nn-EFIT with the imputation method green or with the zero-padding method pink. It is evident that the nn-EFIT with the imputation method performs well at least up to nine missing values. Such a result is, in fact, expected since the imputation method has inferred the missing values well as shown in \reffig{fig:imp_example}(b) in addition to the fact that a well-trained neural network typically has a reasonable degree of resistance on noises. Again, the nn-EFIT with the zero-padding method is not reliable.

\begin{figure}[t]
    \centering
    \includegraphics[width=0.46\linewidth]{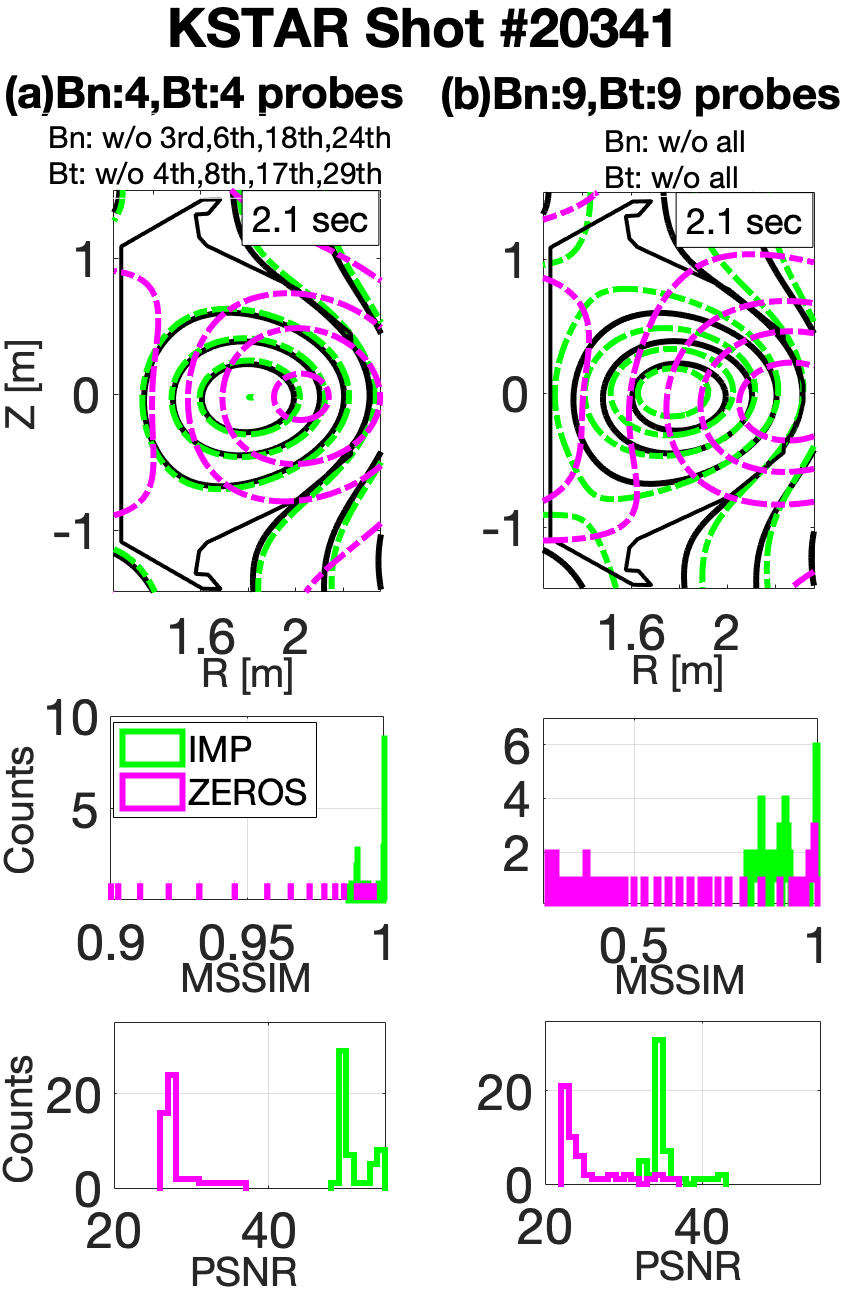}
    \caption{Same color code as in \reffig{fig:imp_two}. Combinations of missing $B_\textrm{n}$ and $B_\textrm{t}$ are examined: (a) four missing $B_\textrm{n}$ and four mssing $B_\textrm{t}$ case and (b) nine missing $B_\textrm{n}$ and nine missing $B_\textrm{t}$ case.}
    \label{fig:imp_many_comb}
\end{figure}

\reffig{fig:imp_many_Bn} (a) and (b) are results with the eight (except only Probe \#37) and nine (all) missing values of $B_\textrm{n}$, respectively. Color codes are same as in \reffig{fig:imp_two}. We find that the nn-EFIT with the eight missing values reconstructs the equilibrium similar to the reference one, while the reconstruction quality becomes notably worse for nine missing values. This is caused mostly due to poor inference of Probe \#$37$ by the imputation method (see \reffig{fig:imp_example}(a)). Nevertheless, the result is still better than the zero-padding method. \reffig{fig:imp_many_comb} shows the reconstruction results with the same color codes as in \reffig{fig:imp_two} when we have (a) $4+4$ and (b) $9+9$ combinations of $B_\textrm{n}$ and $B_\textrm{t}$ missing values simultaneously.

All these results suggest that the nn-EFIT with the imputation method reconstructs equilibria reasonably well except when the imputation infers the true value poorly, e.g., $B_\textrm{n}$ Probe \#37 in \reffig{fig:imp_example}(a) and \reftable{tab:imputation}. In fact, the suggested imputation method \cite{Joung:2018ju} infers the missing values based on the neighboring intact values (using Gaussian processes) while satisfying the Maxwell's equations (using Bayesian probability theory). Consequently, such a method becomes less accurate if (1) the neighboring channels are also missing \textit{AND} (2) the true values change fast from the neighboring values. In fact, $B_\textrm{n}$ Probe \#37 happens to satisfy these two conditions, i.e., Probe \#35 is also missing, and the true values of Probe \#35, \#37 and \#38 are changing fast as one can discern from \reffig{fig:imp_example}(a).

\subsection{Discussions and Conclusions} \label{S6:con}

\noindent
We have developed and presented the neural network based Grad-Shafranov solver constrained with the measured magnetic signals. The networks take the plasma current from a Rogowski coil, 32 normal and 36 tangential components of the magnetic fields from the magnetic pick-up coils, 22 poloidal fluxes from the flux loops, and $\lp R, Z\rp$ position of the interest as inputs. With three fully connected hidden layers consisting of 61 nodes each layer, the network outputs a value of poloidal flux $\psi$. We set the cost function used to train the networks to be a function of not only the poloidal flux $\psi$ but also the Grad-Shafranov equation $\Delta^*\psi$ itself. The networks are trained and validated with $1,118$ KSTAR discharges from 2017 and 2018 campaigns.

Treating the off-line EFIT results as accurate magnetic equilibria to train the networks, our networks fully reconstruct magnetic equilibria, not limited to obtaining selected information such as positions of magnetic axis, X-points or plasma boundaries, more similar to the off-line EFIT results than the rt-EFIT is to the off-line EFIT. Owing to the fact that $\lp R, Z\rp$ position is a part of the input, our networks have adjustable spatial resolution within the first wall. The imputation method supports the networks to obtain the nn-EFIT results even if there exist a few missing inputs. 

As all the necessary computation time is approximately $1$ msec, the networks have potential to be used for real-time plasma control. In addition, the networks can be used to provide large number of automated EFIT results fast for many other data analyses requiring magnetic equilibria.

\newpage
\section{Article \RNum{5}: Plasma reconstruction via unsupervised learning}

\noindent
This approach deals with GS-DeepNet: Mastering tokamak plasma equilibria with deep neural networks and the Grad-Shafranov equation\footnote{\textsc{S. J{\small oung}, Y.{\small -c}. G{\small him}, J. K{\small im}, S. K{\small wak}, S. L{\small ee}, D. K{\small won}, H.S. K{\small im}, J.G. B{\small ak}} \\
\textit{Science Advances}, (2022), in preparation}, which is largely taken from Ref \cite{joung2022deep}, as a part of \textit{reconstruction of plasma equilibria via deep neural networks}.

This article embodies the essence of this thesis, taking advantage of all the principles and methods described previously, in order to reconstruct plasma equilibria in a magnetic confinement fusion experiment via deep neural networks. KSTAR serves as a test bed for an application of this approach. As the networks demonstrated the fact that the plasma equilibrium can be encoded in the network architecture, this article taking one step forward brings up a question that neural networks can directly solve plasma equilibria by themselves unlike Article \RNum{4}. 

Here, solving plasma equilibria denotes obtaining plasma equilibria by solving the Grad-Shafranov equation. Since this approach does not rely on training database containing solutions of the Grad-Shafranov equation, i.e., the EFIT database used in Article \RNum{4}, a large number of different measurement data employed in current reconstruction algorithms are collected for the networks such as the magnetic pick-up coils, the poloidal flux loops, KSTAR Thomson scattering system and KSTAR Charge exchange spectroscopy system as well as Motional Stark Effect system (albeit not being introduced). Since the Grad-Shafranov equation is derived from both the plasma force equation with equilibrium assumption and Maxwell's equations, two neural network architectures take charge of each contribution for reasonably solving the whole equations. Furthermore, two additional modules are introduced to determine a plasma boundary (dividing a plasma region from a vacuum area) since solving the Grad-Shafranov equation is a free-boundary problem, and using only the networks is not enough to resolve it. From the modelling for the relations between tokamak current sources and the measured magnetic signals, the networks can learn various plasma equilibria via unsupervised learning manner and the gradient descent, meanwhile kinetic profiles of plasmas are optimally found by the network themselves. After the training is done, the networks can produce plasma equilibria in real time which can be used for tokamak operations. Note that Article \RNum{5} applies all the techniques suggested in Article \RNum{1}--\RNum{4}.

\subsection{Introduction} \label{S1:gsintro}

\noindent
The ultimate goal of scientific and engineering research in the field of nuclear fusion is to build a power plant producing sustainable and clean electricity through fusion reactions from a confined plasma heated up to $\sim$100 million degrees in Celsius \cite{cite-keyVerberk, Chen:2011wc}. A tokamak is a torus-shape vacuum vessel within which the plasma is confined by magnetic fields directed along the long (toroidal) and short (poloidal) way around the torus. In order to maintain such a high-temperature plasma for a long period of time (for instance, more than 400 seconds \cite{aymar2002iter}), it is necessary to balance the plasma pressure and the Lorentz force within the entire plasma volume \cite{Freidberg:1987} during a tokamak operation. This means that spatial structures of plasma pressure and magnetic fields must be known in real time. 

It is often hard to make direct \textit{in situ} measurements of the plasma structures due to its harsh environments, e.g., high temperature and radiation. Although there are some optics systems directly measuring internal information such as electron temperature and density \cite{cite-key} and magnetic pitch-angle \cite{levinton1989magnetic}, these measurements are spatially localized and require a magnetic field structure for the measured data to be mapped onto a whole plasma volume. Hence, a suite of magnetic diagnostics \cite{Strait:2008}, fundamental measurement devices installed on a tokamak wall far from the plasma, are used to obtain the magnetic field structures indirectly by solving the Grad-Shafranov (GS) equation \cite{grad1958hydromagnetic, shafranov1966plasma}. The GS equation describes a force balanced plasma state conforming to Maxwell’s equations with a toroidal axisymmetry assumption, thus finding a solution to the GS equation is regarded as reconstructing the magnetohydrodynamic (MHD) equilibrium of a toroidal plasma.

The GS equation, resembling the Hicks equation \cite{hicks1899ii} which describes axisymmetric inviscid fluid, is a two-dimensional (poloidal cross-section), nonlinear, elliptic partial differential equation. Owing to its nonlinearity, finding a solution to the GS equation typically requires an iterative numerical approach. Complicating the problem even further, it is an inverse and free boundary problem as only external measurements of magnetic fields are often only available. These difficulties hinder a real-time application of the GS equation. Of course, a simple resolution for the real-time application is to sacrifice accuracy of the solution as in Ref. \cite{Ferron:2002fw}. But, even if accuracy is eschewed, there exist human expert choices for numerical convergence. Current numerical algorithms of the reconstruction, chiefly EFIT \cite{Lao:1985hn}, often require decisions made by human experts in manually choosing measured magnetic data. Those neglected data do not participate in reconstructing a plasma equilibrium, i.e., in finding a solution to the GS equation, as they tend to obstruct finding a converged numerical solution. 

There have been attempts to parallelize the numerical algorithms based on GPUs \cite{Yue:2013cj, huang2020gpu} or to use a supervised deep neural network \cite{joung2019deep}, which fulfil real-time demand but human decisions as they are all based on the EFIT algorithm. Contrarily, reconstruction methods using Bayesian inference \cite{von2013unified,romero2018inference,kwak2020thesis} were introduced to eliminate (or at least explicitly articulate) manual selections, but they are unlikely to be used for real-time purpose due to their required heavy computational time. We note that reconstructing more detailed plasma equilibria in real time using internal information is also an active research area \cite{carpanese2020first, berkery2021kinetic}.

As recent scientific computing is highly supported by deep learning \cite{lecun2015deep}, there have been various approaches for neural networks to learn physics-based differential equations such as solving the many-electron Schr{\"o}dinger equation \cite{pfau2020ab, hermann2020deep}, the Navier-Stokes equation \cite{raissi2020hidden} and an atmospheric model for climate modelling \cite{beucler2021enforcing}. Other examples include interpolating partial differential equations \cite{vanMilligen:1995dv,bar2019learning,kochkov2021machine} and regularizing neural networks with the Kohn-Sham equations \cite{li2021kohn}. These previous works require actual solutions \cite{bar2019learning, kochkov2021machine}, prior knowledge on some of the unknown parameters \cite{raissi2020hidden,beucler2021enforcing,vanMilligen:1995dv}, or approximated solution states \cite{pfau2020ab, hermann2020deep, li2021kohn} of the target governing equations. There were attempts of solving the GS equation using neural networks \cite{vanMilligen:1995dv, kaltsas2022neural}; however, they work only if entire internal profiles over a plasma volume are given with a fixed boundary condition, e.g., from a numerical code, VMEC \cite{vanMilligen:1995dv,hirshman1983steepest}, or prescribed polynomial-based functions \cite{kaltsas2022neural}, which may not be applicable for many of existing tokamaks. Besides, there was an approach \cite{wang2021deep} for neural networks to solve a Stefan problem \cite{stefan1891theorie} which is a free-boundary problem and describes a phase-change between a liquid and a solid state. However, the method assumed that a boundary of the phase-change between the states is already known.

We propose an algorithm, Grad-Shafranov Deep Neural Networks (GS-DeepNet), that learns plasma equilibria solely via unsupervised learning without existing numerical algorithms. GS-DeepNet does not depend on several aspects in both current reconstruction methods and other neural networks solving differential equations. First and foremost, it is trained by self-teaching unsupervised learning, without use of any guess of solutions. Only known information is the GS equation and externally (and locally) measured data with no manual selections. Second, it uses typical fully-connected neural networks known as retaining real-time plausibility. Finally, it uses an auxiliary module that detects boundary information based solely on network outputs. To reach these outcomes, we develop neural networks that are capable of solving a nonlinear elliptic partial differential equation in a free-boundary and inverse condition, i.e., the GS equation. As a simple survey, we introduce that a neural network can solve a first-order linear differential equation, which is discussed in Appendix.


\subsection{Modelling Grad-Shafranov Deep Neural Networks} \label{S2:gsdd}

\begin{figure}[!]
    \centering
    \includegraphics[width=0.835\linewidth]{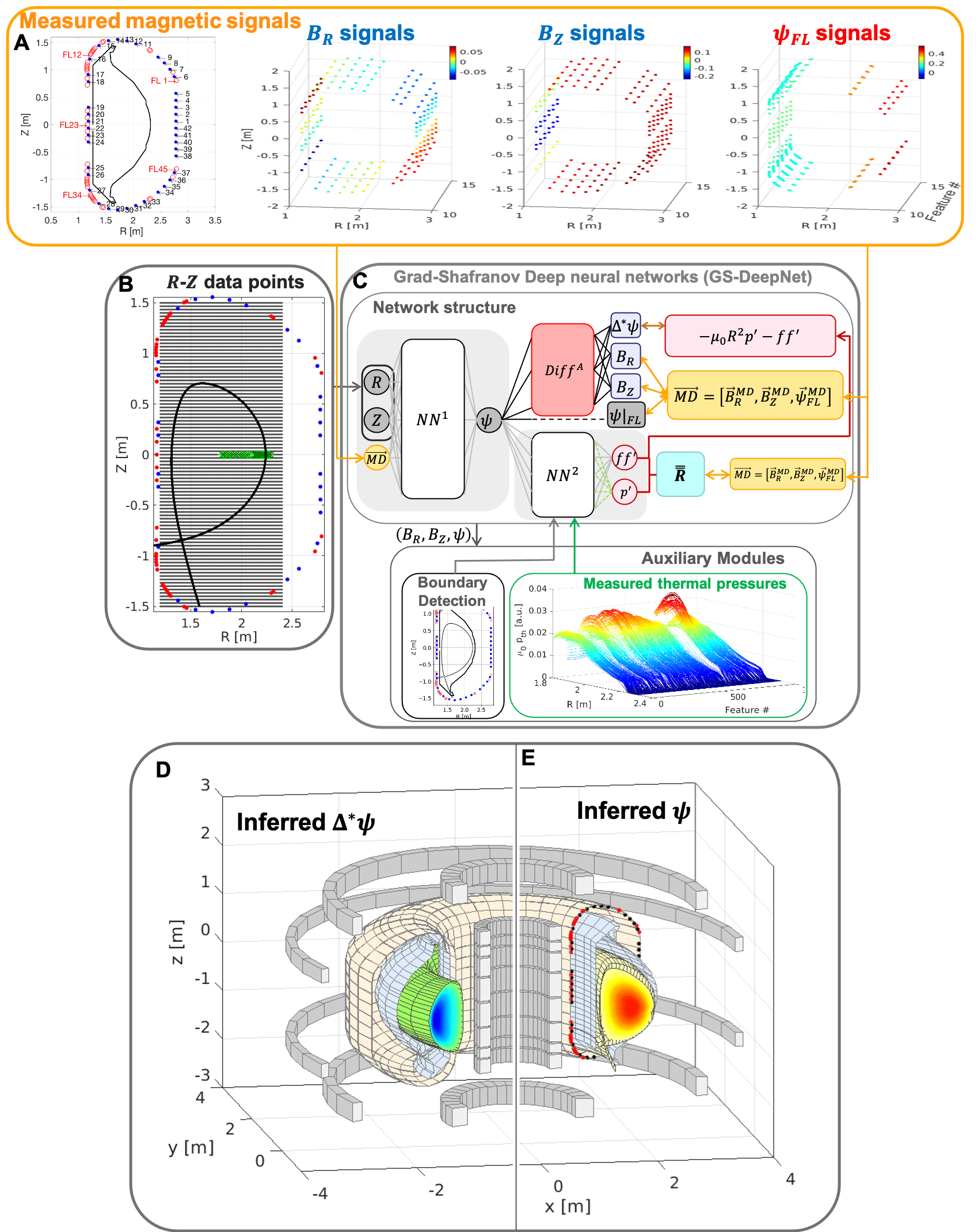}
    \caption[Self-teaching unsupervised learning scheme in GS-DeepNet]{\textbf{Self-teaching unsupervised learning scheme in GS-DeepNet}. (A) The locations and sensor numbers of the KSTAR magnetic diagnostics. Colors represent the magnitude of the data. (B) $41\times41$ grid points where the plasmas are potentially generated. (C) Schematic representation of GS-DeepNet. We call the bundle of $NN_{\Theta}^{1}$ and $Diff^{A}$ the Maxwell-Net, and $NN_{\theta}^{2}$ is named as the Force-Balance (FB) Net. (D--E) The three-dimensional configurations of $\Delta^{*}\psi$ and $\psi$ from GS-DeepNet. KSTAR poloidal field coils (gray), the vacuum vessel wall (orange) and the plasma facing components (blue) are also shown.}
    \label{fig:gs1}
\end{figure}

\noindent
Our novel algorithm GS-DeepNet has two deep neural networks $NN_{\Theta}^{1}$ and $NN_{\theta}^{2}$ with parameters $\Theta$ and $\theta$, respectively. The goal of it is to discover the poloidal flux function, $\psi$, a solution of the GS equation on the spatial positions shown in Fig \ref{fig:gs1}B under a given measurement state as its input.

Figure \ref{fig:gs1}B shows $41\times41$ grid points where a plasma potentially exists and the positions of the magnetic diagnostics which obtains radial and axial components of the poloidal magnetic field ($B_{R}$ and $B_{Z}$) and the poloidal magnetic flux ($\psi_{FL}$). Taking a single spatial point $(R,Z)$ among the spatial positions and a set of the magnetic data $\overrightarrow{MD}=(\vec{B}_{R},\vec{B}_{Z},\vec{\psi}_{FL})$ as an input, $NN_{\Theta}^{1}$ outputs a flux function, $\psi=NN_{\Theta}^{1}(R,Z,\overrightarrow{MD})$. The vector of $\vec{B}_{(R(Z))}$ (or $\psi_{FL}$) means a feature of all obtained $\vec{B}_{(R(Z))}$ (or $\psi_{FL}$) from its measurement locations at a single time slice during a tokamak operation (Fig. \ref{fig:gs1}A). This network output is fed into $NN_{\theta}^{2}$ which outputs both a plasma pressure gradient and a quantity related to the toroidal magnetic field, $(dp/d\psi,f df/d\psi)=(p',ff')=NN_{\theta}^{2}(\psi)$. Here, $p$ represents the plasma pressure, and the toroidal field $B_{\phi}$ can turn into $f=R B_{\phi}$ which is related to the poloidal plasma current. $(p',ff')$ is the variables of the GS equation (see equation 4). Both neural networks have multiple fully-connected layers \cite{lecun2015deep} with dropout \cite{srivastava2014dropout} and swish nonlinear activation functions \cite{ramachandran2017searching} (See Materials and Methods).

Since we do not depend on existing numerical algorithms, there are not any guesses of the solution $\psi$ to train $NN_{\Theta}^{1}$. Instead, GS-DeepNet teaches itself by an unsupervised learning algorithm following the GS equation and the magnetic measurements. With a given measurement set $\overrightarrow{MD}$, $NN_{\Theta}^{1}$ generates $\boldsymbol{\psi}$ all over the spatial positions $(\boldsymbol{R},\boldsymbol{Z})$ where $\boldsymbol{\psi}$ is the vector representation of $\psi$ consistent with the vector $(\boldsymbol{R},\boldsymbol{Z})$, and $(\boldsymbol{R},\boldsymbol{Z})$ is the vector representation for all the points in Fig. \ref{fig:gs1}B. Passing through the automatic differential operator \cite{paszke2017automatic, baydin2018automatic} ($Diff^{A}$ in Fig. \ref{fig:gs1}C), the flux functions $\boldsymbol{\psi}$ turn into $\boldsymbol{B}_{R}(=-1/\boldsymbol{R}\cdot\partial\boldsymbol{\psi}/\partial Z)$, $\boldsymbol{B}_{Z}(=1/\boldsymbol{R}\cdot\partial\boldsymbol{\psi}/\partial R)$ and $\Delta^{*}\boldsymbol{\psi}\{=(\partial^{2}/\partial R^{2}+\partial^{2}/\partial Z^{2} - 1/\boldsymbol{R}\cdot\partial/\partial R)\boldsymbol{\psi}\}$ based on Maxwell’s equations, where the symbol $\cdot$ represents the element-wise product.

A plasma is usually placed inside a boundary called the plasma boundary (Fig. \ref{fig:gs1}B). This boundary dividing the plasma region from the vacuum area cannot be defined until the solution $\psi$ is prepared. The pressure $p$ and the poloidal current function $f$ of the plasma are ideally defined within the plasma region. Thus, after locating the boundary through an auxiliary module for boundary detection (Fig. \ref{fig:gs1}C) based on $(\boldsymbol{B}_{R},\boldsymbol{B}_{Z},\boldsymbol{\psi})$, $NN_{\theta}^{2}$ generates $(\boldsymbol{p}',\boldsymbol{ff}')$ by using the flux functions inside the plasma, $\boldsymbol{\psi}_{in}$, where $(\boldsymbol{p}',\boldsymbol{ff}')$ is also the vector representations for $(p',ff')$. Then, $(\boldsymbol{p}',\boldsymbol{ff}')$ forms $-\mu_{0}\boldsymbol{R}^{2}\cdot\boldsymbol{p}' - \boldsymbol{ff}'(=\Delta^{*}\boldsymbol{\psi}_{in})$ based on the force balance equation. This is also related to the toroidal plasma current density $J_{\phi}=-\Delta^{*}\psi/(\mu_{0}R)$ where $\mu_{0}$ is the vacuum permeability. Since $-\mu_{0}\boldsymbol{R}^{2}\cdot\boldsymbol{p}' - \boldsymbol{ff}'$ is constrained with the given magnetic data $\overrightarrow{MD}$ via a response matrix $\bar{\bar{\boldsymbol{R}}}$ estimated based on the Biot-Savart law, it may be viewed as a potential guess of the GS equation. Thus, the main concept of the unsupervised learning algorithm is that $NN_{\Theta}^{1}$ is repeatedly taught by $NN_{\theta}^{2}$ which takes $\psi$ from $NN_{\Theta}^{1}$ as an input. We call the bundle of $NN_{\Theta}^{1}$ and the differential operator $Diff^{A}$ the Maxwell-Net, and $NN_{\theta}^{2}$ is named as the Force-Balance (FB) Net.

The networks’ parameters are updated to keep their $\Delta^{*}\boldsymbol{\psi}$ matched with each other. Since we theoretically have a vacuum on the outside of the plasma boundary, $NN_{\Theta}^{1}$ is updated to have null current densities for $\Delta^{*}\boldsymbol{\psi}_{out}$ estimated from $\boldsymbol{\psi}_{out}$, the flux functions outside the plasma. Furthermore, the Maxwell-Net outputs corresponding to the measurement locations, $(\boldsymbol{B}_{R},\boldsymbol{B}_{Z},\boldsymbol{\psi})_{md}$, are trained to match $\overrightarrow{MD}$ as the initial value of a differential equation. The pipeline of this self-teaching procedure is presented in Fig. \ref{fig:gs1}C.

The Biot-Savart law explains a relationship between a magnetic field (or flux) and its corresponding current source. This relationship can depend on a single independent variable, i.e., a magnitude of the current source in Ampere if the locations of the magnetic field (or flux) and current source are fixed, and the source carries a constant current at a certain time. Thus, we treat the toroidal current density $J_{\phi}$ of the plasma as a constant current source, and each $J_{\phi}$ on a single $R$-$Z$ grid position is modelled with an arbitrary three-dimensional volumetric current beam having rectangular cross-section (See fig. S4). Thus, we can pre-calculate a contribution of $J_{\phi}$ generating a magnetic field (or flux) at a measurement position as $r_{ij}$ where the subscript $i$ and $j$ represent the indices of the magnetic sensor locations and $\boldsymbol{J}_{\phi}$ over all $41\times41$ grid points, respectively. We estimate $r_{ij}$ for all grids at first due to undefined plasma boundary. This contribution $r_{ij}$ can be contained in a matrix called the response matrix for the plasma, i.e., $\bar{\bar{\boldsymbol{R}}}_{p}$ whose matrix size is $N_{md}\times41^{2}$ where $N_{md}$ is the size of $\overrightarrow{MD}$. After the boundary is detected via the auxiliary module, the matrix $\bar{\bar{\boldsymbol{R}}}_{p}$ is reduced to $\bar{\bar{\boldsymbol{R}}}_{p,in}$ whose size is $N_{md}\times N_{in}$ where $N_{in}$ is the number of the grid points inside the boundary.

A tokamak has external field coils that we control to generate the poloidal magnetic field (see Fig. \ref{fig:gs1}D--E). With the identical procedure with the plasma, we pre-calculate the contributions of the external coils as $\bar{\bar{\boldsymbol{R}}}_{ext}$. We also pre-estimate the contributions of the vessel currents (currents induced in tokamak structures such as the vacuum vessel wall) which cannot be negligible \cite{lee2008magnetic, park2011kstar} as $\bar{\bar{\boldsymbol{R}}}_{VV}$. Therefore, we can relate $\overrightarrow{MD}$ with every current source in the tokamak, i.e., $\overrightarrow{MD}=\bar{\bar{\boldsymbol{R}}} \bar{\bar{\boldsymbol{I}}}_{\phi}$ where $\bar{\bar{\boldsymbol{R}}}=(\bar{\bar{\boldsymbol{R}}}_{p,in}, \bar{\bar{\boldsymbol{R}}}_{ext}, \bar{\bar{\boldsymbol{R}}}_{VV})$ with $\bar{\bar{\boldsymbol{I}}}_{\phi}$ containing $J_{\phi}$, $J_{ext}$ and $J_{VV}$ as a column vector. The current densities of the external coils $J_{ext}$ are known, while the vessel currents $J_{VV}$ are required to be inferred \cite{urano2020breakdown} during the unsupervised training procedure. Further technical details are described in Materials and Methods.

GS-DeepNet includes two auxiliary modules for plasma boundary detection and locally measured plasma pressure. Because solving the GS equation is a free-boundary problem, the plasma boundary cannot be determined until the flux function $\psi$ is defined all over the grid points. The boundary can be regarded as an outermost line which connects $R$-$Z$ positions whose flux functions are identical to one another as well as encloses a whole plasma area. Thus, the boundary module detects the $R$-$Z$ positions by searching $\psi$ at the boundary spots based on the Maxwell-Net outputs (see Methods for details).

As explained before, the FB-Net outputs $(p',ff')$ by taking $\psi_{in}$, and its parameters are updated based on the magnetic data $\overrightarrow{MD}$ by forming $-\mu_{0}R^{2}p' - ff'$. We came up with this update procedure in case that measured data to train the network outputs directly is not available. Thus, the pressure module is used when measured plasma pressure $p^{m}$ is available although it is spatially localized in the fixed $R$-$Z$ positions (see Fig. \ref{fig:gs1}B). To use the measured pressure $p^{m}$ in order to train the FB-Net $p'$, the module performs Gaussian process regressions \cite{Rasmussen:2006} (a non-parametric regression) to the measured pressure $p^{m}$ \cite{chilenski2015improved, ho2019application} and subsequently estimates derivatives of the regressed pressure $p^{m,GP}$ with respect to $\psi$ which is given by the Maxwell-Net. Similarly, the poloidal current function $f$ can be deduced from measurements for the local pitch angle between the toroidal and the poloidal magnetic fields \cite{levinton1989magnetic}, which can be used to train the FB-Net $ff'$ with using Gaussian process regressions. This is left as future works since we here consider a way how to use spatially localized data in GS-DeepNet, and it is presumably sufficient to show the method for the measured pressure.

The neural networks in GS-DeepNet are trained by the self-teaching unsupervised training algorithm fundamentally based on the GS equation. First and foremost, we initialize both networks with random parameters, $\Theta_{0}$ and $\theta_{0}$, respectively. At every iteration $i\geq1$ and each feature $t$, a guess of the solution $\boldsymbol{\psi}=NN_{\Theta_{i-1}}^{1}(\boldsymbol{R},\boldsymbol{Z},\overrightarrow{MD}_{t})$ is generated, and $(\boldsymbol{B}_{R,t},\boldsymbol{B}_{Z,t},\boldsymbol{\psi}_{t},\Delta^{*}\boldsymbol{\psi}_{t})$ is estimated using the automatic differential operator $Diff^{A}$. After a plasma boundary is determined (as well as preprocessing of $p_{t}^{m}$ is done), $(\boldsymbol{p}_{t}',\boldsymbol{ff}_{t}')=NN_{\theta_{i-1}}^{2}(\boldsymbol{\psi}_{in,t})$ is generated, and $(-\mu_{0}\boldsymbol{R}^{2}\cdot\boldsymbol{p}_{t}'-\boldsymbol{ff}_{t}',\bar{\bar{\boldsymbol{I}}}_{\phi,t},\boldsymbol{p}_{t}')$ is prepared. With a randomly sampled feature from the total feature space, renewed network parameters $\Theta_{i}$ are trained by $(\boldsymbol{B}_{R},\boldsymbol{B}_{Z},\boldsymbol{\psi},\Delta^{*}\boldsymbol{\psi})$ compared to $(\overrightarrow{MD}, -\mu_{0}\boldsymbol{R}^{2}\cdot\boldsymbol{p}'-\boldsymbol{ff}' (or \text{nulls}))$, while new network parameters $\theta_{i}$ are updated by $(-\mu_{0}\boldsymbol{R}^{2}\cdot\boldsymbol{p}'-\boldsymbol{ff}',\bar{\bar{\boldsymbol{I}}}_{\phi},\boldsymbol{p}')$ compared to $(\overrightarrow{MD},p^{m,GP})$ using the response matrix $\bar{\bar{\boldsymbol{R}}}$. We use gradient descent for training the parameters $\Theta$ and $\theta$ by means of loss functions $l_{1}$:
\begin{equation}
\label{eq:l1}
l_{1} = \Big\{(\boldsymbol{B}_{R},\boldsymbol{B}_{Z},\boldsymbol{\psi})_{md} - \overrightarrow{MD} \Big\}^{2} + (\Delta^{*}\psi_{in} + \mu_{0}R^{2}p'+ff')^{2} + (\Delta^{*}\psi_{out})^2 + c_{1}||\Theta||^{2}
\end{equation}
and $l_{2}$, respectively:
\begin{equation}
\label{eq:l2}
l_{2} = (\bar{\bar{\boldsymbol{R}}}\bar{\bar{\boldsymbol{I}}}_{\phi} - \overrightarrow{MD})^{2} + \Bigg\{\alpha\sum_{i=1}^{N_{in}}\bigg(R_{i}p_{i}' - \frac{ff_{i}'}{R_{i}\mu_{0}} \bigg) - I_{P}^{m} \Bigg\}^{2} + c_{2}|\theta| + (p' - p^{m, GP})^2
\end{equation}
where $l_{1}$ and $l_{2}$ are averaged over the mean-squared errors, $c_{1}$ and $c_{2}$ are the coefficients for L2 and L1 weight regularizations, respectively, to avoid overfitting, and $\alpha\sum_{i=1}^{N_{in}}(R_{i}p_{i}' - ff_{i}'/R_{i}\mu_{0})$ is the sum of the toroidal current densities multiplied by the area of the rectangular cross-section $\alpha$, which turns out to be the total plasma current $I_{P}$. $I_{P}^{m}$ is the measured total plasma current by the Rogowski coil \cite{Strait:2008}. The last term of $l_{2}$ are only used when the measured pressure $p^{m}$ is usable.

As an example, three-dimensional configurations of $\Delta^{*}\psi$ and $\psi$ from the Maxwell-Net are shown in Fig. \ref{fig:gs1}D and E, respectively, with some tokamak structures of the KSTAR \cite{kim2009design}, one of the first research tokamaks with fully superconducting magnets in the world.

\begin{figure}[!]
    \centering
    \includegraphics[width=0.85\linewidth]{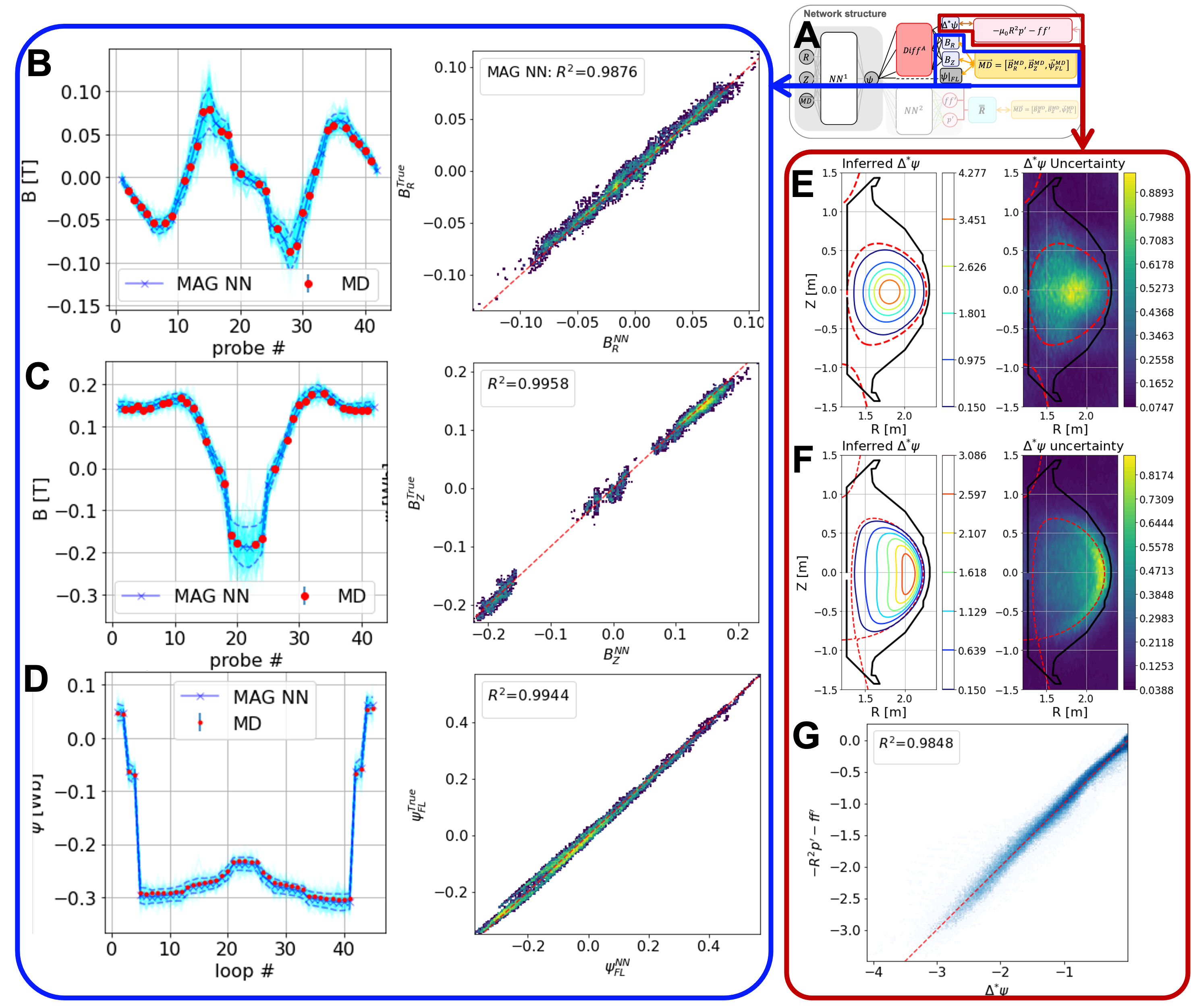}
    \caption[Statistical evaluation of GS-DeepNet training]{\textbf{Statistical evaluation of GS-DeepNet training}. (A) Blue box: the comparison of the initial values of the Maxwell-Net with the magnetic dataset. Red box: the comparison of the Maxwell-Net $\Delta^{*}\psi$ with the FB-Net $\Delta^{*}\psi$. (B--D) The Maxwell-Net $B_{R}$, $B_{Z}$ and $\psi$. (E--F) Left: the configurations of the Maxwell-Net $\Delta^{*}\psi$ for a limited and diverted plasmas, respectively. Right: the uncertainty configurations of the Maxwell-Net $\Delta^{*}\psi$. They show their structural consistencies with $\Delta^{*}\psi$ such that if $\Delta^{*}\psi$ increases, then uncertainty increases with it. (G) $R^{2}$ for $\Delta^{*}\psi$ between the Maxwell-Net (x-axis) and the FB-Net (y-axis).}
    \label{fig:gs2}
\end{figure}

\subsection{Statistical analysis of GS-DeepNet training}

\noindent
Our unsupervised training scheme was performed to train our algorithm GS-DeepNet. From the completely random network parameters, the training continued until it was terminated by the early stopping method (a regularization method to avoid losing generalization for unseen features) \cite{morgan1989generalization}, which approximately took a day.

We collected 50 experimental plasma discharges of the KSTAR. A plasma discharge (fig. S2A--E) represents a series of experimental procedures: first, the external magnetic field from the external coils is built up; then, a plasma is initiated (or discharged) and controlled by the external magnetic field until it is eventually disappeared due to mechanical and physical reasons such as the maximum limit of coil currents. A typical discharge length of the KSTAR is of the order of 10$^{1\text{-}2}$ seconds. Among all time-steps of 50 discharges, we chose $\sim10^{4}$ time slices (features) for the magnetic and pressure measurements. With approximate $2\times10^{2}$ spatial $R$-$Z$ positions, network parameters were dealt with about $2\times10^{6}(=2\times10^{2}\times10^{4})$ dataset for covering combination of the total features as well as all the $R$-$Z$ points. 80\%, 5\% and 15\% of the dataset went to the training, validation and test datasets, respectively.

Figure \ref{fig:gs2} shows the statistical evaluations of GS-DeepNet training. We did not use solutions of the GS equation calculated from an existing numerical algorithm to train the networks, rather we used the magnetic measurements to constrain the initial values of the network solutions and the self-teaching unsupervised learning for the networks to acquire the knowledge of the GS equation. Thus, as instructed in Fig. \ref{fig:gs2}A, we compared the Maxwell-Net $(\boldsymbol{B}_{R},\boldsymbol{B}_{Z},\boldsymbol{\psi})_{md}$ and $\Delta^{*}\boldsymbol{\psi}$ with the obtained magnetic data $\overrightarrow{MD}$ and the FB-Net $\Delta^{*}\boldsymbol{\psi}(=-\mu_{0}\boldsymbol{R}^{2}\cdot\boldsymbol{p}'-\boldsymbol{ff}')$, respectively, to prove whether GS-DeepNet has good understanding for both the initial values and the GS equation. Here, we did not use the measured plasma pressure, meaning that the last term of the loss function $l_{2}$ constraining the form of the pressure gradient did not participate during the self-teaching unsupervised learning.

Figure \ref{fig:gs2}B--D show the comparisons of the Maxwell-Net $(\boldsymbol{B}_{R},\boldsymbol{B}_{Z},\boldsymbol{\psi})_{md}$ and $\overrightarrow{MD}$ over the test dataset. KSTAR has 42 magnetic pick-coil coils each for $\vec{B}_{R}$ and $\vec{B}_{Z}$, and 45 flux loops for $\vec{\psi}_{FL}$ \cite{doi:10.1063/1.3494275} (see Fig. \ref{fig:gs1}A). Among them, 31 intact pick-up coils each were selected, while 45 flux loops were used after their intactness was inferred based on both a deep neural network and the intact pick-up coils \cite{joung2022fl}. It is worth to mention that, often, some of the magnetic measurements are impaired since they are susceptible to damage. The fundamental difference between our GS-DeepNet and an existing numerical algorithm is that we used every magnetic measurement except the one that is fully out of order such that only null signals are measured, while an existing numerical algorithm yet requires human selections among those intact data for its numerial convergence. Moreover, we can also cover the flawed data by invoking an imputation scheme \cite{Joung:2018ju} that estimates the missing magnetic data based on Bayesian inference \cite{Sivia:2006}.

As an example, the left column in Fig. \ref{fig:gs2}B--D shows the initial values of the network compared to their corresponding $\overrightarrow{MD}$ where they qualitatively well agree with each other within one standard deviation (1-$\sigma$) uncertainties of the network given by Monte Carlo (MC) dropout \cite{gal2016dropout}. In addition, with the use of the coefficient of determination $R^{2}$ \cite{rao1973linear} ($B_{R}$, $R^{2}=0.9876$; $B_{Z}$, $R^{2}=0.9958$; $\psi_{FL}$, $R^{2}=0.9944$), these suggest that GS-DeepNet may achieve proper initial values for solutions $\psi$ of the GS equation which were also used as its input.

We presented the examples of configurations of the Maxwell-Net $\Delta^{*}\boldsymbol{\psi}$ under certain features as inputs with their determined plasma boundaries using the boundary module (Fig. \ref{fig:gs2}E--F). We have fundamentally two kinds of plasma boundaries in a tokamak \cite{beghi2005advances}: a limited plasma boundary where the plasma is ‘limited’ by hitting a solid wall (Fig. \ref{fig:gs2}E); a diverted plasma boundary where a magnetic X-point, a null point whose poloidal magnetic field is zero, is created, and only a leg extended from the X-point touches the wall. Figure \ref{fig:gs2}F shows the X-point (the point where the red line crosses) with two legs extended from it in the lower left corner. Note that the uncertainty configurations of the Maxwell-Net show their structural consistencies with $\Delta^{*}\boldsymbol{\psi}$ configurations such that large uncertainty happens around large $\Delta^{*}\psi$.

To assess whether the GS equation was properly learned, we compared the Maxwell-Net $\Delta^{*}\boldsymbol{\psi}$ (the left hand side of the GS equation) with the FB-Net $\Delta^{*}\boldsymbol{\psi}$ (the right hand side of the GS equation) in Fig. 2G. Namely, the Maxwell-Net $\Delta^{*}\boldsymbol{\psi}$ is required to be comparable with the FB-Net $\Delta^{*}\boldsymbol{\psi}$ generated from taking the Maxwell-Net $\psi$ as the network inputs if the self-teaching training reasonably worked. With the coefficient of determination $R^{2}=0.9848$ estimated with the test dataset, this proposes that GS-DeepNet may achieve the knowledge of the GS equation with its solutions $\psi$ as well. It is worth mentioning that an example of comparing the FB-Net $\bar{\bar{\boldsymbol{R}}}\bar{\bar{\boldsymbol{I}}}_{\phi}$ with $\overrightarrow{MD})^{2}$ can be found in Supplementary section S3.

\begin{figure}[t]
    \centering
    \includegraphics[width=0.8765\linewidth]{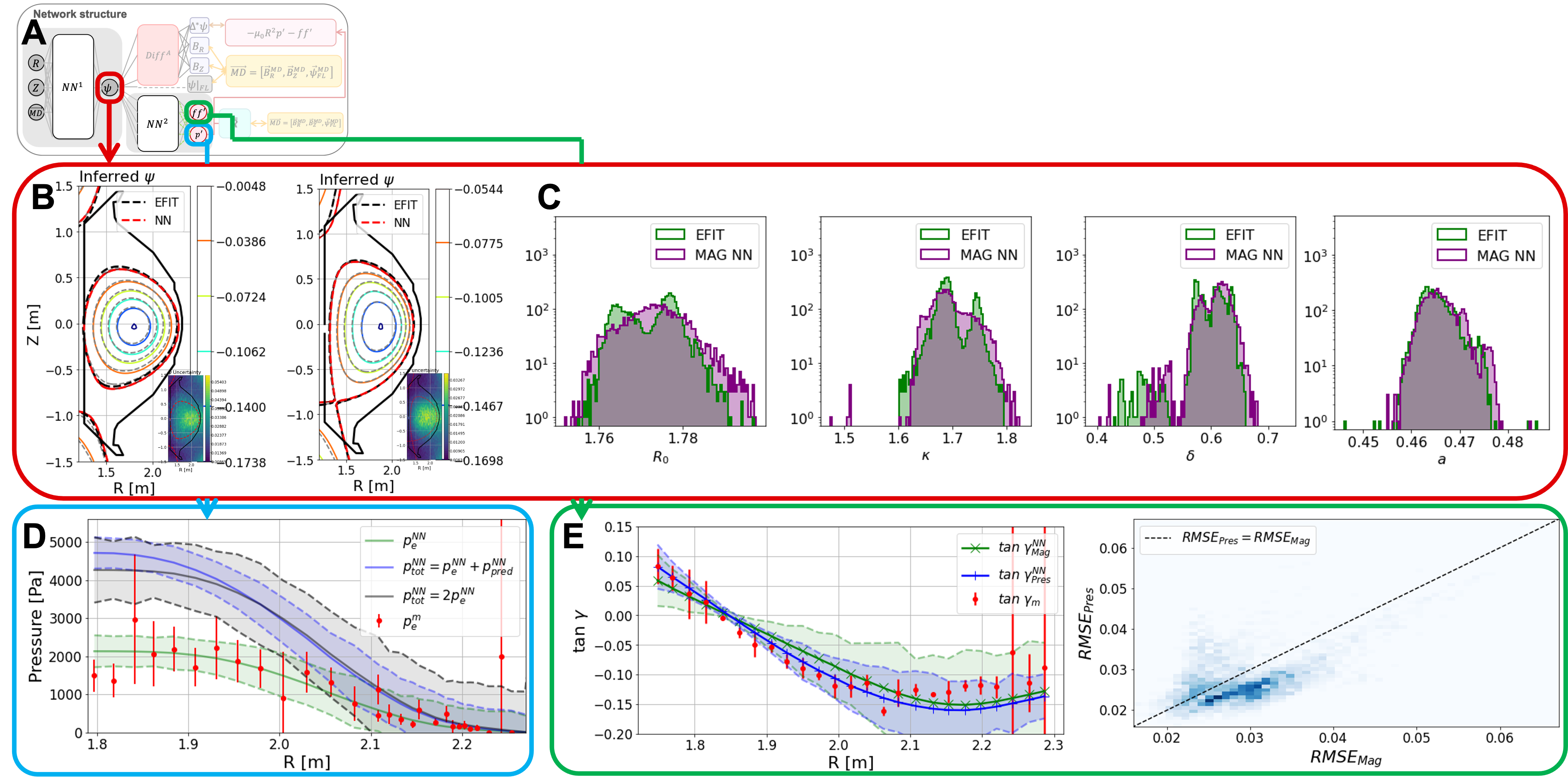}
    \caption[Equilibrium knowledge learned by GS-DeepNet]{\textbf{Equilibrium knowledge learned by GS-DeepNet}. (A) The physical knowledge discovered by GS-DeepNet. (B) The limited and diverted plasma equilibria. (C) The histograms of plasma parameters. (D) The comparison of the plasma pressure from the FB-Net with the measured one. (E) Left: the measured $\tan{\gamma}$ (red), the estimated $\tan{\gamma}$ from GS-DeepNet without the kinetic constraints (green) and that from GS-DeepNet with the kinetic constraints (blue) are presented. Right: the comparison of $RMSE$ of $\tan{\gamma}$ between GS-DeepNet with (y-axis) and without (x-axis) the kinetic constraints. Colors represent the histogram.}
    \label{fig:gs3}
\end{figure}

\begin{figure}[t]
    \centering
    \includegraphics[width=.85\linewidth]{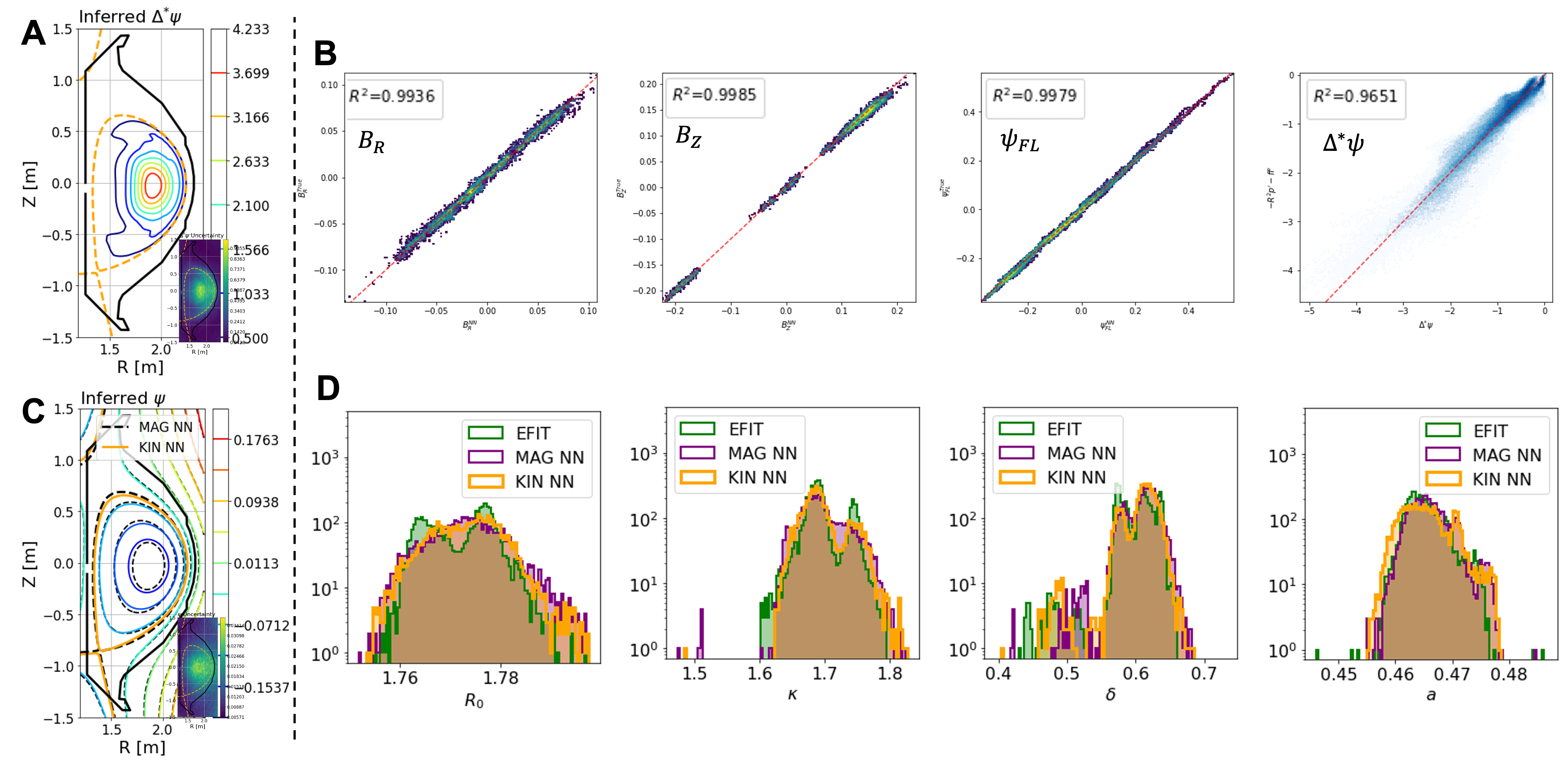}
    \caption[Performance of GS-DeepNet with local pressure constraints]{\textbf{Performance of GS-DeepNet with local pressure constraints}. (A) The Maxwell-Net $\Delta^{*}\psi$ with the kinetic constraints is shown. Compared to Fig. \ref{fig:gs2}F, the configuration become noticeably changed. (B) $R^{2}$ for $B_{R}$, $B_{Z}$, $\psi_{FL}$ and $\Delta^{*}\psi$. (C) Two plasma equilibria with (colored) and without (black) the kinetic constraints is shown. (D) The histograms of the plasma parameters from EFIT (green), GS-DeepNet with (orange) and without (purple) the kinetic constraints are shown.}
    \label{fig:gs4}
\end{figure}

\subsection{Physical knowledge learned by GS-DeepNet}

\noindent
GS-DeepNet understood the knowledge of the GS equation conforming to Maxwell’s equations as well as the force balance from its self-teaching training procedure with the measurement constraints. Here, we presented that GS-DeepNet also discovered the physical knowledge within the GS equation from the Maxwell-Net $\boldsymbol{\psi}$ and the FB-Net $(\boldsymbol{p}_{t}',\boldsymbol{ff}_{t}')$ (Fig. \ref{fig:gs3}A) which were trained indirectly (or partially directly).

Figure \ref{fig:gs3}B shows two plasma equilibria (the solutions $\boldsymbol{\psi}$) discovered by the Maxwell-Net corresponding to the configurations of the Maxwell-Net $\Delta^{*}\boldsymbol{\psi}$ in Fig. \ref{fig:gs2}E--F. We compared these solutions with equilibria reconstructed from an existing numerical algorithm, EFIT. Note that we cannot argue whether or not the plasma equilibria from EFIT are the accurate solutions of the GS equation since they have human decisions for numerical convergence, and most importantly there are no means to measure the equilibria directly to compare. At least, what we may be able to argue is that EFIT is commonly used to reconstruct equilibria in the field of nuclear fusion, and GS-DeepNet is capable of providing such equilibria without influence on several aforementioned aspects in numerical algorithms. In the right corners of Fig. \ref{fig:gs3}B, we presented the uncertainty configurations for the Maxwell-Net $\boldsymbol{\psi}$ that show larger uncertainty at the center region.

For further statistical comparison, we presented the histograms of plasma parameters (Fig. 3C) such as minor radius $a$ (half distance between innermost and outermost R positions), major radius $R_{0}$ (R at the central axis of an equilibrium), elongation $\kappa$ (ratio of vertical to horizontal sizes of a plasma) and triangularity $\delta$ (degree of shape closeness to triangle) (see fig. S2G in Supplementary section S2).

Furthermore, we named and related the FB-Net outputs with the plasma pressure gradient $p'$ and the toroidal magnetic field $B_{\phi}=f/R$, respectively. Thus, we presented the suitability of their terminologies compared with their corresponding measured data.

Figure \ref{fig:gs3}D shows the plasma pressure from the FB-Net compared with the measured one. The plasma pressure $p$ is generally estimated based on the ideal gas law \cite{clapeyron1834memoire}, i.e., $p=nT$ where $n$ is the plasma density, and $T$ is the plasma temperature. A plasma is an ionized gas that contains charged particles: ions and electrons. This plasma is mainly heated up by using its own electrical resistivity \cite{coppi2013compact}, externally injected energetic ions (whose temperature is greater than the plasma) \cite{sweetman1981heating} and external injection of electromagnetic waves \cite{cairns1991radiofrequency}. Therefore, to assess the plasma pressure completely, it is required to know the electron pressure $n_{e}T_{e}$, the ion pressure $n_{i}T_{i}$ as well as the pressure of externally injected energetic ions $p_{ext}$, i.e., $p=n_{e}T_{e}+n_{i}T_{i}+p_{ext}$.

The electron pressure $n_{e}T_{e}$ can be measured by the Thomson Scattering (TS) system \cite{cite-key, doi:10.1063/1.3494275} using the scattered and Doppler-shifted photons from an interaction between high-power laser photons and the plasma electrons. Since the plasma is known as having quasi-neutrality $(n_{e}\approx n_{i})$ \cite{Chen:2011wc}, the ion pressure $n_{i}T_{i}$ can be estimated with the quasi-neutrality and measuring $T_{i}$ based on the energetic ion injection called the Charge Exchange Spectroscopy (CES) system \cite{doi:10.1063/1.3496991} by capturing the Doppler line width and deviation of a spectrum emitted from an interaction between the energetic and the plasma ions. Unfortunately, measurement systems for $p_{ext}$ still need to be developed further \cite{sarwar2018effective,kim2012initial,yoo2021fast}, meaning that we cannot measure $p_{ext}$ yet.

Thus, to check whether or not GS-DeepNet is able to contain the physical knowledge for the FB-Net $p'$, we presented the plasma heated up by solely the plasma resistivity, meaning that there were no externally injected energetic ions to be considered. But, this fact also made measuring $T_{i}$ unavailable, thus the pressure was estimated with an assumption that the electron temperature came into a thermal equilibrium with the ion temperature, i.e., $p=n_{e}T_{e} + n_{i}T_{i} = 2n_{e}T_{e}$ (this assumption has commonly been made in the field of nuclear fusion \cite{goldston2011heuristic,kim2018high,frei2020gyrokinetic}). Here, we devised two different ways of the verification: first, the measured pressure is known by $p=2n_{e}T_{e}$ which is used to train the FB-Net $p'$ based on equation 2; second, the electron pressure is only known, and the FB-Net $p'$ is in charge of inferring the ion pressure. Thus, the plasma pressure $p$ turns out to be $p=n_{e}T_{e}+p_{pred}^{NN}$ where $p_{pred}^{NN}$ is inferred by the FB-Net. To this end, we trained a neural network (using the same architecture of the FB-Net except the output $ff'$) with respect to the electron pressure $p_{e}=n_{e}T_{e}$ by only using the last term of equation 2, while the rest of equation 2 was used to train the FB-Net.

Both ways of the verification yielded the total plasma pressures consistent with each other within their uncertainties (Fig. \ref{fig:gs3}D) although they are not perfectly matched with each other. This suggests that the FB-Net $p'$ may capture some knowledge of the pressure gradient. These uncertainties are quantified via MC dropout.

As mentioned before, the FB-Net $ff'$ can be constrained with a magnetic pitch angle measurement which is called the Motional Stark Effect (MSE) system \cite{levinton1989magnetic, ko2013polarimetric} measuring local pitch angles using the polarization of the motional Stark effect emission signals by the energetic ion injection. Although we left this to future works, we presented the comparison between the FB-Net $ff'$ and the MSE measurements in order to verify the knowledge learned by the FB-Net $ff'$.

KSTAR MSE system \cite{ko2013polarimetric} has 25 channels, and each channel measures the local pitch angle $\gamma$ given by $\tan{\gamma}\cong A_{1}B_{Z}/(A_{2}B_{\phi}+A_{3}B_{R})$ where the $A$ coefficients are the fixed values related to the geometry information of the measurement system. We prepared GS-DeepNet results trained with and without using the last term of equation 2. Here, we collected the measured $\tan{\gamma}$ and $n_{e}T_{e}$ as well as $T_{i}$ from 50 KSTAR discharges and let the FB-Net take charge of inferring $p_{ext}$. The FB-Net was used to generate $\boldsymbol{f}(=\boldsymbol{R}\cdot B_{\phi})$, and the Maxwell-Net was used to generate $(\boldsymbol{B}_{R},\boldsymbol{B}_{Z})$ to estimate the $\tan{\gamma}$ formula. Note that the density and temperature data are often called the kinetic data.

Both GS-DeepNet results with and without using the kinetic information qualitatively show the similarity to the measured pitch angle, $\tan{\gamma_{m}}$, in Fig. \ref{fig:gs3}E even though their poloidal current functions $f$ were trained indirectly. Figure \ref{fig:gs3}E also shows the histogram of Root Mean Square Error ($RMSE$) of $\tan{\gamma}$, i.e., $RMSE=\{\sum_{i=1}^{n} (\tan{\gamma_{GS,i}} - \tan{\gamma_{m,i}})^{2}/n \}^{0.5}$ for each feature in the test dataset where $n$ is the total channel number of the MSE system $(n=25)$, and $\tan{\gamma_{GS,i}}$ is the local pitch angle estimated from GS-DeepNet results. Although some of the $RMSE$ results constrained with the kinetic data, $RMSE_{Pres}$, are greater than those using the magnetic data only, $RMSE_{Mag}$, most of $RMSE_{Pres}$ are smaller than $RMSE_{Mag}$. This is presumably due to the fact that $RMSE_{Pres}$ can contain the internal information of the plasmas even if it is localized. It is worth to mention that the $\tan{\gamma_{GS}}$ profiles in Fig. \ref{fig:gs3}E have $RMSE_{Pres}=0.023$ and $RMSE_{Mag}=0.024$, respectively.

\subsection{Final performance of GS-DeepNet with the kinetic constraints}

\noindent
As mentioned earlier, we applied the kinetic measurement constraints to GS-DeepNet via our unsupervised learning scheme. Since we used the localized internal measurements to train the FB-Net $p'$, the Maxwell-Net $\psi$ and $\Delta^{*}\psi$ were qualitatively altered, compared to Fig. \ref{fig:gs2}F--G and Fig. \ref{fig:gs3}B--C whose results depended only on the magnetic measurements.

Figure \ref{fig:gs4}A shows that the Maxwell-Net $\Delta^{*}\boldsymbol{\psi}$ with the kinetic constraints became noticeably varied compared to Fig. \ref{fig:gs2}F where they used the same input feature. But of course, GS-DeepNet caught the initial values for its solutions well enough with the coefficient of determination $R^{2}$ of 0.9936, 0.9985 and 0.9979 for $B_{R}$, $B_{Z}$ and $\psi_{FL}$, respectively, in Fig. \ref{fig:gs4}B. The result of comparing the Maxwell-Net $\Delta^{*}\boldsymbol{\psi}$ with the FB-Net $\Delta^{*}\boldsymbol{\psi}$ shows the coefficient of determination $R^{2}=0.9651$ which slightly diminishes compared to Fig. \ref{fig:gs2}G. We presume that this might happen because we kept using the same network architecture and the coefficients for L2 and L1 weight regularizations although the shapes of \ref{fig:gs2} became more complex when we adopted the kinetic constraints to the unsupervised training procedure. Thus, increasing complexity of the networks (in other words, decreasing the coefficients for the regularizations for instance) can possibly improve the network performance. In addition, we also expect that using the MSE measurements to constrain the FB-Net $ff'$ directly may help the performance improved, which might bring the histogram of $RMSE$ of $\tan{\gamma}$ in Fig. \ref{fig:gs3}E much closer to zero.

Figure \ref{fig:gs4}C shows the comparison of two plasma equilibria corresponding to Fig. \ref{fig:gs4}A and Fig. \ref{fig:gs2}F, respectively. Unlike the significant structural change in $\Delta^{*}\boldsymbol{\psi}$, there were seemingly slight variation depending on the constraints used in GS-DeepNet. Nevertheless, this slight change made notable difference in $\Delta^{*}\boldsymbol{\psi}$ followed by the GS equation, meaning that careful acquirement of solutions of the GS equation might be required for those who would like to have more complex structures of plasma equilibria. We also prepared the histograms of the plasma parameters (minor radius $a$, major radius $R_{0}$, elongation $\kappa$ and triangularity $\delta$) in Fig. \ref{fig:gs4}D for additional statistical comparisons.

\subsection{Materials and Methods} \label{S3:mm}

\subsubsection{Domain knowledge}

\noindent
The plasma force-balance equation can be regarded as the well-known incompressible Navier-Stokes equations \cite{batchelor2000introduction} under assumptions of the steady-state and no viscosity conditions, together with the Lorentz force as the external force instead of the gravitational force, i.e.,
\begin{equation}
-\nabla p + \vec{J}\times\vec{B} = 0
\end{equation}
where $p$ is the plasma pressure, $\vec{J}$ is the current density, and $\vec{B}$ is the magnetic field. From this force-balance equation, we can derive the GS equation \cite{Freidberg:1987} (which can be explained in a form of the Hicks equation) with the toroidal symmetry assumption in the cylindrical coordinates $(R,\phi,Z)$ by taking advantage of Maxwell’s equations as shown below:
\begin{equation}
\begin{split}
\Delta^{*}\psi &\equiv \lp R\frac{\partial}{\partial R} \frac{1}{R} \frac{\partial}{\partial R} + \frac{\partial^2}{\partial Z^2} \rp \psi \\
& = -\mu_{0}R J_{t} \\
& = - R^2 \mu_0 \frac{d p(\psi)}{d \psi} - f(\psi)\frac{d f(\psi)}{d \psi}
\end{split}
\end{equation}
where combining Maxwell’s equations, i.e., the Gauss’s law for magnetism, $\nabla\cdot\vec{B}=0$, and the Ampere’s law, $\nabla\times\vec{B}=\mu_{0}\vec{J}$, together derives the first two lines of the equation, $R\frac{\partial}{\partial R}\frac{1}{R}\frac{\partial\psi}{\partial R}+\frac{\partial^{2}\psi}{\partial Z^{2}}=-\mu_{0}RJ_{\phi}$, and the force balance equation is used to derive the second to third lines of the equation above, $-R^{2}\mu_{0}\frac{dp}{d\psi}-f\frac{df}{d\psi}=-\mu_{0}RJ_{\phi}$. Here, $\psi$ is the poloidal flux function, $J_{\phi}$ is the toroidal current density, $\mu_{0}$ is the vacuum permeability, $p(\psi)$ is the plasma pressure as a function of $\psi$, and $f(\psi)$ is the poloidal current function as a function of $\psi$. The poloidal current function $f$ has the relation with the toroidal magnetic field $B_{\phi}$ such that $f=RB_{\phi}$.

Technically, the toroidal current density $J_{\phi}$ is required to be known over the whole tokamak area in order to solve the GS equation, which is barely achievable due to the harsh environment of the plasma. Instead, the external magnetic measurements (and the spatially localized pressure measurements) are the only information for the GS equation, which causes an inverse problem. In addition, since we only have $\sim10^{2}$ number of the measurement data to be used for discovering solutions of the GS equation on the $41\times41(\cong10^{3})$ spatial grids for a certain during a tokamak operation (see Fig. \ref{fig:gs1}B), this results in an ill-posed problem. Finally, the plasma boundary which divides the plasma region from the vacuum region can be determined after the solution $\psi$ is found. This corresponds to the definition of a free-boundary problem since a boundary is unknown until we solve the GS equation. Therefore, solving the GS equation is a free-boundary and ill-posed inverse problem.

Again, we tackled this problem by developing the Maxwell-Net, the FB-Net and the auxiliary modules. Starting from scratch, the Maxwell-Net generates a solution $\psi$ of the GS equation and achieves initial values of the solution from the prepared magnetic data. This Maxwell-Net solution is used to determine a plasma boundary via the auxiliary module for boundary detection, and the FB-Net generates a toroidal current density $J_{\phi}$ given by the Maxwell-Net solution as well as the boundary information from the module, constrained with the magnetic data using the response matrix $\bar{\bar{\boldsymbol{R}}}$. Therefore, the Maxwell-Net is taught by the FB-Net output under our self-teaching unsupervised learning algorithm. In case that we can use the local pressure measurements to train the FB-Net, we prepare the auxiliary module for plasma pressure which performs Gaussian process regressions to the measured pressure and takes the derivative of it with respect to the Maxwell-Net $\psi$. Thus, our GS-DeepNet is capable of solving the GS equation, starting \textit{tabula rasa}.

\subsubsection{Data acquisition and preprocessing}

\noindent
From 50 KSTAR experimental discharges that we collect, the measured data from the magnetic pick-up coils, the flux loops and the Rogowski coils for the plasma and external coil currents (the poloidal field coils and in-vessel coils \cite{kim2009design}) as well as the TS, CES and MSE systems is used for training, validation and testing our GS-DeepNet.

The magnetic pick-up coils, in fact, measure the poloidal magnetic field normal and tangential to the vessel wall, $B_{n}$ and $B_{t}$, where the measurements are installed. Thus, we transform $B_{n}$ and $B_{t}$ into $B_{R}$ and $B_{Z}$ based on the following coordinate transformations:
\begin{equation}
\begin{split} \label{eq:corrd}
&B_{R} = -B_{n}\sin{\xi} + B_{t}\cos{\xi} \\
&B_{Z} = B_{n}\cos{\xi} + B_{t}\sin{\xi}
\end{split}
\end{equation}
where $\xi$ is the angle between the direction normal to the wall and the radial direction (see fig. S5 in Supplementary section S5).

A noise reduction technique is applied to the magnetic pick-up coils, the flux loops and the Rogowski coils based on the boxcar average with a time scale of 1 $m$sec which is smaller scale than a typical time scale of KSTAR equilibrium reconstruction. Furthermore, we preprocess the signal drifts in the magnetic measurements \cite{joung2019deep} based on Bayesian inference since the magnetic data tends to suffer from the signal drift in time such that the baseline of the data increases or decreases over time.

\subsubsection{Response matrix}

\noindent
Following a previous approach \cite{Svensson:2008in} that models the plasma and external coils as toroidal current beams with rectangular cross sections \cite{urankar1982vector}, we estimate the response matrix $\bar{\bar{\boldsymbol{R}}}$ with the following expressions for a matrix component $r_{ij}$ where the subscript $i$ is the index for the magnetic measurements:
\begin{equation}
r_{ij}^{(B_{R})}=\frac{\mu_{0}}{2\pi}\int_{Z_{j,1}}^{Z_{j,2}}\int_{R_{j,1}}^{R_{j,2}} \,dRdZ \frac{Z_{i}-Z}{R_{i}} \sqrt{\frac{k}{4RR_{i}}} \Big[K(k)+\frac{R^{2}+R_{i}^{2}+(Z_{i}-Z)^{2}}{(R-R_{i})^{2}+(Z_{i}-Z)^{2}}E(k) \Big]
\end{equation}
where $r_{ij}^{(B_{R})}$ is the component for $B_{R}$, and the subscript $i$ is set to $1\leq i\leq31$,
\begin{equation}
r_{ij}^{(B_{Z})}=\frac{\mu_{0}}{2\pi}\int_{Z_{j,1}}^{Z_{j,2}}\int_{R_{j,1}}^{R_{j,2}} \,dRdZ \sqrt{\frac{k}{4RR_{i}}} \Big[K(k)+\frac{R^{2}-R_{i}^{2}-(Z_{i}-Z)^{2}}{(R-R_{i})^{2}+(Z_{i}-Z)^{2}}E(k) \Big]
\end{equation}
where $r_{ij}^{(B_{Z})}$ is the component for $B_{Z}$, and the subscript $i$ is set to $32\leq i\leq62$,
\begin{equation}
r_{ij}^{(\psi_{FL})}=2\mu_{0}\int_{Z_{j,1}}^{Z_{j,2}}\int_{R_{j,1}}^{R_{j,2}} \,dRdZ \sqrt{\frac{R}{R_{i}}} \frac{1}{\sqrt{k}} \Big[\Big(1-\frac{1}{2}k \Big)K(k)-E(k) \Big]
\end{equation}
where $r_{ij}^{(\psi_{FL})}$ is the component for $\psi_{FL}$, and the subscript $i$ is set to $63\leq i\leq107$. Here, $k$ is the elliptic modulus $k=\frac{4RR_{i}}{(R+R_{i})^{2}+(Z_{i}-Z)^{2}}$, $(R_{i},Z_{i})$ is the location of the magnetic measurement, and $(R_{j,1},R_{j,2},Z_{j,3},Z_{j,4})$ represents the location of the rectangular cross section of the toroidal current beam (see fig. S4 in Supplementary section S4). The subscript $j$ is set to $1\leq j\leq41^{2}(=\text{1,681})$ for $\bar{\bar{\boldsymbol{R}}}_{p}$, $1\leq j\leq30(=14+16)$ for $\bar{\bar{\boldsymbol{R}}}_{ext}$ and $1\leq j\leq18$ for $\bar{\bar{\boldsymbol{R}}}_{VV}$. The KSTAR has 14 poloidal field coils and 16 segments of the in-vessel coils as the external magnetic coils. We divide the tokamak vessel wall into 18 current-carrying segments, following a previous approach used in the KSTAR \cite{park2013investigation}.

In the equations above, $K(k)$ and $E(k)$ are the complete elliptic integral of the first and the second kinds, respectively, as defined below:
\begin{equation}
\begin{split} \label{eq:ke}
&K(k)=\int_{0}^{\frac{\pi}{2}}\,d\theta \frac{1}{\sqrt{1-k^{2}\sin^{2}{\theta}}}, \\
&E(k)=\int_{0}^{\frac{\pi}{2}}\,d\theta \sqrt{1-k^{2}\sin^{2}{\theta}}.
\end{split}
\end{equation}
The vessel currents $J_{VV}$ are treated as free-parameters \cite{Svensson:2008in} which are optimized during the network optimization process based on an approach \cite{urano2020breakdown} suggesting that the vessel currents can be reasonably inferred even though actual conducting structures are not precisely identified.

\subsubsection{Auxiliary module: boundary and pressure modules}

\noindent
A fundamental concept to determine a plasma boundary is finding a X-point by using the fact that the poloidal magnetic field at the X-point is ideally zero. Since a plasma boundary is a connected line of $R$-$Z$ positions enclosing the plasma area where flux functions on the $R$-$Z$ positions are all the same, we estimate a flux function $\psi$ at the X-point and use it to determine a location of the boundary. Thus, the auxiliary module for boundary detection performs the following procedure: first, the module receives the Maxwell-Net $(\boldsymbol{\psi},\boldsymbol{B}_{R},\boldsymbol{B}_{Z})$ over the spatial grids and estimate magnitudes of the poloidal magnetic field, $\boldsymbol{B}_{P}$, via $B_{P}=\sqrt{B_{R}^{2}+B_{R}^{2}}$; second, the module searches a $R$-$Z$ point where magnitude of its $B_{P}$ is the smallest; third, by feeding the $R$-$Z$ point into the Maxwell-Net again, the module can obtain the flux function $\psi$ at the X-point, $\psi_{b}$, and use it to find $R$-$Z$ positions whose flux functions are identical to $\psi_{b}$; finally, in case that the plasma is limited, the module compares $\psi_{b}$ with flux functions on locations of the solid wall by using the Maxwell-Net. If some of the $R$-$Z$ positions possessing $\psi_{b}$ are out of the wall, then the module defines that the plasma is limited or vice versa. It is worth to mention that the plasma boundary is always changed during a tokamak operation (fig. S2F).

The major role of the auxiliary module for plasma pressure module is to perform Gaussian process regression (GPR) to the measured plasma pressure. As mentioned before, this module takes the derivative of the GPR-regressed pressure with respect to the Maxwell-Net $\psi$. Here, we use a finite difference method to calculate the regressed pressure gradient.

\subsubsection{Optimization}

\noindent
Our neural networks $NN_{\Theta_{i}}^{1}$ and $NN_{\theta_{i}}^{2}$ are trained via TensorFlow \cite{tensorflow2015-whitepaper} with one GPU worker and 20 CPU cores. We pass one randomly selected measurement feature to the optimization process at a time, meaning that the batch-size is $41^{2}+107=\text{1,788}$ corresponding to the grid points and the locations of the magnetic measurements. A total mini-batch size is approximately 8,000. Stochastic gradient descent is used to optimized out network parameters with the loss functions in equations 1 and 2. The coefficients for the L2 and L1 regularizations used in the loss functions are set to $c_{1}=10^{-3}$ and $c_{2}=10^{-2}$, respectively. At every iteration, a new checkpoint is saved and used to estimate the validation loss for the early stopping method.

\subsubsection{Neural network architectures}

\noindent
The network $NN_{\Theta}^{1}$ uses 31 $B_{R}$, 31 $B_{Z}$, 45 $\psi_{FL}$ from a certain time slice, and a single $R$-$Z$ point as its input whose size is $1\times109(=31+31+45+2)$. This input passes through three fully connected layers with the swish nonlinearities and a fully connected linear layer, then turns into a scalar. Each layers have 100 hidden neurons and a bias. The output scalar is treated as a solution $\psi$ on the $R$-$Z$ position, and processed by the automatic differential operator $Diff{^A}$ that produces $B_{R}$, $B_{Z}$ and $\Delta^{*}\psi$. Dropout with a rate of 0.05 is applied to all the fully connected layers.

Similarly, the network $NN_{\theta}^{2}$ has a fully connected layer with the swish nonlinearity and a fully connected linear layer. Each layer has 60 and five neurons, respectively, and applies dropout with a rate of 0.10 as well. Taking a scalar (a solution $\psi$ from the network $NN_{\Theta}^{1}$) as an input, this network outputs a vector of size 2, corresponding to the pressure gradient $p'$ and the poloidal current related variable $ff'$.

These networks are initialized to random weights based on Glorot (or Xavier) initialization \cite{pmlr-v9-glorot10a}.

\subsection{Discussion} \label{sd:con}

\noindent
Our results reasonably demonstrate that a self-teaching unsupervised learning scheme is applicable for deep neural networks to learn a second-order nonlinear differential equation such as the GS equation. Without using any guess or precalculated solutions of the GS equation, we prove that it is possible to train deep neural networks to acquire knowledge of the solution of the GS equation which is a free-boundary and inverse problem required numerical algorithms in general. Since our approach do not depend on existing numerical algorithms (or existing reconstruction methods), GS-DeepNet is unfettered with the challenges raised in the existing methods as mentioned earlier. Furthermore, our approach is possible to include not only external measurement constraints but also both external and internal (but localized) measurement constraints in the unsupervised training procedure without any significant changes.

We introduce the method to solve a differential equation via neural networks, and leave using various measurements to constrain the network training as future works. Thus, we plan training neural networks with our unsupervised learning scheme by including other plasma measurements such as the MSE pitch angle measurement. Furthermore, we have plans to search the most suitable architecture of a neural network to improve its performance when internal constraints are involved.

Most physical phenomena can be, in general, expressed in differential equations. Thus, our work might be helpful for other engineering and science fields to support solving their differential equations, starting from scratch. Furthermore, previous researches using GAN \cite{goodfellow2014generative} to simulate complex physical systems such as accelerators \cite{de2017learning,paganini2018calogan,paganini2018accelerating} are likely to be improved based on our approach by giving physical knowledge to them.

\chapter{Conclusions}
\epigraph{\textit{``Farewell... My brave Hobbits. My work is now finished. Here at last, on the shores of the sea...comes the end of our Fellowship. I will not say ``Do not weep"...for not all tears are an evil."}}{--- Gandalf the White,\\The Lord of the Rings}

\noindent
So far, we -- me and the readers of this thesis -- have had a journey with the neural networks which proved the fact that they are really able to solve a governing equation, i.e., the Grad-Shafranov equation in our case which is a second-order non-linear and elliptic partial differential equation for a two-dimensional plasma, by themselves. Here again, I would like to emphasize that the way that the networks solve the Grad-Shafranov equation is indeed general and applicable to many other governing equations in various scientific problems. Only tedious part that one needs to do is switching the Grad-Shafranov equation with other differential equations, then modifying the network architectures slightly to be suitable for their own problems. As we have shown together previously, I have defined the differential equations as the cost functions of the networks. Namely, given measurement data and their corresponding governing equations, the networks are trained automatically based on gradient descents of the cost functions and capture physical behaviors of interest without taking a glance at any man-made solutions of the problems. Furthermore, one can reflect the principle of Occam's razor to the networks by appropriately modifying forms of the cost functions. 

This thesis has showed that the neural networks are able to produce plasma equilibria whose quality is equivalent to EFIT results, which are regarded as well-converged solutions of the GS equation, in real time. In addition, the consistent flux signals can be fitted by the neural networks, which can be available for the plasma reconstruction whose quality is as reasonable as the plasma equilibria attained by selecting some of the measurements arbitrarily based on human (expert) decisions. Bayesian neural networks applied in fusion research have showed that we can quantify the uncertainty of the network models which solve a free-boundary and inverse problem to generate plasma equilibria. Of course, this thesis still needs to show applicability to long pulse discharges of tokamaks, and thorough consistency with the MSE diagnostic system, which is future works. Applying the methods used in this thesis to a method such as GAN that may complement existing physics simulations is also planned to be conducted for tokamak plasmas.

However, we should think about this, i.e., what is better about solving differential equations with neural networks than with conventional methods such as finite difference method? There are numerous conventional methods that exist already to solve differential equations, and these methods also proved that they can be quantitatively evaluated and provide reasonable outcomes of interest. In other words, good (sometimes fantastic) solutions of differential equations can be already obtained by the existing algorithms. At this moment, what I can mention about neural networks is that a lot of trends have appeared and disappeared in the field of neural networks over the past decades, and the method of solving differential equations via networks may also be one of them. Of course, I want to note that networks are powerful to calculate derivatives really conveniently through the back-propagation.

Nevertheless, let me suppose that neural networks are trained only with prepared solutions of governing equations of interests through supervised learning. The results of the trained networks, of course, will be really similar to the solutions used. However, can we affirm that the network results truly satisfy the governing equations used to prepare the solutions? I would say, \textit{no}. I have confirmed through my findings that this may not be true. As dealt with previously, I trained the network with the solutions of the Grad-Shafranov equation calculated from EFIT, but the network results did not follow the equation. Thus, I would argue that if networks are needed to be used for rigorous physics, they should obviously produce physically rigorous results. I would like to emphasize that attempts to learn differential equations through networks can be a stepping stone for neural networks to be able to be used strictly in the field of physics. Finally, I would also like to add that we need to find a way for networks to collaborate with the existing methods rather than persisting in the use of neural networks only as a perspective of taking advantage of each other. 

As one may have noticed, I did not answer the question, ``Why is it better to use neural networks than the conventional methods?" A persuasive answer to this question will be very subjective (at this moment I guess), and I would like to leave it as a future work. \textbf{So, let us simply enjoy our journey now - \textit{and beyond}.} It is unknown what will happen next, but at least it is really fascinating.

\appendix
\chapter{Bayesian Deep Learning: Model uncertainty}\label{ap1}

\noindent
Bayesian neural networks (BNNs) was first suggested in the 1990s \cite{mackay1992practical, neal1995bayesian} where a brief history can be found elsewhere \cite{gal2016uncertainty}. These networks set prior distributions for their weights, offering a probabilistic interpretation of their models. While their formulations are relatively easy, the probabilistic inference is quite tricky. Thus, we would like to approximate $p\big(\boldomega|\boldX,\boldY \big)$ by means of VI. This approximate BNN inference comes out with stochastic regularization techniques (SRTs) like dropout used in VI. This approach is largely taken from Ref. \cite{gal2016uncertainty}.

To approximate the BNN inference, we would like to approximate $p\big(\boldomega|\boldX,\boldY \big)$ in light of VI, i.e.,
\begin{equation} 
\label{eq:bi7}
\mathcal{L}_{VI}(\theta) \equiv -\sum_{i=1}^{N} \int q_{\theta}(\boldomega) \log p\big(\boldy_{i}|\boldsymbol{f}^{\boldomega}(\boldX_{i}) \big) d\boldomega + KL\big(q_{\theta}(\boldomega) \big|\big| p(\boldomega) \big)
\end{equation}
where $\boldsymbol{f}^{\boldomega}$ is the function parameterized by $\boldomega$, and the subscript $i$ is the row or column of a matrix denoted in boldface. This stands for an average (or integration) over the entire dataset, which costs heavy computations for large $N$, thus we apply the \textit{mini-batch} approach to approximate it as:
\begin{equation} 
\label{eq:bi8}
\hat{\mathcal{L}}_{VI}(\theta) \equiv -\frac{N}{M}\sum_{i\in \mathcal{S}} \int q_{\theta}(\boldomega) \log p\big(\boldy_{i}|\boldsymbol{f}^{\boldomega}(\boldX_{i}) \big) d\boldomega + KL\big(q_{\theta}(\boldomega) \big|\big| p(\boldomega) \big)
\end{equation}
where $\mathcal{S}$ is the random index set of size $M$, and Monte Carlo (MC) integration has been applied to this. The data sub-sampling also gives us an optimum \cite{robbins1951stochastic, hoffman2013stochastic}.

After reparameterizing $q_{\theta}(\boldomega)$ and $\boldomega$ as $p(\boldeps)$ and $g(\theta,\boldeps)$, i.e., using the pathwise derivative estimator (the reparametrization trick, infinitesimal perturbation analysis, and stochastic backpropagation \cite{glasserman2004monte, kingma2013auto, titsias2014doubly}) which assumes that $q_{\theta}(\boldomega)$ can be re-parameterized as a parameter-free distribution $p(\boldeps)$ with a deterministic differentiable bivariate transformation $g$, we rewrite the sub-sampling VI as:
\begin{equation} 
\label{eq:bi9}
\hat{\mathcal{L}}_{VI}(\theta) = -\frac{N}{M}\sum_{i\in \mathcal{S}} \int p(\boldeps) \log p\big(\boldy_{i}|\boldsymbol{f}^{g(\theta,\boldeps)}(\boldX_{i}) \big) d\boldeps + KL\big(q_{\theta}(\boldomega) \big|\big| p(\boldomega) \big).
\end{equation}
This expression turns out to be the form below:
\begin{equation} 
\label{eq:bi10}
\hat{\mathcal{L}}_{MC}(\theta) = -\frac{N}{M}\sum_{i\in \mathcal{S}} \log p\big(\boldy_{i}|\boldsymbol{f}^{g(\theta,\boldeps)}(\boldX_{i}) \big) + KL\big(q_{\theta}(\boldomega) \big|\big| p(\boldomega) \big).
\end{equation}
where the log likelihood is replaced with its stochastic estimator. This is a new MC estimator.

Now, we can optimize $\hat{\mathcal{L}}_{MC}(\theta)$ with respect to $\theta$ following a sequence for inference:
\begin{equation}
\begin{split}
&\widehat{\Delta \theta} \leftarrow -\frac{N}{M}\sum_{i\in \mathcal{S}} \frac{\partial}{\partial\theta} \log p\big(\boldy_{i}|\boldsymbol{f}^{g(\theta,\hat{\boldeps}_{i})}(\boldX_{i}) \big) + \frac{\partial}{\partial\theta}KL\big(q_{\theta}(\boldomega) \big|\big| p(\boldomega) \big) \\
&\theta \leftarrow \theta + \eta\widehat{\Delta \theta}
\end{split}
\end{equation}
where $\eta$ is the learning rate, $\hat{\boldeps}_{i}~p(\boldeps)$ is $M$ random variables, and $\theta$ is initialized randomly.

From now on, we relate the approximate inference above to SRTs used in deep learning. Dropout \cite{hinton2012improving, srivastava2014dropout} is the most popular SRT which is easily applicable to any neural networks and used to avoid over-fitting issues. Thus, let us focus mainly on dropout.

Suppose that there is a neural network having a single hidden layer, dropout applied. When we start to estimate the network's outputs through dropout, two binary vectors $\widehat{\boldeps}_{1}$ and $\widehat{\boldeps}_{2}$ whose dimension corresponds to the input and the hidden layer, respectively, are sampled to be assigned with value 0 with probability $0\leq p_{1(\text{or 2})}\leq 1$. Then, we multiply given input $\boldx$ with $\widehat{\boldeps}_{1}$ like $\widehat{\boldx}=\boldx\odot\widehat{\boldeps}_{1}$ making some inputs to zero (turn off activation of some input nodes). $\odot$ is the element-wise product. Similarly, some hidden nodes $\boldh$ are turned off through $\widehat{\boldh} = \boldh\odot\widehat{\boldeps}_{2}$ where $\boldh=f(\widehat{\boldx}\boldM_{1}+\boldb)$ and $f$ is an activation function, and therefore the network's output with given dropout $\widehat{\boldy}=\widehat{\boldh}\boldM_{2}$. Here, $\boldM_{1}$ and $\boldM_{2}$ are the deterministic matrix for the network weights for the input and the hidden layer, respectively.


A stark difference between BNNs and dropout is that BNNs quantify their uncertainty over their model parameters while dropout injects its noise into the feature space, i.e., $\boldx$ and $\boldh$. Thus, let us treat the dropout noise as the parameter space noise from the feature space noise as shown below:
\begin{equation} 
\label{eq:bi11}
\begin{split}
\widehat{\boldy} &= \widehat{\boldh}\boldM_{2} \\
& = (\boldh\odot\widehat{\boldeps}_{2})\boldM_{2} \\
& = (\boldh\cdot\text{diag}(\widehat{\boldeps}_{2}))\boldM_{2} \\
& = \boldh(\text{diag}(\widehat{\boldeps}_{2})\boldM_{2}) \\
& = f\Big(\widehat{\boldx}\boldM_{1}+\boldb \Big)(\text{diag}(\widehat{\boldeps}_{2})\boldM_{2}) \\
& = f\Big((\boldx\odot\widehat{\boldeps}_{1})\boldM_{1}+\boldb \Big)(\text{diag}(\widehat{\boldeps}_{2})\boldM_{2}) \\
& = f\Big(\boldx\big(\text{diag}(\widehat{\boldeps}_{1})\boldM_{1} \big)+\boldb \Big)(\text{diag}(\widehat{\boldeps}_{2})\boldM_{2})
\end{split}
\end{equation}
where we define $\widehat{\boldW}_{1}\equiv \text{diag}(\widehat{\boldeps}_{1})\boldM_{1}$ and $\widehat{\boldW}_{2}\equiv \text{diag}(\widehat{\boldeps}_{2})\boldM_{2}$. $\widehat{\boldW}$ is a realization of $\boldW$, i.e., a random variable defined over the set of real matrices. This turns out to be:
\begin{equation} 
\label{eq:bi12}
\widehat{\boldy} = f\Big(\boldx\widehat{\boldW}_{1}+\boldb \Big)\widehat{\boldW}_{2} \equiv \boldsymbol{F}^{\widehat{\boldW}_{1}, \widehat{\boldW}_{2}, \boldb}(\boldx)
\end{equation}
where we define $\widehat{\omega}=\{\widehat{\boldW}_{1}, \widehat{\boldW}_{2}, \boldb \}$. 

Therefore, we can optimize this dropout network based on:
\begin{equation} 
\label{eq:bi13}
\mathcal{L}_{dropout}(\boldM_{1}, \boldM_{2}, \boldb) \equiv \frac{1}{M} \sum_{i\in \mathcal{S}} E^{\widehat{\boldW}_{1}^{i}, \widehat{\boldW}_{2}^{i}, \boldb}\big(\boldx_{i}, \boldy_{i} \big) + \lambda_{1}||\boldM_{1}||^{2} + \lambda_{2}||\boldM_{2}||^{2} + \lambda_{3}||\boldb||^{2}
\end{equation}
where $\widehat{\boldW}_{1}^{i}$ and $\widehat{\boldW}_{2}^{i}$ are the contributions of $\widehat{\boldeps}_{1}^{i}$ and $\widehat{\boldeps}_{2}^{i}$ sampled from each data point $i$ to the matrix $\boldM_{1\text{or 2}}$, respectively. The \textit{mini-batch} approach is applied here to sub-sample a random index set $\mathcal{S}$ whose size is $M$.

We can rewrite $E^{\boldM_{1}, \boldM_{2}, \boldb}(\boldx, \boldy)$ as follows \cite{tishby1989consistent}:
\begin{equation}
\label{eq:bi14}
\begin{split}
E^{\boldM_{1}, \boldM_{2}, \boldb}(\boldx, \boldy) &= \frac{1}{2} ||\boldy-\boldsymbol{F}^{\boldM_{1}, \boldM_{2}, \boldb}(\boldx)||^{2} \\
&= -\frac{1}{\tau}\log p(\boldy|\boldsymbol{F}^{\boldM_{1}, \boldM_{2}, \boldb}(\boldx)) + \text{const}
\end{split}
\end{equation}
where this is a form of the negative log-likelihood with an offset (constant), and $p(\boldy|\boldsymbol{F}^{\boldM_{1}, \boldM_{2}, \boldb}(\boldx)) = \mathcal{N}(\boldy;\boldsymbol{F}^{\boldM_{1}, \boldM_{2}, \boldb}(\boldx),\tau^{-1}I)$ with observation noise $\tau^{-1}$. 

For the data point $i$ in the range of $1\leq i\leq N$, we can also rewrite $\widehat{\omega}$ as follows:
\begin{equation}
\begin{split}
\widehat{\omega}_{i} &= \{\widehat{\boldW}_{1}^{i}, \widehat{\boldW}_{2}^{i}, \boldb \} = \{\text{diag}(\hat{\boldeps}_{1}^{i})\boldM_{1}, \text{diag}(\hat{\boldeps}_{2}^{i})\boldM_{2}, \boldb \} \\
&\equiv g(\theta, \hat{\boldeps}_{i})
\end{split}
\end{equation}
if we define $\theta=\{\boldM_{1}, \boldM_{2}, \boldb \}$, and $\widehat{\boldeps}_{1}^{i}$ and $\widehat{\boldeps}_{2}^{i}$ are approximately $p(\boldeps_{1})$ and $p(\boldeps_{2})$ where $p(\boldeps_{1(\text{or }2)})$ is products of Bernoulli distributions with probabilities $1-p_{1(\text{or }2)}$. 

Finally, we can re-define $\mathcal{L}_{dropout}(\boldM_{1}, \boldM_{2}, \boldb)$ as follows:
\begin{equation} 
\label{eq:bi15}
\begin{split}
\hat{\mathcal{L}}_{dropout}(\boldM_{1}, \boldM_{2}, \boldb) = &-\frac{1}{M\tau} \sum_{i\in \mathcal{S}} \log p(\boldy|\boldsymbol{F}^{g(\theta, \hat{\boldeps}_{i})}(\boldx)) \\
&+ \lambda_{1}||\boldM_{1}||^{2} + \lambda_{2}||\boldM_{2}||^{2} + \lambda_{3}||\boldb||^{2}
\end{split}
\end{equation}
where $\hat{\boldeps}$ is the realization of $\boldeps$. Thus, taking a differentiation of this with respect to $\theta$ is shown as:
\begin{equation} 
\label{eq:bi16}
\begin{split}
\frac{\partial}{\partial \theta}\hat{\mathcal{L}}_{dropout}(\theta) = &-\frac{1}{M\tau} \sum_{i\in \mathcal{S}} \frac{\partial}{\partial \theta}\log p(\boldy|\boldsymbol{F}^{g(\theta, \hat{\boldeps}_{i})}(\boldx)) \\
&+ \frac{\partial}{\partial \theta} \Big(\lambda_{1}||\boldM_{1}||^{2} + \lambda_{2}||\boldM_{2}||^{2} + \lambda_{3}||\boldb||^{2} \Big)
\end{split}
\end{equation}

Now, we can optimize $\hat{\mathcal{L}}_{dropout}(\theta)$ through the sequence below for inference:
\begin{equation}
\label{eq:bi17}
\begin{split}
&\widehat{\Delta \theta} \leftarrow -\frac{1}{M\tau} \sum_{i\in \mathcal{S}} \frac{\partial}{\partial \theta}\log p(\boldy|\boldsymbol{F}^{g(\theta, \hat{\boldeps}_{i})}(\boldx)) + \frac{\partial}{\partial \theta} \Big(\lambda_{1}||\boldM_{1}||^{2} + \lambda_{2}||\boldM_{2}||^{2} + \lambda_{3}||\boldb||^{2} \Big) \\
&\theta \leftarrow \theta + \eta\widehat{\Delta \theta}
\end{split}
\end{equation}
where $\eta$ is the learning rate, $\hat{\boldeps}_{i}~p(\boldeps)$ is $M$ random variables, and $\theta$ is initialized randomly. As can be easily noticed, except a constant scale of $1/N\tau$, this is greatly identical to the approximate inference of a BNN if we let $KL(q_{\theta}(\boldomega)||p(\boldomega))$ become:
\begin{equation}
\label{eq:bi18}
\frac{\partial}{\partial\theta}KL\big(q_{\theta}(\boldomega) \big|\big| p(\boldomega) \big) = \frac{\partial}{\partial \theta} N\tau \Big(\lambda_{1}||\boldM_{1}||^{2} + \lambda_{2}||\boldM_{2}||^{2} + \lambda_{3}||\boldb||^{2} \Big)
\end{equation}
toward the optimization as follows:
\begin{equation}
\label{eq:bi19}
\frac{\partial}{\partial \theta} \hat{\mathcal{L}}_{dropout}(\theta) = \frac{1}{N\tau} \frac{\partial}{\partial \theta} \hat{\mathcal{L}}_{MC}(\theta).
\end{equation}

This result shows that if we optimize a network's weights by means of dropout, this optimization process is exactly identical to the variational optimization of a Bayesian neural network. Thus, this dropout neural network is the Bayesian neural network, allowing us to express the network uncertainty quantitatively with given observed data. I would like to stress that we now possess the tool to construct a reliable neural network being able to provide its confidence to given inputs, which is tested with the typical sine regressions again in chapter \ref{ch3-1-2}.

\chapter{Neural Network Differentiation}\label{ap2}

\noindent
In this chapter, I would like to introduce a method to learn physical theories by a neural network itself. Theories are generally represented in the form of differential equations, i.e., \textit{differentiable physics}, thus, if I can show that the network is able to learn differential equations, then it would become true that the network can be trained with the theories directly. Learning differential equations is eventually related to setting cost functions in terms of those equations. Therefore, we will confirm how to model a cost function with a differential equation through taking derivatives of the neural networks. To this end, suppose that we have a simple neural network where this has only a single hidden layer with a single input and output node.

\begin{figure}[t]
    \centerline{\includegraphics[width=0.375\columnwidth]{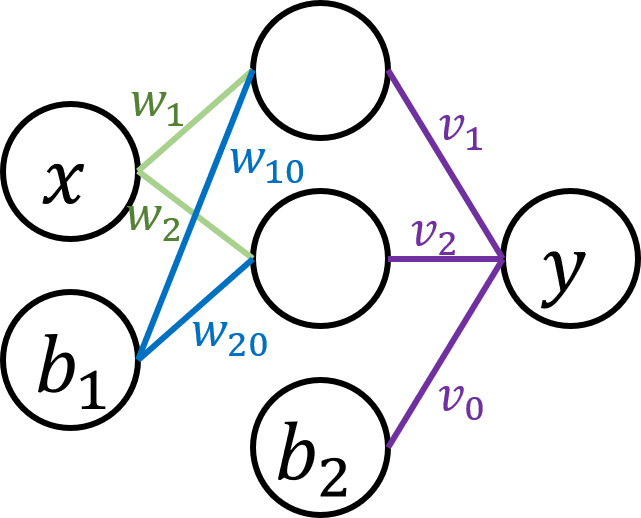}}
    \caption[Simple neural network 2]{A simple neural network having only one input node (except the input bias), hidden layer and output node. There are two nodes in the hidden layer.
    } \label{ch2-fig4}
\end{figure}

Figure \ref{ch2-fig4} shows the structure of the basic neural network. With two nodes in the hidden layer, there are biases in the input and the hidden layers. This is expressed as a basic algebra as follows:
\begin{equation} 
\label{eq:nd1}
y = v_{0} + \left[v_{1} f\big(w_{10}+w_{1}x\big) + v_{2}f\big(w_{20}+w_{2}x\big) \right]
\end{equation}
where it is worth to note that there should be $b_{1}$ and $b_{2}$ multiplied to $w_{10}$ and $w_{20}$, respectively, in the equation above. Instead, I simply prescribe unity to the biases in order to ignore the explicit expressions. I set an activation function in the hidden layer to the linear function for simplicity.

From the fact that the network can be represented analytically, this then gives us the ability to express a derivative of the network outputs with respect to the input $x$, together with an arbitrary activation function, $f$. Therefore, if one take first order derivative of the network, this can be represented as follows:
\begin{equation} 
\label{eq:nd2}
\frac{\partial y}{\partial x} = v_{1}w_{1} f'\big(w_{10}+w_{1}x\big) + v_{2}w_{2} f'\big(w_{20}+w_{2}x\big)
\end{equation}
where $f'(X, ...)$ refers to the first order partial derivative of the function $f$ with respect to $X$, i.e., $\partial f(X, ...) / \partial X$. 

This shows how the network output changes according to the input. Note that there are no restrictions for this input and output to be any quantities. This means that those can be any physical parameters of interest being able to be expressed in the analytic expression through the network derivative. Namely, if optimal weights of the network can be found to explain related physical phenomena appropriately, then we are now capable of calculating all possible changes of the phenomena in, e.g., time and space. This is quite powerful since we treat the network as solutions of physics. However, an important issue left is how we can obtain the proper weights for the physics. Here, this can be simply resolved if a cost function of the network contains the differential equations themselves with a given training dataset.

For instance, suppose that we are interested in a simple differential equation below:
\begin{equation}
\label{eq:rn1}
\frac{d\hat{t}}{dx} - 14\pi \cos{(14\pi x)} = 0
\end{equation}
where $14\pi \cos{(14\pi x)}$ is the observed quantity. If we replace $\hat{t}$ with $y$, the network output, in the equation and use the equation itself as a cost function with given $14\pi \cos{(14\pi x)}$, then we can eventually acquire the network whose output satisfies the equation above such that the output $y=\sin{(14\pi x)}$, i.e., the solution $\hat{t}$. Likewise, we can generalize the cost function as follows:
\begin{equation} 
\label{eq:nd3}
\epsilon = \left(\frac{\partial y_i}{\partial x_i} - t_i \right)^{2}
\end{equation}
where subscript $i$ refers to an arbitrary feature to be used for training the network, and $t$ is a single training data point. 

Therefore, the training of the network is possible through the gradient descent method with the cost function above. From this, the gradients of the weights in the hidden layer are expressed as follows:
\begin{equation} 
\label{eq:nn4}
\begin{split}
\frac{\partial \epsilon}{\partial v_{1}} &= \frac{\partial}{\partial v_{1}} \left(\frac{\partial y_i}{\partial x_i} - t_i \right)^{2} \\
&= 2\sqrt{\epsilon}\:\: \frac{\partial}{\partial v_{1}} \left[v_{1}w_{1} f'\big(w_{10}+w_{1}x_{i}\big) + v_{2}w_{2} f'\big(w_{20}+w_{2}x_{i}\big) - t_{i} \right] \\
&= 2\sqrt{\epsilon}\:\: w_{1} f'\big(w_{10}+w_{1}x_{i}\big)
\end{split}
\end{equation}
\begin{equation} 
\label{eq:nd5}
\frac{\partial \epsilon}{\partial v_{2}} = 2\sqrt{\epsilon}\:\: w_{2} f'\big(w_{20}+w_{2}x_{i}\big)
\end{equation}
where $v_{1}$ and $v_{2}$ are the weights between the hidden and the output layer. Note that the cost function (Equation \ref{eq:nd2}) does not have $v_{0}$ dependency, meaning that prescribed $v_{0}$ cannot be trained. Thus, one can see an error message while training a neural network with a bias in the last hidden layer, together with the cost function defined before.

Similarly, we can analytically represent the gradients of the weights between the input and the hidden layer as follows:
\begin{equation} 
\label{eq:nd6}
\begin{split}
&\frac{\partial \epsilon}{\partial w_{1}} = 2\sqrt{\epsilon}\:\: v_{1} \left[f'\big(w_{10}+w_{1}x_{i}\big) + w_{1}x_{i} f''\big(w_{10}+w_{1}x_{i}\big) \right] \\
&\frac{\partial \epsilon}{\partial w_{2}} = 2\sqrt{\epsilon}\:\: v_{2} \left[f'\big(w_{20}+w_{2}x_{i}\big) + w_{2}x_{i} f''\big(w_{20}+w_{2}x_{i}\big) \right] \\
&\frac{\partial \epsilon}{\partial w_{10}} = 2\sqrt{\epsilon}\:\: v_{1}w_{1} f''\big(w_{10}+w_{1}x_{i}\big) \\
&\frac{\partial \epsilon}{\partial w_{20}} = 2\sqrt{\epsilon}\:\: v_{2}w_{2} f''\big(w_{20}+w_{2}x_{i}\big)
\end{split}
\end{equation}
where it is worth to mention that $f''(X)$ stands for the second order partial derivative of the function $f$ with respect to $X$. In contrast to $v_{0}$, the cost function $\epsilon$ is a function of $w_{i0}$ which is updated during the process.

So far, we have identified how the first order derivative of the network with respect to the input can be trained. To check a tendency of high order derivatives with respect to the input, I introduce the second order derivatives regarding the input and show corresponding training procedure as an example. The following equation is the second order derivative:
\begin{equation}
\label{eq:nd7}
\begin{split}
\frac{\partial^{2} y}{\partial x^{2}} &= \frac{\partial}{\partial x} \left[v_{1}w_{1} f'\big(w_{10}+w_{1}x\big) + v_{2}w_{2} f'\big(w_{20}+w_{2}x\big) \right] \\
&= v_{1}w_{1}^{2} f''\big(w_{10}+w_{1}x\big) + v_{2}w_{2}^{2} f''\big(w_{20}+w_{2}x\big)
\end{split}
\end{equation}
where the only difference with the first order derivative is that we now have the square terms of $w$. From the cost function for the second order derivative, i.e.,
\begin{equation} 
\label{eq:nd8}
\epsilon_{2} = \left(\frac{\partial^{2} y_i}{\partial x_i^{2}} - t_i \right)^{2},
\end{equation}
the gradients of the neural network weights can be expressed as follows:
\begin{equation} 
\label{eq:nd9}
\begin{split}
&\frac{\partial \epsilon_{2}}{\partial v_{1}} = 2\sqrt{\epsilon_{2}}\:\: w_{1}^{2} f''\big(w_{10}+w_{1}x_{i}\big) \\
&\frac{\partial \epsilon_{2}}{\partial v_{2}} = 2\sqrt{\epsilon_{2}}\:\: w_{2}^{2} f'\big(w_{20}+w_{2}x_{i}\big) \\
&\frac{\partial \epsilon_{2}}{\partial w_{1}} = 2\sqrt{\epsilon_{2}}\:\: v_{1} \left[2W_{1} f''\big(w_{10}+w_{1}x_{i}\big) + w_{1}^{3}x_{i} f'''\big(w_{10}+w_{1}x_{i}\big) \right] \\
&\frac{\partial \epsilon_{2}}{\partial w_{2}} = 2\sqrt{\epsilon_{2}}\:\: v_{2} \left[2W_{2} f''\big(w_{20}+w_{2}x_{i}\big) + w_{2}^{3}x_{i} f'''\big(w_{20}+w_{2}x_{i}\big) \right] \\
&\frac{\partial \epsilon_{2}}{\partial w_{10}} = 2\sqrt{\epsilon_{2}}\:\: v_{1}w_{1}^{2} f''\big(w_{10}+w_{1}x_{i}\big) \\
&\frac{\partial \epsilon_{2}}{\partial w_{20}} = 2\sqrt{\epsilon_{2}}\:\: v_{2}w_{2}^{2} f''\big(w_{20}+w_{2}x_{i}\big)
\end{split}
\end{equation}
where $v_{0}$ is also not updated at all during the training, and we can train the network output according to the second (or high) order changes in space (or time) of certain physical phenomena that we are concerned with. I collect the simple formulas for the neural network output, the first order, and the second order derivatives as shown below:
\begin{equation} 
\label{eq:nd10}
\begin{split}
y &= v_{0} + \left[v_{1} f\big(w_{10}+w_{1}x\big) + v_{2}f\big(w_{20}+w_{2}x\big) \right] \\
\frac{\partial y}{\partial x} &= v_{1}w_{1} f'\big(w_{10}+w_{1}x\big) + v_{2}w_{2} f'\big(w_{20}+w_{2}x\big) \\
\frac{\partial^{2} y}{\partial x^{2}} &= v_{1}w_{1}^{2} f''\big(w_{10}+w_{1}x\big) + v_{2}w_{2}^{2} f''\big(w_{20}+w_{2}x\big).
\end{split}
\end{equation}

Could one argue that what if the activation function $f$ is an exponential function? If the function $f(X)$ is $\exp{(X)}$ where $X$ is a certain variable, all the derivatives of the function $f$ will be identical, leading us to estimate high order derivatives of the network easily. Then, one can raise a question like ``is $\exp{(X)}$ able to play a role of non-linearity of the neural network as the sigmoid function, tanh and ReLU?" The answer is \textit{yes}, however, I would like to leave this fact that $\exp{(\exp{(\exp{(\exp{(1)})})})} \approx e^{3814279}$. This means that the output given by $\exp{(X)}$ will be literally exponentially skyrocketing, making the training procedure unstable.

Then what if we simplify $w$ to be unity. The major difference in Equation \ref{eq:nd10} is the power of $w$. Furthermore, it is reasonable that although $w$ is fixed, non-linearity of the network can be achievable from the hidden layer as well as a non-linear activation function. I would like to leave this argument as a future work, however it is clear that if we can find proper weights, then derivatives may easily be estimated.

\begin{figure}[t]
    \centerline{\includegraphics[width=0.575\columnwidth]{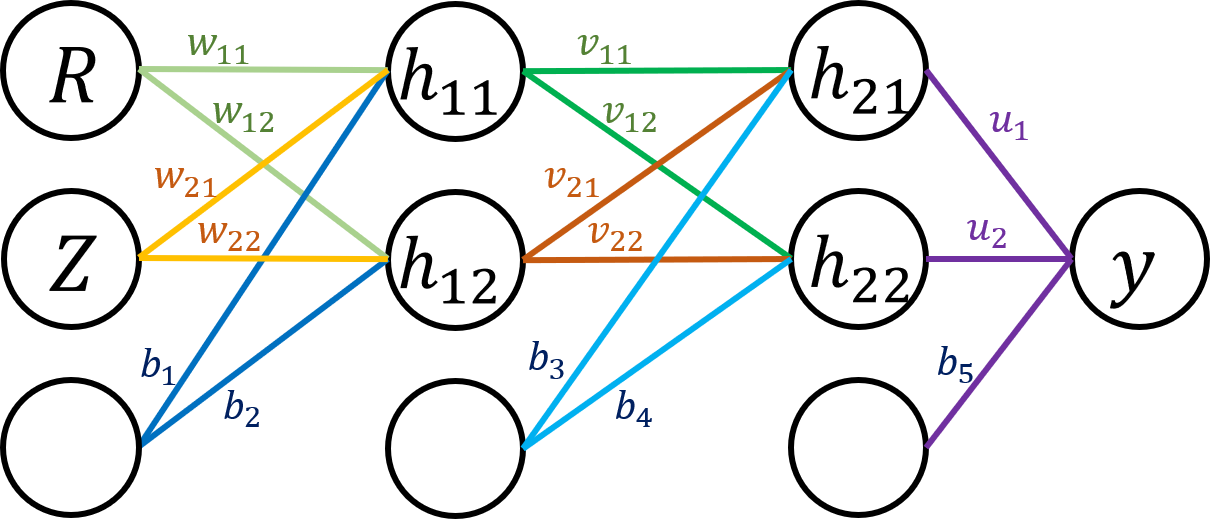}}
    \caption[Simple neural network 3]{A simple neural network having two input nodes, two hidden layers and one output node. There are two nodes in each hidden layer.
    } \label{ch2-fig5}
\end{figure}

Here, I would like to introduce an excursion into the topic that a network has more than one hidden layer with more input nodes. Let me take a look at what kind of complexity occurs when I add another hidden layer to the network used before. For simplicity, the hidden layers contain two hidden nodes each with one bias, and there are two input nodes. The functional form of the network displayed in Figure \ref{ch2-fig5} is presented as follows: 
\begin{equation} 
\label{eq:nd11}
\begin{split}
&h_{11} = f\big(w_{11}R + w_{21}Z + b_{1}\big) \\
&h_{12} = f\big(w_{12}R + w_{22}Z + b_{2}\big) \\
&h_{21} = f\big(v_{11}h_{11} + v_{21}h_{12} + b_{3}\big) \\
&h_{22} = f\big(v_{12}h_{11} + v_{22}h_{12} + b_{4}\big)
\end{split}
\end{equation}
where $b$ is the bias. Based on the expressions above, the output of the network can be expressed as follows:
\begin{equation}
\label{eq:nd12}
y = u_{1}h_{21} + u_{2}h_{22} + b_{5}.
\end{equation}

I would like to present here $\partial y/ \partial R$ only for instance, which is
\begin{equation}
\label{eq:nd13}
\begin{split}
\frac{\partial y}{\partial R} &= u_{1}\frac{\partial h_{21}}{\partial R} + u_{2}\frac{\partial h_{22}}{\partial R} \\
&= u_{1} \biggl[v_{11}w_{11}\: f'\big(w_{11}R + w_{21}Z + b_{1}\big) \\
&+ v_{21}w_{12}\: f'\big(w_{12}R + w_{22}Z + b_{2}\big) \biggl] f'\big(v_{11}h_{11} + v_{21}h_{12} + b_{3}\big) \\
&+ u_{2} \biggl[v_{12}w_{11}\: f'\big((w_{11}R + w_{21}Z + b_{1}\big) \\
&+ v_{22}w_{12}\: f'\big(w_{12}R + w_{22}Z + b_{2}\big) \biggl] f'\big(v_{12}h_{11} + v_{22}h_{12} + b_{4}\big) \\
&\equiv u_{1} \Big[v_{11}w_{11}\: \tilde{h}_{11} + v_{21}w_{12}\: \tilde{h}_{12} \Big] \: \tilde{h}_{21} + u_{2} \Big[v_{12}w_{11}\: \tilde{h}_{11} + v_{22}w_{12} \: \tilde{h}_{12} \Big] \: \tilde{h}_{22}
\end{split}
\end{equation}
The derivatives of the network look like becoming more complex than those of the previous network which has one hidden layer. However, if we define $\Delta_{21} \equiv v_{11}\: \tilde{h}_{11} + v_{21}\: \tilde{h}_{12}$ and $\Delta_{22} \equiv v_{12}\: \tilde{h}_{11} + v_{22}\: \tilde{h}_{12}$ with $w\equiv w_{11}=w_{12}$ assumption and assume $h = \tilde{h}$ by using $\exp{(X)}$ as the activation function, then the first order derivative turns out to be:
\begin{equation}
\label{eq:nd15}
\frac{\partial y}{\partial R} \approx u_{1}\: w\Delta_{21}h_{21} + u_{2}\: w\Delta_{22}h_{22}
\end{equation}
where this is somewhat identical to Equation \ref{eq:nd10} if we make such strict constraints, e.g., $\Delta$ should be unity, meaning that stacking more layers is futile.

We have seen so far the fact that how we train a neural network holds a differential equation itself. Recall that a differential equation is a fundamental representative of a physical phenomenon, which is often called differentiable physics as well, and inculcating that kind of equation in the network through the procedures above means that we possibly think that the network \textit{have knowledge of} corresponding physics that we are interested in. This might lead us to step forward to believe the network can be used for the scientific fields. In section \ref{ch3-1-3}, I show practical results of this scheme with the usual sine function problem.

\bibliographystyle{unsrt}
\bibliography{jpbib_thesis}

\begin{thebibliography}{100}

\bibitem{lee2001design}
GS~Lee, M~Kwon, CJ~Doh, BG~Hong, K~Kim, MH~Cho, W~Namkung, Choong-Seock Chang,
  YC~Kim, JY~Kim, et~al.
\newblock \href{https://doi.org/10.1088/0029-5515/41/10/318}{Design and
  construction of the KSTAR tokamak}.
\newblock {\em Nuclear Fusion}, 41(10):1515, 2001.

\bibitem{Lee:2008cl}
S~G Lee, J~G Bak, E~M Ka, J~H Kim, and S~H Hahn.
\newblock \href{https://doi.org/10.1063/1.2953587}{Magnetic diagnostics for the
  first plasma operation in Korea Superconducting Tokamak Advanced Research}.
\newblock {\em Review of Scientific Instruments}, 79(10):10F117--4, 2008.

\bibitem{Joung:2018ju}
Semin Joung, Jaewook Kim, Sehyun Kwak, Kyeo-reh Park, S~H Hahn, H~S Han, H~S
  Kim, J~G Bak, S~G Lee, and Y~c Ghim.
\newblock {Imputation of faulty magnetic sensors with coupled Bayesian and
  Gaussian processes to reconstruct the magnetic equilibrium in real time}.
\newblock {\em Review of Scientific Instruments}, 89(10):10K106, 2018.

\bibitem{lister1990implementation}
Jonathan~B Lister, HARRY Schnurrenberger, and Ph~Marmillod.
\newblock Implementation of a multi-layer perceptron for a non-linear control
  problem.
\newblock Technical report, LRP 398/90, CRPP-Lausanne, 1990.

\bibitem{Lister:1991gx}
J~B Lister and H~Schnurrenberger.
\newblock \href{https://doi.org/10.1088/0029-5515/31/7/005}{Fast non-linear
  extraction of plasma equilibrium parameters using a neural network mapping}.
\newblock {\em Nuclear Fusion}, 31(7):1291--1300, 1991.

\bibitem{LAGIN19931057}
L.~LAGIN, R.~BELL, S~DAVIS, T.~ECK, S.~JARDIN, C.~KESSEL, J.~MCENERNEY,
  M.~OKABAYASHI, J.~POPYACK, and N.~SAUTHOFF.
\newblock \href{https://doi.org/10.1016/B978-0-444-89995-8.50205-8}{APPLICATION
  OF NEURAL NETWORKS FOR REAL-TIME CALCULATIONS OF PLASMA EQUILIBRIUM
  PARAMETERS FOR PBX-M}.
\newblock In C.~FERRO, M.~GASPAROTTO, and H.~KNOEPFEL, editors, {\em Fusion
  Technology 1992}, pages 1057--1061. North-Holland, Oxford, 1993.

\bibitem{Coccorese:1994jt}
E~Coccorese, C~Morabito, and R~Martone.
\newblock \href{https://doi.org/10.1088/0029-5515/34/10/I05}{Identification of
  noncircular plasma equilibria using a neural network approach}.
\newblock {\em Nuclear Fusion}, 34(10):1349--1363, 1994.

\bibitem{vanMilligen:1995dv}
B~Ph van Milligen, V~Tribaldos, and J~A Jim{\'e}nez.
\newblock \href{https://doi.org/10.1103/PhysRevLett.75.3594}{Neural Network
  Differential Equation and Plasma Equilibrium Solver}.
\newblock {\em Physical Review Letters}, 75(20):3594--3597, 1995.

\bibitem{Bishop10.1162}
Chris~M. Bishop, Paul~S. Haynes, Mike E.~U. Smith, Tom~N. Todd, and David~L.
  Trotman.
\newblock \href{https://doi.org/10.1162/neco.1995.7.1.206}{Real-Time Control of
  a Tokamak Plasma Using Neural Networks}.
\newblock {\em Neural Computation}, 7(1):206--217, 01 1995.

\bibitem{windsor1997real}
Colin~George Windsor, Thomas~Noel Todd, David~Leonard Trotman, and Michael
  Edward~Underhill Smith.
\newblock \href{https://doi.org/10.13182/FST97-A5}{Real-time electronic neural
  networks for iter-like multiparameter equilibrium reconstruction and control
  in compass-d}.
\newblock {\em Fusion technology}, 32(3):416--430, 1997.

\bibitem{mc1998realtime}
Patrick~J Mc~Carthy and Francesco~C Morabito.
\newblock \href{https://doi.org/10.1007/978-1-4615-5353-3_66}{Realtime
  Identification of Plasma Shape and Position on ITER Using an Equilibrium
  Database}.
\newblock In {\em Diagnostics for Experimental Thermonuclear Fusion Reactors
  2}, pages 541--544. Springer, 1998.

\bibitem{Bishop10}
Chris~M. Bishop.
\newblock \href{https://doi.org/10.1063/1.1144830}{Neural networks and their
  applications}.
\newblock {\em Review of Scientific Instruments}, 65(6):1803--1832, 1994.

\bibitem{Wang:JFE2016}
Bo~Wang, Bingjia Xiao, Jiangang Li, Yong Guo, and Zhengping Luo.
\newblock \href{https://doi.org/10.1007/s10894-015-0044-z}{Artificial Neural
  Networks for Data Analysis of Magnetic Measurements on East}.
\newblock {\em Journal of Fusion Energy}, 35(2):390--400, Apr 2016.

\bibitem{carpanese2020first}
F~Carpanese, F~Felici, C~Galperti, A~Merle, JM~Moret, O~Sauter, et~al.
\newblock \href{https://doi.org/10.1088/1741-4326/ab81ac}{First demonstration
  of real-time kinetic equilibrium reconstruction on TCV by coupling LIUQE and
  RAPTOR}.
\newblock {\em Nuclear Fusion}, 60(6):066020, 2020.

\bibitem{degrave2022magnetic}
Jonas Degrave, Federico Felici, Jonas Buchli, Michael Neunert, Brendan Tracey,
  Francesco Carpanese, Timo Ewalds, Roland Hafner, Abbas Abdolmaleki, Diego
  de~Las~Casas, et~al.
\newblock \href{https://doi.org/10.1038/s41586-021-04301-9}{Magnetic control of
  tokamak plasmas through deep reinforcement learning}.
\newblock {\em Nature}, 602(7897):414--419, 2022.

\bibitem{wai2022neural}
JT~Wai, MD~Boyer, and E~Kolemen.
\newblock \href{ https://doi.org/10.48550/arXiv.2202.13915}{Neural net modeling
  of equilibria in NSTX-U}.
\newblock {\em arXiv preprint arXiv:2202.13915}, 2022.

\bibitem{markusdisruption}
Markus Svens{\'e}n.
\newblock Neural network prediction of disruptions in a tokamak plasma.
\newblock Master's thesis, Aston University, 1994.

\bibitem{windsor2005cross}
CG~Windsor, G~Pautasso, C~Tichmann, RJ~Buttery, TC~Hender, JET~EFDA
  Contributors, et~al.
\newblock \href{https://doi.org/10.1088/0029-5515/45/5/004}{A cross-tokamak
  neural network disruption predictor for the JET and ASDEX Upgrade tokamaks}.
\newblock {\em Nuclear fusion}, 45(5):337, 2005.

\bibitem{Tang:2019Nature}
{Kates-Harbeck, Julian}, {Svyatovskiy, Alexey}, and {Tang, William}.
\newblock \href{https://doi.org/10.1038/s41586-019-1116-4}{Predicting
  disruptive instabilities in controlled fusion plasmas through deep learning}.
\newblock {\em Nature}, 568(7753):526--531, 2019.

\bibitem{hernandez1996neural}
JV~Hernandez, A~Vannucci, T~Tajima, Z~Lin, W~Horton, and SC~McCool.
\newblock \href{https://doi.org/10.1088/0029-5515/36/8/I05}{Neural network
  prediction of some classes of tokamak disruptions}.
\newblock {\em Nuclear fusion}, 36(8):1009, 1996.

\bibitem{wroblewski1997tokamak}
D~Wroblewski, GL~Jahns, and JA~Leuer.
\newblock \href{https://doi.org/10.1088/0029-5515/37/6/I02}{Tokamak disruption
  alarm based on a neural network model of the high-beta limit}.
\newblock {\em Nuclear Fusion}, 37(6):725, 1997.

\bibitem{tresset2002transport}
G~Tresset, CD~Challis, X~Garbet, X~Litaudon, M~Mantsinen, D~Mazon, and
  D~Moreau.
\newblock Transport identification by neural network in jet itb regimes.
\newblock In {\em Proc. 29th EPS Conference on Controlled Fusion and Plasma
  Physics P}, volume~2, page 039, 2002.

\bibitem{wakasa2004development}
Arimitsu Wakasa, Sadayoshi MURAKAMI, Hiroshi YAMADA, Masayuki YOKOYAMA, Hening
  MAASSBERG, Craig BEIDLER, Kiyomasa WATANABE, and Shun-ichi OIKAWA.
\newblock Development of a neoclassical transport database by neural network
  fitting in lhd.
\newblock {\em J. Plasma Fusion Res. SERIES}, 6:203, 2004.

\bibitem{wakasa2008study}
Arimitsu Wakasa, Sadayoshi Murakami, and Shun-ichi Oikawa.
\newblock Study of neoclassical transport in lhd plasmas by applying the
  dcom/nnw neoclassical transport database.
\newblock {\em Plasma and Fusion Research}, 3:S1030--S1030, 2008.

\bibitem{Meneghini:2014ic}
{Meneghini, O}, {Luna, C J}, {Smith, S P}, and {Lao, L L}.
\newblock \href{https://doi.org/10.1063/1.4885343}{Modeling of transport
  phenomena in tokamak plasmas with neural networks}.
\newblock {\em Physics of Plasmas}, 21(6):060702--5, June 2014.

\bibitem{Citrin:2015fj}
{Citrin, J}, {Breton, S}, {Felici, F}, {Imbeaux, F}, {Aniel, T}, {Artaud, J F},
  {Baiocchi, B}, {Bourdelle, C}, {Camenen, Y}, and {Garcia, J}.
\newblock \href{https://doi.org/10.1088/0029-5515/55/9/092001}{Real-time
  capable first principle based modelling of tokamak turbulent transport}.
\newblock {\em Nuclear Fusion}, 55(9):092001--7, July 2015.

\bibitem{Meneghini:2017kp}
O~Meneghini, S~P Smith, P~B Snyder, G~M Staebler, J~Candy, E~Belli, L~Lao,
  M~Kostuk, T~Luce, T~Luda, J~M Park, and F~Poli.
\newblock \href{https://doi.org/10.1088/1741-4326/aa7776}{Self-consistent
  core-pedestal transport simulations with neural network accelerated models}.
\newblock {\em Nuclear Fusion}, 57(8):086034, 2017.

\bibitem{van2020fast}
Karel~Lucas van~de Plassche, Jonathan Citrin, Clarisse Bourdelle, Yann Camenen,
  Francis~J Casson, Victor~I Dagnelie, Federico Felici, Aaron Ho, Simon
  Van~Mulders, and JET Contributors.
\newblock \href{https://doi.org/10.1063/1.5134126}{Fast modeling of turbulent
  transport in fusion plasmas using neural networks}.
\newblock {\em Physics of Plasmas}, 27(2):022310, 2020.

\bibitem{gauss1877theoria}
Carl~Friedrich Gauss.
\newblock {\em Theoria motus corporum coelestium in sectionibus conicis solem
  ambientium}, volume~7.
\newblock FA Perthes, 1877.

\bibitem{bishop2006pattern}
Christopher~M Bishop and Nasser~M Nasrabadi.
\newblock {\em Pattern recognition and machine learning}, volume~4.
\newblock Springer, 2006.

\bibitem{rumelhart1985learning}
David~E Rumelhart, Geoffrey~E Hinton, and Ronald~J Williams.
\newblock Learning internal representations by error propagation.
\newblock Technical report, California Univ San Diego La Jolla Inst for
  Cognitive Science, 1985.

\bibitem{krizhevsky2012imagenet}
Alex Krizhevsky, Ilya Sutskever, and Geoffrey~E Hinton.
\newblock Imagenet classification with deep convolutional neural networks.
\newblock {\em Advances in neural information processing systems}, 25, 2012.

\bibitem{szegedy2015going}
Christian Szegedy, Wei Liu, Yangqing Jia, Pierre Sermanet, Scott Reed, Dragomir
  Anguelov, Dumitru Erhan, Vincent Vanhoucke, and Andrew Rabinovich.
\newblock Going deeper with convolutions.
\newblock In {\em Proceedings of the IEEE conference on computer vision and
  pattern recognition}, pages 1--9, 2015.

\bibitem{kalchbrenner2013recurrent}
Nal Kalchbrenner and Phil Blunsom.
\newblock Recurrent continuous translation models.
\newblock In {\em Proceedings of the 2013 conference on empirical methods in
  natural language processing}, pages 1700--1709, 2013.

\bibitem{sundermeyer2012lstm}
Martin Sundermeyer, Ralf Schl{\"u}ter, and Hermann Ney.
\newblock Lstm neural networks for language modeling.
\newblock In {\em Thirteenth annual conference of the international speech
  communication association}, 2012.

\bibitem{mikolov2010recurrent}
Tomas Mikolov, Martin Karafi{\'a}t, Lukas Burget, Jan Cernock{\`y}, and Sanjeev
  Khudanpur.
\newblock Recurrent neural network based language model.
\newblock In {\em Interspeech}, volume~2, pages 1045--1048. Makuhari, 2010.

\bibitem{cite-keyVerberk}
Bart Verberck and Andrea Taroni.
\newblock Nuclear fusion.
\newblock {\em Nature Physics}, 12(5):383--383, 2016.

\bibitem{Chen:2011wc}
Francis~F Chen.
\newblock {\em {An Indispensable Truth}}.
\newblock How Fusion Power Can Save the Planet. Springer, apr 2011.

\bibitem{federici2019overview}
G~Federici, L~Boccaccini, F~Cismondi, M~Gasparotto, Y~Poitevin, and I~Ricapito.
\newblock \href{https://doi.org/10.1016/j.fusengdes.2019.01.141}{An overview of
  the EU breeding blanket design strategy as an integral part of the DEMO
  design effort}.
\newblock {\em Fusion Engineering and Design}, 141:30--42, 2019.

\bibitem{oh2009commissioning}
Yeong-Kook Oh, WC~Kim, KR~Park, MK~Park, HL~Yang, YS~Kim, Y~Chu, YO~Kim,
  JG~Bak, EN~Baang, et~al.
\newblock \href{https://doi.org/10.1016/j.fusengdes.2008.12.099}{Commissioning
  and initial operation of KSTAR superconducting tokamak}.
\newblock {\em Fusion Engineering and Design}, 84(2-6):344--350, 2009.

\bibitem{han2022sustained}
H~Han, SJ~Park, C~Sung, J~Kang, YH~Lee, J~Chung, TS~Hahm, B~Kim, J-K Park,
  JG~Bak, et~al.
\newblock \href{https://doi.org/10.1038/s41586-022-05008-1}{A sustained
  high-temperature fusion plasma regime facilitated by fast ions}.
\newblock {\em Nature}, 609(7926):269--275, 2022.

\bibitem{Freidberg:1987}
Jeffrey~P. Freidberg.
\newblock {\em {Ideal Magnetohydrodynamics}}.
\newblock Plenum Press, New York, 1987.

\bibitem{grad1958hydromagnetic}
Harold Grad and Hanan Rubin.
\newblock Hydromagnetic equilibria and force-free fields.
\newblock {\em Journal of Nuclear Energy (1954)}, 7(3-4):284--285, 1958.

\bibitem{shafranov1966plasma}
VD~Shafranov.
\newblock Plasma equilibrium in a magnetic field.
\newblock {\em Reviews of plasma physics}, 2:103, 1966.

\bibitem{kim2009design}
HK~Kim, HL~Yang, GH~Kim, Jin-Yong Kim, Hogun Jhang, JS~Bak, and GS~Lee.
\newblock \href{https://doi.org/10.1016/j.fusengdes.2009.01.082}{Design
  features of the KSTAR in-vessel control coils}.
\newblock {\em Fusion Engineering and Design}, 84(2-6):1029--1032, 2009.

\bibitem{kim2005status}
K~Kim, HK~Park, KR~Park, BS~Lim, SI~Lee, Y~Chu, WH~Chung, YK~Oh, SH~Baek,
  SJ~Lee, et~al.
\newblock \href{https://doi.org/10.1088/0029-5515/45/8/003}{Status of the KSTAR
  superconducting magnet system development}.
\newblock {\em Nuclear Fusion}, 45(8):783, 2005.

\bibitem{urankar1982vector}
Laxmikant Urankar.
\newblock \href{https://doi.org/10.1109/TMAG.1982.1062166}{Vector potential and
  magnetic field of current-carrying finite arc segment in analytical form,
  Part III: Exact computation for rectangular cross section}.
\newblock {\em IEEE Transactions on Magnetics}, 18(6):1860--1867, 1982.

\bibitem{Strait:2008}
E~J Strait, E~D Fredrickson, and Takechi~M Moret, J~M.
\newblock \href{https://doi.org/10.13182/FST08-A1674}{Chapter 2: Magnetic
  diagnostics}.
\newblock {\em Fusion Science and Technology}, 53:304, 2008.

\bibitem{Lee:2001cp}
S~G Lee and J~G Bak.
\newblock \href{https://doi.org/10.1063/1.1310576}{Magnetic diagnostics for
  Korea superconducting tokamak advanced research}.
\newblock {\em Review of Scientific Instruments}, 72(1):439--4, 2001.

\bibitem{bak2001performance}
JG~Bak and SG~Lee.
\newblock \href{https://doi.org/10.1063/1.1309010}{Performance test of sample
  coils in the Korea superconducting tokamak advanced research magnetic
  diagnostics test chamber}.
\newblock {\em Review of Scientific Instruments}, 72(1):435--438, 2001.

\bibitem{bak2004performance}
JG~Bak, SG~Lee, KSTAR~Project Team, and DeRac Son.
\newblock \href{https://doi.org/10.1063/1.1789620}{Performance of the magnetic
  sensor and the integrator for the KSTAR magnetic diagnostics}.
\newblock {\em Review of scientific instruments}, 75(10):4305--4307, 2004.

\bibitem{lee2006fabrication}
SG~Lee and JG~Bak.
\newblock \href{https://doi.org/10.1063/1.2227437}{Fabrication details,
  calibrations, and installation activities of magnetic diagnostics for Korea
  Superconducting Tokamak Advanced Research}.
\newblock {\em Review of scientific instruments}, 77(10):10E306, 2006.

\bibitem{Bak:2007jj}
{Bak, J G}, {Lee, S G}, {Son, D}, and {Ga, E M}.
\newblock \href{https://doi.org/10.1063/1.2721405}{Analog integrator for the
  Korea superconducting tokamak advanced research magnetic diagnostics}.
\newblock {\em Review of Scientific Instruments}, 78(4):043504--6, 2007.

\bibitem{bak2008diamagnetic}
JG~Bak, SG~Lee, and EM~Ka.
\newblock \href{https://doi.org/10.1063/1.2956963}{Diamagnetic loop for the
  first plasma in the KSTAR machine}.
\newblock {\em Review of Scientific Instruments}, 79(10):10F118, 2008.

\bibitem{bak2011diamagnetic}
JG~Bak, SG~Lee, and HS~Kim.
\newblock \href{https://doi.org/10.1063/1.3600455}{Diamagnetic loop measurement
  in korea superconducting tokamak advanced research machine}.
\newblock {\em Review of Scientific Instruments}, 82(6):063504, 2011.

\bibitem{lee2010diagnostics}
JH~Lee, HK~Na, SG~Lee, JG~Bak, DC~Seo, SH~Seo, ST~Oh, WH~Ko, J~Chung, YU~Nam,
  et~al.
\newblock \href{https://doi.org/10.1063/1.3429942}{Diagnostics for first plasma
  and development plan on KSTAR}.
\newblock {\em Review of Scientific Instruments}, 81(6):063502, 2010.

\bibitem{kim2017improvements}
Heung-Su Kim, Jun-Gyo Bak, and Sang-hee Hahn.
\newblock \href{https://doi.org/10.1016/j.fusengdes.2017.02.023}{Improvements
  of magnetic measurements for plasma control in KSTAR tokamak}.
\newblock {\em Fusion Engineering and Design}, 123:641--645, 2017.

\bibitem{ali1993long}
S~Ali-Arshad and L~De~Kock.
\newblock \href{https://doi.org/10.1063/1.1143855}{Long-pulse analog
  integration}.
\newblock {\em Review of scientific instruments}, 64(9):2679--2682, 1993.

\bibitem{strait1997hybrid}
EJ~Strait, JD~Broesch, RT~Snider, and ML~Walker.
\newblock \href{https://doi.org/10.1063/1.1147835}{A hybrid digital--analog
  long pulse integrator}.
\newblock {\em Review of scientific instruments}, 68(1):381--384, 1997.

\bibitem{werner2006w7}
Andreas Werner.
\newblock \href{https://doi.org/10.1063/1.2220073}{W7-X magnetic diagnostics:
  Performance of the digital integrator}.
\newblock {\em Review of scientific instruments}, 77(10):10E307, 2006.

\bibitem{seo2010development}
Seong-Heon Seo, Andreas Werner, and M~Marquardt.
\newblock \href{https://doi.org/10.1063/1.3519303}{Development of a digital
  integrator for the KSTAR device}.
\newblock {\em Review of Scientific Instruments}, 81(12):123507, 2010.

\bibitem{wang2014digital}
Y~Wang, ZS~Ji, GJ~Xu, F~Wang, S~Li, XY~Sun, and ZC~Zhang.
\newblock \href{https://doi.org/10.1016/j.fusengdes.2014.02.019}{A digital long
  pulse integrator for EAST Tokamak}.
\newblock {\em Fusion Engineering and Design}, 89(5):618--622, 2014.

\bibitem{seo2009plasma}
Seong-Heon Seo.
\newblock \href{https://doi.org/10.1063/1.3079781}{Plasma current position
  measurements in the Korea Superconducting Tokamak Advanced Research device}.
\newblock {\em Physics of Plasmas}, 16(3):032501, 2009.

\bibitem{Moreau:2009dg}
{Moreau, Ph}, {Saint-Laurent, F}, and {Lister, J B}.
\newblock \href{https://doi.org/10.1016/j.fusengdes.2009.02.023}{Drift-free
  magnetic equilibrium reconstruction using the response to plasma position
  modulation}.
\newblock {\em Fusion Engineering and Design}, 84(7-11):1339--1343, June 2009.

\bibitem{doi:10.1063/1.3494275}
J.~H. Lee, S.~T. Oh, and H.~M. Wi.
\newblock \href{https://doi.org/10.1063/1.3494275}{Development of KSTAR Thomson
  scattering system}.
\newblock {\em Review of Scientific Instruments}, 81(10):10D528, 2010.

\bibitem{doi:10.1063/1.3496991}
Won-Ha Ko, Hyungho Lee, Dongcheol Seo, and Myeun Kwon.
\newblock \href{https://doi.org/10.1063/1.3496991}{Charge exchange spectroscopy
  system calibration for ion temperature measurement in KSTAR}.
\newblock {\em Review of Scientific Instruments}, 81(10):10D740, 2010.

\bibitem{bae2012commissioning}
YS~Bae, YM~Park, JS~Kim, WS~Han, SW~Kwak, YB~Chang, HT~Park, NH~Song, HL~Yang,
  SW~Yoon, et~al.
\newblock \href{https://doi.org/10.1016/j.fusengdes.2012.05.011}{Commissioning
  of the first KSTAR neutral beam injection system and beam experiments}.
\newblock {\em Fusion Engineering and Design}, 87(9):1597--1610, 2012.

\bibitem{sarwar2018effective}
S~Sarwar, HK~Na, and JM~Park.
\newblock \href{https://doi.org/10.1063/1.5004217}{Effective ion charge (Zeff)
  measurements and impurity behavior in KSTAR}.
\newblock {\em Review of Scientific Instruments}, 89(4):043504, 2018.

\bibitem{kim2012initial}
Junghee Kim, Jun~Young Kim, SW~Yoon, M~Garc{\'\i}a-Mu{\~n}oz, M~Isobe, and
  WC~Kim.
\newblock \href{https://doi.org/10.1063/1.4733550}{Initial measurements of fast
  ion loss in KSTAR}.
\newblock {\em Review of Scientific Instruments}, 83(10):10D305, 2012.

\bibitem{yoo2021fast}
JW~Yoo, Junghee Kim, MW~Lee, J~Kang, W-H Ko, SG~Oh, J~Ko, JH~Lee, YU~Nam,
  L~Jung, et~al.
\newblock \href{https://doi.org/10.1063/5.0040559}{Fast-ion D$\alpha$
  spectroscopy diagnostic at KSTAR}.
\newblock {\em Review of Scientific Instruments}, 92(4):043504, 2021.

\bibitem{levinton1989magnetic}
FM~Levinton, RJ~Fonck, GM~Gammel, R~Kaita, HW~Kugel, ET~Powell, and DW~Roberts.
\newblock \href{https://doi.org/10.1103/PhysRevLett.63.2060}{Magnetic field
  pitch-angle measurments in the PBX-M tokamak using the motional Stark
  effect}.
\newblock {\em Physical review letters}, 63(19):2060, 1989.

\bibitem{ko2013polarimetric}
J~Ko, J~Chung, AGG Lange, and MFM De~Bock.
\newblock \href{https://doi.org/10.1088/1748-0221/8/10/C10022}{Polarimetric
  spectra analysis for tokamak pitch angle measurements}.
\newblock {\em Journal of Instrumentation}, 8(10):C10022, 2013.

\bibitem{Lao:1985hn}
L~L Lao, H~St John, R~D Stambaugh, A~G Kellman, and W~Pfeiffer.
\newblock \href{https://doi.org/10.1088/0029-5515/25/11/007}{Reconstruction of
  current profile parameters and plasma shapes in tokamaks}.
\newblock {\em Nuclear Fusion}, 25(11):1611--1622, 1985.

\bibitem{Lao:2005kd}
L~L Lao, H~E~St John, Q~Peng, and J~R Ferron.
\newblock \href{https://doi.org/10.13182/FST48-968}{MHD Equilibrium
  Reconstruction in the DIII-D Tokamak}.
\newblock {\em Fusion Science and Technology}, 48(2):968--977, 2005.

\bibitem{Ferron:2002fw}
J~R Ferron, M~L Walker, L~L Lao, H~E~St John, D~A Humphreys, and J~A Leuer.
\newblock \href{https://doi.org/10.1088/0029-5515/38/7/308}{Real time
  equilibrium reconstruction for tokamak discharge control}.
\newblock {\em Nuclear Fusion}, 38(7):1055--1066, May 1998.

\bibitem{Yue:2013cj}
{Yue, X N}, {Xiao, B J}, {Luo, Z P}, and {Guo, Y}.
\newblock \href{https://doi.org/10.1088/0741-3335/55/8/085016}{Fast equilibrium
  reconstruction for tokamak discharge control based on GPU}.
\newblock {\em Plasma Physics and Controlled Fusion}, 55(8):085016--10, June
  2013.

\bibitem{berkery2021kinetic}
JW~Berkery, SA~Sabbagh, L~Kogan, D~Ryan, JM~Bialek, Y~Jiang, DJ~Battaglia,
  S~Gibson, and C~Ham.
\newblock \href{https://doi.org/10.1088/1361-6587/abf230}{Kinetic equilibrium
  reconstructions of plasmas in the MAST database and preparation for
  reconstruction of the first plasmas in MAST upgrade}.
\newblock {\em Plasma Physics and Controlled Fusion}, 63(5):055014, 2021.

\bibitem{joung2019deep}
Semin Joung, Jaewook Kim, Sehyun Kwak, JG~Bak, SG~Lee, HS~Han, HS~Kim, Geunho
  Lee, Daeho Kwon, and Y-C Ghim.
\newblock \href{https://doi.org/10.1088/1741-4326/ab555f}{Deep neural network
  Grad--Shafranov solver constrained with measured magnetic signals}.
\newblock {\em Nuclear Fusion}, 60(1):016034, 2019.

\bibitem{Park:2011go}
Y~S Park, S~A Sabbagh, J~W Berkery, J~M Bialek, Y~M Jeon, S~H Hahn, N~Eidietis,
  T~E Evans, S~W Yoon, J~W Ahn, J~Kim, H~L Yang, K~I You, Y~S Bae, J~Chung,
  M~Kwon, Y~K Oh, W~C Kim, J~Y Kim, S~G Lee, H~K Park, H~Reimerdes, J~Leuer,
  and M~Walker.
\newblock \href{https://doi.org/10.1088/0029-5515/51/5/053001}{KSTAR
  equilibrium operating space and projected stabilization at high normalized
  beta}.
\newblock {\em Nuclear Fusion}, 51(5):053001, 2011.

\bibitem{mcculloch1943logical}
Warren~S McCulloch and Walter Pitts.
\newblock A logical calculus of the ideas immanent in nervous activity.
\newblock {\em The bulletin of mathematical biophysics}, 5(4):115--133, 1943.

\bibitem{hebb1949organization}
Donald~Olding Hebb.
\newblock {\em The organization of behavior: a neuropsychological theory}.
\newblock Science editions, 1949.

\bibitem{rosenblatt1958perceptron}
Frank Rosenblatt.
\newblock The perceptron: a probabilistic model for information storage and
  organization in the brain.
\newblock {\em Psychological review}, 65(6):386, 1958.

\bibitem{minsky1960learning}
Marvin Minsky and Oliver~G Selfridge.
\newblock {\em Learning in random nets}, volume~46.
\newblock MIT Lincoln Laboratory, 1960.

\bibitem{minsky1969perceptrons}
Marvin Minsky and Seymour Papert.
\newblock Perceptrons an introduction to computational geometry.
\newblock {\em MIT Press}, 1969.

\bibitem{minsky1988perceptrons}
Marvin Minsky and Seymour Papert.
\newblock {\em Perceptrons: expanded edition}.
\newblock MIT press, 1988.

\bibitem{Sivia:2006}
D~S Sivia and J~Skilling.
\newblock {\em {Data Analysis: A Bayesian Tutorial}}.
\newblock Oxford: Oxford University Press, 2006.

\bibitem{kwak2017bayesian}
Sehyun Kwak, J~Svensson, M~Brix, Y-c Ghim, and JET Contributors.
\newblock \href{https://doi.org/10.1088/1741-4326/aa5072}{Bayesian electron
  density inference from JET lithium beam emission spectra using Gaussian
  processes}.
\newblock {\em Nuclear Fusion}, 57(3):036017, 2017.

\bibitem{kwak2020bayesian}
Sehyun Kwak, Jakob Svensson, S~Bozhenkov, Joanne Flanagan, Mark Kempenaars,
  Alexandru Boboc, Y-C Ghim, and JET Contributors.
\newblock \href{https://doi.org/10.1088/1741-4326/ab686e}{Bayesian modelling of
  Thomson scattering and multichannel interferometer diagnostics using Gaussian
  processes}.
\newblock {\em Nuclear Fusion}, 60(4):046009, 2020.

\bibitem{kwak2021bayesian}
Sehyun Kwak, U~Hergenhahn, U~H{\"o}fel, M~Krychowiak, A~Pavone, J~Svensson,
  O~Ford, R~K{\"o}nig, S~Bozhenkov, G~Fuchert, et~al.
\newblock \href{https://doi.org/10.1063/5.0043777}{Bayesian inference of
  spatially resolved Z eff profiles from line integrated Bremsstrahlung
  spectra}.
\newblock {\em Review of Scientific Instruments}, 92(4):043505, 2021.

\bibitem{chilenski2015improved}
MA~Chilenski, M~Greenwald, Y~Marzouk, NT~Howard, AE~White, JE~Rice, and
  JR~Walk.
\newblock \href{https://doi.org/10.1088/0029-5515/55/2/023012}{Improved profile
  fitting and quantification of uncertainty in experimental measurements of
  impurity transport coefficients using Gaussian process regression}.
\newblock {\em Nuclear Fusion}, 55(2):023012, 2015.

\bibitem{chilenski2017experimentally}
MA~Chilenski, MJ~Greenwald, AE~Hubbard, JW~Hughes, JP~Lee, YM~Marzouk, JE~Rice,
  and AE~White.
\newblock \href{https://doi.org/10.1088/1741-4326/aa8387}{Experimentally
  testing the dependence of momentum transport on second derivatives using
  Gaussian process regression}.
\newblock {\em Nuclear Fusion}, 57(12):126013, 2017.

\bibitem{ho2019application}
Aaron Ho, Jonathan Citrin, Fulvio Auriemma, Clarisse Bourdelle, Francis~J
  Casson, Hyun-Tae Kim, Pierre Manas, Gabor Szepesi, Henri Weisen, and JET
  Contributors.
\newblock \href{https://doi.org/10.1088/1741-4326/ab065a}{Application of
  Gaussian process regression to plasma turbulent transport model validation
  via integrated modelling}.
\newblock {\em Nuclear fusion}, 59(5):056007, 2019.

\bibitem{linder2018flux}
O~Linder, J~Citrin, GMD Hogeweij, C~Angioni, Clarisse Bourdelle, Francis~J
  Casson, E~Fable, A~Ho, Florian Koechl, M~Sertoli, et~al.
\newblock \href{https://doi.org/10.1088/1741-4326/aae875}{Flux-driven
  integrated modelling of main ion pressure and trace tungsten transport in
  ASDEX Upgrade}.
\newblock {\em Nuclear Fusion}, 59(1):016003, 2018.

\bibitem{leddy2022single}
Jarrod Leddy, Sandeep Madireddy, Eric Howell, and Scott Kruger.
\newblock \href{https://doi.org/10.48550/arXiv.2202.11557}{Single Gaussian
  Process Method for Arbitrary Tokamak Regimes with a Statistical Analysis}.
\newblock {\em arXiv preprint arXiv:2202.11557}, 2022.

\bibitem{gal2016uncertainty}
Yarin Gal et~al.
\newblock {\em Uncertainty in deep learning}.
\newblock PhD thesis, 2016.

\bibitem{goodfellow2014generative}
Ian Goodfellow, Jean Pouget-Abadie, Mehdi Mirza, Bing Xu, David Warde-Farley,
  Sherjil Ozair, Aaron Courville, and Yoshua Bengio.
\newblock Generative adversarial nets.
\newblock {\em Advances in neural information processing systems}, 27, 2014.

\bibitem{creswell2018generative}
Antonia Creswell, Tom White, Vincent Dumoulin, Kai Arulkumaran, Biswa Sengupta,
  and Anil~A Bharath.
\newblock Generative adversarial networks: An overview.
\newblock {\em IEEE Signal Processing Magazine}, 35(1):53--65, 2018.

\bibitem{de2017learning}
Luke de~Oliveira, Michela Paganini, and Benjamin Nachman.
\newblock Learning particle physics by example: location-aware generative
  adversarial networks for physics synthesis.
\newblock {\em Computing and Software for Big Science}, 1(1):1--24, 2017.

\bibitem{paganini2018calogan}
Michela Paganini, Luke de~Oliveira, and Benjamin Nachman.
\newblock Calogan: Simulating 3d high energy particle showers in multilayer
  electromagnetic calorimeters with generative adversarial networks.
\newblock {\em Physical Review D}, 97(1):014021, 2018.

\bibitem{paganini2018accelerating}
Michela Paganini, Luke de~Oliveira, and Benjamin Nachman.
\newblock Accelerating science with generative adversarial networks: an
  application to 3d particle showers in multilayer calorimeters.
\newblock {\em Physical review letters}, 120(4):042003, 2018.

\bibitem{kim2020generative}
Sungwon Kim, Juhwan Noh, Geun~Ho Gu, Alan Aspuru-Guzik, and Yousung Jung.
\newblock Generative adversarial networks for crystal structure prediction.
\newblock {\em ACS central science}, 6(8):1412--1420, 2020.

\bibitem{kim2020inverse}
Baekjun Kim, Sangwon Lee, and Jihan Kim.
\newblock Inverse design of porous materials using artificial neural networks.
\newblock {\em Science advances}, 6(1):eaax9324, 2020.

\bibitem{tensorflow2015-whitepaper}
Mart\'{\i}n Abadi, Ashish Agarwal, Paul Barham, Eugene Brevdo, Zhifeng Chen,
  Craig Citro, Greg~S. Corrado, Andy Davis, Jeffrey Dean, Matthieu Devin,
  Sanjay Ghemawat, Ian Goodfellow, Andrew Harp, Geoffrey Irving, Michael Isard,
  Yangqing Jia, Rafal Jozefowicz, Lukasz Kaiser, Manjunath Kudlur, Josh
  Levenberg, Dandelion Man\'{e}, Rajat Monga, Sherry Moore, Derek Murray, Chris
  Olah, Mike Schuster, Jonathon Shlens, Benoit Steiner, Ilya Sutskever, Kunal
  Talwar, Paul Tucker, Vincent Vanhoucke, Vijay Vasudevan, Fernanda Vi\'{e}gas,
  Oriol Vinyals, Pete Warden, Martin Wattenberg, Martin Wicke, Yuan Yu, and
  Xiaoqiang Zheng.
\newblock {TensorFlow}: Large-scale machine learning on heterogeneous systems,
  2015.
\newblock Software available from tensorflow.org.

\bibitem{Sakakibar2010}
S.~Sakakibara, H.~Yamada, and LHD~Experiment Group.
\newblock Magnetic measurements in lhd.
\newblock {\em Fusion Science and Technology}, 58(1):471--481, 2010.

\bibitem{Edlington:2001jg}
{Edlington, T}, {Martin, R}, and {Pinfold, T}.
\newblock {MAST magnetic diagnostics}.
\newblock {\em Review of Scientific Instruments}, 72(1):421--6, 2001.

\bibitem{Strait:2006kh}
{Strait, E J}.
\newblock {Magnetic diagnostic system of the DIII-D tokamak}.
\newblock {\em Review of Scientific Instruments}, 77(2):023502--15, 2006.

\bibitem{Moret:1998fh}
J~M Moret, F~Buhlmann, D~Fasel, F~Hofmann, and G~Tonetti.
\newblock {Magnetic measurements on the TCV Tokamak}.
\newblock {\em Review of Scientific Instruments}, 69(6):2333--17, 1998.

\bibitem{Liu:2013dn}
{Liu, G J}, {Wan, B N}, {Sun, Y W}, {Xiao, B J}, {Wang, Y}, {Luo, Zh P}, {Qian,
  J P}, and {Liu, D M}.
\newblock {Analysis of uncertainty in equilibrium reconstruction in the EAST
  superconducting tokamak}.
\newblock {\em Review of Scientific Instruments}, 84(7):073502--7, 2013.

\bibitem{Peruzzo:2009et}
Simone Peruzzo, Raffaele Albanese, Giovanni Artaserse, Vincenzo Coccorese,
  Sergei Gerasimov, Norman Lam, Francesco Maviglia, Ian Pearson, Phil Prior,
  Antonio Quercia, and Luca Zabeo.
\newblock {Installation and commissioning of the JET-EP magnetic diagnostic
  system}.
\newblock {\em Fusion Engineering and Design}, 84(7-11):1495--1498, 2009.

\bibitem{Chavan:2009gk}
R~Chavan, G~Chitarin, R~S Delogu, A~Encheva, A~Gallo, E~R Hodgson, L~C
  Ingesson, A~Le-Luyer, J~B Lister, Ph~Moreau, J~M Moret, S~Peruzzo, J~Romero,
  D~S Testa, M~Toussaint, G~Vayakis, and R~Vila.
\newblock {The magnetics diagnostic set for ITER}.
\newblock {\em Fusion Engineering and Design}, 84(2-6):295--299, June 2009.

\bibitem{Orlinskij:191982}
D~V Orlinskij and G~Magyar.
\newblock {Plasma diagnostics on large Tokamaks}.
\newblock {\em Nuclear Fusion}, 28(JET-P-88-33):611--697, 1988.

\bibitem{Shiraki:2015jp}
D~Shiraki, N~Commaux, L~R Baylor, N~W Eidietis, E~M Hollmann, V~A Izzo, R~A
  Moyer, and C~Paz-Soldan.
\newblock {Characterization of MHD activity and its influence on radiation
  asymmetries during massive gas injection in DIII-D}.
\newblock {\em Nuclear Fusion}, 55(7):073029, 2015.

\bibitem{Snipes:1988cq}
J~A Snipes, D~J Campbell, P~S Haynes, et~al.
\newblock {Large amplitude quasi-stationary MHD modes in JET}.
\newblock {\em Nuclear Fusion}, 28(6):1085--1097, 1988.

\bibitem{Gerasimov:2014dp}
S~N Gerasimov, T~C Hender, J~Morris, V~Riccardo, L~E Zakharov, and JET~EFDA
  Contributors.
\newblock {Plasma current asymmetries during disruptions in JET}.
\newblock {\em Nuclear Fusion}, 54(7):073009, 2014.

\bibitem{Buttery:2002im}
R~J Buttery, O~Sauter, R~Akers, M~Gryaznevich, R~Martin, C~D Warrick, H~R
  Wilson, and the~MAST Team.
\newblock {Neoclassical Tearing Physics in the Spherical Tokamak MAST}.
\newblock {\em Physical Review Letters}, 88(12), 2002.

\bibitem{Park:2013ex}
Y~S Park, S~A Sabbagh, J~M Bialek, et~al.
\newblock {Investigation of MHD instabilities and control in KSTAR preparing
  for high beta operation}.
\newblock {\em Nuclear Fusion}, 53(8):083029, 2013.

\bibitem{Svensson:2008in}
{Svensson, J}, {Werner, A}, and {JET EFDA Contributors}.
\newblock {Current tomography for axisymmetric plasmas}.
\newblock {\em Plasma Physics and Controlled Fusion}, 50(8):085002--26, May
  2008.

\bibitem{Lazerson:2015bb}
S~A Lazerson and the DIII-D Team.
\newblock {Three-dimensional equilibrium reconstruction on the DIII-D device}.
\newblock {\em Nuclear Fusion}, 55(2):023009, 2015.

\bibitem{Romero:2013eg}
{Romero, J A} and {Svensson, J}.
\newblock {Optimization of out-vessel magnetic diagnostics for plasma boundary
  reconstruction in tokamaks}.
\newblock {\em Nuclear Fusion}, 53(3):033009--26, February 2013.

\bibitem{Qian:2017gj}
J~P Qian, L~L Lao, C~T Holcomb, B~N Wan, Y~W Sun, D~Moreau, E~Li, L~Zeng,
  K~Hanada, A~M Garofalo, X~Z Gong, B~Shen, and B~J Xiao.
\newblock {An efficient technique for magnetic equilibrium reconstruction with
  q profile constraints and its application on the EAST tokamak}.
\newblock {\em Nuclear Fusion}, 57(8):084001, 2017.

\bibitem{Xia:2015cv}
{Xia, Yu-Jun}, {Zhang, Zhong-Dian}, {Xia, Zhen-Xin}, {Zhu, Shi-Liang}, and
  {Zhang, Rui}.
\newblock {A precision analogue integrator system for heavy current measurement
  in MFDC resistance spot welding}.
\newblock {\em Measurement Science and Technology}, 27(2):025104--12, December
  2015.

\bibitem{Shikama:2003jn}
T~Shikama, T~Nishitani, T~Kakuta, S~Yamamoto, S~Kasai, M~Narui, E~Hodgson,
  R~Reichle, B~Brichard, A~Krassilinikov, R~Snider, G~Vayakis, A~Costley,
  S~Nagata, B~Tsuchiya, and K~Toh.
\newblock {Irradiation test of diagnostic components for ITER application in
  the Japan Materials Testing Reactor}.
\newblock {\em Nuclear Fusion}, 43(7):517--521, 2003.

\bibitem{Ariola:2008}
Marco Ariola and Alfredo Pironti.
\newblock {\em {Magnetic Control of Tokamak Plasmas}}.
\newblock Springer, 2008.

\bibitem{Bishop:1994kr}
Chris~M Bishop, Paul~S Haynes, Mike E~U Smith, Tom~N Todd, and David~L Trotman.
\newblock {Fast feedback control of a high temperature fusion plasma}.
\newblock {\em Neural Computing {\&} Applications}, 2(3):148--159, 1994.

\bibitem{Arshad:1993ev}
S~Ali Arshad and L~de~Kock.
\newblock {Long-pulse analog integration}.
\newblock {\em Review of Scientific Instruments}, 64(9):2679--2682, 1993.

\bibitem{Strait:1997ds}
E~J Strait, J~D Broesch, R~T Snider, and M~L Walker.
\newblock {A hybrid digital{\textendash}analog long pulse integrator}.
\newblock {\em Review of Scientific Instruments}, 68(1):381--5, 1997.

\bibitem{Werner:2006hz}
{Werner, Andreas}.
\newblock {W7-X magnetic diagnostics: Performance of the digital integrator}.
\newblock {\em Review of Scientific Instruments}, 77(10):10E307--6, 2006.

\bibitem{Wang:2014el}
Y~Wang, Z~S Ji, G~J Xu, F~Wang, S~Li, X~Y Sun, and Z~C Zhang.
\newblock {A digital long pulse integrator for EAST Tokamak}.
\newblock {\em Fusion Engineering and Design}, 89(5):618--622, 2014.

\bibitem{Kwon:2011dq}
M~Kwon et~al.
\newblock {Overview of KSTAR initial operation}.
\newblock {\em Nuclear Fusion}, 51(9), September 2011.

\bibitem{Ka:2008jg}
{Ka, E M}, {Lee, S G}, {Bak, J G}, and {Son, D}.
\newblock {Performance test of the integrator system for magnetic diagnostics
  in KSTAR}.
\newblock {\em Review of Scientific Instruments}, 79(10):10F119--4, 2008.

\bibitem{Rasmussen:2006}
C~E Rasmussen and C~K~I Williams.
\newblock {\em {Gaussian Processes for Machine Learning}}.
\newblock The MIT Press, 2006.

\bibitem{vanLint:2005cn}
J~W~C van Lint, S~P Hoogendoorn, and H~J van Zuylen.
\newblock {Accurate freeway travel time prediction with state-space neural
  networks under missing data}.
\newblock {\em Transportation Research Part C: Emerging Technologies},
  13(5-6):347--369, October 2005.

\bibitem{Neto:2014cm}
Andre~C Neto, Diogo Alves, Bernardo~B Carvalho, Gianmaria De~Tommasi, Robert
  Felton, Horacio Fernandes, Peter~J Lomas, Francesco Maviglia, Fernanda~G
  Rimini, Filippo Sartori, Adam~V Stephen, Daniel~F Valcarcel, and Luca Zabeo.
\newblock {A Real-Time Architecture for the Identification of Faulty Magnetic
  Sensors in the JET Tokamak}.
\newblock {\em IEEE Transactions on Nuclear Science}, 61(3):1228--1235, 2014.

\bibitem{Nouailletas:2012ho}
R{\'e}my Nouailletas, Philippe Moreau, and Sylvain Bremond.
\newblock {A generic method for real time detection of magnetic sensor failure
  on tokamaks}.
\newblock {\em Fusion Engineering and Design}, 87(3):289--297, March 2012.

\bibitem{Freidberg:2014dt}
Jeffrey Freidberg.
\newblock {\em {Ideal MHD}}.
\newblock Cambridge University Press, 2014.

\bibitem{Tsaun:2007il}
S~Tsaun and Hogun Jhang.
\newblock {Real-time plasma boundary reconstruction in the KSTAR tokamak using
  finite element method}.
\newblock {\em Fusion Engineering and Design}, 82(2):163--174, February 2007.

\bibitem{Mises:1964}
R~von Mises.
\newblock {\em {Mathematical Theory of Probability and Statistics}}.
\newblock Academic Press, 1964.

\bibitem{Kwak:2017gy}
Sehyun Kwak, J~Svensson, M~Brix, and Y.-c Ghim.
\newblock {Bayesian electron density inference from JET lithium beam emission
  spectra using Gaussian processes}.
\newblock {\em Nuclear Fusion}, 57(3), March 2017.

\bibitem{Foster:2009rr}
Leslie Foster et~al.
\newblock {Stable and Efficient Gaussian Process Calculations}.
\newblock {\em Journal of Machine Learning Research}, 10:857, April 2009.

\bibitem{Kwak:2016kv}
Sehyun Kwak, J~Svensson, M~Brix, Y~c Ghim, and JET Contributors.
\newblock {Bayesian modelling of the emission spectrum of the Joint European
  Torus Lithium Beam Emission Spectroscopy system}.
\newblock {\em Review of Scientific Instruments}, 87(2):023501, 2016.

\bibitem{Li:2013apa}
Dong Li, J~Svensson, H~Thomsen, F~Medina, A~Werner, and R~Wolf.
\newblock {Bayesian soft X-ray tomography using non-stationary Gaussian
  Processes}.
\newblock {\em Rev. Sci. Instrum.}, 84:083506, 2013.

\bibitem{joung2022fl}
Semin Joung, Jaewook Kim, HS~Kim, JG~Bak, and Y-C Ghim.
\newblock A deep learning approach to recover hidden consistency of kstar flux
  loop signals.
\newblock {\em arXiv}, 2022.

\bibitem{worswick2018deep}
Steven~G Worswick, James~A Spencer, Gunnar Jeschke, and Ilya Kuprov.
\newblock Deep neural network processing of deer data.
\newblock {\em Science advances}, 4(8):eaat5218, 2018.

\bibitem{bar2019learning}
Yohai Bar-Sinai, Stephan Hoyer, Jason Hickey, and Michael~P Brenner.
\newblock Learning data-driven discretizations for partial differential
  equations.
\newblock {\em Proceedings of the National Academy of Sciences},
  116(31):15344--15349, 2019.

\bibitem{raissi2020hidden}
Maziar Raissi, Alireza Yazdani, and George~Em Karniadakis.
\newblock Hidden fluid mechanics: Learning velocity and pressure fields from
  flow visualizations.
\newblock {\em Science}, 367(6481):1026--1030, 2020.

\bibitem{pfau2020ab}
David Pfau, James~S Spencer, Alexander~GDG Matthews, and W~Matthew~C Foulkes.
\newblock Ab initio solution of the many-electron schr{\"o}dinger equation with
  deep neural networks.
\newblock {\em Physical Review Research}, 2(3):033429, 2020.

\bibitem{mills2017deep}
Kyle Mills, Michael Spanner, and Isaac Tamblyn.
\newblock Deep learning and the schr{\"o}dinger equation.
\newblock {\em Physical Review A}, 96(4):042113, 2017.

\bibitem{torlai2018neural}
Giacomo Torlai, Guglielmo Mazzola, Juan Carrasquilla, Matthias Troyer, Roger
  Melko, and Giuseppe Carleo.
\newblock Neural-network quantum state tomography.
\newblock {\em Nature Physics}, 14(5):447--450, 2018.

\bibitem{li2021kohn}
Li~Li, Stephan Hoyer, Ryan Pederson, Ruoxi Sun, Ekin~D Cubuk, Patrick Riley,
  Kieron Burke, et~al.
\newblock Kohn-sham equations as regularizer: Building prior knowledge into
  machine-learned physics.
\newblock {\em Physical review letters}, 126(3):036401, 2021.

\bibitem{joung2022deep}
Semin Joung and Y-C Ghim.
\newblock Mastering plasma equilibria with deep neural networks and
  grad-shafranov equation.
\newblock {\em Science Advances}, 2022, submitted.

\bibitem{pmlr-v9-glorot10a}
Xavier Glorot and Yoshua Bengio.
\newblock Understanding the difficulty of training deep feedforward neural
  networks.
\newblock In Yee~Whye Teh and Mike Titterington, editors, {\em Proceedings of
  the Thirteenth International Conference on Artificial Intelligence and
  Statistics}, volume~9 of {\em Proceedings of Machine Learning Research},
  pages 249--256, Chia Laguna Resort, Sardinia, Italy, 13--15 May 2010. PMLR.

\bibitem{ramachandran2017searching}
Prajit Ramachandran, Barret Zoph, and Quoc~V Le.
\newblock Searching for activation functions.
\newblock {\em arXiv preprint arXiv:1710.05941}, 2017.

\bibitem{gal2016dropout}
Yarin Gal and Zoubin Ghahramani.
\newblock Dropout as a bayesian approximation: Representing model uncertainty
  in deep learning.
\newblock In {\em international conference on machine learning}, pages
  1050--1059. PMLR, 2016.

\bibitem{rao1973linear}
Calyampudi~Radhakrishna Rao, Calyampudi~Radhakrishna Rao, Mathematischer
  Statistiker, Calyampudi~Radhakrishna Rao, and Calyampudi~Radhakrishna Rao.
\newblock {\em Linear statistical inference and its applications}, volume~2.
\newblock Wiley New York, 1973.

\bibitem{Sabbagh:2001hh}
S~A Sabbagh, S~M Kaye, J~Menard, F~Paoletti, M~Bell, R~E Bell, J~M Bialek,
  M~Bitter, E~D Fredrickson, D~A Gates, A~H Glasser, H~Kugel, L~L Lao, B~P
  LeBlanc, R~Maingi, R~J Maqueda, E~Mazzucato, D~Mueller, M~Ono, S~F Paul,
  M~Peng, C~H Skinner, D~Stutman, G~A Wurden, W~Zhu, and NSTX~Research Team.
\newblock {Equilibrium properties of spherical torus plasmas in NSTX}.
\newblock {\em Nuclear Fusion}, 41(11):1601--1611, 2001.

\bibitem{Jinping:2009jn}
Qian Jinping, Wan Baonian, L~L Lao, Shen Biao, S~A Sabbagh, Sun Youwen, Liu
  Dongmei, Xiao Bingjia, Ren Qilong, Gong Xianzu, and Li~Jiangang.
\newblock {Equilibrium Reconstruction in EAST Tokamak}.
\newblock {\em Plasma Science and Technology}, 11(2):142--145, 2009.

\bibitem{OBrien:1992gm}
D~P O'Brien, L~L Lao, E~R Solano, M~Garribba, T~S Taylor, J~G Cordey, and J~J
  Ellis.
\newblock {Equilibrium analysis of iron core tokamaks using a full domain
  method}.
\newblock {\em Nuclear Fusion}, 32(8):1351--1360, 1992.

\bibitem{Joung:2019}
Semin Joung and Y.-c. Ghim.
\newblock {Bayesian based real-time numerical method to correct signal drifts
  in magnetic measurements from tokamaks}.
\newblock 2019.
\newblock unpublished.

\bibitem{VanHoutte:1993}
D~Van~Houtte and Equipe~TORE SUPRA.
\newblock {One minute pulse operation in the Tore Supra Tokamak}.
\newblock {\em Nuclear Fusion}, 33(1):137--141, 1993.

\bibitem{Ekedahl_2010}
A.~Ekedahl, L.~Delpech, M.~Goniche, D.~Guilhem, J.~Hillairet, M.~Preynas, P.K.
  Sharma, J.~Achard, Y.S. Bae, X.~Bai, C.~Balorin, Y.~Baranov, V.~Basiuk,
  A.~B{\'{e}}coulet, J.~Belo, G.~Berger-By, S.~Br{\'{e}}mond, C.~Castaldo,
  S.~Ceccuzzi, R.~Cesario, E.~Corbel, X.~Courtois, J.~Decker, E.~Delmas,
  X.~Ding, D.~Douai, C.~Goletto, J.P. Gunn, P.~Hertout, G.T. Hoang, F.~Imbeaux,
  K.K. Kirov, X.~Litaudon, R.~Magne, J.~Mailloux, D.~Mazon, F.~Mirizzi,
  P.~Mollard, P.~Moreau, T.~Oosako, V.~Petrzilka, Y.~Peysson, S.~Poli, M.~Prou,
  F.~Saint-Laurent, F.~Samaille, and B.~Saoutic.
\newblock Validation of the {ITER}-relevant passive-active-multijunction {LHCD}
  launcher on long pulses in tore supra.
\newblock {\em Nuclear Fusion}, 50(11):112002, sep 2010.

\bibitem{Itoh_1999}
S~Itoh, K~N Sato, K~Nakamura, H~Zushi, M~Sakamoto, K~Hanada, E~Jotaki,
  K~Makino, S~Kawasaki, H~Nakashima, and A~Iyomasa.
\newblock Recent progress in the superconducting tokamak {TRIAM}-1m.
\newblock {\em Plasma Physics and Controlled Fusion}, 41(3A):A587--A594, jan
  1999.

\bibitem{Zushi_2003}
H~Zushi, S~Itoh, K~Hanada, K~Nakamura, M~Sakamoto, E~Jotaki, M~Hasegawa, Y.D
  Pan, S.V Kulkarni, A~Iyomasa, S~Kawasaki, H~Nakashima, N~Yoshida, K~Tokunaga,
  T~Fujiwara, M~Miyamoto, H~Nakano, M~Yuno, A~Murakami, S~Nakamura, N~Sakamoto,
  K~Shinoda, S~Yamazoe, H~Akanishi, K~Kuramoto, Y~Matsuo, A~Iwamae,
  T~Fuijimoto, A~Komori, T~Morisaki, H~Suzuki, S~Masuzaki, Y~Hirooka,
  Y~Nakashima, and O~Mitarai.
\newblock Overview of steady state tokamak plasma experiments in {TRIAM}-1m.
\newblock {\em Nuclear Fusion}, 43(12):1600--1609, dec 2003.

\bibitem{Saoutic_2002}
B~Saoutic.
\newblock Status of long pulse experiments in magnetic fusion devices.
\newblock {\em Plasma Physics and Controlled Fusion}, 44(12B):B11--B22, nov
  2002.

\bibitem{Park:2019}
H.K. Park, M.J. Choi, S.H. Hong, Y.~In, Y.M. Jeon, J.S. Ko, W.H. Ko, J.G. Kwak,
  J.M. Kwon, J.~Lee, J.H. Lee, W.~Lee, Y.B. Nam, Y.K. Oh, B.H. Park, J.K. Park,
  Y.S. Park, S.J. Wang, M.~Yoo, S.W. Yoon, J.G. Bak, C.S. Chang, W.H. Choe,
  Y.~Chu, J.~Chung, N.~Eidietis, H.S. Han, S.H. Hahn, H.G. Jhang, J.W. Juhn,
  J.H. Kim, K.~Kim, A.~Loarte, H.H. Lee, K.C. Lee, D.~Mueller, Y.S. Na, Y.U.
  Nam, G.Y. Park, K.R. Park, R.A. Pitts, S.A. Sabbagh, and G.S.~Yun and.
\newblock Overview of {KSTAR} research progress and future plans toward {ITER}
  and k-{DEMO}.
\newblock {\em Nuclear Fusion}, 59(11):112020, jul 2019.

\bibitem{Wan:2019}
B.N. Wan, Y.~Liang, X.Z. Gong, N.~Xiang, G.S. Xu, Y.~Sun, L.~Wang, J.P. Qian,
  H.Q. Liu, L.~Zeng, L.~Zhang, X.J. Zhang, B.J. Ding, Q.~Zang, B.~Lyu, A.M.
  Garofalo, A.~Ekedahl, M.H. Li, F.~Ding, S.Y. Ding, H.F. Du, D.F. Kong, Y.~Yu,
  Y.~Yang, Z.P. Luo, J.~Huang, T.~Zhang, Y.~Zhang, G.Q. Li, T.Y. Xia, and and.
\newblock Recent advances in {EAST} physics experiments in support of
  steady-state operation for {ITER} and {CFETR}.
\newblock {\em Nuclear Fusion}, 59(11):112003, jun 2019.

\bibitem{Park_NP:2018}
Jong-Kyu Park, YoungMu Jeon, Yongkyoon In, Joon-Wook Ahn, Raffi Nazikian,
  Gunyoung Park, Jaehyun Kim, HyungHo Lee, WonHa Ko, Hyun-Seok Kim, Nikolas~C.
  Logan, Zhirui Wang, Eliot~A. Feibush, Jonathan~E. Menard, and Michael~C.
  Zarnstroff.
\newblock 3d field phase-space control in tokamak plasmas.
\newblock {\em Nature Physics}, 14:1223, 2018.

\bibitem{Reimold:2015}
F.~Reimold, M.~Wischmeier, M.~Bernert, S.~Potzel, A.~Kallenbach, H.W.
  MÃŒller, B.~Sieglin, and U.~Stroth and.
\newblock Divertor studies in nitrogen induced completely detached h-modes in
  full tungsten {ASDEX} upgrade.
\newblock {\em Nuclear Fusion}, 55(3):033004, feb 2015.

\bibitem{Jaervinen:2016}
A.E. Jaervinen, C.~Giroud, M.~Groth, P.~Belo, S.~Brezinsek, M.~Beurskens,
  G.~Corrigan, S.~Devaux, P.~Drewelow, D.~Harting, A.~Huber, S.~Jachmich,
  K.~Lawson, B.~Lipschultz, G.~Maddison, C.~Maggi, C.~Marchetto, S.~Marsen,
  G.F. Matthews, A.G. Meigs, D.~Moulton, B.~Sieglin, M.F. Stamp, and S.~Wiesen
  and.
\newblock Comparison of h-mode plasmas in {JET}-{ILW} and {JET}-c with and
  without nitrogen seeding.
\newblock {\em Nuclear Fusion}, 56(4):046012, mar 2016.

\bibitem{Huang:2016gz}
Yao Huang, BJ~Xiao, ZP~Luo, QP~Yuan, XF~Pei, and XN~Yue.
\newblock Implementation of gpu parallel equilibrium reconstruction for plasma
  control in east.
\newblock {\em Fusion Engineering and Design}, 112:1019--1024, 2016.

\bibitem{Barana:2002RSI}
O.~Barana, A.~Murari, P.~Franz, L.~C. Ingesson, and G.~Manduchi.
\newblock Neural networks for real time determination of radiated power in jet.
\newblock {\em Review of Scientific Instruments}, 73(5):2038--2043, 2002.

\bibitem{Murari:2013cm}
{Murari, A}, {Arena, P}, {Buscarino, A}, {Fortuna, L}, {Iachello, M}, and
  {contributors, JET-EFDA}.
\newblock {On the identification of instabilities with neural networks on JET}.
\newblock {\em Nuclear Inst. and Methods in Physics Research, A}, 720(C):2--6,
  August 2013.

\bibitem{Boyer_2019}
M.D. Boyer, S.~Kaye, and K.~Erickson.
\newblock Real-time capable modeling of neutral beam injection on {NSTX}-u
  using neural networks.
\newblock {\em Nuclear Fusion}, 59(5):056008, mar 2019.

\bibitem{Murari:2012fl}
Andrea Murari, Didier Mazon, N~Martin, Guido Vagliasindi, and Michela Gelfusa.
\newblock {Exploratory Data Analysis Techniques to Determine the Dimensionality
  of Complex Nonlinear Phenomena: The L-to-H Transition at JET as a Case
  Study}.
\newblock {\em IEEE Transactions on Plasma Science}, 40(5):1386--1394, April
  2012.

\bibitem{MURARI2010850}
A.~Murari, J.~Vega, D.~Mazon, D.~Patan, G.~Vagliasindi, P.~Arena, N.~Martin,
  N.F. Martin, G.~Ratt, and V.~Caloone.
\newblock Machine learning for the identification of scaling laws and dynamical
  systems directly from data in fusion.
\newblock {\em Nuclear Instruments and Methods in Physics Research Section A},
  623(2):850 -- 854, 2010.

\bibitem{Gaudio_2014}
P~Gaudio, A~Murari, M~Gelfusa, I~Lupelli, and J~Vega.
\newblock An alternative approach to the determination of scaling law
  expressions for the l{\textendash}h transition in tokamaks utilizing
  classification tools instead of regression.
\newblock {\em Plasma Physics and Controlled Fusion}, 56(11):114002, oct 2014.

\bibitem{Cannas:NF2010}
B.~Cannas, A.~Fanni, G.~Pautasso, G.~Sias, and P.~Sonato.
\newblock An adaptive real-time disruption predictor for {ASDEX} upgrade.
\newblock {\em Nuclear Fusion}, 50(7):075004, jun 2010.

\bibitem{Pau:NF2019}
A.~Pau, A.~Fanni, S.~Carcangiu, B.~Cannas, G.~Sias, A.~Murari, and F.~Rimini
  and.
\newblock A machine learning approach based on generative topographic mapping
  for disruption prevention and avoidance at {JET}.
\newblock {\em Nuclear Fusion}, 59(10):106017, aug 2019.

\bibitem{Felici:2018db}
F~Felici, J~Citrin, A~A Teplukhina, J~Redondo, C~Bourdelle, F~Imbeaux,
  O~Sauter, JET Contributors, and the EUROfusion~MST1 Team.
\newblock {Real-time-capable prediction of temperature and density profiles in
  a tokamak using RAPTOR and a first-principle-based transport model}.
\newblock {\em Nuclear Fusion}, 58(9):096006, 2018.

\bibitem{Matos:2017kl}
{Matos, Francisco A}, {Ferreira, Diogo R}, and {Carvalho, Pedro J}.
\newblock {Deep learning for plasma tomography using the bolometer system at
  JET}.
\newblock {\em Fusion Engineering and Design}, 114:18--25, January 2017.

\bibitem{Ferreira:2018}
Diogo~R. Ferreira, Pedro~J. Carvalho, Horcio Fernandes, and JET Contributors.
\newblock Full-pulse tomographic reconstruction with deep neural networks.
\newblock {\em Fusion Science and Technology}, 74(1-2):47--56, 2018.

\bibitem{Bockenhoff:2018hl}
Daniel B{\"o}ckenhoff, Marko Blatzheim, Hauke H{\"o}lbe, Holger Niemann, Fabio
  Pisano, Roger Labahn, Thomas~Sunn Pedersen, and The W7-X Team.
\newblock {Reconstruction of magnetic configurations in W7-X using artificial
  neural networks}.
\newblock {\em Nuclear Fusion}, 58(5):056009, 2018.

\bibitem{Cannas:2019gr}
B.~Cannas, S.~Carcangiu, A.~Fanni, T.~Farley, F.~Militello, A.~Montisci,
  F.~Pisano, G.~Sias, and N.~Walkden.
\newblock Towards an automatic filament detector with a faster r-cnn on mast-u.
\newblock {\em Fusion Engineering and Design}, 146:374 -- 377, 2019.

\bibitem{Clayton:2013dx}
{Clayton, D J}, {Tritz, K}, {Stutman, D}, {Bell, R E}, {Diallo, A}, {LeBlanc, B
  P}, and {Podest{\`a}, M}.
\newblock {Electron temperature profile reconstructions from multi-energy SXR
  measurements using neural networks}.
\newblock {\em Plasma Physics and Controlled Fusion}, 55(9):095015--9, August
  2013.

\bibitem{Cacciola:2006bq}
{Cacciola, Matteo}, {Greco, Antonino}, {Morabito, Francesco Carlo}, and
  {Versaci, Mario}.
\newblock {An Exhaustive Employment of Neural Networks to Search the Better
  Configuration of Magnetic Signals in ITER Machine}.
\newblock In {\em Neural Information Processing}, pages 353--360. Springer
  Berlin Heidelberg, Berlin, Heidelberg, 2006.

\bibitem{Jeon:2001bd}
Young-Mu Jeon, Yong-Su Na, Myung-Rak Kim, and Y~S Hwang.
\newblock {Newly developed double neural network concept for reliable fast
  plasma position control}.
\newblock {\em Review of Scientific Instruments}, 72(1):513--5, 2001.

\bibitem{Rodrigues:2005PRL}
Paulo Rodrigues and Jo\~ao P.~S. Bizarro.
\newblock Grad-shafranov equilibria with negative core toroidal current in
  tokamak plasmas.
\newblock {\em Phys. Rev. Lett.}, 95:015001, Jun 2005.

\bibitem{Rodrigues:2007PRL}
Paulo Rodrigues and Jo\~ao P.~S. Bizarro.
\newblock Tokamak equilibria with toroidal-current reversal in the plasma core
  consistent with experimental data.
\newblock {\em Phys. Rev. Lett.}, 99:125001, Sep 2007.

\bibitem{Ludwig:2013NF}
G.O. Ludwig, Paulo Rodrigues, and Jo{\~{a}}o~P.S. Bizarro.
\newblock Tokamak equilibria with strong toroidal current density reversal.
\newblock {\em Nuclear Fusion}, 53(5):053001, apr 2013.

\bibitem{Haykin:2008}
Simon Haykin.
\newblock {\em {Neural Networks and Learning Machines}}.
\newblock Pearson, 2008.

\bibitem{DBLP:journals/corr/KingmaB14}
Diederik~P. Kingma and Jimmy Ba.
\newblock Adam: {A} method for stochastic optimization.
\newblock {\em CoRR}, abs/1412.6980, 2014.

\bibitem{DBLP:journals/corr/GeHJY15}
Rong Ge, Furong Huang, Chi Jin, and Yang Yuan.
\newblock Escaping from saddle points - online stochastic gradient for tensor
  decomposition.
\newblock {\em CoRR}, abs/1503.02101, 2015.

\bibitem{Bottou10large-scalemachine}
L{\'e}on Bottou.
\newblock Large-scale machine learning with stochastic gradient descent.
\newblock In {\em Proceedings of COMPSTAT'2010}, pages 177--186. Springer,
  2010.

\bibitem{HuynhThu:2008fm}
Q~Huynh-Thu and M~Ghanbari.
\newblock {Scope of validity of PSNR in image/video quality assessment}.
\newblock {\em Electronics Letters}, 44(13):800, 2008.

\bibitem{Ebrahimi:2004fz}
{\em {JPEG vs. JPEG 2000: an objective comparison of image encoding quality}},
  2004.

\bibitem{Wang:2004gj}
Z~Wang, A~C Bovik, H~R Sheikh, and E~P Simoncelli.
\newblock {Image Quality Assessment: From Error Visibility to Structural
  Similarity}.
\newblock {\em IEEE Transactions on Image Processing}, 13(4):600--612, 2004.

\bibitem{aymar2002iter}
R~Aymar, P~Barabaschi, and Y~Shimomura.
\newblock The iter design.
\newblock {\em Plasma physics and controlled fusion}, 44(5):519, 2002.

\bibitem{cite-key}
N.~J. PEACOCK, D.~C. ROBINSON, M.~J. FORREST, P.~D. WILCOCK, and V.~V.
  SANNIKOV.
\newblock Measurement of the electron temperature by thomson scattering in
  tokamak t3.
\newblock {\em Nature}, 224(5218):488--490, 1969.

\bibitem{hicks1899ii}
William~Mitchinson Hicks.
\newblock Ii. researches in vortex motion.—part iii. on spiral or gyrostatic
  vortex aggregates.
\newblock {\em Philosophical Transactions of the Royal Society of London.
  Series A, Containing Papers of a Mathematical or Physical Character},
  (192):33--99, 1899.

\bibitem{huang2020gpu}
Y~Huang, ZP~Luo, BJ~Xiao, LL~Lao, A~Mele, A~Pironti, M~Mattei, G~Ambrosino,
  QP~Yuan, YH~Wang, et~al.
\newblock Gpu-optimized fast plasma equilibrium reconstruction in fine grids
  for real-time control and data analysis.
\newblock {\em Nuclear Fusion}, 60(7):076023, 2020.

\bibitem{von2013unified}
GT~Von~Nessi, MJ~Hole, MAST team, et~al.
\newblock A unified method for inference of tokamak equilibria and validation
  of force-balance models based on bayesian analysis.
\newblock {\em Journal of Physics A: Mathematical and Theoretical},
  46(18):185501, 2013.

\bibitem{romero2018inference}
JA~Romero, SA~Dettrick, E~Granstedt, T~Roche, and Y~Mok.
\newblock Inference of field reversed configuration topology and dynamics
  during alfvenic transients.
\newblock {\em Nature communications}, 9(1):1--10, 2018.

\bibitem{kwak2020thesis}
Sehyun Kwak.
\newblock {\em Bayesian modelling of nuclear fusion experiments}.
\newblock PhD thesis, Technische Universit{\"a}t Berlin, Berlin, Germany, 2020.

\bibitem{lecun2015deep}
Yann LeCun, Yoshua Bengio, and Geoffrey Hinton.
\newblock Deep learning.
\newblock {\em nature}, 521(7553):436--444, 2015.

\bibitem{hermann2020deep}
Jan Hermann, Zeno Sch{\"a}tzle, and Frank No{\'e}.
\newblock Deep-neural-network solution of the electronic schr{\"o}dinger
  equation.
\newblock {\em Nature Chemistry}, 12(10):891--897, 2020.

\bibitem{beucler2021enforcing}
Tom Beucler, Michael Pritchard, Stephan Rasp, Jordan Ott, Pierre Baldi, and
  Pierre Gentine.
\newblock Enforcing analytic constraints in neural networks emulating physical
  systems.
\newblock {\em Physical Review Letters}, 126(9):098302, 2021.

\bibitem{kochkov2021machine}
Dmitrii Kochkov, Jamie~A Smith, Ayya Alieva, Qing Wang, Michael~P Brenner, and
  Stephan Hoyer.
\newblock \href{https://www.pnas.org/doi/abs/10.1073/pnas.2101784118}{Machine
  learning--accelerated computational fluid dynamics}.
\newblock {\em Proceedings of the National Academy of Sciences}, 118(21), 2021.

\bibitem{kaltsas2022neural}
DA~Kaltsas and GN~Throumoulopoulos.
\newblock Neural network tokamak equilibria with incompressible flows.
\newblock {\em Physics of Plasmas}, 29(2):022506, 2022.

\bibitem{hirshman1983steepest}
Steven~P Hirshman and JC~Whitson.
\newblock Steepest-descent moment method for three-dimensional
  magnetohydrodynamic equilibria.
\newblock {\em The Physics of fluids}, 26(12):3553--3568, 1983.

\bibitem{wang2021deep}
Sifan Wang and Paris Perdikaris.
\newblock Deep learning of free boundary and stefan problems.
\newblock {\em Journal of Computational Physics}, 428:109914, 2021.

\bibitem{stefan1891theorie}
Johan Stefan.
\newblock {\"U}ber die theorie der eisbildung, insbesondere {\"u}ber die
  eisbildung im polarmeere.
\newblock {\em Annalen der Physik}, 278(2):269--286, 1891.

\bibitem{srivastava2014dropout}
Nitish Srivastava, Geoffrey Hinton, Alex Krizhevsky, Ilya Sutskever, and Ruslan
  Salakhutdinov.
\newblock Dropout: a simple way to prevent neural networks from overfitting.
\newblock {\em The journal of machine learning research}, 15(1):1929--1958,
  2014.

\bibitem{paszke2017automatic}
Adam Paszke, Sam Gross, Soumith Chintala, Gregory Chanan, Edward Yang, Zachary
  DeVito, Zeming Lin, Alban Desmaison, Luca Antiga, and Adam Lerer.
\newblock Automatic differentiation in pytorch.
\newblock {\em NIPS Workshop}, 2017.

\bibitem{baydin2018automatic}
Atilim~Gunes Baydin, Barak~A Pearlmutter, Alexey~Andreyevich Radul, and
  Jeffrey~Mark Siskind.
\newblock Automatic differentiation in machine learning: a survey.
\newblock {\em Journal of Marchine Learning Research}, 18:1--43, 2018.

\bibitem{lee2008magnetic}
SG~Lee, JG~Bak, EM~Ka, JH~Kim, and SH~Hahn.
\newblock Magnetic diagnostics for the first plasma operation in korea
  superconducting tokamak advanced research.
\newblock {\em Review of Scientific Instruments}, 79(10):10F117, 2008.

\bibitem{park2011kstar}
YS~Park, SA~Sabbagh, JW~Berkery, JM~Bialek, YM~Jeon, SH~Hahn, N~Eidietis,
  TE~Evans, SW~Yoon, J-W Ahn, et~al.
\newblock Kstar equilibrium operating space and projected stabilization at high
  normalized beta.
\newblock {\em Nuclear Fusion}, 51(5):053001, 2011.

\bibitem{urano2020breakdown}
H~Urano, Y~Miyata, T~Suzuki, and K~Kurihara.
\newblock Breakdown optimization method based on inverse reconstruction of
  magnetic fluxes in jt-60sa.
\newblock {\em Nuclear Fusion}, 60(6):066002, 2020.

\bibitem{morgan1989generalization}
Nelson Morgan and Herv{\'e} Bourlard.
\newblock Generalization and parameter estimation in feedforward nets: Some
  experiments.
\newblock {\em Advances in neural information processing systems}, 2, 1989.

\bibitem{beghi2005advances}
Alessandro Beghi and Angelo Cenedese.
\newblock Advances in real-time plasma boundary reconstruction: From gaps to
  snakes.
\newblock {\em IEEE Control Systems Magazine}, 25(5):44--64, 2005.

\bibitem{clapeyron1834memoire}
{\'E}mile Clapeyron.
\newblock M{\'e}moire sur la puissance motrice de la chaleur.
\newblock {\em Journal de l'{\'E}cole polytechnique}, 14:153--190, 1834.

\bibitem{coppi2013compact}
B~Coppi.
\newblock Compact experiments for o-particle heating.
\newblock {\em Tokamak Reactors for Breakeven: A Critical Study of the
  Near-Term Fusion Reactor Program}, page 303, 2013.

\bibitem{sweetman1981heating}
DR~Sweetman, JG~Cordey, and TS~Green.
\newblock Heating and plasma interactions with beams of energetic neutral
  atoms.
\newblock {\em Philosophical Transactions of the Royal Society of London.
  Series A, Mathematical and Physical Sciences}, 300(1456):589--598, 1981.

\bibitem{cairns1991radiofrequency}
RA~Cairns.
\newblock {\em Radiofrequency heating of plasmas}.
\newblock CRC Press, 1991.

\bibitem{goldston2011heuristic}
Robert~James Goldston.
\newblock Heuristic drift-based model of the power scrape-off width in
  low-gas-puff h-mode tokamaks.
\newblock {\em Nuclear Fusion}, 52(1):013009, 2011.

\bibitem{kim2018high}
Hyun-Tae Kim, ACC Sips, M~Romanelli, CD~Challis, F~Rimini, L~Garzotti,
  E~Lerche, J~Buchanan, X~Yuan, S~Kaye, et~al.
\newblock High fusion performance at high ti/te in jet-ilw baseline plasmas
  with high nbi heating power and low gas puffing.
\newblock {\em Nuclear Fusion}, 58(3):036020, 2018.

\bibitem{frei2020gyrokinetic}
Baptiste~Jimmy Frei, Rog{\'e}rio Jorge, and Paolo Ricci.
\newblock A gyrokinetic model for the plasma periphery of tokamak devices.
\newblock {\em Journal of Plasma Physics}, 86(2), 2020.

\bibitem{batchelor2000introduction}
G~K Batchelor.
\newblock {\em An introduction to fluid dynamics}.
\newblock Cambridge university press, 2000.

\bibitem{park2013investigation}
YS~Park, SA~Sabbagh, JM~Bialek, JW~Berkery, SG~Lee, WH~Ko, JG~Bak, YM~Jeon,
  JK~Park, J~Kim, et~al.
\newblock Investigation of mhd instabilities and control in kstar preparing for
  high beta operation.
\newblock {\em Nuclear Fusion}, 53(8):083029, 2013.

\bibitem{mackay1992practical}
David~JC MacKay.
\newblock A practical bayesian framework for backpropagation networks.
\newblock {\em Neural computation}, 4(3):448--472, 1992.

\bibitem{neal1995bayesian}
RM~Neal.
\newblock Bayesian learning for neural networks [phd thesis].
\newblock {\em Toronto, Ontario, Canada: Department of Computer Science,
  University of Toronto}, 1995.

\bibitem{robbins1951stochastic}
Herbert Robbins and Sutton Monro.
\newblock A stochastic approximation method.
\newblock {\em The annals of mathematical statistics}, pages 400--407, 1951.

\bibitem{hoffman2013stochastic}
Matthew~D Hoffman, David~M Blei, Chong Wang, and John Paisley.
\newblock Stochastic variational inference.
\newblock {\em Journal of Machine Learning Research}, 2013.

\bibitem{glasserman2004monte}
Paul Glasserman.
\newblock {\em Monte Carlo methods in financial engineering}, volume~53.
\newblock Springer, 2004.

\bibitem{kingma2013auto}
Diederik~P Kingma and Max Welling.
\newblock Auto-encoding variational bayes.
\newblock {\em arXiv preprint arXiv:1312.6114}, 2013.

\bibitem{titsias2014doubly}
Michalis Titsias and Miguel L{\'a}zaro-Gredilla.
\newblock Doubly stochastic variational bayes for non-conjugate inference.
\newblock In {\em International conference on machine learning}, pages
  1971--1979. PMLR, 2014.

\bibitem{hinton2012improving}
Geoffrey~E Hinton, Nitish Srivastava, Alex Krizhevsky, Ilya Sutskever, and
  Ruslan~R Salakhutdinov.
\newblock Improving neural networks by preventing co-adaptation of feature
  detectors.
\newblock {\em arXiv preprint arXiv:1207.0580}, 2012.

\bibitem{tishby1989consistent}
Naftali Tishby, Esther Levin, and Sara~A Solla.
\newblock Consistent inference of probabilities in layered networks:
  Predictions and generalization.
\newblock In {\em International Joint Conference on Neural Networks}, volume~2,
  pages 403--409. IEEE New York, 1989.

\end{thebibliography}


\acknowledgment[4]

\noindent
감사했던 분들의 성함은 다 적지 못할까 염려되어 선뜻 남기기 어려웠습니다. 그럼에도 이야기하고자 합니다. 지난 2015년 1월을 시작으로 제가 감히 이해할 수 없는 마음으로 여러 가르침을 주신 김영철 교수님께 감사드립니다. 아직 더 배움이 필요함에도 불구하고 미숙한 저의 학위 심사에 참여해주시고 많은 가르침을 주셨던 최원호 교수님, 윤시우 박사님, 성충기 교수님, 미숙한 석사과정 학생에 불과했던 저에게 여러 가르침과 격려를 해주셨던 김현석 박사님께 감사드립니다. 연구의 시작점이며 방향이 되었던 곽세현 박사님께 감사드립니다. 언제나 말도 안 되는 이야기와 주장을 해도 깊은 이해로 진지하게 받아주셨던 김재욱 박사님께도 감사드립니다. 좌충우돌 어디로 튈지 저 조차도 모르던 미숙함을 언제나 높은 마음으로 이해해주셨던 김건희, 때로는 본이 되는 연구자로서 때로는 깊은 토론을 할 수 있는 동료로서 때로는 함께 즐길 수 있는 친구며 동생이 되어준 오태석, 마찬가지로 깊은 연구 토론과 때로는 삶을 공유할 수 있었던 임예건, 언제나 여러 힘이 되어준 권대호, 미숙한 모습에도 내색없이 바라봐주었던 이정진, 정충순, 해준 것 없는 데도 많은 도움을 주었던 유용성, 이원준, 박사 마지막 시기에 많은 도움이 되었던 김동욱, 연구에 영어에 많은 도움을 준 Alvin 모두 다 감사드립니다. 언제나 저를 어여삐 여겨주신 Scott, Mandy, SeongOk, Kidoo 그리고 힘든 시기 도움이 되어주신 여러 선생님들, 학교에서 까지 의지할 곳이 되어준 이신의 모두 감사드립니다. 언제나 툴툴대고 선택 못 하고 하고 싶은 말이 아니라 감정에 치우친 말이 나와도 이해해주었던 신찰범 고맙습니다. 혹여 급히 작성하느라 미처 다 언급하지 못한 분들 및 연구실 일원들에게도 깊은 감사드립니다.

지난 날을 되돌아 보면 기억은 빛을 바래도 추억은 보다 더 또렷해지는 순간들이 있습니다. 이미 한참 지났지만 성남 구 시가지에서 친구들과 뛰어놀던 기억에 더욱 추억이 깊습니다. 그리고 아쉽습니다. 흐려져 가는 기억들이 아깝습니다. 일기라도 적어놓을 것 그랬습니다. 어찌 하다보니 지금 이 순간을 맞이하게 되었습니다. 어렸을 때의 저는 이런 공부를 하는 제가 되리라고는 전혀 생각도 못 했었겠지요. 돌아보니 석사 졸업 시기와는 사뭇 다른 감정을 지금은 느끼는 것 같습니다. 보다 명확해진 점은 이전에는 제가 제 길을 만들려고 부단히 애를 썼지만 지금은 이끄시는 대로 그저 따라 가려 합니다. 이미 많이 미숙했기에 앞으로 성숙해질 것만이 남아 있다는 점은 참 감사하게 여겨집니다. 그래서 어느 방향의 길이든 그 끝은 다 감사하지 않을까 생각합니다. 지금 이 순간도 후에 돌아보면 기억은 바랬지만 추억은 보다 살아남아 있을 것이라 생각합니다. 언제나 제 삶을 지지해주시는 아버지 어머니 감사드립니다 동생들도 미안하고 사랑합니다. 마지막으로 혜린아 고생 정말 많았어요, 덕분에 나무 아래 누워 쉬던 날에서 일어나 걸을 수 있었습니다. 잡은 손 놓지 않겠습니다. 감사합니다.

1. 내 그대를 생각함은 항상 그대가 앉아 있는 배경에서 해가 지고 바람이 부는 일처럼 사소한 일일 것이나 언젠가 그대가 한없이 괴로움 속을 헤매일 때에 오랫동안 전해오던 그 사소함으로 그대를 불러보리라. // 2. 진실로 진실로 내가 그대를 사랑하는 까닭은 내 나의 사랑을 한없이 잇닿은 그 기다림으로 바꾸어 버린 데 있었다. 밤이 들면서 골짜기엔 눈이 퍼붓기 시작했다. 내 사랑도 어디쯤에선 반드시 그칠 것을 믿는다. 다만 그때 내 기다림의 자세를 생각하는 것뿐이다. 그 동안에 눈이 그치고 꽃이 피어나고 낙엽이 떨어지고 또 눈이 퍼붓고 할 것을 믿는다. (황동규, `즐거운 편지', 현대문학, 1958)

\curriculumvitae[3]

    \begin{personaldata}
        \name       {정 세 민 (SEMIN JOUNG)}
        \email      {smjoung@kaist.ac.kr, smjoungsemail@gmail.com}
    \end{personaldata}

    \begin{education}
        \item[2017. 8.\ --\ 2022. 8.] KAIST {\small (Korea Advanced Institute of Science and Technology)}\\ Department of Nuclear and Quantum Engineering (PhD)
        \item[2015. 2.\ --\ 2017. 2.] KAIST\\ Department of Nuclear and Quantum Engineering (MS)
        \item[2011. 3.\ --\ 2015. 2.] KYUNG HEE University\\ Department of Nuclear Engineering (Bachelor's degree)
    \end{education}


    \begin{publication}
        \item {\bfseries\scshape S.} {\bfseries\scshape J}{\bfseries\scshape\small oung}, \textsc{Y.{\small -c}. G{\small him}, J. K{\small im}, S. K{\small wak}, S. L{\small ee}, D. K{\small won}, H.S. K{\small im} and J.G. B{\small ak}}. `GS-DeepNet: Mastering tokamak plasma equilibria with deep neural networks and the Grad-Shafranov equation'. In: \textit{Science Advances}, (2022), In preparation.
        \item \textsc{\textbf{S. J{\small oung}}, J. K{\small im}, H.S. H{\small an}, J.G. B{\small ak}} and \textsc{Y.{\small -c}. G{\small him}}. `A deep learning approach to recover hidden consistency of KSTAR flux loop signals'. In: \textit{Scientific Reports}, (2022), In preparation.
        \item \textsc{\textbf{S. J{\small oung}}, J. K{\small im}, S. K{\small wak}, J.G. B{\small ak}, S.G. L{\small ee}, H.S. H{\small an}, H.S. K{\small im}, G. L{\small ee}, D. K{\small won}} and \textsc{Y.{\small -c}. G{\small him}}. `Deep neural network Grad–Shafranov solver constrained with measured magnetic signals'. In: \textit{Nuclear Fusion}, Vol.60.1 (3$^{rd}$ Dec. 2019), \textsc{{\small DOI:}}\href{https://doi.org/10.1088/1741-4326/ab555f}{10.1088/1741-4326/ab555f}
        \item \textsc{\textbf{S. J{\small oung}}, J. K{\small im}, S. K{\small wak}, K. P{\small ark}, S.H. H{\small ahn}, H.S. H{\small an}, H.S. K{\small im}, J.G. B{\small ak}, S.G. L{\small ee}} and \textsc{Y.{\small -c}. G{\small him}}. `Imputation of faulty magnetic sensors with coupled Bayesian and Gaussian processes to reconstruct the magnetic equilibrium in real time'. In: \textit{Review of Scientific Instruments}, Vol.89.10 (7$^{th}$ May 2018), \textsc{{\small DOI:}}\href{https://doi.org/10.1063/1.5038938}{10.1063/\linebreak1.5038938}
    \end{publication}

    \begin{activity}
        \item \textsc{\textbf{S. J{\small oung}}, J. K{\small im}, S. K{\small wak}, M. K{\small im}, H.S. K{\small im}, J.G. B{\small ak}} and \textsc{Y.{\small -c}. G{\small him}}. `Learning plasma equilibria from scratch with deep neural network Grad-Shafranov solver'. 63rd Annual Meeting of the APS Division of Plasma Physics. Pittsburgh, PA, USA, 8$^{th}$ Nov. 2021.
        \item Jiheon S{\small ong}, \textsc{\textbf{S. J{\small oung}}, Y.{\small -c}. G{\small him}, J{\small ungpyo}} and \textsc{S.H. H{\small ahn}}. `Implement of machine learning U-net model for automatic ELM-burst detection in the KSTAR tokamak'. Korean Physical Society Fall Conference. Remote e-conference, 20$^{th}$ Oct. 2021.
        \item \textsc{\textbf{S. J{\small oung}}, J. K{\small im}, S. K{\small wak}, H.S. H{\small an}, H.S. K{\small im}, J.G. B{\small ak}, S.G. L{\small ee}} and \textsc{Y.{\small -c}. G{\small him}}. `Learning tokamak equilibria from scratch with deep neural network Grad-Shafranov solver'. 3rd International Conference on Data Driven Plasma Science. Remote e-conference, 29$^{th}$ May 2021.
        \item \textsc{\textbf{S. J{\small oung}}, J. K{\small im}, S. K{\small wak}, Y.M. J{\small eon}, S.H. H{\small ahn}, H.S. K{\small im}, H.S. H{\small an}, J.G. B{\small ak}, S.G. L{\small ee}, G. L{\small ee}, D. K{\small won}} and \textsc{Y.{\small -c}. G{\small him}}. `Data-driven Grad-Shafranov solver with KSTAR EFIT data based on neural network and Bayesian inference'. 3rd IAEA Technical Meeting on Fusion Data Processing, Validation and Analysis. Vienna, Austria, 27$^{th}$ May 2019.
        \item \textsc{Y.{\small -c}. G{\small him}, J. K{\small im}, S. K{\small wak}} and \textsc{\textbf{S. J{\small oung}}}. `Bayesian based data analysis to infer plasma parameters and to generate synthetic data in tokamaks'. Korean Physical Society Fall Conference. Changwon, South Korea, 24$^{th}$ Oct. 2018.
        \item \textsc{\textbf{S. J{\small oung}}, S. K{\small wak}, Y.M. J{\small eon}, S.H. H{\small ahn}, H.S. K{\small im}, H.S. H{\small an}, J.G. B{\small ak}, S.G. L{\small ee}} and \textsc{Y.{\small -c}. G{\small him}}. `Neural network magnetic equilibrium reconstruction with Bayesian based preprocessor in KSTAR'. 11th IAEA Technical Meeting on Control, Data Acquisition, and Remote Participation for Fusion Research. Greifswald, Germany, 8$^{th}$ May 2017.
    \end{activity}

    \begin{activity2}
        \item \textsc{\textbf{S. J{\small oung}}, J. K{\small im}} and \textsc{Y.{\small -c}. G{\small him}}. `Bayesian modelling of magnetic pick-up coils and flux loops using Gaussian processes for real-time control of plasmas'. 23rd Topical Conference on High Temperature Plasma Diagnostics. Remote e-conference, 13$^{th}$ Dec. 2020.
        \item \textsc{\textbf{S. J{\small oung}}, H.S. K{\small im}} and \textsc{Y.{\small -c}. G{\small him}}. `Real-time compensation of KSTAR magnetic signal drifts based on Bayesian inference'. KSTAR Conference 2019. COEX, Seoul, South Korea, 20$^{th}$ Feb. 2019.
        \item \textsc{\textbf{S. J{\small oung}}, J. K{\small im}, S. K{\small wak}, K. P{\small ark}, Y.M. J{\small eon}, S.H. H{\small ahn}, H.S. H{\small an}, H.S. K{\small im}, J.G. B{\small ak}, S.G. L{\small ee}} and \textsc{Y.{\small -c}. G{\small him}}. `Bayesian based missing input imputation scheme for neural network reconstructing magnetic equilibria in real time'. 22nd Topical Conference on High Temperature Plasma Diagnostics. San Diego, CA, USA, 16$^{th}$ Apr. 2018.
        \item \textsc{\textbf{S. J{\small oung}}, S. K{\small wak}, Y.M. J{\small eon}, S.H. H{\small ahn}, H.S. K{\small im}, H.S. H{\small an}, J.G. B{\small ak}, S.G. L{\small ee}} and \textsc{Y.{\small -c}. G{\small him}}. `Neural network based real-time reconstruction of KSTAR magnetic equilibria with Bayesian-based preprocessing'. 59th Annual Meeting of the APS Division of Plasma Physics. Milwaukee, WI, USA, 22$^{nd}$ Oct. 2017.
        \item \textsc{\textbf{S. J{\small oung}}, S. K{\small wak}} and \textsc{Y.{\small -c}. G{\small him}}. `Plasma equilibrium reconstruction for real-time control using artificial neural network in KSTAR'. KSTAR Conference 2016. Daejeon, South Korea, 24$^{th}$ Feb. 2016.
    \end{activity2}

  \label{paperlastpagelabel}     
\end{document}